\documentclass[namedreferences,hyperref,optionalrh,solaromanenum]{spr-sola}

\usepackage{graphicx}                    
\usepackage{amssymb}                    
\usepackage{amsmath}
\usepackage{color}                       
\usepackage{natbib}
\usepackage{lineno}
\graphicspath{{Figures/}}
\usepackage{overpic}
\newcommand{\ca}{\mbox{Ca\,{\textsc{ii}}~K\,}}
\newcommand{\lwig}{{\leavevmode\kern0.3em\raise.3ex\hbox{$<$}
		\kern-0.8em\lower.7ex \hbox{$\sim$}\kern0.3em}}

\newcommand{\degr}{^\circ}
\newcommand\degree{\mbox{$^\circ$}}

\newcommand{\aap}{{\it Astron. Astrophys.}}

\newcommand{\actaa}{{\it Acta Astron.}}

\newcommand{\aj}{{\it Astron. J.}} 
\newcommand{\apj}{{\it Astrophys. J.}}
\newcommand{\apjl}{{\it Astrophys. J. Lett.}}
\newcommand{\apjs}{{\it Astrophys. J. Suppl.}}

\newcommand{\grl}{{\it Geophys. Res. Lett.}}

\newcommand{\jcap}{{\it J. Cosm. Astropart. Phys.}}
\newcommand{\mnras}{{\it Mon. Not. R. Astron. Soc.}}
\newcommand{\nat}{{\it Nature}}

\newcommand{\pasj}{{\it Publ. Astron. Soc. Japan}}

\newcommand{\physrep}{{\it Phys. Rep.}}
\newcommand{\prl}{{\it Phys. Rev. Lett.}}
\newcommand{\solphys}{{\it Sol. Phys.}}
 
\newcommand{\ssr}{{\it Space Sci. Rev.}} 
\chardef\us=`\_
\newcommand{\sunrad}{\mbox{\ensuremath{R_\odot}}}

\newcommand{\ms}{m\,s$^{-1}$}
\newcommand\sun{\odot}%
\begin{document}

\begin{frontmatter}

\title{Structure and Dynamics of the Sun’s Interior Revealed by the Helioseismic and Magnetic \\Imager}

%
\author[addressref={aff1},corref,email={alexander.g.kosovichev@njit.edu}]{\inits{A.G.}\fnm{Alexander}~\snm{Kosovichev}\orcid{0000-0003-0364-4883}}

\author[addressref=yale,email={sarbani.basu@yale.edu}]{\inits{S.}\fnm{Sarbani}~\snm{Basu}\orcid{0000-0002-6163-3472}}

\author[addressref=affMPS,email={bekki@mps.mpg.de}]{\inits{Y.B.}\fnm{Yuto}~\snm{Bekki}\orcid{0000-0002-5990-013X}}

\author[addressref=affUCB,email={juan@berkeley.edu}]{\inits{J.C.}\fnm{Juan Camilo}~\snm{Buitrago-Casas}\orcid{0000-0002-8203-4794}}

\author[addressref=affMPS,
email={chatzistergos@mps.mpg.de}]{\inits{T.}\fnm{Theodosios}
~\snm{Chatzistergos}\orcid{0000-0002-0335-9831}}

\author[addressref=aff2,email={rzchen@stanford.edu}]{\inits{R.}\fnm{Ruizhu}~\snm{Chen}\orcid{}}

\author[addressref=affAU,
email={jcd@phys.au.dk}]{\inits{J.}\fnm{J{\o}rgen}
~\snm{Christensen-Dalsgaard}\orcid{0000-0001-5137-0966
}}

\author[addressref=affMON,email={alina.donea@monash.edu}]{\inits{A.D.}\fnm{Alina }~\snm{Donea}\orcid{0000-0002-4111-3496}}

\author[addressref=affESA,email={bfleck@esa.nascom.nasa.gov}]{\inits{B.}\fnm{Bernhard}~\snm{Fleck}\orcid{0000-0001-5777-9121}}

\author[addressref=affMPS,email={fournier@mps.mpg.de}]{\inits{D.F.}\fnm{Damien}~\snm{Fournier}\orcid{0000-0001-7749-8650}}

\author[addressref=affCEA,email={rafael.garcia@cea.fr}]{\inits{R.A.}\fnm{Rafael A.}~\snm{García}
\orcid{0000-0002-8854-3776}}

\author[addressref=affMSU,email={a.getling@mail.ru}]{\inits{A.V.}\fnm{Alexander}~\snm{Getling}\orcid{0000-0002-1766-2620}}

\author[addressref=affMPS,email={gizon@mps.mpg.de}]{\inits{L.G.}\fnm{Laurent}~\snm{Gizon}}\orcid{0000-0001-7696-8665}

\author[addressref=affIOA,email={douglas@ast.cam.ac.uk}]{\inits{D.O.}\fnm{Douglas O.}~\snm{Gough}}\orcid{0000-0002-6086-2636}

\author[addressref=affTIFR,email={shravan.hanasoge@tifr.res.in}]{\inits{S.}\fnm{Shravan}~\snm{Hanasoge}\orcid{0000-0003-2896-1471}}

\author[addressref=affNYUAD,email={hanson@nyu.edu}]{\inits{C.S.}\fnm{Chris S.}~\snm{Hanson}\orcid{0000-0003-2536-9421}}

\author[addressref=aff2,email={shessweb@stanford.edu}]{\inits{S.A.}\fnm{Shea A.}~\snm{Hess~Webber}\orcid{0000-0002-3631-6491}}

\author[addressref=aff2,email={todd@sun.stanford.edu}]{\inits{J.T.}\fnm{J.~Todd}~\snm{Hoeksema}\orcid{0000-0001-9130-7312}}

\author[addressref=affBHAM, email={r.howe@bham.ac.uk}]{\inits{R.}\fnm{Rachel}~\snm{Howe}\orcid{0000-0002-3834-8585}}

\author[addressref=aff3,email={kjain@nso.edu}]
{\inits{K.}\fnm{Kiran}~\snm{Jain}\orcid{0000-0002-1905-1639}}

\author[addressref=aff4,email={spiridon.kasapis@nasa.gov}]{\inits{S.}\fnm{Spiridon}~\snm{Kasapis}
\orcid{0000-0002-0972-8642}}

\author[addressref=affMPS,
email={kashyap@mps.mpg.de}]{\inits{S. G.}\fnm{Samarth G.}~\snm{Kashyap}
\orcid{0000-0001-5443-5729}}

\author[addressref=aff4,email={irina.n.kitiashvili@nasa.gov}]{\inits{N.}\fnm{Irina}~\snm{Kitiashvili}
\orcid{0000-0003-4144-2270}}

\author[addressref=aff3,email={rkomm@nso.edu}]{\inits{R.}\fnm{Rudolf}~\snm{Komm}
\orcid{}}

\author[addressref=affCfA,email={skorzennik@cfa.harvard.edu}]{\inits{S.G.}\fnm{Sylvain}~\snm{Korzennik}\orcid{0000-0003-1531-1541}}

\author[addressref=affMPS,
email={natalie@mps.mpg.de}]{\inits{N. A.}\fnm{Natalie A.}
~\snm{Krivova}\orcid{0000-0002-1377-3067}}

\author[addressref=affIfA,email={jeff.reykuhn@yahoo.com}]{\inits{J.R.}\fnm{Jeff}~\snm{Kuhn}\orcid{0000-0003-1361-9104}}

\author[addressref=affMPS,email={zhichao@mps.mpg.de}]{\inits{Z.L.}\fnm{Zhi-Chao}~\snm{Liang}}

\author[addressref=affNWRA,email={lindsey@nwra.com}]{\inits{C.}\fnm{Charles}~\snm{Lindsey}\orcid{0000-0002-5658-5541}}

\author[addressref=aff2,email={mahajans@stanford.edu}]{\inits{S.S.}\fnm{Sushant S.}~\snm{Mahajan}\orcid{0000-0003-1753-8002}}

\author[addressref=aff1,email={krishnendu.mandal@njit.edu}]{\inits{K.}\fnm{Krishnendu }~\snm{Mandal}\orcid{0000-0003-3067-288X}}

\author[addressref=affUSYD, email={pman0946@sydney.edu.au}]{\inits{P.}\fnm{Prasad}~\snm{Mani}\orcid{0000-0002-8707-201X}}

\author[addressref=affUCB,email={oliveros@ssl.berkeley.edu}]{\inits{J.C.}\fnm{Juan Carlos}~\snm{Martinez Oliveros}}

\author[addressref=affIAC,email={smathur@iac.es}]{\inits{S.}\fnm{Savita}~\snm{Mathur}
\orcid{0000-0002-0129-0316}}

\author[addressref=aff2,email={csoares@stanford.edu}]{\inits{M.C.}\fnm{M. Cristina}~\snm{Rabello Soares}\orcid{0000-0003-0172-3713}}

\author[addressref=affIIA,email={rajaguru@iiap.res.in}]{\inits{S.P.}\fnm{S. Paul}~\snm{Rajaguru}\orcid{0000-0003-0003-4561}}

\author[addressref=afftum,
email={jreiter@tum.de}]{\inits{J.}\fnm{Johann}
~\snm{Reiter}\orcid{0009-0000-0970-634X}}

\author[addressref=aff5,email={erhodes@usc.edu}]{\inits{E.J.}\fnm{Edward J.}~\snm{Rhodes, Jr.}\orcid{0009-0007-8009-8018}}

\author[addressref=affCA,email={jp.rozelot@orange.fr}]{\inits{J.P}\fnm{Jean-Pierre}~\snm{Rozelot}\orcid{0000-0002-5369-1381}}

\author[addressref=aff2,
email={pscherrer@solar.stanford.edu}]{\inits{P.H.}\fnm{Philip H.}
~\snm{Scherrer}\orcid{0000-0002-6937-6968}}

\author[addressref=affMPS,
email={solanki@mps.mpg.de}]{\inits{S. K.}\fnm{Sami K.}
~\snm{Solanki}\orcid{0000-0002-3418-8449 }}

\author[addressref=aff1,email={john.stefan@njit.edu}]{\inits{J.T.}\fnm{John}~\snm{Stefan}\orcid{0000-0002-5519-8291}}

\author[addressref=affCU,
email={jtoomre@lcd.colorado.edu}]{\inits{J.}\fnm{Juri}~\snm{Toomre}\orcid{0000-0002-3125-4463}}

\author[addressref=aff3,email={stripathy@nso.edu}]{\inits{S.C.}\fnm{Sushanta C.}~\snm{Tripathy}\orcid{0000-0002-4995-6180}}
\author[addressref=affswri,email={lisa.upton@swri.org}]{\inits{L.}\fnm{Lisa A.}~\snm{Upton}\orcid{0000-0003-0621-4803}}
\author[addressref=aff2,email={junwei@sun.stanford.edu}]{\inits{J.}\fnm{Junwei}~\snm{Zhao}\orcid{0000-0002-6308-872X}}

\runningtitle{\textit{Sol. Phys.} Structure and Dynamics of the Sun’s Interior}

%

\address[id=aff1]{New Jersey Institute of Technology, Newark, NJ 07102, U.S.A.}
\address[id=yale]{Dept. of Astronomy, Yale University, New Haven, CT 06517, U.S.A.}
\address[id=affMPS]{Max Planck Institute for Solar System Research, 37077 G\"ottingen, Germany}
\address[id=affUCB]{Space Sciences Laboratory, University of California Berkeley, Berkeley, California, U.S.A.}
\address[id=aff2]{W. W. Hansen Experimental Physics Laboratory, Stanford University, Stanford, CA 94305-4085, U.S.A.}
\address[id=affAU]{Department of Physics and Astronomy, Aarhus University, Ny Munkegade 120, 8000 Aarhus C, Denmark}
\address[id=affMON]{School of Mathematical Sciences, Monash University, 9 Rainforest Walk, Clayton, Victoria 3800, Australia}
\address[id=affESA]{ESA Science and Operations Department, c/o NASA/GSFC Code 671, Greenbelt, MD 20071, U.S.A.}
\address[id=affCEA]{Université Paris-Saclay, Université Paris Cité, CEA, CNRS, AIM, 91191, Gif-sur-Yvette, France}
\address[id=affMSU]{Skobeltsyn Institute of Nuclear Physics, Lomonosov Moscow State University, Moscow, 119991, Russia}
\address[id=affIOA]{Institute of Astronomy, and Department of Applied Mathematics and Theoretical Physics, University of Cambridge, Madingley Road, Cambridge, CB3 0HA, UK}
\address[id=affTIFR]{Department of Astronomy and Astrophysics, Tata Institute of Fundamental Research, Mumbai, India}
\address[id=affNYUAD]{Center for Astrophysics and Space Science, NYUAD Institute, New York University Abu Dhabi, Abu Dhabi, UAE}
\address[id=affBHAM]{School of Physics and Astronomy, University of Birmingham, Edgbaston, Birmingham B15 2TT, UK}
\address[id=aff3]{National Solar Observatory, Boulder, CO 80303, U.S.A.}
\address[id=aff4]{NASA Ames Research Center, Moffett Field, CA 94035, U.S.A.}
\address[id=affCfA]{Center for Astrophysics $|$ Harvard \& Smithsonian, Cambridge, MA 02138, U.S.A.}
\address[id=affIfA]{Institute for Astronomy, University of Hawaii, Honolulu, HI, U.S.A.}
\address[id=affNWRA]{North West Research Associates, Boulder, CO 80301, U.S.A.}
\address[id=affUSYD]{Sydney Institute for Astronomy, School of Physics, University of Sydney, Sydney NSW 2006, Australia}
\address[id=affIAC]{Instituto de Astrof\'isica de Canarias (IAC), E-38205 La Laguna, Tenerife, Spain \& Universidad de La Laguna (ULL), Departamento de Astrof\'isica, E-38206 La Laguna, Tenerife, Spain}
\address[id=affIIA]{Indian Institute of Astrophysics, Bangalore 560034, India}
\address[id=afftum]{TUMAM17 Lehrstuhl f\"ur Optimalsteuerung (Prof. Vexler),
Technische Universit\"at M\"unchen,
D-85748 Garching bei M\"unchen, Germany}
\address[id=aff5]{Dept. of Physics and Astronomy, University of Southern California, Los Angeles, CA 90089,U.S.A.}
\address[id=affCA]{Universit\'e de la C\^ote d'Azur, Grasse 06130, France}
\address[id=affCU]{JILA and Dept Astrophysical \& Planetary Sciences
University of Colorado Boulder, Boulder CO 80309-0440,
U.S.A.}
\address[id=affswri]{Southwest Research Institute, 1301 Walnut St, Suite 400, Boulder, CO 80302, U.S.A.\vspace*{2cm}}

\begin{abstract}
High-resolution helioseismology observations with the Helioseismic and Magnetic Imager (HMI) onboard Solar Dynamics Observatory (SDO) provide a unique three-dimensional view of the solar interior structure and dynamics, revealing a tremendous complexity of the physical processes inside the Sun. We present an overview of the results of the HMI helioseismology program and discuss their implications for modern theoretical models and simulations of the solar interior.
\end{abstract}
%
\keywords{Active Regions -- Flares -- Helioseismology -- Interior -- Magnetic Fields -- Oscillations -- Rotation -- Sunspots -- Turbulence -- Velocity Field}

\end{frontmatter}
\setcounter{section}{0}
\section*{Introduction}
\label{sec:Introduction}

The Solar Dynamics Observatory (SDO) is the first mission for NASA's Living With a Star (LWS) Program and was launched on 11 February 2010. The SDO science goals are to determine how the Sun’s magnetic field is generated and structured and how this stored magnetic energy is released into the heliosphere and geospace as the solar wind, energetic particles, and variations in solar irradiance  \citep{2012SoPh..275....3P}. 
The Helioseismic and Magnetic Imager (HMI) instrument is one of three instruments onboard SDO with the primary goal of studying the origin of solar variability and characterizing and understanding the Sun’s interior and the various components of magnetic activity, as described in the LWS Science Architecture Team Report \citep{Mason2001} and the SDO Science Definition Team Report \citep{Hathaway2001}.
The HMI instrument is designed to study convection-zone dynamics and the solar dynamo, the origin and evolution of sunspots, active regions and complexes of activity, the sources and drivers of solar magnetic activity and disturbances, links between the internal processes and dynamics of the corona and heliosphere, and precursors of solar disturbances for space-weather forecasts \citep{2012SoPh..275..207S}.

With 14 years of SDO/HMI data analyzed by the heliophysics community, we provide an overview of the results of the SDO/HMI program focusing on the dynamics and structure of the Sun's interior.  The review describes results that are publicly available, mainly in refereed journals, and does not report on work in progress.  The review paper is organized into 14 sections, as follows: 
1. HMI Observables and Data Products, 
2. Global Helioseismology Measurements,
3. Local Helioseismology Techniques,
4. Solar Structure, 
5. Convection and Large-Scale Flows, 
6. Differential Rotation, 
7. Meridional Circulation,
8. Solar Inertial Waves, 
9. Solar Cycle Variation and Dynamo, 
10. Seismology of Sunspots and Active Regions,
11. Helioseismic Imaging of the Solar Far-Side Active Regions,
12. Early Detection of Emerging Magnetic Flux, 
13. Sunquakes, and
14. Solar Irradiance.

Finally, we summarize the results presented in this review and discuss current challenges and unresolved problems.

\section{HMI Observables and Data Products}
\label{sec:HMI_Observables}

Calibrated filtergrams from the first {HMI} camera collected over an interval of 180\,s are interpolated in both time and space and analyzed to compute full-disk Dopplergrams, line-of-sight magnetograms, and images in continuum intensity and line depth every 45\,s. These mapped quantities are termed \texttt{observables}. Spatial interpolation is required for coalignment and elimination of smearing due to solar rotation. Temporal interpolation reduces the effects of acceleration during the time the filtergrams are collected. The timing of the observables is chosen so the images are evenly spaced in time for an observer at a constant Sun-Earth distance of 1 AU. A separate pipeline uses filtergrams from both cameras to generate a full set of 24 Stokes-parameter filtergrams every 720\,s; this {\texttt hmi.S\_720s} time series is used to compute the vector magnetic field and other quantities averaged over 12 minutes to reduce the helioseismic signals \citep{2014Hoeksema_Vector,2016Couvidat_Obs, 2018Hoeksema_Performance}. From these observables, the HMI pipeline derives a variety of higher-level data products, including those used for helioseismology.

\subsection{Helioseismic Data Products}

The time series of 45\,s Dopplergrams is used to produce higher-level data products that can be used to derive the subsurface structure and dynamics of the Sun using helioseismic techniques.
\begin{figure} 
	\centerline{\includegraphics[width=0.9\textwidth,clip=]{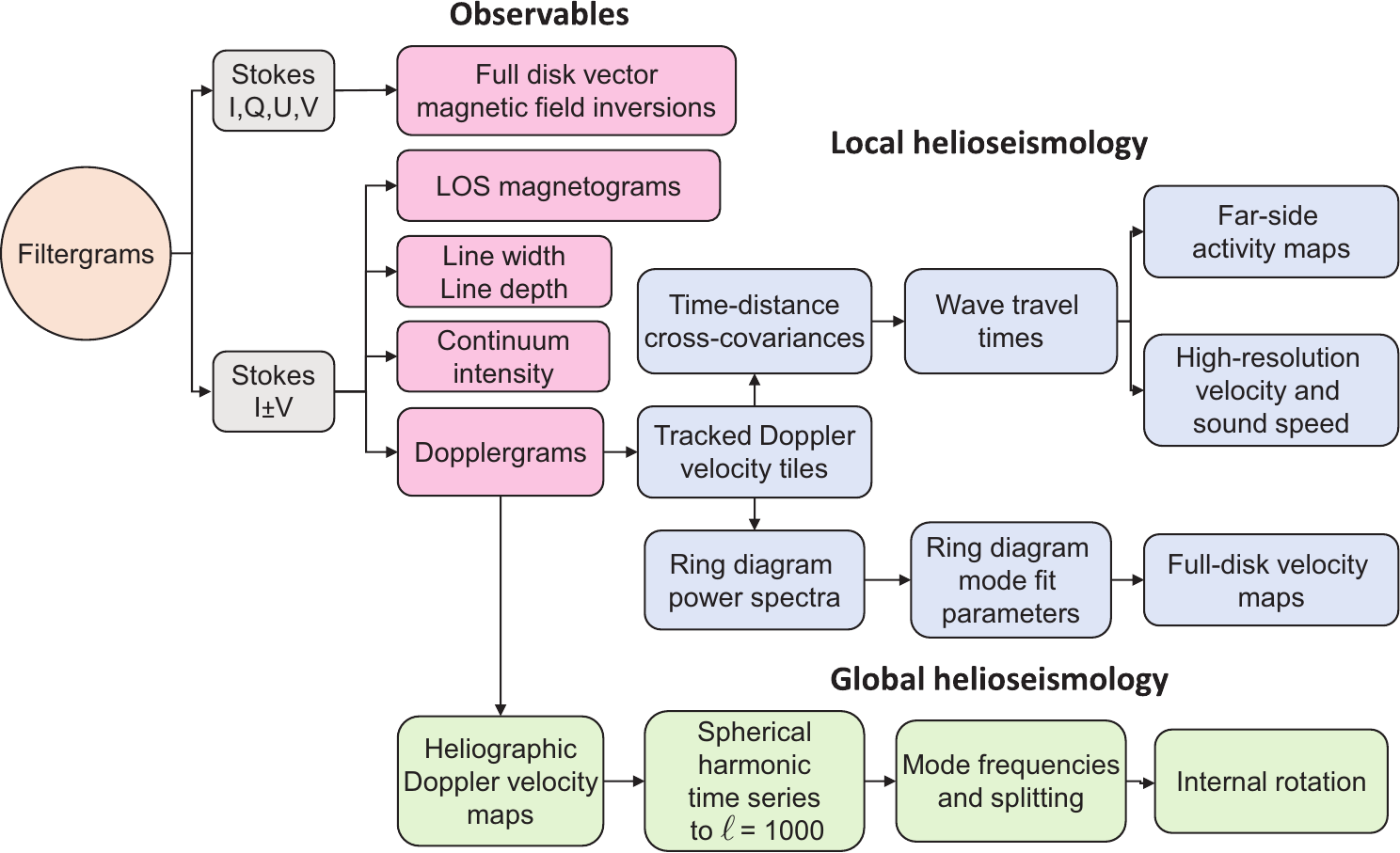}}
	\caption{Simplified HMI Science Pipeline. Local helioseismology data products are shown in {\it blue boxes}, and global helioseismology products are shown in {\it green}.}\label{fig:pipeline}
\end{figure}
Figure~\ref{fig:pipeline} shows the simplified HMI data pipeline emphasizing data products related to helioseismology. Boxes on the right edge show final high-level data products. Users of HMI data can access any of the products via the SDO Joint Science Operations Center (JSOC)\footnote{\url{jsoc.stanford.edu}}.

The 45\,s Dopplergrams feed two helioseismology pipelines: global and local. The global pipeline remaps the full-disk Dopplergrams to heliographic coordinates, computes time series of spherical harmonics that reveal resonant acoustic-wave oscillations, determines mode frequencies and splittings, and performs inversions to determine internal rotation and sound speed profiles. Boxes in {green} near the bottom of the figure indicate the global-mode helioseismic data products.
The HMI global pipeline analysis \citep{Larson2018} is similar to that performed using data from the Michelson Doppler Imager \citep[MDI:][]{1995Scherrer_MDI,1997Kosovichev_Medl,Larson2015} on board the Solar and Heliospheric Observatory \citep[SOHO:][]{1995Domingo_SOHO} in Solar Cycle 23. Others have extended the analysis to higher order modes \citep{2015ApJ...803...92R,2020Reiter_Modes} and have applied alternative methods for inversion to determine rotation profiles \citep{2024Korzennik_Modes}.

The local-helioseismology pipeline starts with tracked surface-velocity tiles that are then used in two analysis techniques: Ring Diagrams and Time-Distance. 
The {blue} boxes in the middle show local-helioseismology products.
The tiles in the synoptic pipeline are centered on a regular Carrington grid and tracked at a fixed rotation rate \citep{2011Bogart_Rings1, 2011Bogart_Rings2}. Pipeline tiles have three sizes: $5^\circ$, $15^\circ$, and $30^\circ$, and are spatially sampled at half the size of the tile out to $\sim80^\circ$ from the disk center. The velocities are corrected for the motion of the spacecraft relative to that point on the Sun and for instrument sensitivity. Tiles are remapped from heliographic coordinates to  Postel's azimuthal equidistant projection. The data cubes formed by time series of tracked tiles form the basis for subsequent analysis.

The Ring-Diagram technique, {initially suggested by \citet{Gough_DetectionSubphotosphericConvective_1983},} was first developed by \citet{1988Hill_Rings} to measure local flows at moderate depths beneath the solar surface, $\sim$~15 - 30 Mm, depending on the tile size. Tiles for the HMI Rings synoptic pipeline \citep{2011Bogart_Rings1, 2011Bogart_Rings2} are tracked at the Carrington rate. The software performs a 3D Fourier spectral analysis of the cubes and fits the resulting shape and location of the rings of mode power at each frequency. The ring parameters are then inverted to resolve the depth dependence of the horizontal velocity.
Additional details can be found in Section~\ref{sec:Ring-diagram Method} and the \href{hmi.stanford.edu/teams/rings/}{HMI Rings Team} website\footnote{hmi.stanford.edu/teams/rings/}.

The Time-Distance helioseismology technique was pioneered by \citet{Duvall1993}. Acoustic waves of a particular wavelength and frequency penetrate to a certain depth (wave turning point). By applying appropriate temporal and spatial filters, waves traveling in opposite directions between surface points can be identified, and the differences in travel times can be used to determine the velocity along the ray path. With assumptions about the ray propagation, an appropriate inversion for many points results in a map of subsurface motions \citep{Kosovichev1996}. The method has been developed over the years and implemented in the HMI pipeline (see \citealp{Zhao2012} for an introduction and a detailed description of the HMI analysis pipeline). The pipeline uses $30^\circ$ tiles tracked at the differential rotation rate to build up a map of flow vs. depth for the near side of the Sun to a depth of $\sim20$ Mm every 8 hours (see Section~\ref{sec:Time-Distance Method}). 


\citet{Lindsey2000Farside} first realized that the largest spatial-scale waves could be used to detect perturbations on the far side of the Sun. Particular acoustic waves that originate at one point on the far side will follow well-defined paths whose reflection points can be detected in predictable patterns on the visible side of the Sun. By isolating specific combinations of such waves that originate at different locations on the far side, increasingly sophisticated maps of propagation delays can be constructed \citep{Duvall2001,Zhao2007,Zhao2019}. Delays of a few seconds in the several-hour travel time of waves reveal locations on the far side where the magnetic field is strong \citep{2022Chen_Farside}.

Table 1 lists some of the most relevant helioseismology data series available in the HMI pipeline. The rest of this report discusses scientific results based on these data.

\begin{table}[ht!]
	\centering
	\begin{tabular}{|r|p{0.17\textwidth}|p{0.43\textwidth}|}
		\hline
		JSOC Data Series & Method & Description \\ 
		\hline
		\texttt{hmi.V\_45s} & &Time series of full-disk Dopplergrams at 45~s cadence \\
		\texttt{hmi.V\_sht\_gf\_72d} & Global & Spherical harmonic time series of HMI velocity data, gap-filled and detrended \\
		\texttt{hmi.V\_sht\_pow} & Global & Power spectra of detrended and gap-filled time series\\
		\texttt{hmi.V\_sht\_modes} & Global & Mode fits to the spherical harmonic power spectra, symmetric mode profile \\
		\texttt{hmi.V\_sht\_modes\_asym} & Global & Same as above but asymmetric profile \\
		\texttt{hmi.V\_sht\_2drls} & Global & 2D RLS inversion for rotation profile from symmetric mode parameters \\
		\texttt{hmi.V\_sht\_2drls\_asym} & Global & Same as above using asymmetric mode parameters \\
		\texttt{hmi.rdVpspec\_fd30} & Rings & Power spectra of $30^\circ$ tracked regions\\
		\texttt{hmi.rdVpspec\_fd15} & Rings & Power spectra of $15^\circ$ tracked regions\\
		\texttt{hmi.rdVflows\_fd30\_frame} & Rings & Inverted flows for tracked $30^{\circ}$  regions \\
		\texttt{hmi.rdVflows\_fd15\_frame} & Rings & Inverted flows for $15^{\circ}$ regions \\
		\texttt{hmi.rdVfitsc\_fd05} & Rings & Ring-diagram fits of flows and mode parameters for $5^{\circ}$ regions \\
		\texttt{hmi.tdVtimes\_synopHC} & Time-Distance & Travel times for 25 areas of $30\times30^\circ$\\
		\texttt{hmi.tdVinvrt\_synopHC} & Time-Distance & Subsurface flow fields \\
		\texttt{hmi.td\_fsi\_12h} & Time-Distance & Time-distance far-side images \\
		\texttt{hmi.fsi\_phase\_lon\_lat} & Holography & Holography-method far-side images \\
		\hline
	\end{tabular}
	\caption{Selected HMI pipeline helioseismology data series available at the JSOC.}
	\label{tab:JSOCProducts}
\end{table}

\subsection{JSOC Data Access}

The SDO Joint Science Operations Center (JSOC) Science Data Processing facility at Stanford receives SDO data in near-real time from the satellite downlink station. JSOC processes both HMI and Atmospheric Imaging Assembly \citep[AIA: ][]{2012Lemen_AIA} data for use by the community. Quick-look data, such as magnetograms and solar far-side activity maps, are produced for space-weather studies. Definitive science data are generated with some delay, {which can take several days,} to ensure that all of the relevant calibrations that depend on time history are complete and that time series of sufficient duration are available.

Each data product is stored in a \texttt{data\_series} that consists of a series of data \texttt{records}. The record is the basic unit of data and consists of \texttt{keywords}, \texttt{segments}, and \texttt{links}. Each \texttt{data\_series} is organized around one or more \texttt{prime} keywords, such as time. Other keywords contain metadata describing the series or a particular record or segment. A segment is a collection of files with the actual data, typically one or more files in the FITS format. A link is a pointer to a segment in another data series associated with the same prime keywords.

For example, the HMI Dopplergrams are in the data series \texttt{hmi.V\_45s} with records organized by the prime keywords \textsc{t\_rec} and \textsc{camera}. Many other keywords provide information about the data in that record; for example, \textsc{obs\_vr} provides the radial velocity of the SDO satellite relative to the disk center at the time of observation. This series has a single segment called \textsc{Dopplergram}, which is a FITS file with the 4096$\times$4096 line-of-sight velocity image in m\,s$^{-1}$. 

By contrast, the $15^\circ$ ring diagram flow series, \texttt{hmi.rdVflows\_fd15\_frame} has prime keywords \textsc{CarrRot} and \textsc{CMLon} indicating the Carrington Rotation and central meridian longitude for a set of computed $15^\circ$ rings; there is one record for each $15^\circ$ of Carrington Longitude. For each record, there are two segments, \textsc{Ux} and \textsc{Uy}, that contain velocities in the zonal and meridional directions, respectively. In this series, each segment consists of 284 ASCII tables, one for each of the rings spaced every $7.5^\circ$ in longitude from $75^\circ$E to $75^\circ$W and in latitude between +$60^\circ$ and $-60^\circ$ within $80^\circ$ of the disk center. For each ring pixel, the table provides the velocity computed at 31 target depths from 0.970 to 1.000 solar radii, along with uncertainties and other information.

The JSOC data series are managed by a Data Record Management System (DRMS). Various tools are available to query the database and export data. On export, the keywords, segments, and links are combined. A general introduction to SDO data access is available in the \href{www.lmsal.com/sdodocs/doc/dcur/SDOD0060.zip/zip/entry/}{Guide to SDO Data Analysis}\footnote{www.lmsal.com/sdodocs/doc/dcur/SDOD0060.zip/zip/entry/}. For exploratory work, the web-based tool \href{jsoc.stanford.edu/ajax/lookdata.html}{lookdata}\footnote{jsoc.stanford.edu/ajax/lookdata.html} is recommended. The export data tool provides various options for preprocessing and refining, extracting, or projecting data. The \href{jsoc.stanford.edu/}{JSOC website}\footnote{jsoc.stanford.edu/} provides more information. There are also Python tools for \href{docs.sunpy.org/en/stable/tutorial/acquiring\_data/jsoc.html}{sunpy}\footnote{docs.sunpy.org/en/stable/tutorial/acquiring\_data/jsoc.html} and \href{www.lmsal.com/sdodocs/doc/dcur/SDOD0060.zip/zip/entry/sdoguidese6.html#x11-390006.3}{solarsoft}\footnote{www.lmsal.com/sdodocs/doc/dcur/SDOD0060.zip/zip/entry/sdoguidese6.html\#x11-390006.3} tools for IDL.

 
 \section{Global Helioseismology Measurements}\label{sec:global}

The primary goal of global helioseismology measurements is to provide estimates of resonant mode frequencies, amplitudes, line widths, and asymmetry, as well as parameters of the background convective noise. Mode frequencies are primarily used to determine the solar interior structure and differential rotation and their variations using helioseismic inversion procedures. In addition, the mode amplitudes, line widths, asymmetry, and background are valuable for studying the physics of solar oscillations and characterizing solar-cycle variations.

Continuous high-resolution HMI observations substantially extended the capabilities of global helioseismology measurements from the Global Oscillation Network Group (GONG) and Michelson Doppler Imager (MDI) and led to the development of new data analysis methods. Because of the telemetry restrictions for transmitting the observational data from SoHO, MDI's full-disk $1024\times 1024$-pixel Dopplergram images with a 1-min temporal cadence were transmitted only two months a year during the Dynamics campaigns. All other times the data were transmitted only after a `vector-weighted' binning of the full-disk images, which reduced the spatial resolution by a factor of 4. These data were used to measure the frequencies of the oscillation modes in the range of the spherical harmonic degrees, $\ell$, from 0 to 300 \citep[aka the Medium-$\ell$ Program; ][]{1997Kosovichev_Medl}. SDO orbit allows for the continuous and nearly uninterrupted transmission of HMI full-resolution $4096\times 4096$-pixel Doppler images with a 45-sec cadence.

The HMI data analysis pipeline continues the MDI Medium-$\ell$ program. The data are processed {by performing the spherical harmonic decomposition for the full-resolution $4096\times 4096$-pixel Dopplergrams, and also by simulating the MDI vector-weighted binning to investigate potential systematic effects caused by the binning scheme}. In addition, the frequency measurement procedure was substantially improved and thoroughly tested \citep{Larson2018,Larson2024SoPh299a}. These improvements made it possible to investigate potential systematic errors. In particular, it was determined that a signature of a high-latitude jet, which appeared in some inversions of the MDI data \citep{1998ApJ...505..390S}, was an artifact. 

The global helioseismology analysis of the HMI data in the medium-degree range
measures the mean frequencies of mode multiplets and the coefficients of a polynomial expansion to
represent the mode frequency splitting as a function of azimuthal order, $m$, using
orthogonal polynomials (the so-called $a$-coefficients). This frequency splitting results from the
differential rotation and departures of the solar structure from spherical symmetry. The global helioseismology analysis also measures other properties of the oscillation modes,  such as the mode amplitude, line width, and background noise level \citep{Larson2015}. Two
models are used for fitting the oscillation line profiles, a symmetrical Lorentzian profile describing a damped harmonic oscillator and a
non-symmetrical profile \citep[based on the formulation of ][]{1998ApJ...505L..51N}, describing
oscillations excited by acoustic sources localized in the subsurface
layers. The mode fitting results and the inferred differential
rotation profiles are available via the JSOC for these two models.

\begin{figure}
	\begin{center}
		\includegraphics[width=0.9\linewidth]{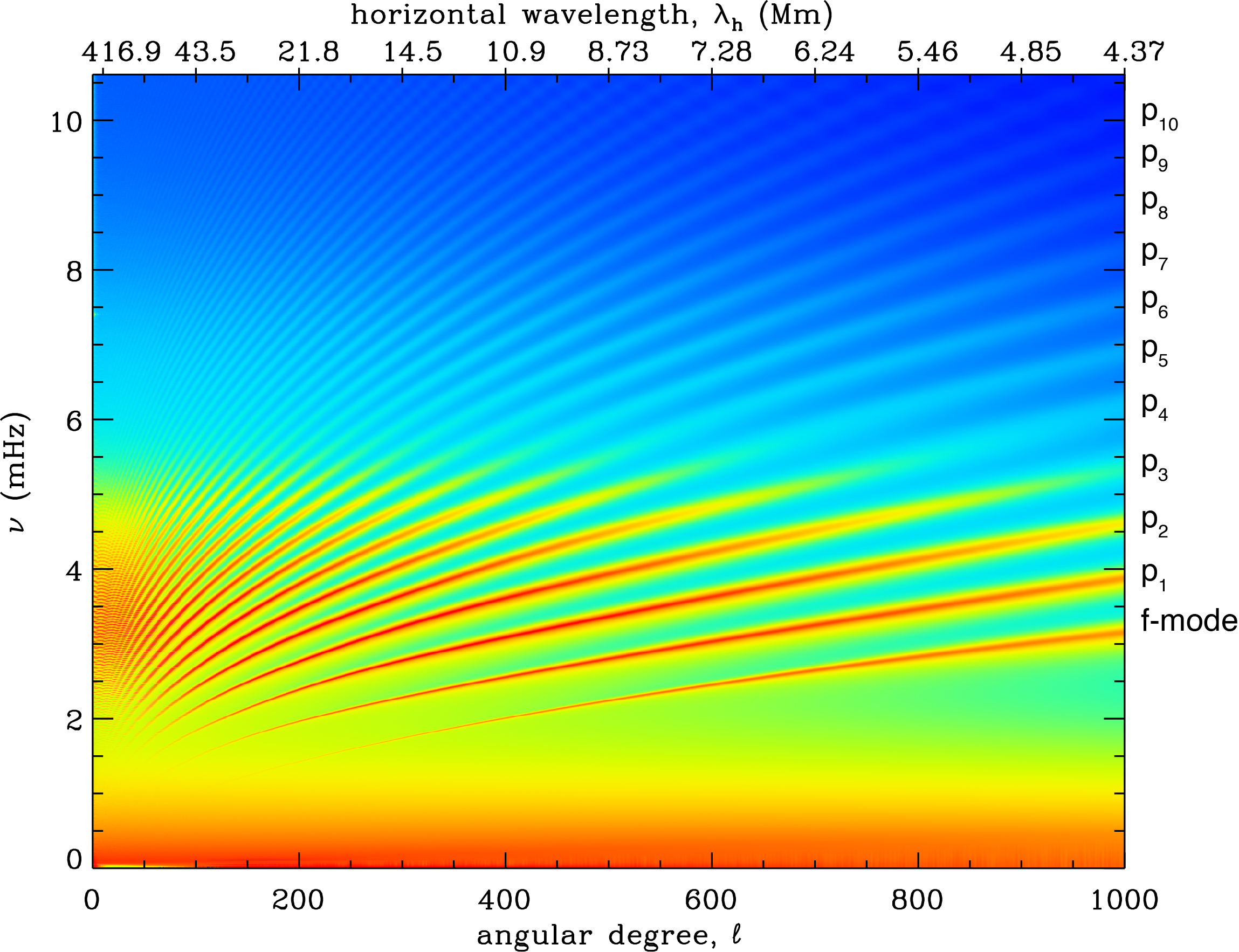}
		\caption{The oscillation power spectrum (on a logarithmic scale) as a function of the angular degree $\ell$ and frequency, $\nu$ (so-called $\ell-\nu$ diagram) obtained from a 72-day time series of the Doppler velocity images from {the HMI data series \texttt{hmi.V\_sht\_pow}}. The left vertical axis shows the identification of the surface gravity (\textit{f}) and acoustic (\textit{p}) modes in the range of the radial order, $n$, from 1 to 10. The top axis shows the horizontal wavelength, $\lambda_h$.
		}\label{HMI_l-nu_1000}
	\end{center}
\end{figure}

The precision of global helioseismology measurements is higher for longer observing periods. However, the analysis has to be performed for relatively short intervals to capture variations in the solar structure and rotation associated with solar magnetic activity. Hence the JSOC  global heliosesimology pipeline fits time series that are 72-day and 360-day long. The oscillation power spectrum corresponding to a 72-day long time series is illustrated in Figure~\ref{HMI_l-nu_1000}. This power spectrum is obtained by projecting the Doppler velocity images onto spherical harmonics, performing the Fourier transform for the spherical harmonic coefficients time series, and averaging the power over the mode multiplets after applying corrections for the rotational frequency splitting.

\begin{figure}
	\begin{center}
		\includegraphics[width=\linewidth]{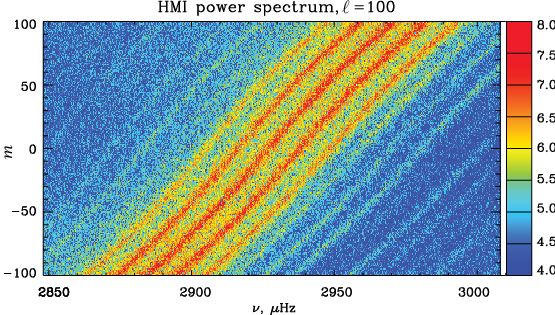}
		\caption{Small portion power spectrum (on a logarithmic scale) shown as a function of frequency and azimuthal order, $m$, in the vicinity of the $\ell=100, n=6$ mode. The mode frequencies vary with $m$ due to the internal solar rotation. The multiple nearly parallel mode ridges around the central target mode ridge are caused by spatial leakage.
		}\label{HMI_m-nu_l100}
	\end{center}
\end{figure}

The power spectral density (on a logarithmic scale) 
as a function of frequency and the azimuthal order, $m$, for a mode of the angular degree, $\ell=100$, is shown in Figure~\ref{HMI_m-nu_l100}. The change of slope in this diagram reflects the latitudinal changes in the solar rotation. The multiple lines (`spatial leaks') are caused by the leakage of the power of modes with the angular degrees in the vicinity of the target mode (mostly in the range of $\ell=100 \pm 2$) because the spherical harmonic transform is performed only in a visible hemisphere of the Sun. The mode leakage is calculated and taken into account in the data analysis 
\citep[e.g.][]{Korzennik2013}.

The measurements (fitting the mode peaks in the power spectrum) are performed approximating the rotational splitting by polynomials of three orders, 6, 18, and 36, {and available in the HMI data products \texttt{hmi.V\_sht\_modes} and \texttt{hmi.V\_sht\_modes\_asym} (Table~\ref{tab:JSOCProducts})}.  The odd $a$-coefficients in the  6-order approximation provide estimates of only the first three terms in the differential rotation, but the high-degree approximations resolve the evolving zonal flows (so-called `torsional oscillations') and investigate their relationship to the solar activity. {The 36-order approximation provides better latitudinal resolution than the 18-order approximation; however, the inversion results for the 36-order approximation may be noisier. Therefore, both approximations should be used for analysis.} The even-degree $a$-coefficients are used to study the aspherical structure of the Sun.      

\begin{figure}
	\begin{center}
		\includegraphics[width=0.9\linewidth]{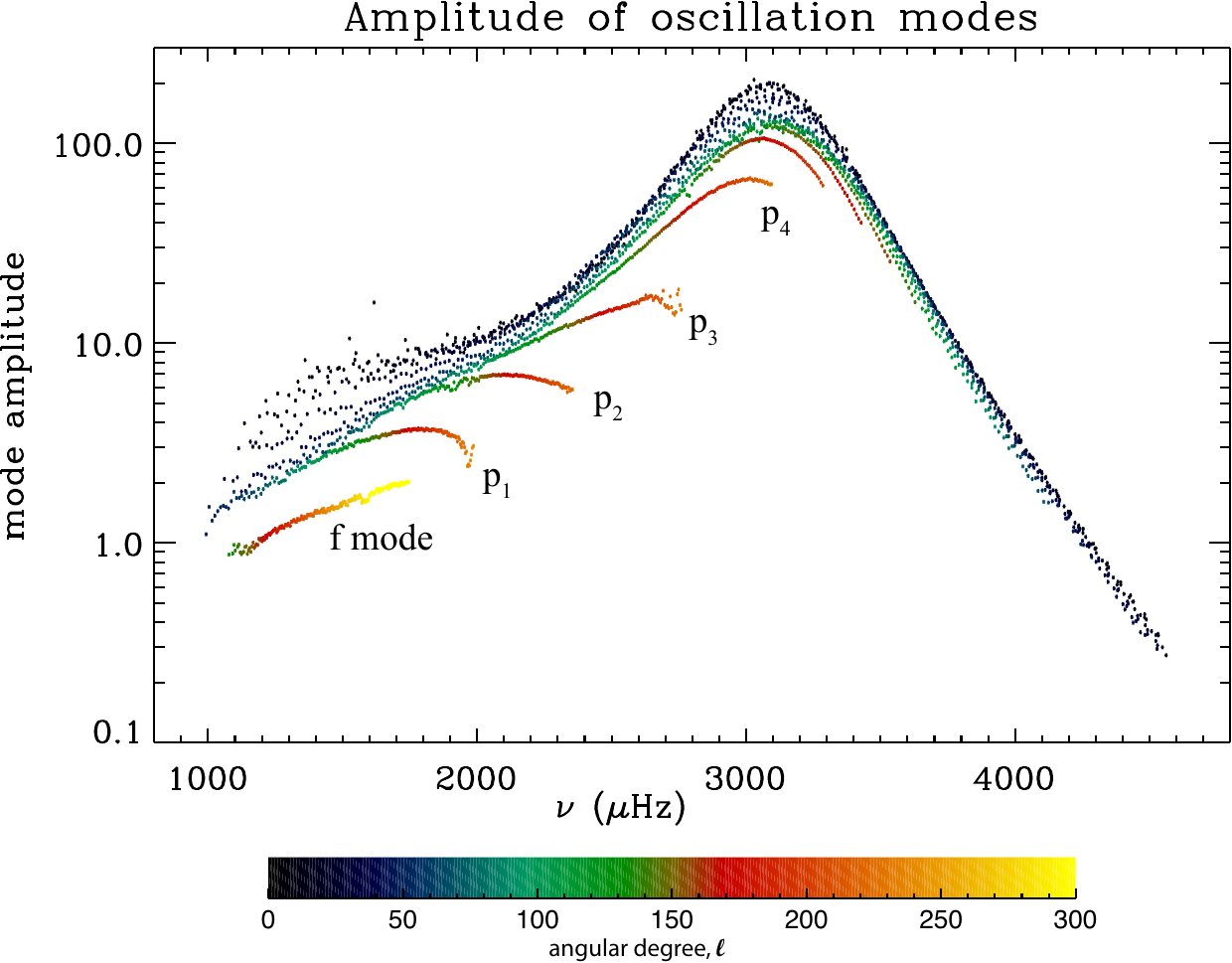}
		\caption{The amplitudes of the mean multiplet frequencies for modes $\ell \leq 300$ obtained from a 2304-day time series of observations of Doppler velocity, with the start date of 31 October 2016 from the SDO/HMI instrument. {The data points are colored by the mode angular degree.}}\label{HMI_amplitude}
	\end{center}
\end{figure}

\begin{figure}
	\begin{center}
		\includegraphics[width=0.9\linewidth]{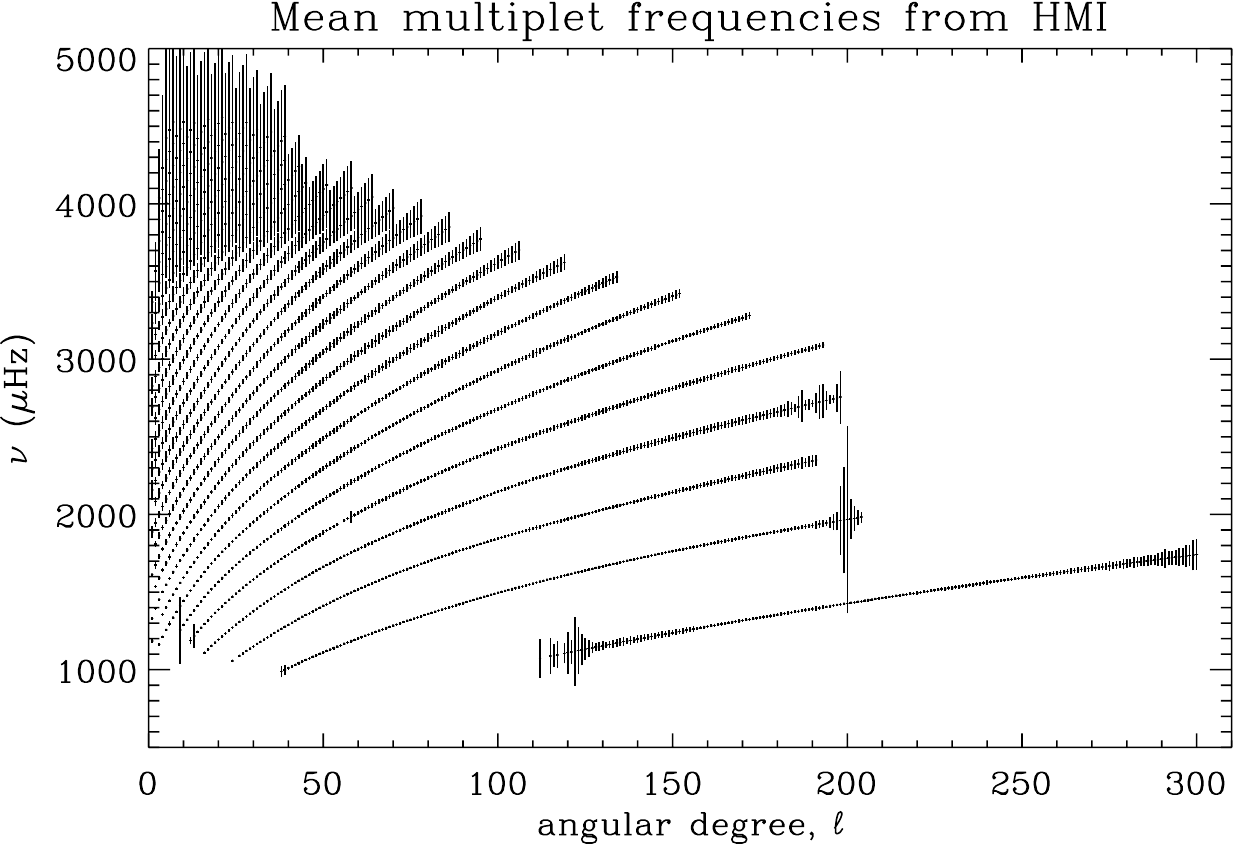}
		\caption{The $\ell-\nu$ diagram of the mean multiplet frequencies for modes with $\ell \leq 300$ obtained from a 2304-day long time series of observations of Doppler velocity from the SDO/HMI instrument \citep{Korzennik2023FrASS}. The error bars show 2000$\sigma$ uncertainties. }\label{HMI_lnu}
	\end{center}
\end{figure}

In addition, the data are analyzed at the Center for Astrophysics, Harvard \& Smithsonian (CfA) for various time intervals using a novel technique \citep{Korzennik2023FrASS}. This technique uses {optimal} multi-taper power spectrum estimators to reduce the noise and fit the individual multiplets that are visible in the power spectrum \citep{Korzennik2005ApJ}. The rotational frequency coefficients are calculated from the individual frequencies for each multiplet. The frequencies and frequency splitting obtained by this technique from the HMI data are available at the CfA website\footnote{\url{lweb.cfa.harvard.edu/~sylvain/research/tables/}} \citep{Korzennik2023FrASS} and {the JSOC data products \texttt{su\_sylvain.hmi\_V\_sht\_modes\_sym\_v7} for fitting symmetric mode profiles and \texttt{su\_sylvain.hmi\_V\_sht\_modes\_asym\_v7}.} for the asymmetrical profile. In particular, the HMI provides an opportunity to obtain high-precision measurements by using very long time series. Figures~\ref{HMI_amplitude} and \ref*{HMI_lnu} show the amplitude and frequency measurements obtained from a 2304-day (6.3 years) long time series of observations of Doppler velocity. {The mode amplitude is colored according to the mode angular degree. The amplitude of higher-degree modes is reduced because of stronger dissipation in the near-surface layers.} The error bars in Figure~\ref{HMI_lnu} show the error estimates magnified by a factor of 2000. 
The work on identifying potential systematic uncertainties that may affect the inversion results for the solar rotation and structure is still actively pursued \citep{Korzennik2023FrASS}.

The oscillation modes of high angular degree ($\ell>300$) carry important information about the near-surface layers of the Sun. 
 At high degrees, however, the spatial leaks get closer in frequency due to a smaller mode separation, and the peaks get wider and eventually overlap since the mode's lifetime gets smaller. As a result, high-degree modes blend into broad ridges of power. Unfortunately, the properties of such ridges do not correspond to those of the underlying individual peaks. This is because the amplitudes of the spatial leaks are asymmetrical with respect to the targeted mode, and so the central frequency of the ridge is significantly shifted away from the underlying individual mode frequency. Therefore, fitting a ridge requires detailed knowledge of the distribution of power that leaks into the sidelobes adjacent to the targeted spectral peak. 

\begin{figure}
\begin{center}
\includegraphics[width=\linewidth]{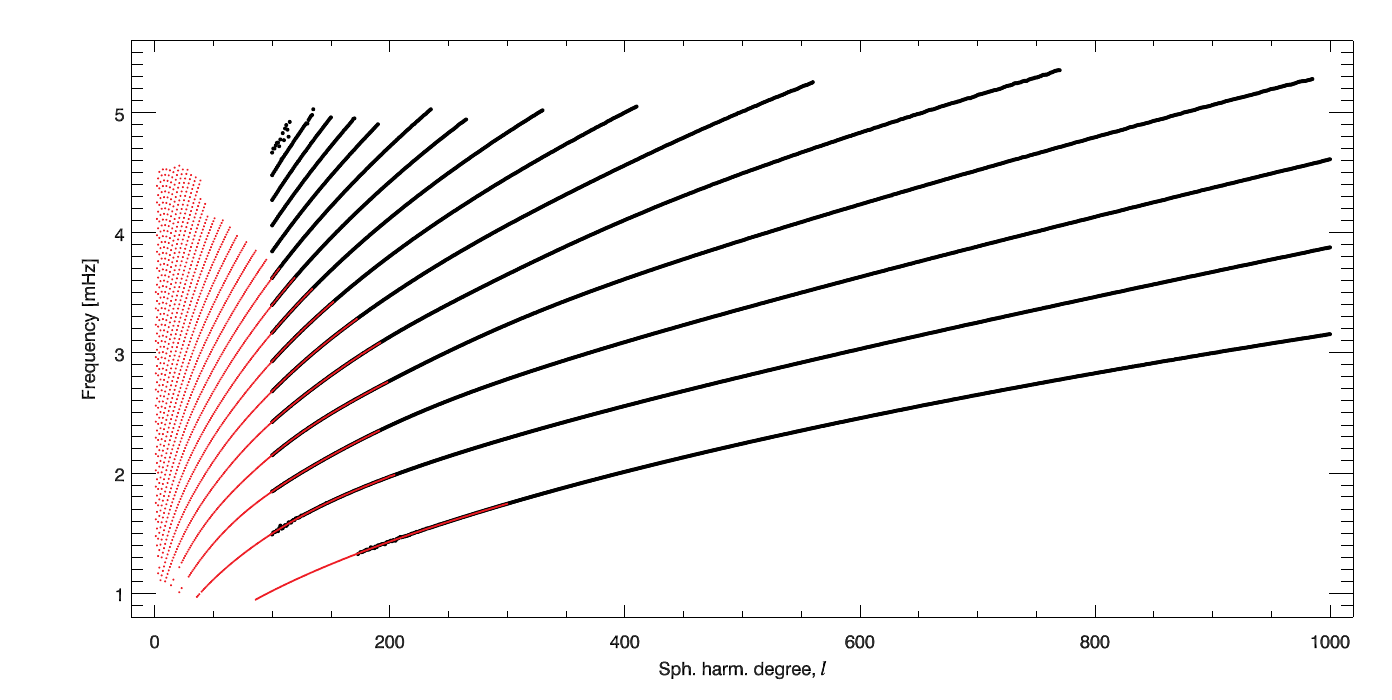}
\caption{Coverage in the $\ell-\nu$ space of the high degree (\textit{black dots}) and low and intermediate degree (\textit{red}) modes.}
\label{modes-medium+high-l}
\end{center}
\end{figure}

\begin{figure}
	\begin{center}
 \includegraphics[width=0.85\linewidth]{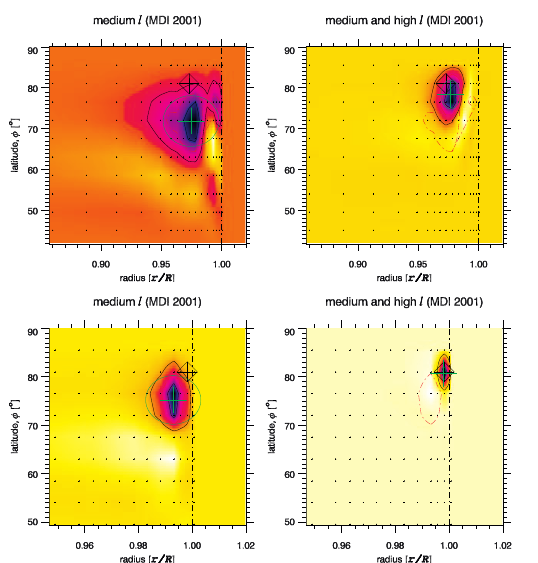}
\caption{{Comparison of the resolution of solar rotation inversion achieved when not including (\textit{left panels}) and including high degree modes (\textit{right panels}). The resulting averaging kernels are shown near the surface at either $r/R_\odot$ = 0.973 (\textit{top panels}) or $r/R_\odot$ = 0.998 (\textit{bottom panels}) and for a latitude of $81^\circ$. The value of the averaging kernel is shown using a color map (and contours at 0.25 and 0.75 of the kernel’s maximum value). The \textit{black dots} indicate the inversion grid, the \textit{black cross-diamond symbols} indicate the inversion target location, while the \textit{green crosses} and \textit{circles} show the respective averaging kernel’s center of gravity and a single measure of width. The \textit{red dot-dashed contour} on the \textit{right panels} indicates the location of the corresponding averaging kernel half height when the high-degree modes are not included (i.e., from the averaging kernel on the left panels).}} \label{ak-hil-soho28}
\end{center}
\end{figure}

Techniques for measuring properties of the high-degree modes have been developed \citep{Korzennik2013,2020Reiter_Modes}, but they have yet to be routinely applied for analyzing the HMI data. One approach consists of fitting the power ridges and correcting the resulting values by using a sophisticated model of the underlying power distribution of the modes that blend into such a ridge. The offsets between the underlying mode characteristics and the resulting modeled ridge properties are then used in a parametric formulation to derive underlying mode characteristics from measured ridge characteristics. 
The other approach consists of fitting a multimodal extended model to the observed ridge, which incorporates some a priori knowledge of the power distribution. 

The extent and potential of these high-degree modes are illustrated in Figures \ref{modes-medium+high-l} and \ref{ak-hil-soho28}, where we show the coverage in an $\ell-\nu$ diagram of the high-degree modes compared to the medium-degree ones, while Figure \ref{ak-hil-soho28} compares the expected resolution, as indicated by averaging kernels, near the surface when including or not high degree modes in a rotation inversion. It has also been shown that the high-degree mode measurements can substantially improve the resolution of the solar structure in the near-surface shear layer \citep{2015ApJ...803...92R,2020Reiter_Modes}.

\section{Local Helioseismology Techniques}\label{sec:Local Helioseismology}


Several techniques have been developed to measure and characterize large-scale mass flows on the Sun. Doppler imaging uses the Dopplergrams to measure the line-of-sight velocities of plasma motions on the solar surface. Correlation and Structure Tracking methods involve tracking the motion of magnetic features or intensity patterns over time. Local helioseismology methods enabled the inference of near-surface and deep subsurface flows from observations of solar oscillations. 
The Time--Distance method measures the travel times of acoustic waves between different points on the solar surface, while the Ring Diagram technique analyzes the power spectra of solar oscillations in localized regions; both methods determine flow velocities at various depths through which the observed waves propagate. The Helioseismic Holography method provides important information about acoustic sources of solar flares and quiet Sun regions and is routinely used for detecting and tracking sunspot regions on the far side of the Sun. In addition, new methods of Mode Coupling enabled the discovery and analysis of Rossby waves and inertial modes on the Sun. 

\subsection{Ring-Diagram Method}\label{sec:Ring-diagram Method}

\begin{figure}
	\begin{center}
		\includegraphics[width=\linewidth]{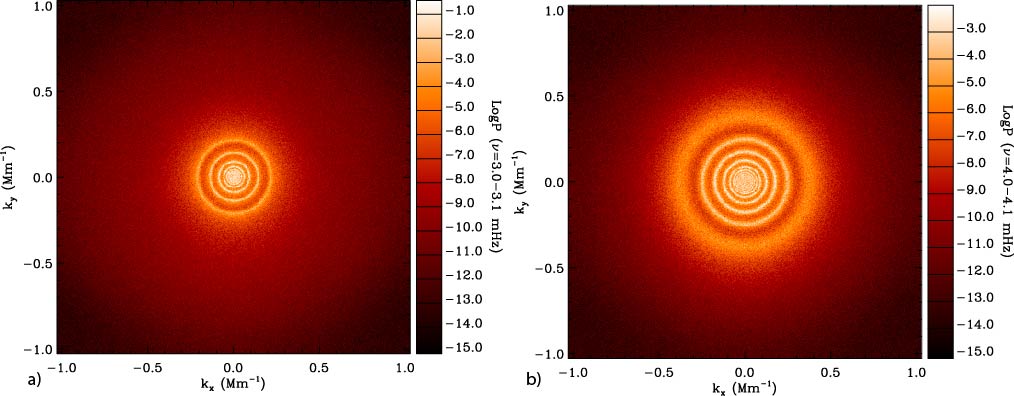}
		\caption{{The ring-diagram power spectrum calculated for a $15^\circ \times 15^\circ$ tile located at the disk center from the HMI observations with the start date of 2024.07.31~01:02:15~TAI	and the end date of 2024.08.01~05:49:30~TAI: the power spectrum as a function of wavenumbers, $k_x$ and $k_y$, in the frequency ranges: \textit{a)} 3.0\,--\,3.1~mHz and \textit{b)} 4.0\,--\,4.1~mHz. The data used in these images are obtained from the JSOC data series \texttt{hmi.rdVpspec\_fd15} (Table~\ref{tab:JSOCProducts})}.}\label{HMI_rings_power_spec}
	\end{center}
\end{figure}

In the Ring-Diagram technique, small regions of the solar disk, typically with diameters of 5$^\circ$, 15$^\circ$, and 30$^\circ$, are tracked at the Carrington rotation rate for durations corresponding to their angular size (approximately 9.6, 28.8, and 57.6 hours, respectively). These regions are then remapped into solar latitude and longitude coordinates {using the azimuthal equidistant (Postel's) projection}. A plane-wave approximation can be applied because the horizontal wavelengths of acoustic waves observed at these scales are significantly smaller than the solar radius. This enables us to calculate the power spectrum using three-dimensional Fourier transforms as a function of the horizontal wavenumber components, $k_x$ and $k_y$ (corresponding to longitude and latitude directions), and temporal frequency. The resulting spectrum exhibits a characteristic structure resembling nested ``trumpets" along the frequency axis. When examining horizontal slices of this power distribution at constant frequencies, distinct rings for each radial order $n$ are observed (Figure~\ref{HMI_rings_power_spec}). 

This distinctive pattern gives rise to the term ``ring diagram analysis" for this technique.
The presence of a velocity field, $\mathbf{U}$, introduces perturbations to the wave frequency producing an apparent frequency shift, which can be expressed as:
$\Delta \omega = k_x U_x + k_y U_y$,
where $U_x$ and $U_y$ represent the components of the velocity vector $\mathbf{U}$ along the longitude ($x$) and latitude ($y$) directions. 
{
	\begin{figure}
		\begin{center}
			\includegraphics[width=0.7\linewidth]{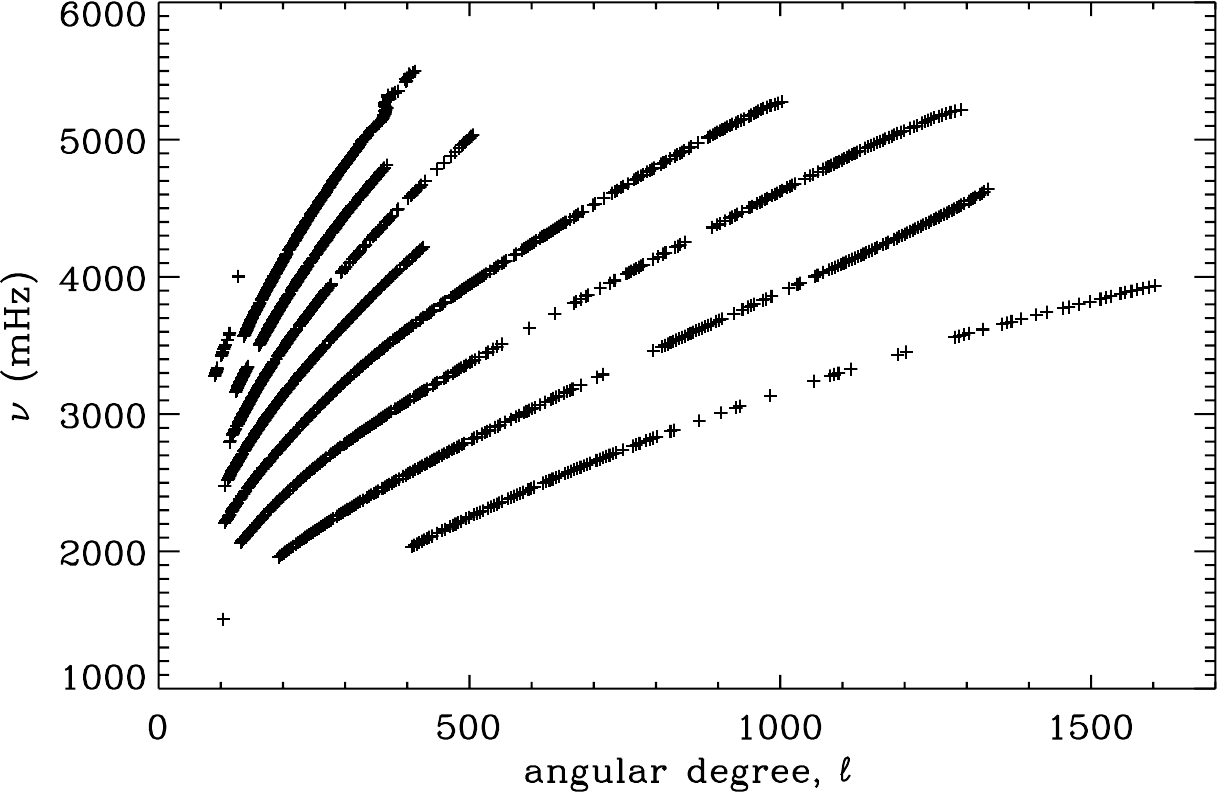}
			\caption{{The angular degree - frequency ($\ell-\nu$) diagram illustrating a set of selected frequencies used in the ring-diagram method.  The data used in this plot are obtained from the JSOC data series \texttt{hmi.rdVfitsc} (Table~\ref{tab:JSOCProducts})}.}\label{HMI_rings_lnu}
		\end{center}
	\end{figure}
}

The 3D power spectrum is analyzed using two fitting methods to estimate the horizontal velocity field. In the first approach \citep{BasuAntia1999}, implemented in the HMI pipeline as \texttt{hmi.rdVfitsc} data series (Table 1, HMI Observables), the spectrum is fitted at a set of selected frequencies (Figure~\ref{HMI_rings_lnu}). 


The second method, developed by \citet{SchouBogart1998}, involves unwrapping the 3D power spectrum. For each wavenumber $k$  
(where $k = \sqrt{k_x^2 + k_y^2}$),
a cylindrical section is extracted by interpolating the original three-dimensional spectrum onto a uniform grid in azimuth angle $\theta$ at each frequency. This spectrum is then Fourier filtered in $\theta$ and subsampled to reduce computational time. While this approach is significantly faster than the first method, it results in a smaller number of measurements. The velocity field is fitted for each $k$, and in the HMI pipeline, the resulting fitted parameters are provided in the \texttt{hmi.rdVfitsf} data series. It's worth noting that this second method is also employed in the GONG pipeline \citep{Corbard2003}.

{
\begin{figure}
	\begin{center}
		\includegraphics[width=\linewidth]{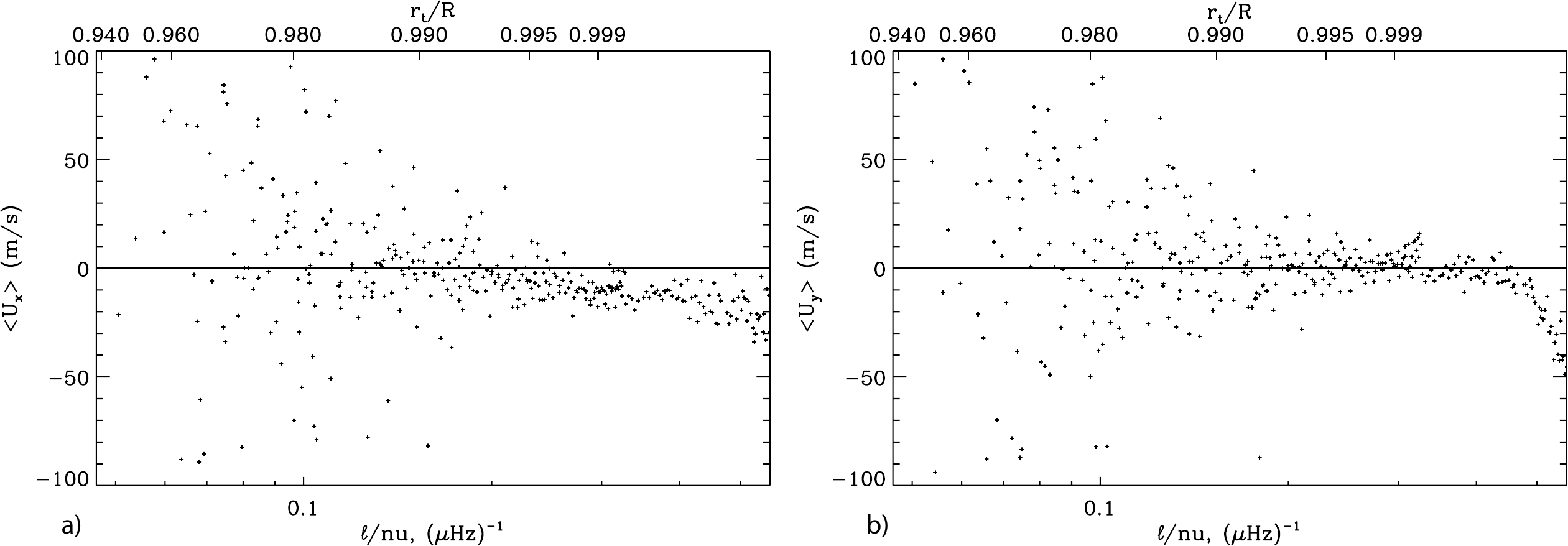}
		\caption{{An example of the ring-diagram fitted parameters $U_x$ and $U_y$ from the HMI observations with the start and end dates indicated in the caption to Figure~\ref{HMI_rings_power_spec}, plotted as a function of $\log(\ell/\nu)$ and the radius of the mode turning points, $r_t/R_\odot$. The data used in this plot are obtained from the JSOC data series \texttt{hmi.rdVfitf} (Table~\ref{tab:JSOCProducts})}.}\label{HMI_rings_ux_uy_f}
	\end{center}
\end{figure}
}
 In ring-diagram analysis, we infer the horizontal velocity variation with depth using the relation $k = \sqrt{\ell(\ell+1)}/R_\odot$, derived from spherical harmonics, to estimate an effective angular degree, $\ell$. These estimated $\ell$ values are then interpolated to the nearest integral value. Subsequently, we apply inversion methods from global helioseismology to analyze the data. The ring-diagram fitted parameters, $U_x$ and $U_y$, plotted as a function of $\log(\ell/\nu)$ and the radius of the mode inner turning points, $r_t/R_\odot$, in Figure~\ref{HMI_rings_ux_uy_f}, illustrate the depth coverage of the ring-diagram method. The inner turning point defines the deepest point of wave propagation into the Sun's interior from the surface. It depends on the ratio of the mode angular degree, $\ell$, and frequency $\nu$.

The HMI pipeline employs the Optimally Localized Average (OLA) technique, as described by \citet{BackusGilbert1968} and further elaborated by \citet{BasuAntia1999}. Another inversion method commonly used is the regularized least squares (RLS). In the HMI pipeline, the OLA method is used to calculate the inverted flows, which are provided in the data series \texttt{hmi.rdVflows\_fd15\_frame} and \texttt{hmi.rdVflows\_fd30\_frame} (Table~\ref{tab:JSOCProducts} in Section~\ref{sec:HMI_Observables}: HMI Observables). These calculations are performed by the \texttt{rdvinv} module \citep{2011Bogart_Rings1}.

{
\begin{figure}
	\begin{center}
		\includegraphics[width=\linewidth]{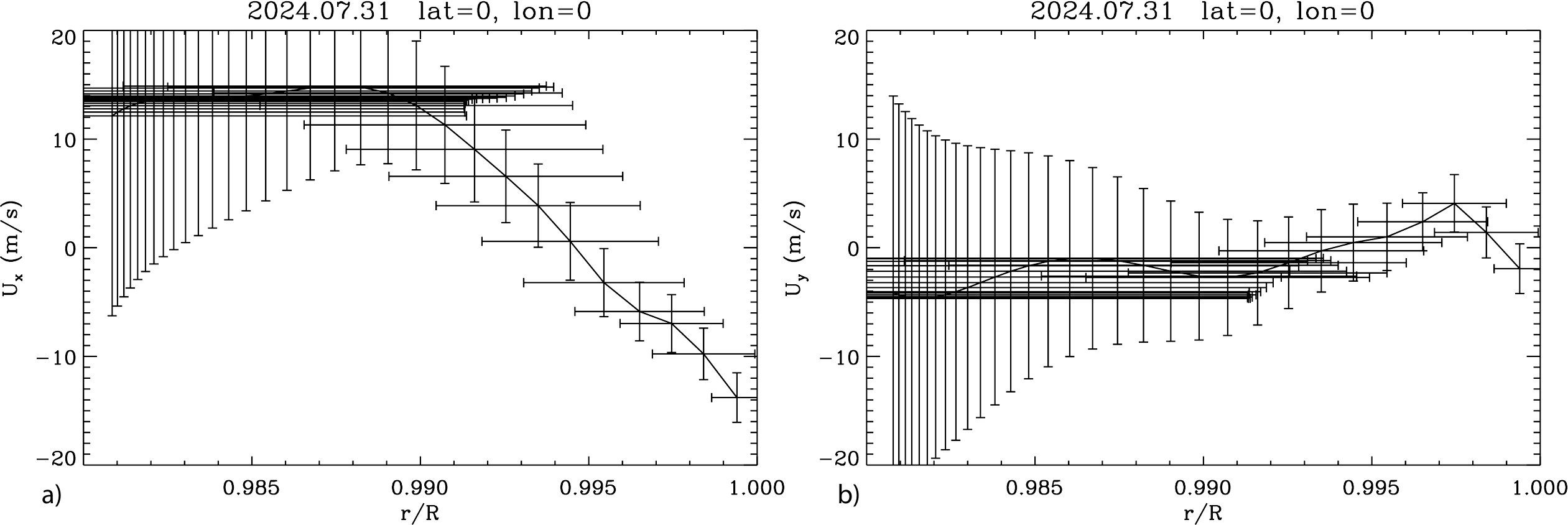}
		\caption{{The azimuthal, $U_x$, and meridional, $U_y$, velocity components as a function of radius obtained from the HMI observations with the start and end dates indicated in Figure~\ref{HMI_rings_power_spec}. The \textit{vertical bars} show $1\sigma$ error estimates, and the \textit{horizontal bars} show the horizontal resolution (a width of averaging kernels). The data used in this plot are obtained from the JSOC data series \texttt{hmi.rdVfitf} (Table~\ref{tab:JSOCProducts})}.}\label{HMI_rings_inver_radial_ux_uy_f}
	\end{center}
\end{figure}
}
{
	\begin{figure}
		\begin{center}
			\includegraphics[width=0.7\linewidth]{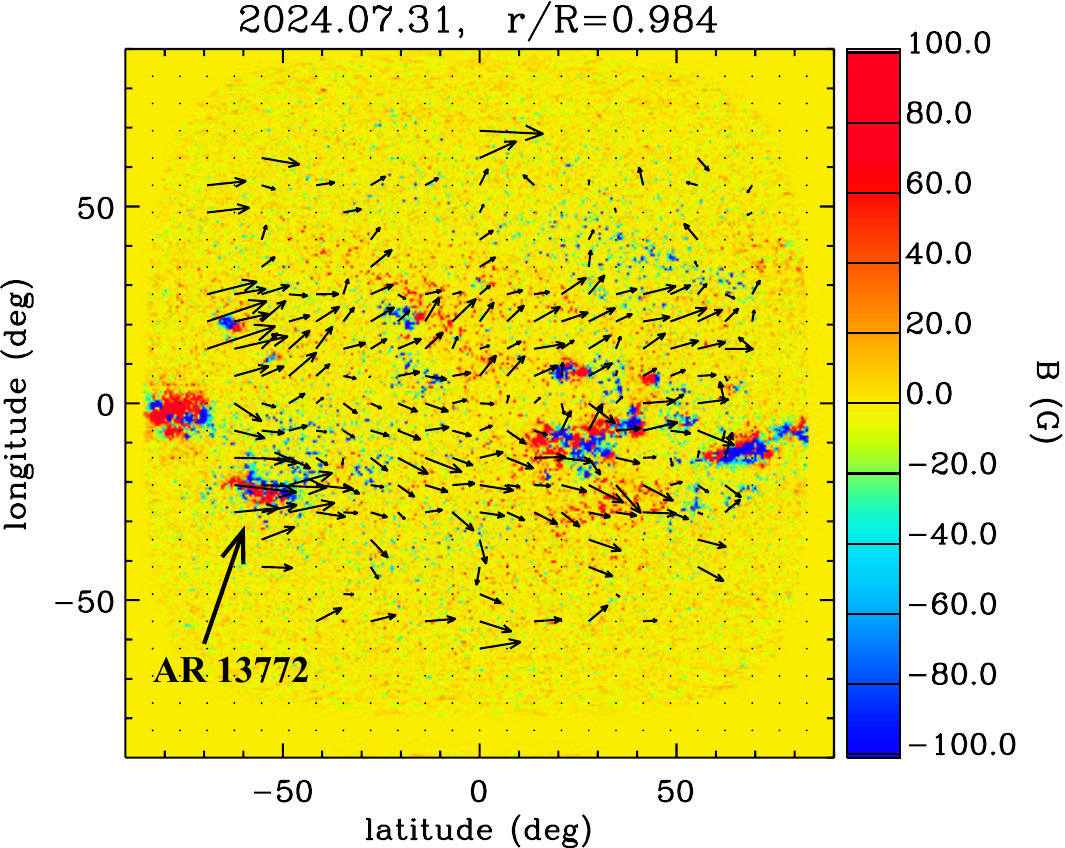}
			\caption{{The horizontal flow field (\textit{arrows}) at a radius of 0.984$R_\odot$ (the corresponding depth is $\approx 11$~Mm) and the surface magnetogram (\textit{color image}), obtained from the HMI observations with the start and end dates indicated in the caption to Figure~\ref{HMI_rings_power_spec}. The data used in this plot are obtained from the JSOC data series \texttt{hmi.rdVfitf} (Table~\ref{tab:JSOCProducts})}.}\label{HMI_rings_inver_mag_f}
		\end{center}
	\end{figure}
}

The maximum depth achievable in this analysis is approximately equal to the size of the tile under examination. For instance, when analyzing a 15-degree (30-degree) region, we can infer flows from very close to the surface down to depths of about 15 Mm (30 Mm). It's important to note that as fewer waves penetrate deeper layers, the uncertainty in the horizontal location increases with depth (Figure~\ref{HMI_rings_inver_radial_ux_uy_f}).

The analysis of smaller tiles presents additional challenges due to the reduced number of measurements. This limitation makes the inversion process more difficult. As a result, the HMI pipeline currently does not provide inversion results for 5-degree tiles \citep{Bogart2023}.

A useful data product for ring-diagram analysis is the Magnetic Activity Index (MAI), which quantifies the level of magnetic activity in each analyzed region. The MAI is calculated by the HMI processing pipeline. Using the same mappings and temporal and spatial apodizations as the tracked regions, the absolute values of all pixels in corresponding HMI line-of-sight magnetograms with an absolute value greater than 50 G ($|B_z| >$ 50~G) are averaged. The magnetograms are sampled once every 48 minutes for this calculation, as described by \citet{2011Bogart_Rings1, 2011Bogart_Rings2}. {Arrows in  Figure~\ref{HMI_rings_inver_mag_f} illustrate a typical horizontal flow map at a depth of about 11~Mm plotted over the corresponding surface magnetogram. Results of the ring-diagram method are discussed in Sections~\ref{sec:convection}, \ref{sec:dynamo}, \ref{sec:meridional_vary}, and \ref{sec:zonal}.}

\subsection{Time-Distance Method}
\label{sec:Time-Distance Method}

Acoustic waves are believed to travel between the Sun's surface locations through the solar interior along curved paths. 
The acoustic travel times from one surface location to the other can be measured, as a function of the distance between the two locations, by time-distance helioseismology \citep{Duvall1993} by cross-correlating two time-sequences of oscillations observed at those surface locations. 
One travel time can be measured from both the positive lag and the negative lag of the cross-correlation function, corresponding to the waves traveling from one location to the other and traveling in the opposite direction, respectively. 

	\begin{figure}
	\begin{center}
		\includegraphics[width=0.65\linewidth]{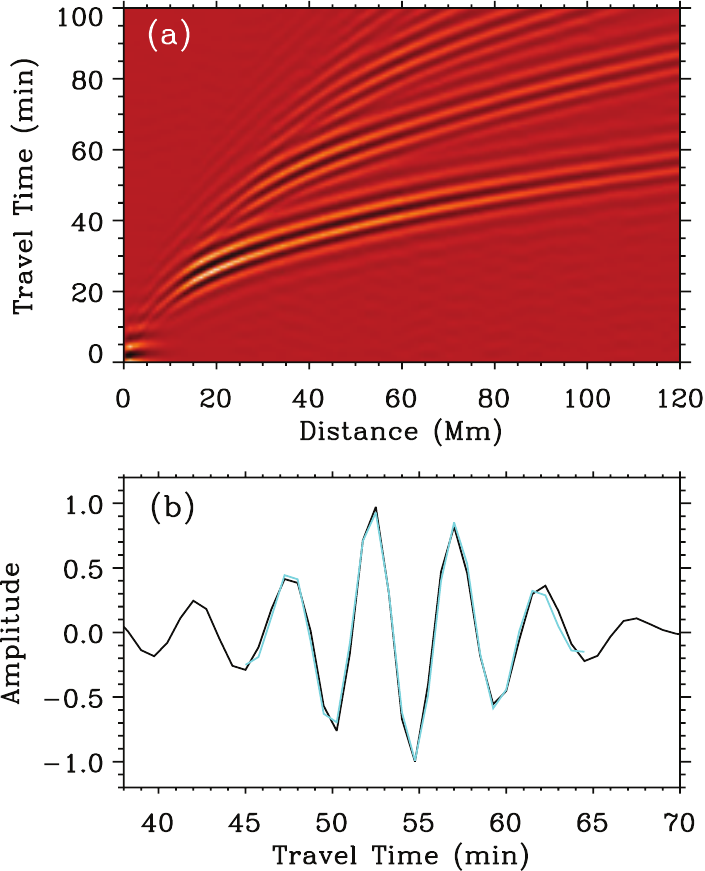}
		\caption{{\textit{a)} Time-distance diagram made using the SDO/HMI Dopplergrams. \textit{b)} As an example, Gabor wavelet fitting (\textit{cyan}) was applied on the cross-covariance function (\textit{dark}) taken at the distance of 109 Mm, getting a phase travel time of 52.28 min}.}\label{HMI_td_fit}
	\end{center}
\end{figure}

\begin{figure}
	\begin{center}
		\includegraphics[width=\linewidth]{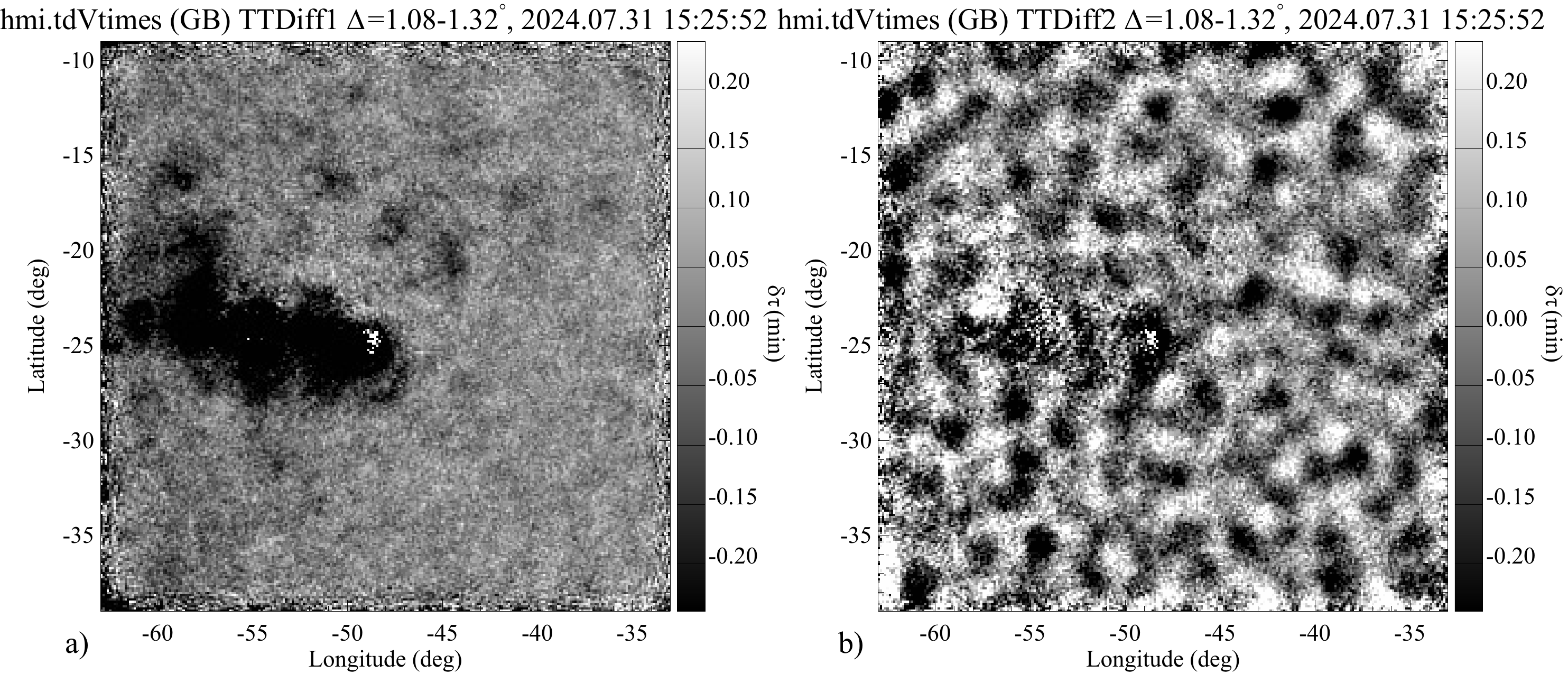}
		\caption{{Illustration of acoustic travel data obtained in the \texttt{hmi.tdVt\textit{}imes\_synopHC} dataset by the `GB' method \citep{Gizon2002} for the range of travel time distances of $1.08-1.32$ heliographic degrees: \textit{a)} variations of the mean travel times; \textit{b)} variations of the travel difference sensitive to the vertical flow component and flow divergence.}}\label{hmi.tdVtimes_synopHC_ttdiff1-2}
	\end{center}
\end{figure}

{A combination of the cross-covariance functions calculated for different distances and lag times forms a time-distance diagram (Figure~\ref{HMI_td_fit}\textit{a}). It shows ridges corresponding to wave packets arriving at a particular distance after multiple reflections. The lowest ridge represents direct waves (called the first skip). Figure~\ref{HMI_td_fit}\textit{b} illustrates a cross-covariance function of the first skip (black) for a distance of 109 Mm and a Gabor-wavelet fitting function (green) that measures a phase travel time of 52.28 min.}

Small perturbations in the acoustic wave travel times reflect changes in the physical conditions of the solar interior along the wave paths. For instance, a plasma flow along the wave's propagation direction will shorten the travel time, while a flow in the opposite direction will lengthen it. Thus, the travel-time shifts between waves traveling in opposite directions along the same path serve as indicators of interior flows and can be measured from observation. Thus, the mean of these travel time shits is related to the sound-speed variations, and the difference between the shifts is caused by subsurface flows.

A time-distance helioseismic analysis pipeline was developed using the HMI Doppler-velocity data \citep{Zhao2012,Couvidat2012}. The primary data product from this pipeline is subsurface flow fields of 25 patches of $30\times 30$ heliographic degrees, covering the near-full-disk area of $120\degr\times120\degr$, with a spatial sampling rate of $0.12\degr$ pixel$^{-1}$ and a depth coverage from the surface up to around 20\,Mm. In this analysis, each patch is tracked and remapped using Postel's projection.

The pipeline employs two methods to fit for the travel times, Gabor wavelet fitting \citep{Kosovichev1997} and linear fitting \citep{Gizon2002}. {The HMI data product \texttt{hmi.tdVtimes\_synopHC} contains fitted travel times, with data segments `Gabor' from the Gabor wavelet fitting and `GB' from the travel times defined by \citet{Gizon2002}. {The travel times are calculated for 11 travel distance ranges \citep[Table 1 in ][]{Zhao2012}, which correspond to the range of depths of acoustic wave turning points from 0 to $\simeq 35$~Mm.  Figure~\ref{hmi.tdVtimes_synopHC_ttdiff1-2} illustrates (a) the mean travel times (defined as \texttt{TTDiff1} in data product \texttt{hmi.tdVtimes\_synopHC}), which are sensitive to the sound-speed variations, and (b) the travel time difference (\texttt{TTDiff2}), which is mostly sensitive to the vertical flow velocity and flow divergence. The travel times are calculated for a $30\times 30^\circ$ patch centered at $-24^\circ$ latitude and $-48^\circ$ longitude, and for an 8-hour time interval centered at 15:32:52 UT on 2024.07.31. This is the same date as in Figure~\ref{HMI_rings_inver_mag_f} illustrating the flow map from the ring-diagram pipeline. The interval of the travel distances, $\Delta = 1.08-1.32$ heliographic degrees, covers a depth range of up to 5.1~Mm. In panel (\textit{a}), the negative travel time variations of \texttt{TTDiff1} (dark areas) reveal the travel time reductions beneath active region AR~13772 (indicated in Figure~\ref{HMI_rings_inver_mag_f}). In panel (\textit{b}), the negative travel-time patches are caused by the diverging supergranulation flows.}     

\begin{figure}
	\begin{center}
		\includegraphics[width=0.8\linewidth]{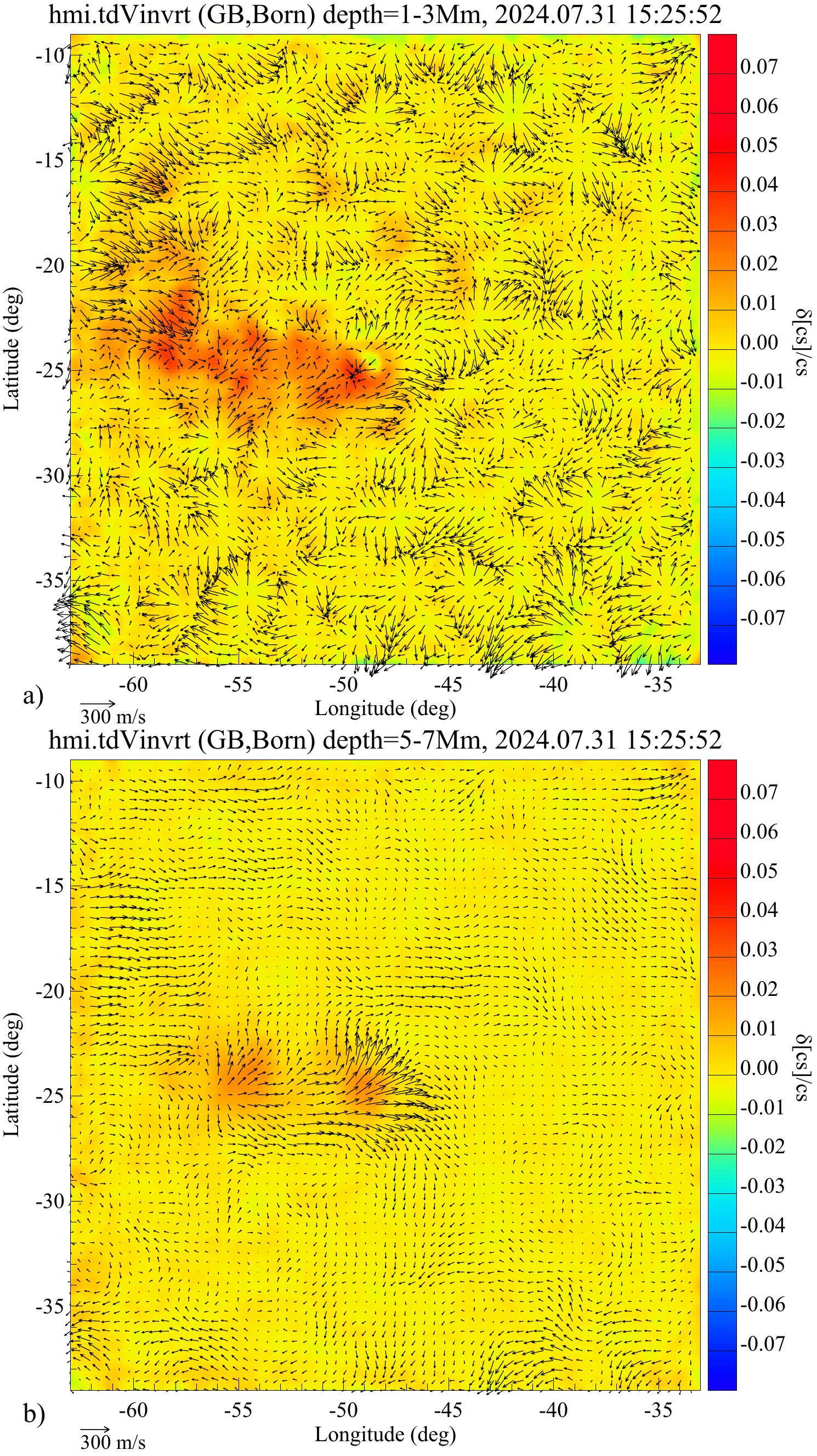}
		\caption{{The maps of subsurface relative sound-speed variations horizontal flow fields from the \texttt{hmi.tdVinvrt\_synopHC} dataset, obtained by inversion of the travel times shown in Figure~\ref{hmi.tdVtimes_synopHC_ttdiff1-2} using the Born-approximation sensitivity kernels in the depth ranges: \textit{a)} 1\,--\,3~Mm and \textit{b)} 5\,--\,7~Mm.}}\label{hmi.tdVinvrt_synopHC_GB_Born}
	\end{center}
\end{figure}

 With theoretically derived sensitivity kernels \citep[e.g.,][]{Birch_TravelTimeSensitivity_2000,Boening2016, Hartlep2021}, which relate the subsurface properties to the surface measurements using a solar model, one can set up linearized equations and solve them for the solar interior flows and sound-speed variations {using inversion techniques \citep[e.g.][]{Kosovichev1996,Jensen_MCDInversionSound_1998,Couvidat_TimeDistanceHelioseismology:_2005}}. 



Two sets of sensitivity kernels, ray-approximation kernels \citep{Kosovichev1997} and Born-approximation kernels \citep{Birch2004,Birch_Linearsensitivityhelioseismic_2007}, are used in the inversion codes. The data product \texttt{hmi.tdVinvrt\_synopHC}  contains inversion results, with `vx' for subsurface east-west direction velocity, `vy' for subsurface north-south direction velocity, `vz' for vertical velocity, and `cs' for subsurface wave-speed perturbations. Keywords `Gabor' and `GizonBirch' define which fitted travel times are used, and `Born' and `Raypath' define which sensitivity kernels are used (see \href{jsoc.stanford.edu/HMI/TimeDistance.html}{HMI Time-Distance Products} website\footnote{jsoc.stanford.edu/HMI/TimeDistance.html}). Figure~\ref{hmi.tdVinvrt_synopHC_GB_Born} illustrates the subsurface sound-speed variations and the horizontal flows at two depth ranges, 1\,--\,3~Mm and 5\,--\,7~Mm. In the shallower layer, the flow pattern is dominated by supergranulation, while in the deeper layer, the supergranulation flows diminish, and the flow map reveals East-West flows associated with the active region. Both cases show enhancements in the sound speed beneath the active region.}

The flow fields become available every 8 hours, several days after the observing date. From these subsurface flow maps, one can compute the divergence and horizontal vorticity at various depths, which then can be used to study, e.g., properties of supergranules \citep{DeGrave2015}. These maps can be used to study the long-term evolution of near-surface zonal and meridional flows \citep{Zhao2014, Getling2021, Mahajan2023}.

Time-distance helioseismology is used not only to infer local subsurface flow fields but also to study global flows. Inference of the Sun's deeper meridional circulation has been studied by various authors in the past decade using HMI observations (see Section~\ref{sec_MC}). Time-distance helioseismology has also been developed to map active regions on the far side of the Sun (see Section~\ref{sec_FS}). {The   \href{jsoc.stanford.edu/data/timed/}{Time-Distance Helioseismology}\footnote{jsoc.stanford.edu/data/timed/} website provides the far-side images and also full-disk merged subsurface flow maps obtained using the Gabor wavelet travel-time fitting and the Raypath sensitivity kernels.}

\subsection{Mode Coupling}
Normal-mode coupling, a seismic technique with a long history in geophysics \citep[see, e.g.][]{DT98}, has seen limited use in helioseismology. This method is similar to global helioseismology, but instead of interpreting eigenfrequencies, it is focused on analyzing the mode eigenfunctions. Non-spherically symmetric flows and structure variations in the Sun perturb the eigenfunctions calculated for a reference solar model. These perturbations can be considered in terms of coupling of the theoretical eigenfunctions. Therefore, this technique is called the Mode Coupling. Unlike eigenfrequencies, which are scalar quantities, eigenfunctions are vectors and require a different measurement technique to analyze observed data. 

The key measurement involves analyzing cross-correlated Fourier components of the Sun's wavefield across different frequencies and spatial scales. This approach is somewhat analogous to estimating the power spectrum of solar acoustic waves. However, instead of focusing on the power spectrum of individual modes, the method considers the cross-spectra of distinct acoustic modes characterized by the radial order $n$, angular degree $\ell$ and order $m$ and frequency $\omega_{n\ell m}$ with the modes with different $(n', \ell', m')$ and frequency $\omega_{n'\ell'm'}$. In the absence of non-spherically symmetric perturbations, the stochastic excitation of acoustic waves would yield zero correlation. However, the presence of flows and perturbations in the Sun with specific spatial and temporal scales result in enhanced power in the cross-correlation measurements interpreted as the mode coupling. The analysis of the enhanced power at certain scales reveals perturbations in the Sun's structure.  
\citet{woodard_89} was the first to describe how the Sun's differential rotation across latitudes distorts its oscillation modes. Since then, this method has been applied to various topics such as meridional circulation and convection \citep[e.g.][]{lavely92,roth_stix_2008,schad_2013,woodard16,hanasoge17,hanasoge18}. 
 The method of mode-coupling has been successfully used to detect various types of inertial waves in the Sun, including Equatorial Rossby waves \citep[][]{Mandal2020,Mandal2021}, high-latitude inertial waves \citep[][]{Mandal2024}, and high-frequency retrograde modes \citep[][]{Hanson2022}. 
 
\section{Solar Structure}
\label{sec:structure}

Helioseismology has provided detailed inferences of the solar internal structure by applying sophisticated techniques for the so-called inversion of the data \citep{Kosovichev2011a,2016LRSP...13....2B}.
This has led to strong constraints on our understanding of solar evolution \citep{2021LRSP...18....2C}.

\begin{figure}
	\centering
	\includegraphics[width=0.95\textwidth]{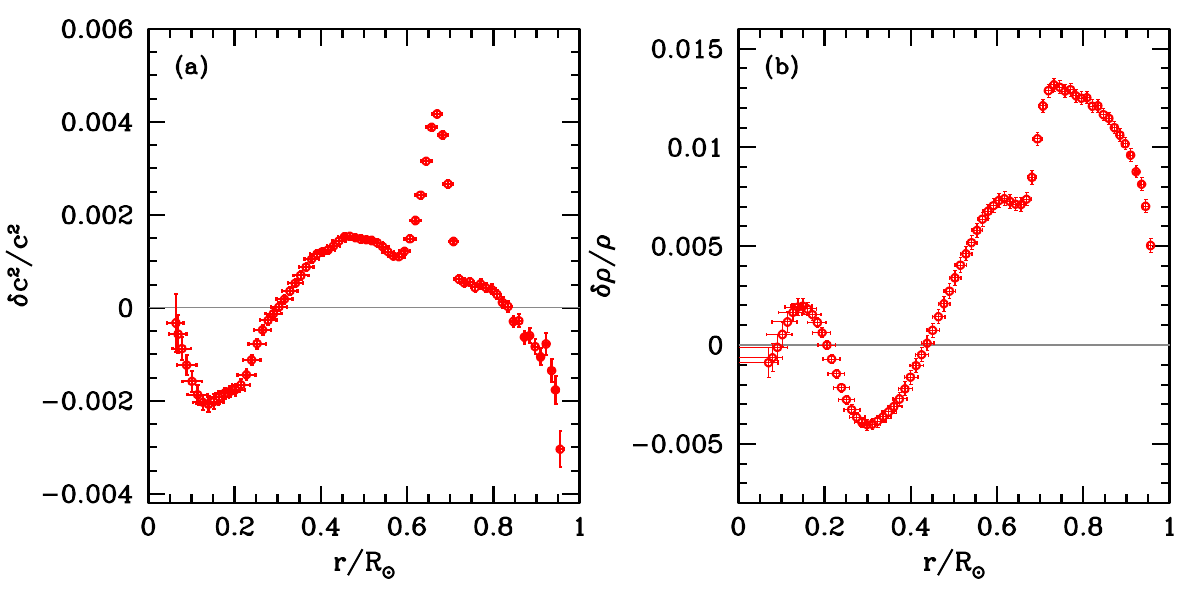}
	\caption{Relative differences between the squared sound speed (\textit{panel a}) and density (\textit{panel b}) between the Sun and Model S of \citet{1996Sci...272.1286C}. The inversion was performed with a mode-set obtained with a 2304-day time series with a start date of 2020.07.11 analyzed as described by \citet{Larson2018}. The differences are in the sense (Sun~$-$~Model)/Sun.
	} 
	\label{fig:csq}
\end{figure}

Early helioseismology data allowed us to determine the structure of the Sun by inverting the available frequencies \citep[e.g.][]{1996Sci...272.1296G}. The inversions showed that the structure of standard solar models is in very good agreement with the structure of the Sun; the relative sound-speed differences are less than 0.5\%, while relative density differences are a few percent. These results have not been changed by the HMI data, as can be seen in Figure~\ref{fig:csq}.

Early inversions, including the one shown in Figure~\ref{fig:csq} were usually done with \textit{p}-modes that ranged in spherical harmonic degree from $\ell=0$ to $\ell=150$ or 250, and hence, the results are not reliable for radii larger than about 0.98\,R$_\odot$. The high spatial resolution data of HMI have allowed the frequencies of higher $\ell$ modes to be determined, and these inversions show that closer to the surface, the sound speed of our models is significantly higher than that of the Sun, with sound-speed differences exceeding 1\% \citep{2015ApJ...803...92R, 2020Reiter_Modes}.

\subsection{The Issue of Solar Abundances}
\label{subsec:abun}

Perhaps the most pressing uncertainty when it comes to solar structure is that the solar heavy element abundance $Z/X$ is in dispute. Model~S, a standard solar model that showed that the structure of the Sun quite well \citep{1996Sci...272.1296G}, was constructed with $Z/X=0.0245$ \citep[henceforth GN93: ][]{1993oee..conf...15G}. Within a few years, updated analyses decreased the estimated solar metallicity to $Z/X=0.023$ \citep[henceforth GS98: ][]{1998SSRv...85..161G}. This decrease in metallicity worsened the match between solar models and the Sun --- the decrease in metallicity decreased opacities, which in turn made the convection zone shallower. This, however, is not the end of the story. Using 3D model atmospheres and non-LTE calculations for abundances, \citet{2005ASPC..336...25A} claimed that the solar metallicity is even lower, $Z/X=0.0165$. 
This group recently revised the abundances upward to $Z/X=0.0187$ \citep{2021A&A...653A.141A}.  Earlier, an independent group of \citet{2010A&A...514A..92C}, using a different 3D model atmosphere and non-LTE effects, obtained a higher metallicity, $Z/X=0.0211$, similar to that of GS98. {More recently, \citet{2022A&A...661A.140M} derived an even higher metallicity of $Z/X=0.0225$. The potential systematic uncertainties in these measurements and the directions for further studies were discussed by \citet{2024arXiv240100697L}.}

The issue with the uncertainty in the solar chemical composition is that it makes the structure of standard solar models uncertain. Low metallicities imply lower opacities, which in turn makes the convection zone more shallow; this introduces large differences between the sound speed and density profiles of low-metallicity models and the Sun. Among other issues is the fact that the convection-zone helium abundance of low-metallicity models is lower than the helioseismically determined one. The small frequency separations of low-metallicity models do not match observations either \citep[][and references therein]{2008PhR...457..217B}. However, all these issues are ultimately connected to the lowered opacity. If opacities are higher than the calculated opacities, and experimental results suggest that iron opacities are higher than calculated \citep{2015Natur.517...56B}, then these discrepancies can go away. However, the experimental results are not easy to interpret, and additional experiments to determine chromium and nickel opacities  \citep{2019PhRvL.122w5001N} have yielded results that were not expected. Thus, the experimental issue of opacities remains unresolved.
On the other hand, we note that opacity investigations can be guided by inferences of opacity from helioseismic inversion \citep[e.g.][]{Buldgen_Nature_25}.

The flux of CNO neutrinos gives an independent measure of abundances, and results from the BOREXINO collaboration \citep{2020Natur.587..577B} reject the low-metallicity models at about a $2.5\sigma$ level. The neutrino results, however, are sensitive to the metallicity of the solar core, and consequently, there have been suggestions that the Sun may have a low-metallicity convection zone but a higher-metallicity core as a result of planet formation \citep{2022A&A...667L...2K}. While most helioseismic results appear to support the high-metallicity solution, there are some inversion results, \citep[e.g.][]{2017MNRAS.472..751B, 2024A&A...681A..57B} that find a low-metallicity solution in the convection zone. Note that these inversions explicitly use the equation of state, which may be uncertain, and also lack precision. However, it is clear that even high metallicity models do not match the Sun perfectly, and much more work needs to be done to understand the physical process inside the Sun \citep{2024A&A...686A.108B}.

\subsection{Seismic Solar Radius}

The MDI and HMI data provided the capability to observe with unprecedented accuracy the surface gravity oscillation (\textit{f}) modes, the frequency of which is not sensitive to sound-speed variations that can be caused by the surface magnetic activity. These data have been used to infer the seismic radius of Sun \citep{Schou1997} and its variations with solar cycle \citep{2000SoPh..192..459A,2001ApJ...553..897D,Kosovichev2018}. 

The \textit{f}-mode frequencies are sensitive to the sharp density gradient in the near-surface layers, occupying about the top 5\% of the solar radius. The low-frequency medium degree \textit{f}-modes in the range of $\ell$ = 137,\--\,300 observed by the HMI concentrate the kinetic energy within a layer of approximately 15 Mm, or about 2\% of the whole solar radius. That means that these \textit{f}-mode radii are well below the subphotospheric superadiabatic boundary layer, which is located about 0.08 Mm below the photospheric surface. The comparison of the observed \textit{f}-mode frequencies with those calculated for standard solar models provided an accurate estimate of the solar radius used as a reference in astronomy. So far, the resulting scaled \textit{f}-mode radius values are lower than the photospheric radius deduced from optical observations by about 310 km \citep{Schou1997} and 203 km \citep{Antia1998}. This discrepancy was explained as a difference of 333 $\pm$ 8 km between the height at the disk center where the optical depth at $\tau_{5000}$ = 1, and the inflection point of the intensity profile on the limb \citep{Haberreiter2008}. They concluded that the standard radius must be lowered by this quantity to be 695.66 Mm, close to the value adopted by IAU in 2015. 

\citet{Takata2024} have proposed inverting the \textit{p}-modes, which react differently from the \textit{f}-modes. They assert that the method is more robust albeit more-or-less, consistent with what is suggested by the \textit{f}-mode analyses, estimating the solar photospheric radius to be 695.78 $\pm$ 0.16 Mm. This accounts for a difference of 210 km compared with the reference model of $R_{\odot}$ = 695.99 Mm used by the authors. 

The analysis of the low-frequency medium-degree \textit{f}-modes in the range of $\ell$ = 137\,--\,299, the data covering nearly two solar cycles, from April 30, 1996, to June 4, 2017, showed that the mean seismic radius was reduced by 1\,--\,2 km during the solar maxima and that most
significant variations of the solar radius occur beneath the
visible surface of the Sun at a depth of about $5 \pm 2$~Mm, where the radius is reduced by 5\,--\,8 km. These variations can be interpreted as changes in the solar subsurface structure caused by the predominately radial $\sim 10$~kG magnetic field \citep{Kosovichev2018}.

\subsection{Solar Asphericity}

\subsubsection{Asphericity of the solar surface}

Most inversions have been done assuming that the structure of the Sun follows spherical symmetry. There have been few studies of the asphericity in structure, mainly because the signal of asphericity, which lies in the even-order splitting coefficients,   is very small. Most previous results \citep[e.g.,][]{Kuhn1988,2001MNRAS.327.1029A, 2003A&A...399..329A} seem to imply that there are small deviations from sphericity close to the surface. It is possible to extract information concerning the coefficients of rotational frequency splitting, $a_n$, {discussed in Section~\ref{sec:global}}. The odd $a_n$ coefficients ($n$ = 1, 3 ...), $a_1$, $a_3$, $a_5$, ..., 
measure the latitudinal differential rotation, whilst the even coefficients $a_2$, $a_4$, $a_6$, ..., 
 are a sensitive probe of the symmetrical (about the equator) part of the distortion described by Legendre polynomials $P_{n}(\cos\theta)$. 

According to \citet{vonZeipel1924} theorem, solar limb contours of temperature, density, or pressure should be nearly coincident near the photosphere. Rotation, magnetic fields, and turbulent pressure are the largest local acceleration sources that violate the von Zeipel theorem \citep{Dicke1970}. Since (geometrical) asphericities are relatively small in the Sun, we may describe the distance from
the center (for instance, for a level of isodensity $\rho$, but the same would happen to a level of isotemperature or isogravity) by:

\begin{equation}
	R(\cos\theta)|_{\rho={\rm constant}} = R_{sp} \left[ 1 + \sum_{n} c_n(R_{sp}) P_n (\cos\theta) \right],
	\label{eq.rdep}
\end{equation}
where $R_{sp}$ is the mean limb contour radius, $\theta$ the angle to the symmetry axis (colatitude) and $P_n$ is the Legendre polynomial of degree $n$. The asphericity shape coefficients, $c_n$, are called quadrupole for $n$ = 2 ($c_2$) and hexadecapole for $n = 4$ ($c_4 $). Terms of higher orders are conventionally named by adding `pole' to the degree numeral prefix.
{It is straightforward to determine the oblateness parameter (the relative difference between the equatorial and polar radii), $f=(R(0)-R(1))/R_{sp}$, from Equation~\ref{eq.rdep} in terms of the asphericity coefficients, $c_2$, $c_4$, and $c_6$, using the standard definition of Legendre polynomials,}
as $f= -(3/2)c_2 - (5/8)c_4 - ({21}/{16})c_6 $; higher-order splitting coefficients are too uncertain to be useful.

The measured splitting coefficients $a_n$ are related to the shape coefficients $c_n$ through a normalization factor $K_n$. An efficient method for calculating this factor was developed by \citet{Kuhn1988}, who showed that it was possible to invert the splitting data to obtain the structural asphericity using the relation:
$a_n   = K_n c_n R_n (\ell)$.
Assuming $R_n$ = $R_{sp}$, as this analysis is conducted only very close to the surface (i.e., the seismic radius at the surface), the corresponding scaling factors are:
$$a_2 = -6 \times 10^{-4} c_2 R_{sp};~~~ a_4 = 1 \times 10^{-4} c_4 R_{sp};~~~ a_6 = -14 \times 10^{-4}  c_6 R_{sp}, ~~~ $$
where the $a_n$ are measured in Hz.

It was shown that the asphericity of the Sun changes from the solar minimum to the maximum. During the solar minimum (from 1996 to 1998) the asphericity was dominated by the $P_2$ and $P_4$ terms, the $P_6$ contribution being negligible. At the time of high activity, all these components have a significant effect. The asphericity of the Sun is strongly affected during the solar cycle progression \citep{Rozelot2020}.

\begin{figure}
	\centering
	\includegraphics[width=2.5 true in]{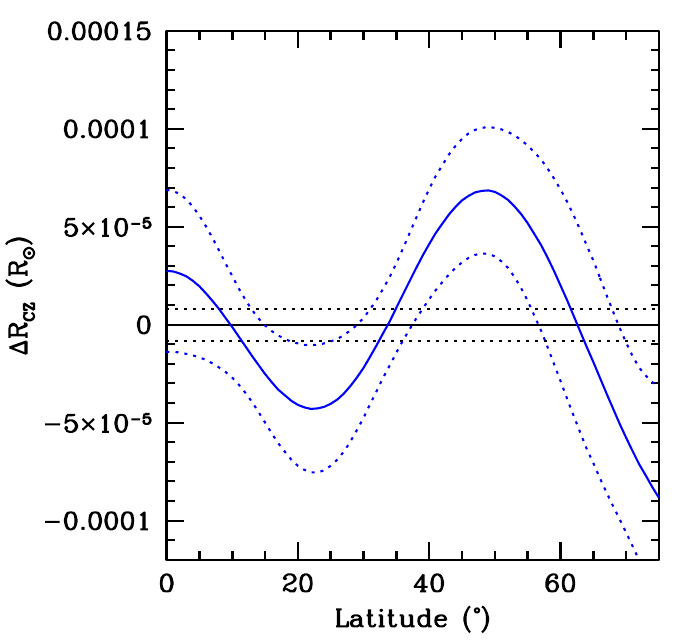}
	\caption{ The deviation of the base of the convection zone from spherical symmetry as a function of latitude is shown as the \textit{blue line}. The \textit{blue dotted lines} show the $1\sigma$ uncertainty. The \textit{black line} shows what the deviation would be in the spherically symmetric case; the \textit{dotted black lines} show the $1\sigma$ uncertainty in the estimate of the spherically symmetric position of the convection-zone base \citep{2024ApJ...964....8B}.
	} 
	\label{fig:asph}
\end{figure}

\subsubsection{Asphericity of the convection-zone base}

Many early attempts to determine the asphericity in the solar structure had been focused on determining whether or not the base of the convection zone has any asphericity. The base of the convection zone (CZ) is where the tachocline is located. The tachocline is the thin shear layer that separates the differentially rotating convection zone from the uniformly rotating interior (see Section~\ref{sec:diffrot}). The tachocline is known to be prolate in shape \citep{1995ESASP.376b..47G,1998MNRAS.298..543A, 1999ApJ...527..445C}, raising the question of whether the base of the convection zone is spherically symmetric.  It was argued that the tachocline could drive meridional flows \citep{1992A&A...265..115Z} and that these flows must affect the thermal balance of the layers and could influence where the layer becomes unstable to convection, thereby causing the base of the CZ to deviate from spherical symmetry \citep{1995ESASP.376b..47G}. 
These authors claimed a difference of no more than $0.02 R_\odot$ between the convection zone base at the pole and the equator, while others derived an upper limit of $0.0005 R_\odot$ on the asphericity of the CZ base \citep{2001MNRAS.324..498B}. 
A more recent attempt at determining the asphericity that used HMI data \citep{2024ApJ...964....8B} gives an upper limit of $\lwig 0.0001$ R$_\odot$ departure from the spherically symmetric position of the base of the convection zone. Analysis of the $a_2$ -- $a_8$ splitting coefficients indicates that deviation from sphericity is more complicated than a simple prolate or oblate spheroid, as can be seen from Figure~\ref{fig:asph}. If one assumes that the cause of the asphericity is the effects of magnetic fields on solar structure then the shift in the base of the convection zone implies a strong hypothetical magnetic field of about $\sim 300$ kG. This is consistent with earlier upper limits on the magnetic field at the base of the convection zone \citep[e.g.][]{1993A&A...272..621D, 1997MNRAS.288..572B, 2003A&A...399..329A}.

\subsection{The Sun as a Physics Laboratory}

The high precision of the helioseismic inferences makes them sensitive to even quite subtle effects in the physics of the solar interior, allowing sensitive tests of the physics used in solar-model computations.

The first full inference of the solar internal sound speed \citep{1985Natur.315..378C} indicated a need to increase the opacity in the radiative interior compared with the then current opacity tables, as later confirmed by new opacity calculations \citep[e.g.][]{1992ApJS...79..507R}.
Helioseismic inferences of opacity were further explored by \citet{2002css1.book.1035G} and \citet{Buldgen_Nature_25}, in the latter case also related to the issues caused by the revisions in the solar abundances (see also Section~\ref{subsec:abun}).

The solar sound speed depends sensitively on the thermodynamic properties of the solar plasma, including the adiabatic compressibility $\Gamma_1 = (\partial \ln p/\partial \ln \rho)_{\rm ad}$.
From the helioseismic inference of $\Gamma_1$, \citet{1998ApJ...500L.199E} inferred that the then-current advanced tables of thermodynamic properties had ignored significant relativistic effects on the electrons at the conditions of the solar core. 
This is corrected in the tables now used; however, the example illustrates the sensitivity of helioseismology to detailed aspects of solar interior physics.
\citet{1999ApJ...518..985B} made a comparison of the inferred $\Gamma_1$ in the solar convection zone with two independent sets of thermodynamic tables, finding significant but different discrepancies between the Sun and the models in both cases. 
Such investigations have the potential to further improve our understanding of the thermodynamic properties of plasmas under extreme conditions; they would be greatly assisted by the availability of reliable helioseismic data for modes of higher degree than currently available \citep{2002A&A...384..666D}.
A closely related issue is the helioseismic determination of the solar envelope helium abundance \citep{1993MNRAS.265.1053K,1995MNRAS.276.1402B}.

\citet{2022MNRAS.517.5281B} considered the constraints on nuclear-reaction parameters provided by solar observations. They found that helioseismic data provided a strong constraint, greatly improving the theoretical or experimental uncertainty on the basic $pp$ and the ${}^3{\rm He} + {}^3{\rm He}$ reaction rates, while the neutrino observations, in particular, constrained the ${}^7{\rm Be} + p$ reaction rate.

Moving beyond standard solar models, helioseismology has also been used to constrain more exotic physics. An important issue is the nature of the dark matter that seems to dominate the dynamics of galaxies. One candidate is the so-called weakly interacting massive particles (WIMPS). 
If present in the solar interior, even in small numbers, their expected long mean free paths could provide an efficient source of energy transport, hence cooling the solar core. 
In the early days of helioseismology and solar neutrino studies, this was seen as a way to account for the apparently low flux of solar neutrinos \citep[e.g.][]{1985ApJ...299..994F, 1985ApJ...294..663S}.
Although the apparent solar neutrino deficit has now been identified as caused by neutrino flavor transitions, the Sun still provides constraints on the properties of potential dark-matter particles \citep[e.g.][]{2014ApJ...780L..15L, 2015JCAP...10..015V}.

Structure inversion results are also being used for more speculative studies. For instance,  \citet{2023ApJ...959..113B} examined what would happen if there were a black hole in the core of the Sun. They find that models of the Sun born around a black hole whose mass has since grown to approximately $10^{-6}M_\odot$ are compatible with current helioseismic results. In other investigations, \citet{Hamerly_DarkMatterand_2012} and Bellinger et al. (submitted) examined what would happen if the Sun had a dark matter core. Interestingly, they find that the sound-speed profile of a solar model
with a $10^{-3}M_\odot$ dark core appears to have a better agreement with the Sun than standard models. However, this improved agreement is by no means evidence of dark matter since a metal-rich core would give a similar result; more precise measurements of frequencies of low-degree \textit{p}-modes and \textit{g}-mode observations may constrain a dark-matter core inside the Sun.

It should be kept in mind that the use of helioseismology to constrain the physics of the solar interior depends critically on constraining other uncertainties in solar modeling than the one under study. Even so, it is clear that the Sun offers valuable information on matter under extreme conditions, supporting basic physics and its use in modeling other stars.

\newcommand{\CSH}[1]{\textcolor{blue}{#1}}

\section{Convection and Large-Scale Flows}
\label{sec:convection}

\subsection{Brief Introduction}

Thermal convection is the primary mechanism for energy transport in the Sun's outer envelope, which encompasses roughly one-third of its radius. This process is responsible for energy transport and drives the complex fluid motions and magnetic fields observed at the solar surface. Deciphering the intricate dynamics within the Sun is crucial for our understanding of many solar phenomena, including the solar cycle, the formation and evolution of active regions, and their influence on space weather.

Previous studies revealed three primary components of solar convection: granulation, supergranulation, and giant cells \citep[e.g.,][]{Nordlund2009}. While the structure and dynamics of granulation are studied in detail using high-resolution observations of the solar surface and numerical simulations, the origin of supergranulation is still a puzzle \citep{Rieutord_SunsSupergranulation_2010}, and the existence of the giant cells is a subject of debate.

The initial studies of supergranulation by time-distance helioseismology using the MDI data revealed that the horizontal outflows that are observed on the surface by tracking the motion of granules or observing the Doppler shift extend several Mm below the surface \citep{Duvall1997,Kosovichev1997}. Evidence of converging flows was at a depth of about 10 Mm, which gave an estimate of the average depth of supergranules of $\simeq 15$~Mm \citep{Zhao2003}. The typical horizontal velocity of the supergranular flows is about 300-500~m\,s$^{-1}$. However, the vertical velocity is substantially lower, about 10~m\,s$^{-1}$, which is rather difficult to measure. To overcome this difficulty, \citet{Duvall2010} suggested averaging the signals from a large set of supergranules and found 10~m\,s$^{-1}$ upflows at the center and downflows of about 5 m\,s$^{-1}$ at the boundaries of such `averaged supergranule.' 

In addition, studies of the supergranulation dynamics deduced from time-distance helioseismology of surface gravity waves (\textit{f}-mode) revealed vorticity caused by the Coriolis force and a wave-like motion of the supergranulation pattern. The latter result added a new element to the long-standing puzzle of the super-rotation of supergranulation \citep{Duvall1980,Meunier2007}. This phenomenon could be explained by the interaction of convection modes with the near-surface rotational shear layer \citep{Green2006}. However, the calculated phase speed was lower than the observed super-rotation; therefore, some other effects must be involved, such as subsurface magnetic field \citep{Green2007} or non-linear dynamics \citep{Rincon2018}. The photospheric convection spectrum constructed from the MDI observations showed distinct peaks representing granules and supergranules but no features at wavenumbers representative of mesogranules or giant cells \citep{Hathaway_etal_2000}.

The uninterrupted high-resolution full-disk observations provided by HMI have significantly enhanced our ability to investigate these internal processes, particularly in the areas of convection and large-scale flows. This section reviews key findings from HMI observations, discusses the methods used to measure solar flows, and examines the challenges in modeling and interpreting these phenomena.

\subsection{Structure and Dynamics of Supergranulation}

\begin{figure} \centering \includegraphics[width=0.95\textwidth]{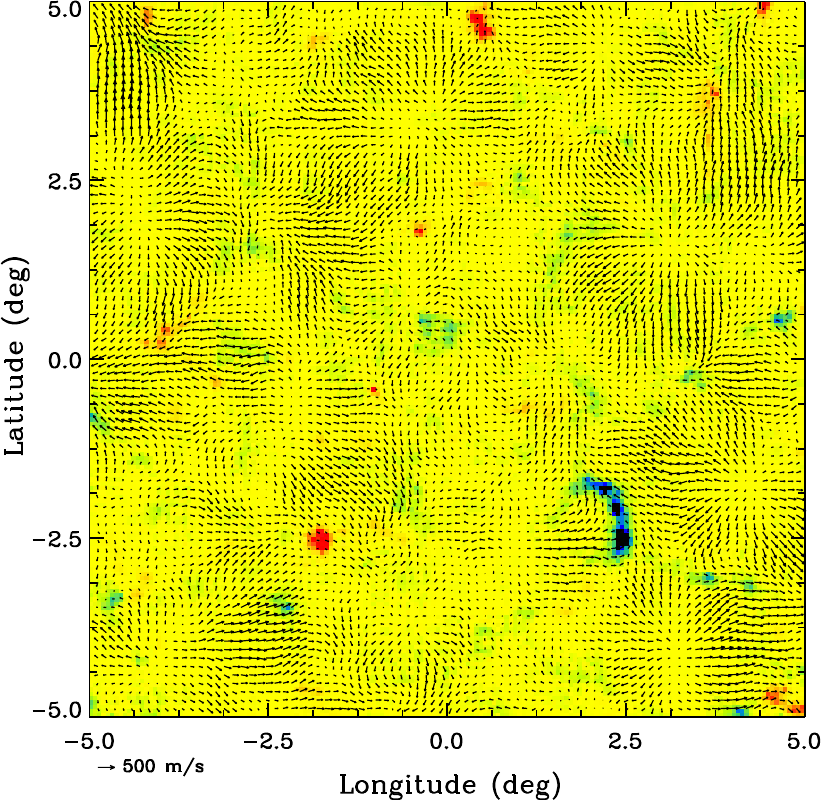} \caption{A sample of subsurface horizontal flow fields with full spatial resolution at a depth of 0\,--\,1 Mm.  The background image shows the line-of-sight magnetic field measured by HMI in the range from -80 to 80 G \citep{Zhao2012}.} \label{Zhao_supergranulation_map} \end{figure}

\begin{figure} \centering \includegraphics[width=0.9\textwidth]{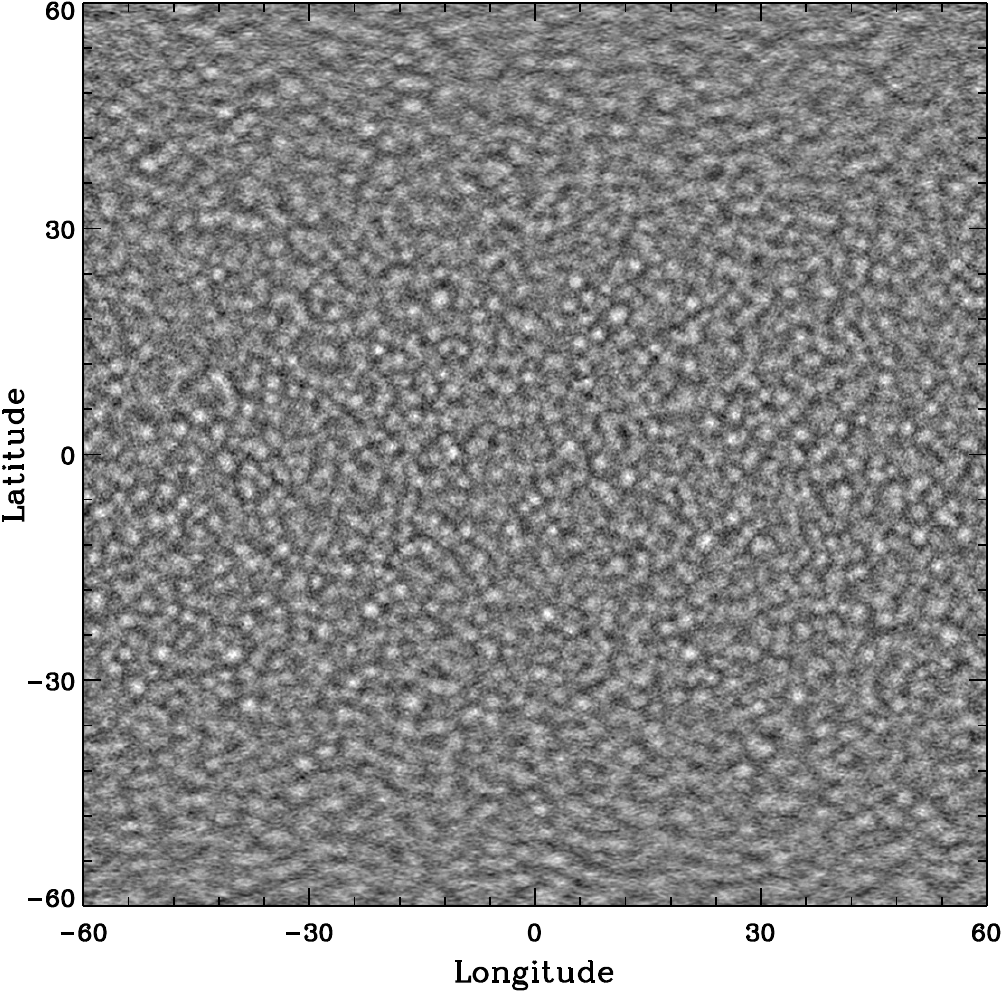} \caption{A map of horizontal flow divergence obtained by the time-distance helioseismology method for a depth range of 1\,--\,3 Mm and a time period of 00:00-07:59 UT 19 May 2010 \citep{Zhao2012}. The display scale is from $-6.2\times 10^{-4}$ to $6.2\times 10^{-4}$~s$^{-1}$. \textit{White areas} with positive divergence correspond to supergranulation. Note that supergranules appear larger at high latitudes because of the rectangular longitude–latitude map projection.} \label{Zhao_supergranulation_divergence} \end{figure}

HMI observations have refined our understanding of supergranular flow patterns, their depth structure, and their relationship to magnetic field distributions. The time-distance helioseismology pipeline developed for processing the HMI data provides the subsurface flow maps covering $120^\circ$ in latitude and longitudes in the range of depth from 0.5 to 19 Mm every 8 hours \citep{Zhao2012}. A sample of a small area of horizontal flow maps is illustrated in Figure~\ref{Zhao_supergranulation_map}, and the flow divergence over the whole 
$120^\circ \times 120^\circ$ area is shown in Figure~\ref{Zhao_supergranulation_divergence}. These maps reveal the interaction of supergranular convection with magnetic fields,  particularly in the areas of active regions, where the supergranular convection is suppressed, and the flow pattern is dominated by larger-scale converging flows \citep{Kosovichev2016}. However, these maps do not show the organization of supergranules in large-scale coherent structures.

\begin{figure} \centering \includegraphics[width=\textwidth] {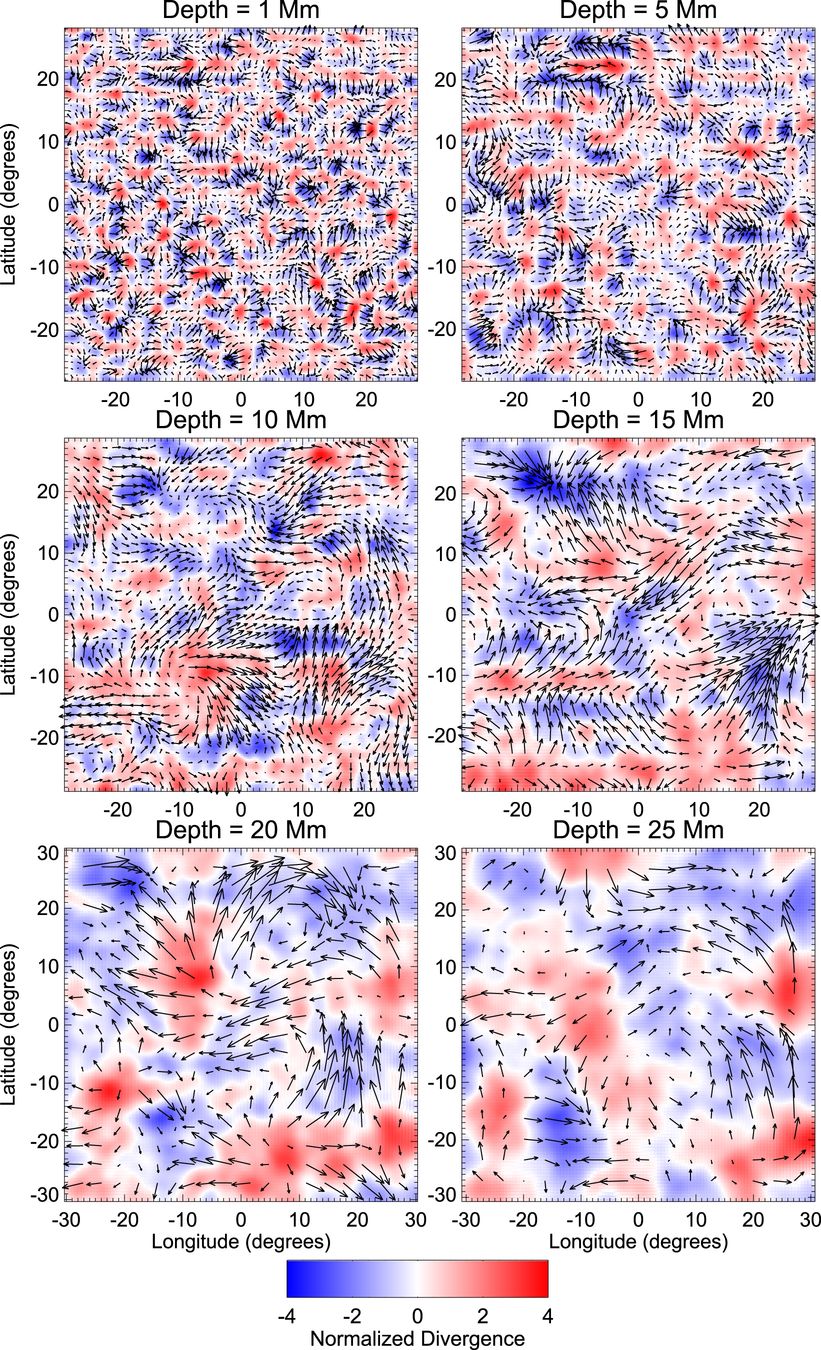} \caption{Horizontal flow field and horizontal divergence maps obtained by the ring-diagram method for a few sample depths \citep{Greer2016}. The vector field at each depth has been sub-sampled to the horizontal resolution of the averaging kernels. For the divergence map,\textit{ red} indicates positive divergence, \textit{blue} is negative, and \textit{white} is zero. The \textit{color map} has been scaled individually at each depth by the RMS average.} \label{fig:greer2016} \end{figure}

In addition, the horizontal flow maps are obtained by the ring-diagram technique. Figure~\ref{fig:greer2016} shows the horizontal flow field and corresponding horizontal divergence maps at various depths \citep{Greer2016}. This study also suggested that supergranulation does not form cell-like convective structures with a well-defined circular flow. Instead, the cell-like nature of supergranulation is only displayed in the horizontal dimensions at the photosphere.  It was suggested that since the lifetime of supergranules is only a day or two, the convective pattern in a deep layer may not resemble the supergranulation at the surface.

The ubiquity of supergranulation in the HMI images opened new opportunities for studying the average supergranule by stacking and averaging thousands of individual supergranules. It was found that the magnetic field that congregates in the downflow lanes was significantly stronger (0.3 G) on the west (prograde) side compared to the east and that the intensity minimum to the east of supergranules was deeper than to the west, suggesting that this anisotropy was related to the wave-like properties of supergranules \citep{Langfellner2018}.

\begin{figure} \centering \includegraphics[width=\textwidth] {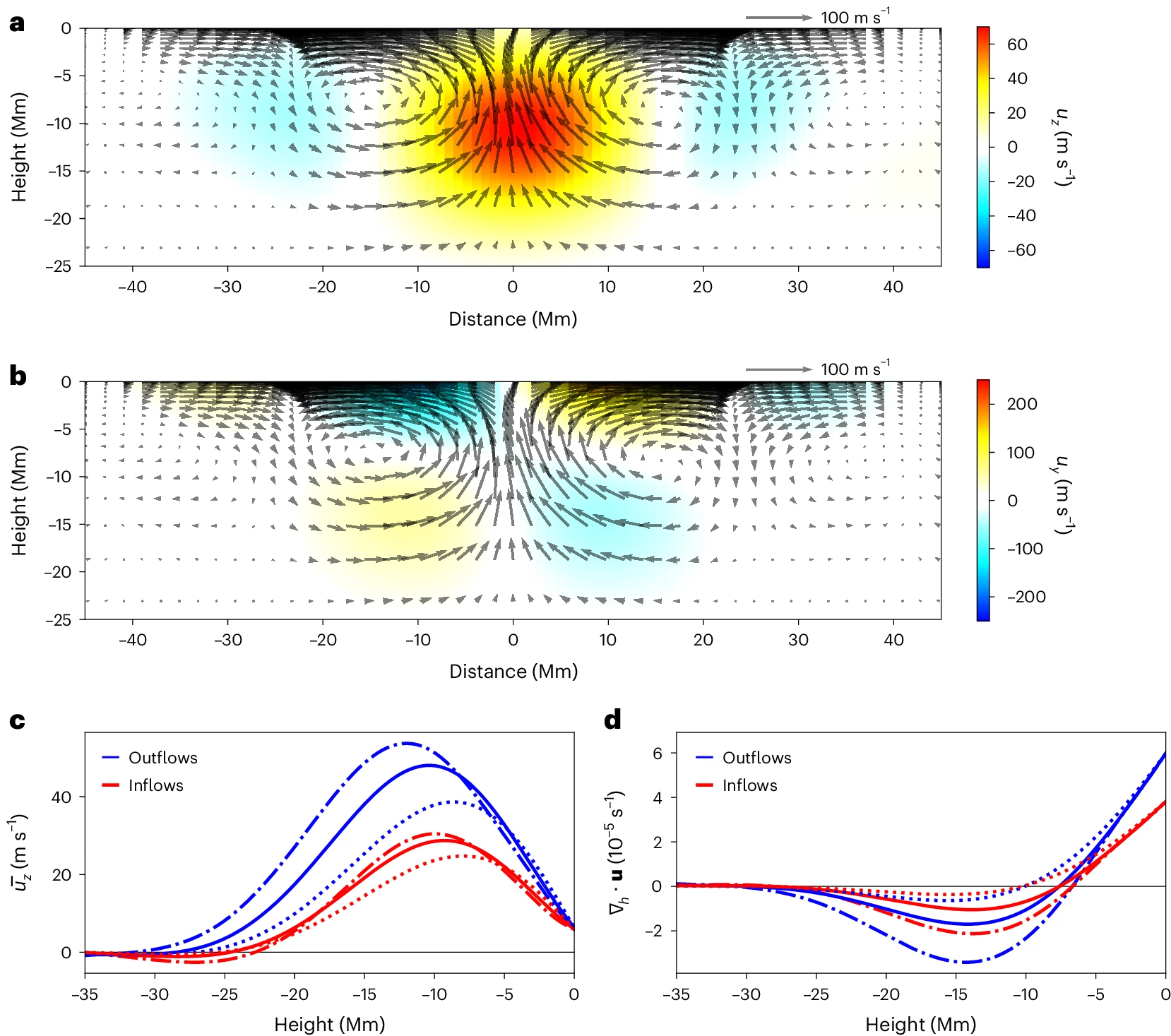} \caption{\textit{a)} Vertical flow component of the average supergranule in the plane of the central meridian. \textit{b)} Flow in the direction of solar north in the plane of the central meridian. Superimposed on both panels is the vector flow field in the plane parallel to the central meridian. The average supergranule's upward flow peaks at 10.5 Mm below the surface, the feature itself extending down to 25 Mm. The horizontal flow is directed away from the center at the surface, and the peak inward flow of $50\pm 25$ m\,s$^{-1}$ occurs 15 Mm below the surface. The horizontal-to-vertical aspect ratio of the overall cell is 3:2. \textit{c)} Mean vertical-flow component, computed over a circular area of radius 10 Mm, at the center of the supergranule outflow (\textit{blue}) and inflow (\textit{red}) for different regularization parameters. \textit{d)} Slices of the horizontal divergence at the center of the supergranular outflow and inflow  \citep{Hanson2024}.} \label{Hanson_supergranulation} \end{figure}

With the high signal-to-noise ratio that HMI provides, many helioseismic studies have attempted to probe the interior flows of supergranulation. Due to the weak vertical flows, many studies have relied upon mass-conservation arguments to infer the vertical flows that become more dominant at depth. In particular, analysis of time-distance measurements of the average supergranule suggested that supergranules were shallow, with peak upflows occurring a few Mm below the surface \citep{Duvall2013,Duvall2014}, and that supergranules extended at least to 7~Mm \citep{Korda2021}. 
A new recent analysis of 23,000 supergranules, including full 3D flow inversions without relying on the mass-conservation constraint, found a peak upflow occurring at 10~Mm below the surface \citep[Figure~\ref{Hanson_supergranulation}; ][]{Hanson2024a}. It was also noted that the peak upflow locations were invariant with the horizontal extent of supergranular-scale convection cells and that the downflow appeared to be 40\% weaker than the upflows. 

\subsection{Detection of Giant Cells}

The detection of convective giant cells with characteristic sizes of $\sim 10^2$ Mm is a much more challenging task. The velocities associated with the giant cells are small compared with granular and supergranular speeds and can hardly be isolated against the background of smaller-scale motions. The existence of such cells was predicted by \citet{Simon_Weiss_1968}. Even before that, \citet{Bumba_etal_1964} noted indications of the presence of giant structures in the distribution of weak magnetic fields. However, giant cells were considered hypothetical 
for more than three decades. The earliest observational detection of giant cells was reported by  \citet{Beck_1998}, who computed correlation functions of Doppler images from the MDI instrument over many months. They reported the existence of large-scale east-west flows that persisted for months, which is in line with theoretical descriptions of giant-cell convection. However, the HMI ring diagram data showed that these signals might be dominated by active region inflows \citep{Hanson2020}. 

Later, the presence of giant flow cells was revealed by tracking the motion of supergranules over many years \citep{Hathaway_etal_2013,Hathaway2021}. It was found that the low-latitude cells have roughly circular shapes and lifetimes of about one month; they rotate nearly rigidly, do not drift in latitude, and do not exhibit any correlation between the longitudinal and latitudinal flow. The high-latitude cells have long extensions that spiral inward toward the poles and can wrap nearly completely around the Sun. They have lifetimes of several months, rotate differentially with latitude, drift poleward at speeds approaching $2\,$m\,s$^{-1}$, and have a strong correlation between prograde and equatorward flows. Spherical harmonic spectral analyses of maps of the divergence and curl of the flows confirm that the flows are dominated by the curl component with RMS velocities of about 12\,m\,s$^{-1}$ at wavenumber $\ell = 10$. It was noted that these large-scale flow patterns can be linked to the inertial convective modes. The structure of these flows with depth is currently unknown. However, similar large-scale spiral flow patterns were detected in 1~Mm deep subsurface layers by time-distance helioseismology \citep{Zhao2016a}.

\subsection{Spectrum of Solar Convection}

Given the diverse spatial scales observed in solar surface flows, from granulation to giant cells, power spectra analysis has emerged as one of the most crucial tools for characterizing the properties of both surface and subsurface dynamics in the solar photosphere. This technique allows researchers to quantify the energy distribution across different scales, providing insights into the complex, multi-scale nature of solar convection \citep{Rincon2018}.

\begin{figure} \centering \includegraphics[width=0.7\textwidth] {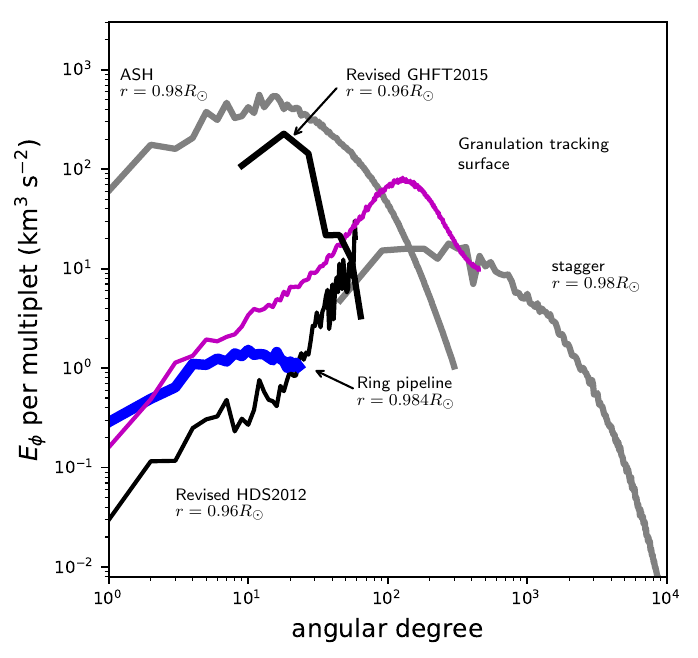} \caption{The estimates of the spectral power density for the azimuthal convective velocity, $E_\phi$, in terms of the spherical harmonic degree. The \textit{black curves} show the revised estimates from \citet{Hanasoge_etal_2012} (Revised HDS2012) and \citet{Greer2015} (Revised GFFT2015). The figure also shows the granulation-tracking measurements (\textit{magenta}) and measurements from the HMI ring-diagram pipeline (\textit{blue}). The curves corresponding to the ASH and Stagger numerical simulations of solar convection are shown in \textit{grey} \citep{Proxauf2021}.} \label{Proxauf_spectrum} \end{figure}

{Figure \ref{Proxauf_spectrum} shows constraints obtained through a variety of techniques, reflecting the state of the art in both observations and numerical simulations, with the following caveats. The flow spectra in Figure~\ref{Proxauf_spectrum} are at different depths and, depending on the type of measurement, may represent averages of flows over a depth range. For instance, supergranulation tracking uses the proper motions of supergranules to measure larger-scale flows; however, supergranules have finite extents in depth, implying that the inferred flow speeds are averaged over the supergranular depth range. Similarly, acoustic waves have radially delocalized eigenfunctions, which implies that the inferred flow speed is an integral of true convective flows with the wave eigenfunctions.}

{\citet{Hanasoge_etal_2012} employed time-distance helioseismology to investigate convective velocity magnitudes in the near-surface shear layer at $r = 0.96R_{\odot}$, examining them as a function of spherical-harmonic degree $\ell$ by representing the mean squared East-West convective velocity in terms of the parameter $E_\phi(\ell)$, defined as $\sum_\ell E_\phi(\ell)=r/2 \langle f^2\rangle$, where $r$ is the radius at which the flow field $f = v_\phi$ is measured. The curve labeled ‘Revised HDS2012’ in Figure \ref{Proxauf_spectrum} revealed smaller horizontal velocities at $r = 0.96R_\odot$ compared to the estimates from the numerical models of solar convection obtained from ASH simulations at $r = 0.98R_\odot$ \citep{Miesch2012}. These two results do not agree at any angular degree, though there is some overlap between the ASH simulation's convective power spectrum and that of the Stagger simulations \citep{Stein2006}, also obtained at $r = 0.98R_\odot$, for intermediate $\ell$. The magnitude of the convective flows as derived from supergranulation tracking on the surface \citep{Roudier2012} also overlap the ASH and Stagger spectra at certain points in the same range of angular degree and with the spectrum derived from the HMI ring diagram pipeline \citep{Bogart2015} at low $\ell$. At the lower end of the intermediate $\ell$ range---around $\ell\approx65$---there does appear to be some agreement between the `Revised HDS2012' and Stagger spectra, in addition to the spectra derived from the novel ring diagram pipeline developed by \citet{Greer2015}. Despite some small overlap at specific angular degrees, the reason for the disagreements between each of these curves is not yet established and is currently a topic of debate and investigation.}

Systematic comparisons between different observational techniques and simulations have highlighted the excess power in large convection scales. For instance, ring-diagram analysis has shown that the peak of the horizontal-velocity spectrum shifts to lower spherical-harmonic degrees with increasing depth, suggesting the presence of larger-scale velocity structures at greater depths \citep{Greer2015}. 

An upper limit on the convective velocity at each spherical harmonic degree of $\ell <60$ at $r = 0.96R_\odot$ that follows from the energy spectra obtained by the time-distance technique and ring pipeline (Revised HDS2012 and Ring pipeline curves in Figure~\ref{Proxauf_spectrum}) is less than 3~m/s for convective motions. When summed over all spherical harmonic modes $\ell < 60$, this corresponds to an upper limit on velocity of less than 10~ m\,s$^{-1}$. This upper limit is more than an order of magnitude lower than the convective velocities predicted by mixing-length theory and found in global convection simulations. 

This led to \textit{the convective conundrum} statement: `The convective velocities required to transport the solar luminosity in global models of solar convection appear to be systematically larger than those needed to maintain the solar differential rotation and those inferred from solar observations \citep{OMara2016}. One suggested solution is the suppression of large-scale convective motions caused by solar rotation \citep{Vasil2021}. This theory predicted that the dominant convection scale throughout the convection zone is about 30~Mm, which is the scale of supergranulation (the corresponding $\ell$ value is 100\,--\,120). It also suggested that supergranulation might be a manifestation of deep convection.

In this respect, a study of the spatial spectra of convective motions and their variation in the solar activity cycle showed an intriguing transition of the convective scales from supergranular to giant scales in the near-surface layer \citep{Getling_Kosovichev_2022}.

\begin{figure} \centering \includegraphics[width=0.9\textwidth] {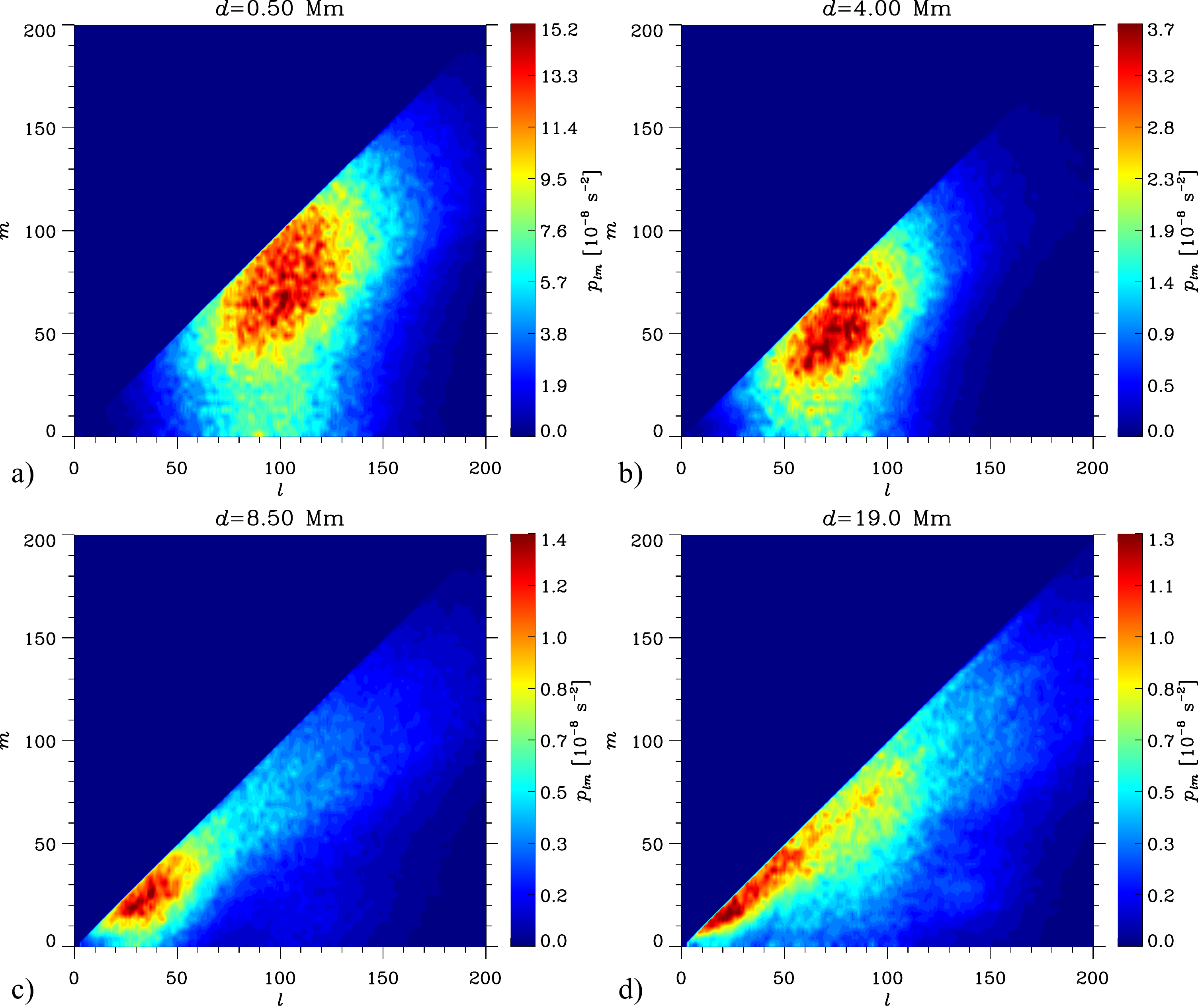} 
\caption{Depth variation of a sample power spectrum of the divergence of the velocity field smoothed with a 17.5\,Mm window, $p_{\ell m}$, with $\ell_{\max}=200$ obtained by 45-day averaging over the low-activity period from 21 December 2019 to 4 February 2020. The depth values are indicated at the top of each panel \citep{Getling_Kosovichev_2022}.} \label{spectra200} \end{figure}

Figure~\ref{spectra200} shows typical power spectra of the divergence field at various depths. In the upper layers, $d \sim 0.5-2$~Mm, the spectrum peaked near degrees of $\ell\sim 100$, which corresponds to wavelengths of $\lambda \sim 20-40$~Mm; this is definitely the supergranulation scale. The peak is fairly wide in these shallow layers. As $d$ increases, the main spectral peak narrows and shifts toward low $\ell$ values or long wavelengths. In the deep layers, harmonics with $\ell\sim 15-30$, of scales $\lambda \sim 150-300$~Mm, are prominent, which appears to be indicative of the presence of giant cells.

It is remarkable that the spectral peak is very close to the line $\ell=m$ in the deep layer. Therefore, with the increase of $d$, the most powerful harmonics become elongated in the meridional direction, approaching the sectorial structure (with a monotonic increase from the pole to the equator in each hemisphere at a given longitude). This seems to be an argument in favor of the repeatedly expressed idea of a banana shape of the global-scale convection structures \citep[][etc.]{Glatzmaier_Gilman_II_1981, Busse_2002}.

The supergranular-scale flows in the upper layers ($\ell \sim 70-130$, $\lambda \sim 30-60$ Mm) coexist with the upper parts of giant convection cells, whose presence is not pronounced there. However, the power values for the largest scales at the upper and deep levels are comparable. 

The variation of the integrated spectral power of the flow during the activity cycle shows that while the total power is anti-correlated with the sunspot number at depths less than 4~Mm, it is positively correlated in the deeper layers. 
Such a behavior of the spectral power in the activity cycle can be interpreted based on the idea that the magnetic field, suppressing convection in the upper layers, redistributes its energy in favor of deeper layers. This agrees with the fact that converging flows observed around active regions may affect the convection spectra, as shown using the ring-diagram \citep{Haber_etal_2004} and time-distance \citep{Zhao2004} techniques.

\subsection{Flow Divergence and Vorticity}

A non-spectral technique has been used to identify supergranules and measure the horizontal scale of supergranulation by estimating the supergranulation scale from horizontal flow divergence maps \citep[see][and references within]{Rincon2017}.  
According to the results of \citet{Getling_Kosovichev_2022} using HMI data, the horizontal flow divergence ranges from $\pm~4 \times 10^{-5}$ s$^{-1}$ at a depth of 0.5 Mm to $\pm~1 \times 10^{-5}$ s$^{-1}$ at a depth of 11.5 Mm. 

The supergranular power spectrum of the divergence signal exhibits wave-like properties, indicating the presence of supergranular waves in surface horizontal flows near the solar equator \citep{gizon2003, schou2003}. Extension of the analysis to higher wavenumbers, corresponding to a spherical-harmonic degree larger than 120, showed an oscillation of the power spectrum with frequency of approximately 2 $\mu$Hz, corresponding to waves with a period of about six days \citep{Langfellner2018}.

Due to the Coriolis force acting on diverging horizontal flows, solar supergranules exhibit a hemisphere-dependent preferred sense of rotation. Consequently, supergranules behave like weak anticyclones, with their vertical vorticity changing sign at the equator \citep{gizon2003}. Using HMI data, it was further demonstrated that the correlation between vertical vorticity and horizontal divergence displays a latitudinal dependence consistent with the Coriolis force's action on convective flows \citep{Langfellner2014}. In addition, a linear relationship between non-magnetic vorticities and divergences was reported in regions within $\pm 30^\circ$ latitude \citep{Sangeetha2016}.

The analysis of both time–distance helioseismology and local correlation tracking to SDO/HMI data found that the typical vortical flow component is of the order of 10 m\,s$^{-1}$ in the diverging core of supergranules, significantly weaker than the diverging horizontal flow component itself \citep{Langfellner2015}. The kinetic helicity proxy, defined as the dot product of velocity and its curl, exhibits a hemispheric pattern similar to vorticity. These flows possess negative kinetic helicity in the northern hemisphere and positive kinetic helicity in the southern hemisphere throughout most of the convection zone \citep{Kosovichev2016,Hathaway2021}.

\subsection{Current Challenges and Future Directions}

Significant advancements in understanding and simulating solar convection have occurred over the past three decades. This progress stems from two main factors: significantly improved computational resources and continuous high-cadence observations of the full solar disk. These observations come from the ground-based Global Oscillation Network Group (GONG), and notably, space-based instruments like the Michelson Doppler Imager (MDI) and Helioseismic and Magnetic Imager (HMI). Since 2010, HMI has provided nearly continuous high-precision data with enhanced spatial resolution. These new data and the remarkable developments in helioseismology they spurred have enabled further insights into solar convective flows, as briefly presented in this section.

The Sun's convective region extends approximately 200 Mm below the surface, where density varies by several orders of magnitude. Solar convection manifests across a broad spectrum of spatial dimensions and temporal scales. This complex and dynamic zone is characterized by intricate interactions between plasma motions and thermal gradients, which in turn interact with magnetic fields. This multi-scale complexity explains why many challenges remain in fully characterizing and understanding convective flows.

An outstanding challenge is `the convective conundrum' - the discrepancy of one or several orders of magnitude in the kinetic energy as a function of angular degree, observed between different helioseismic techniques and types of analysis (Figure~\ref{Proxauf_spectrum}). The disagreement also exists between these observational results and the numerical simulations, and it varies with depth \citep{Lord2014}.

New emerging mode-coupling helioseismic techniques further highlighted this issue by reporting results that contradict numerical models \citep{Woodard2006, Hanson2024}. They also found that downflows are significantly weaker than upflows, suggesting the presence of small, undetectable descending plumes that maintain mass-flux balance. This finding further challenges the validity of the mixing-length theory in explaining solar convection.

Giant cells are prominent in most global simulations, but another puzzling issue is that they appear to have a very weak signal in observations, making them difficult to detect. The recent discovery that many of these large-scale structures are actually inertial waves raises the question of which, if any, are actual giant convective cells. Interestingly, inertial modes are expected to provide additional constraints on the physics of the deep convection zone \citep{Hotta2023}.

Despite all the progress achieved, much remains to be done in both theoretical analysis and observational studies. The continuation of HMI data is highly desirable. Moreover, a new, improved HMI-like instrument with full-disk coverage, high cadence, high spatial resolution, and capability for long-term observations would significantly enhance our understanding of convective flows.

 \section{Differential Rotation}
 \label{sec:diffrot}

The main features of the solar rotation profile -- the near-surface shear layer, differential rotation in the convection zone, and the shear layer or tachocline between the convection zone and the largely rigidly-rotating radiative interior -- were already established before HMI was launched \citep{1996Sci...272.1300T,1998ApJ...505..390S}. There were some discrepancies between the inferred rotation profiles from the two instruments, including a jet-like feature at 75 degrees latitude that was seen in inversions of MDI but not GONG data. These discrepancies were caused by differences in the mode fitting pipelines rather than the instruments; specifically, an anomaly in the MDI odd-order splitting coefficients around a frequency of 3.5~mHz and an underestimation of the low-degree rotational splittings in the GONG data \citep{2002ApJ...567.1234S}.

Inversions of the first months of HMI rotational-splitting data showed a profile that resembled that seen in the MDI full-disk data, confirming that the high-latitude jet in the rotational inversions of the MDI data was an artifact \citep{Howe2011JPhCS271}. Further studies using an improved algorithm to estimate mode parameters from MDI and HMI determined that the systematic effects and the high-latitude jet feature were caused by the `vector-weighted' binning of the MDI data, which was performed onboard the SoHO spacecraft to satisfy the telemetry requirements \citep{Larson2018,Larson2024SoPh299a}. This was confirmed by  a rotational inversion that employed the high-degree modes in the range of angular degree $\ell$ from 0 to
1000 and the frequency range from 965 to $4600\,\mu$Hz, obtained from the 66-day-long 2010 MDI Dynamics Run performed with full $1024\times 1024$-pixel resolution \citep{2020Reiter_Modes}.

The picture of the global solar rotation given by HMI, therefore, is relatively simple, with the rotation rate varying rather smoothly with latitude. 

\subsection{Features of the Solar Rotation Profile}

\begin{figure}
	\centering
	\includegraphics[width=\linewidth]{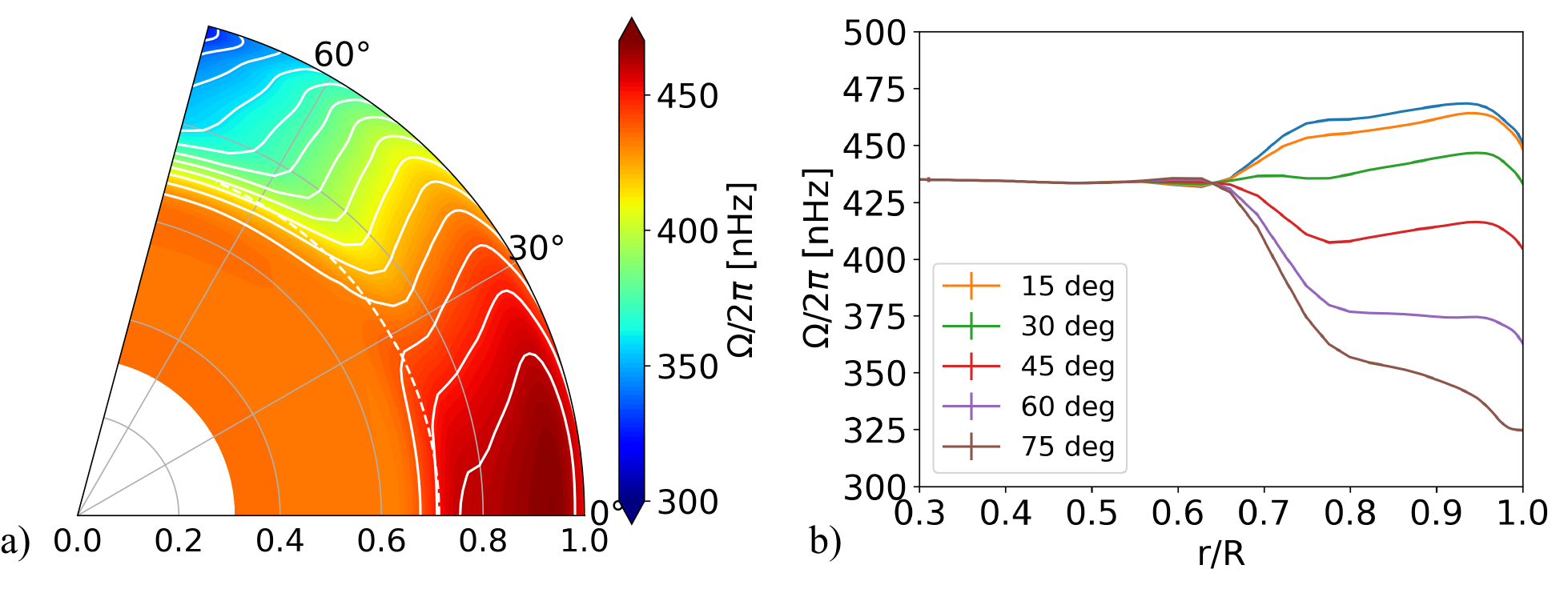}
	\caption{The solar rotation profile from the average of 72 two-dimensional Regularized Least Squares (2dRLS) inversions of 72-day HMI datasets covering the period May 2010 to July 2024. a) Contour map of the rotation as a function of radius and latitude, with the \textit{white contours} at 10\,nHz intervals. The \textit{white dashed line} represents the base of the convection zone at {0.71\,\sunrad}. b) Plots of rotation rate as a function of radius at selected latitudes. Note that the 1-sigma error bars are smaller than the thickness of the curves, but these reflect only the random errors, not systematic effects such as the smoothing of sharp features due to the resolution of the inversion.}
	\label{fig:rh_rotation}
\end{figure}

Figure~\ref{fig:rh_rotation} shows the mean rotation profile derived from the default HMI two-dimensional Regularized Least-Squares (2dRLS) inversions of rotational splittings up to $\ell=300$ that can be downloaded from the JSOC (data series \texttt{hmi.V\_sht\_2drls}). Below we discuss what had been learned from HMI observations about the main features of the rotation. 

\subsubsection{Near-Surface Shear Layer}

The sharp decrease in the rotation rate in a shallow subsurface region (called the Near-Surface Shear Layer or NSSL), which is believed to play a crucial role in the solar dynamo \citep{2010LRSP....7....3C,Pipin2011}, has been in the focus of global and local helioseismology since its initial discovery  \citep{1988ESASP.286...73R,1988ESASP.286..117K}. Of particular interest was a question whether the radial gradient of the rotation rate changes (and even changes sign) at higher latitudes, for which there was some evidence \citep[e.g.][] 
{1996Sci...272.1300T,1997Kosovichev_Medl,2002SoPh..205..211C,2014A&A...570L..12B,Rozelot2025}. 

Some other measurements found no indication of a reversal in the sign of the radial gradient \citep{1990ApJ...351..687R,2013ASPC..478..151C,2020Reiter_Modes}. Although the HMI results in Figure~\ref{fig:rh_rotation} seem to indicate a change in the rotational gradient at about 70$^\circ$, this controversy has not been resolved because of the difficulties of resolving the structure of the solar differential rotation at high latitudes. It is worth noting that the analysis of \citet{2020Reiter_Modes} is the only one using global modes (from MDI) with degrees greater than 300.

\begin{figure}
\centering
	\includegraphics[width=0.6\columnwidth]{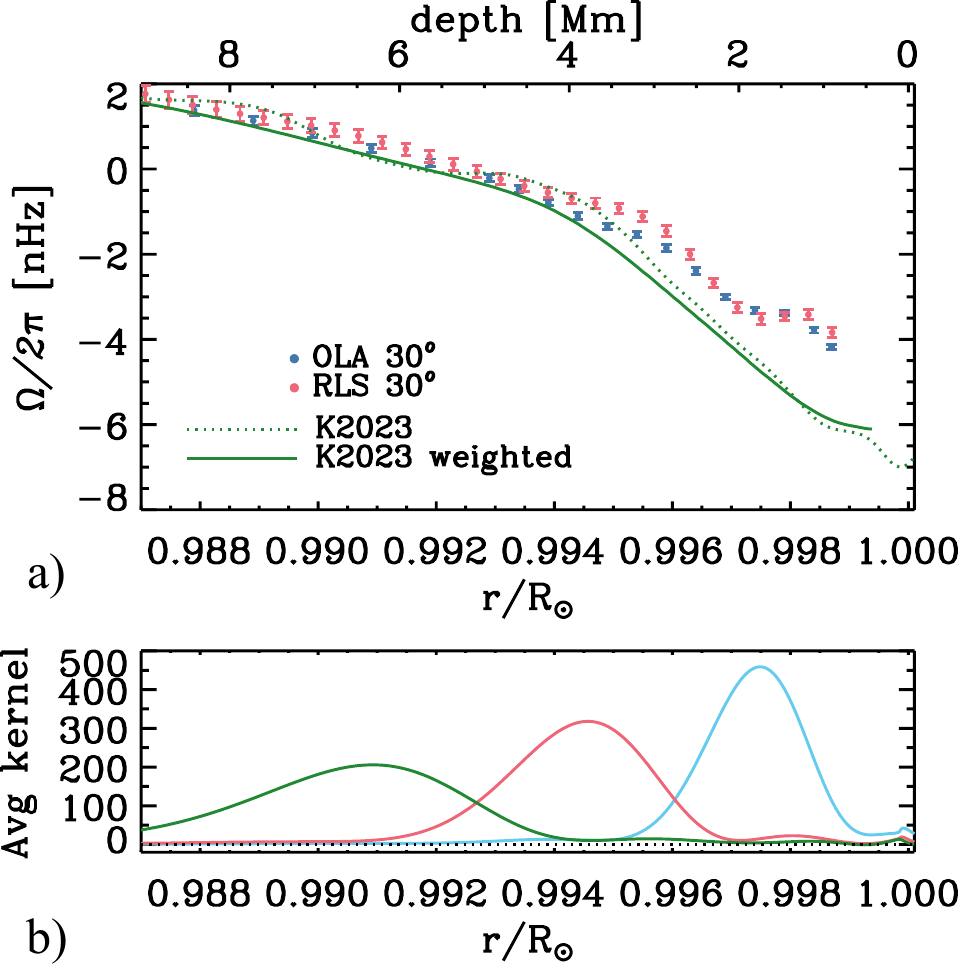}
	\caption{a) A comparison of solar rotation rates, obtained by inversion results for the 30-degree tiles using Optimal Localized Averaging (OLA, \textit{blue circles}) and Regularized Least Squared (RLS, \textit{red circles}) is shown after subtracting the 
		Carrington rotation rate. The rotation rate estimated by \citet[]{2023MNRAS.518..504K} using 3D radiative hydrodynamic simulations is represented by the \textit{dotted green line}, marked K2023. The \textit{full green line} shows the simulation results weighted by the OLA averaging kernels. b) Three examples of OLA averaging kernels are shown to illustrate the challenge of reproducing the model in the inversion results. The \textit{horizontal dotted line} indicates a zero value for the averaging kernel\citep{2024ApJ...967..143R}.
	}
	\label{fig:uxmodel}
\end{figure}

The ring-diagram analysis of \citet{2024ApJ...967..143R} focusing on the NSSL regions along the central meridian and averaging over the 12-year observing period showed that the NSSL can be divided into three regions: a large layer below 0.994~\sunrad\  (4~Mm) with a small shear where the radial gradient of the logarithmic rotation rate increases in amplitude from $-0.5$ at 21\,Mm to $-1$ at 4\,Mm;  a narrow middle layer centered around 0.9965~\sunrad\ (2.4\,Mm) with a strong shear and a gradient of approximately $-3$; and a narrow layer close to the surface from about 0.9977\,\sunrad\ (1.6\,Mm) where the radial gradient is close to zero but becomes negative and steeper again toward the surface (Figure ~\ref{fig:uxmodel}).
A similar fine structure of the NSSL, which is about 10\,Mm-thick (the so-called ``leptocline''), was found in 3D radiative hydrodynamic simulations \citep{2023MNRAS.518..504K} and also in global helioseismology data \citep{Rozelot2025}.

In addition, a combined analysis of data from MDI, HMI, and GONG by \citet{2022ApJ...924...19A} revealed that the shear layer is shallower at higher latitudes than close to the equator and varies with the solar cycle.

\subsubsection{Tachocline}

{The thickness of the tachocline is much smaller than the horizontal resolution of inversions, sometimes estimated to be smaller by a factor of 10. Consequently, estimates of the thickness are done through forward modeling of the splitting coefficients rather than through inversions. Attempts are being made to improve the resolution of the inversions (Korzennik et al. in preparation) by using iterative methods to measure tachocline properties through inversions.} Early results were obtained by modeling the second odd-order splitting coefficient ($a_3$ or $c_3$), which shows the largest signature of the tachocline \citep{1996ApJ...469L..61K,1997MNRAS.288..572B}. Later, a variety of optimization techniques were used to determine the tachocline parameters as a function of latitude \citep{1998MNRAS.298..543A,1999ApJ...527..445C}. These results and later ones showed that (i) the tachocline is prolate in shape, with the midpoint of the change in rotation occurring below the convection-zone base at low latitudes and in the convection zone at mid and high latitudes, and (ii) the change in the rotation rate occurs over a very narrow range of radii, and that this thickness is latitude-dependent. More recent results obtained with HMI data \citep{2019ApJ...883...93B} confirm this.

\subsubsection{Radiative Interior}
The rotation profile of the region below the tachocline cannot be resolved in as much detail as the regions near the surfaces. In general, the observations are believed to be consistent with uniform rotation. It is important to be aware of systematic errors when looking for deviations from this uniformity. The uncertainties in measuring parameters of low-angular-degree modes that are sensitive to the solar core rotation are reduced when longer time series and better mode-fitting methods are used \citep{Korzennik2005ApJ,Korzennik2008AN}. While the multi-instrument analysis of observational data from HMI, MDI, GONG, and the Birmingham Solar-Oscillations Network (BiSON) indicates consistency with a rigidly rotating solar core at 431 nHz, the results did not rule out a faster or a slower rotation rate for $r\le 0.2\,R_\odot$ \citep{EffDarwich2013SoPh287}. Although the radiative zone rotates rigidly, a systematic dip evolving with time was noted at $r=0.4\,R_\odot$ and 60${}^\circ$ latitude in both MDI and GONG data. This feature has not been confirmed by the HMI data.

\subsection{Summary}

The application of both local and global techniques to HMI data has improved our understanding of the Sun's rotation profile, particularly in the near-surface layers where recent studies have uncovered the detailed structure of the near-surface shear layer in both depth and latitude \citep{2024ApJ...967..143R}. Developments in high-degree global helioseismic fitting \citep{Korzennik2013,2020Reiter_Modes} hold promise for further exploring these shallow layers.
At greater depths, inversions of HMI data confirm the prolate shape of the tachocline \citep{2019ApJ...883...93B} and the near-uniform rotation in the radiative interior \citep{EffDarwich2013SoPh287}.

\section{Meridional Circulation}
\label{sec_MC}

\subsection{Brief Introduction}
The Sun's meridional circulation is the global-scale interior flow in the Sun's meridional plane.
 Although it is an order of magnitude weaker than differential rotation -- about $10-20$ \ms\  at the surface and only a few \ms\  in the deeper convection zone -- it plays an important role in solar dynamo and interior dynamics.  The meridional transports magnetic flux and redistributes angular momentum \citep[e.g.,][]{Wang89, Miesch05, Upton14b, Upton14a, Featherstone15}. Its speed was also reported to correlate with the duration and strength of solar cycles \citep{Hathaway2010, Dikpati2010}. 

Over the past three decades, advancements in helioseismology, using data from GONG, MDI, and HMI, have significantly improved our understanding of the Sun's meridional circulation. While early studies focused primarily on meridional flow near the surface, the detection of deep meridional flow throughout the solar convection zone became more feasible with the advent of HMI observations. High-resolution continuous data from HMI have provided valuable information on the flow's structure, depth, and temporal variations, although the profile of the deep meridional flow remains a subject of debate.
This section presents a brief review of recent studies on the structure of meridional circulation. The solar-cycle variations are discussed in Section~\ref{sec:dynamo}.

\subsection{Surface and Sub-Surface Meridional Flow}

The meridional flow on the solar surface and subsurface (a few tens of megameters below the surface) has been intensively investigated and well-recognized by early studies before HMI.  
Various studies have consistently measured a poleward flow on the order of $10 - 20$ \ms, achieved by methods such as surface Doppler measurements \citep[e.g.][]{Duvall1979, Hathaway1996}, surface feature tracking \citep[e.g.][]{Howard1986, Komm1993, Svanda2007}, time-distance helioseismology \citep[e.g.][]{Giles1997, Zhao2004}, ring-diagram analysis \citep[e.g.][]{Gonzalez1999, Haber2000}, and Fourier-Legendre analysis \citep[e.g.][]{Braun1998, Krieger2007,Doerr2010}. HMI, with its higher resolution, has enabled measurements of surface meridional flow up to higher latitudes using local helioseismology methods and extended the study of the long-term evolution of meridional circulation during the solar cycles.

\subsubsection{Observational Results from HMI}

\begin{figure}[t]
	\centering
	\includegraphics[width=0.6\textwidth]{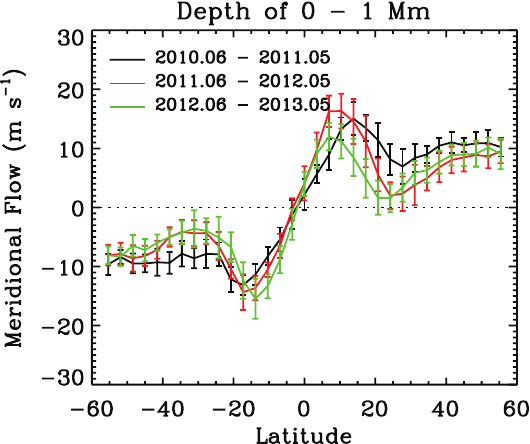}

	\caption{Mean meridional-flow profiles averaged from three consecutive years (represented by different colors) for 0\,--\,1 Mm \citep{Zhao2014}. } 
\label{residual_MF}
\end{figure}

HMI pipeline products provide subsurface flow maps derived from ring-diagram methods \citep{2011Bogart_Rings1, 2011Bogart_Rings2} and time-distance method \citep{Zhao2012} (see Section~\ref{sec:Time-Distance Method} for more details). 
From these data, the meridional flow was obtained and its long-term temporal evolution was studied \citep{Zhao2014,Komm2015,Komm2018,Lekshmi2019,Komm2020,Getling2021}. 
Figure~\ref{residual_MF} shows the one-year-average meridional flow profiles at $0 - 1$ Mm for the first three years of HMI observations, where the largest meridional flow speed occurs at low latitudes near 20\degree. The latitude of the faster meridional flow is observed to migrate toward lower latitudes, as depicted in Figure~\ref{fig:LHMeridional} in Section~\ref{sec:meridional_vary}
, which shows the residual meridional flow after removing a long-term mean profile. In mid-latitude areas,  an anti-correlation between the meridional flow speed and the magnetic flux was reported for Solar Cycle 23. The temporal evolution of the near-surface meridional flow will be further discussed in Section~\ref{sec:meridional_vary}. 

\subsubsection{Polar Meridional Flow}

Surface and subsurface meridional flows are routinely measured by HMI observations up to about $75^\circ$ latitudes, but the polar meridional flow remains a missing piece in helioseismology. 
Observational evidence of counter cells (equator-ward flow) in the polar regions was reportedly found using non-helioseismology methods, such as correlation tracking with MDI and HMI data \citep{ulrich2010,Hathaway2010,hathaway2022,Ulrich2022} and surface flux transport simulations with Hinode data \citep{yang2024}. However, the counter-cell detection is sporadic and shows no clear solar-cycle dependence. 

\subsection{Deep Meridional Flow}

While the surface and subsurface meridional flows have been widely studied with consensus, determining the deep meridional flow remains challenging, as the p- and \textit{f}-mode oscillations are more sensitive to the outer layers of the convection zone than the deeper interiors.  Studies have provided inconsistent results, leaving key questions—such as the number of circulation cells in the Sun's radial direction and the depth of the equatorward return flow—still open for debate.

Although the interior rotation profile across the convection zone has long been established, the meridional flow cannot be determined using the same global helioseismology techniques of measuring frequency splitting since the frequencies of global \textit{p}-modes are not sensitive to meridional circulation to the first order. Instead, the deep meridional circulation has primarily been studied using local helioseismology techniques, particularly the time-distance method (Section~\ref{sec:Time-Distance Method}). Multiple research groups have measured the meridional flow throughout the entire convection zone, but their findings remain inconsistent, largely due to many systematic effects complicating these measurements. In this subsection, we will discuss the results and the ongoing challenges in time-distance measurements. Additionally, a few other techniques, such as Fourier-Legendre decomposition and normal-mode coupling, have been explored to measure deep meridional flow, and we will briefly review these next.

\subsubsection{Deep Meridional Flow Measured by Time-Distance Method}

\begin{figure}[h]
\centering
\includegraphics[width=1\textwidth]{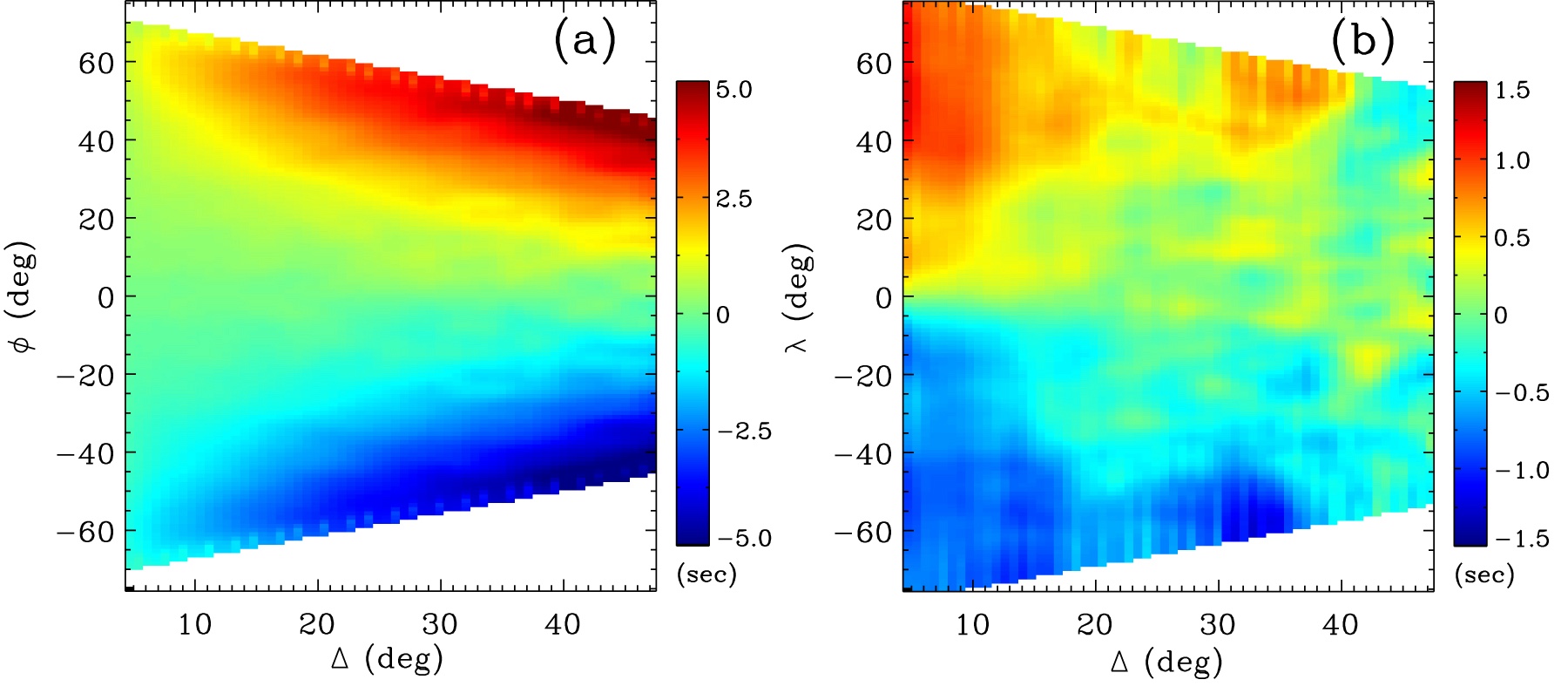}
\caption{\textit{a}) The travel-time shift caused by the center-to-limb effect as a function of travel distance $\Delta$ and disk-centric distance $\phi$.\textit{ b}) The meridional-flow-induced travel-time shifts after removing the CtoL effect, as a function of travel distance $\Delta$ and latitude $\lambda$ \citep{Chen2017}}.
\label{CtoL}
\end{figure}

Analysis of HMI data showed that the travel time measurements for large travel distances that are required for measuring flows in the deep convection zone are affected by systematic errors that could be misinterpreted as a flow along the limb-to-center direction \citep{Zhao2012a,Chen2018}. Known as the center-to-limb (CtoL) effect, it increases with both measurement depth and the center-to-limb position. Figure~\ref{CtoL} illustrates this effect from the HMI data in comparison with the meridional-flow-induced signals \citep{Chen2017}. The travel time shift caused by this effect can be 5\,--\,10 times larger than the flow-induced travel times. Therefore, its removal is critical for determining the deep meridional flow.

\begin{figure}
\centering
\includegraphics[width=1\textwidth]{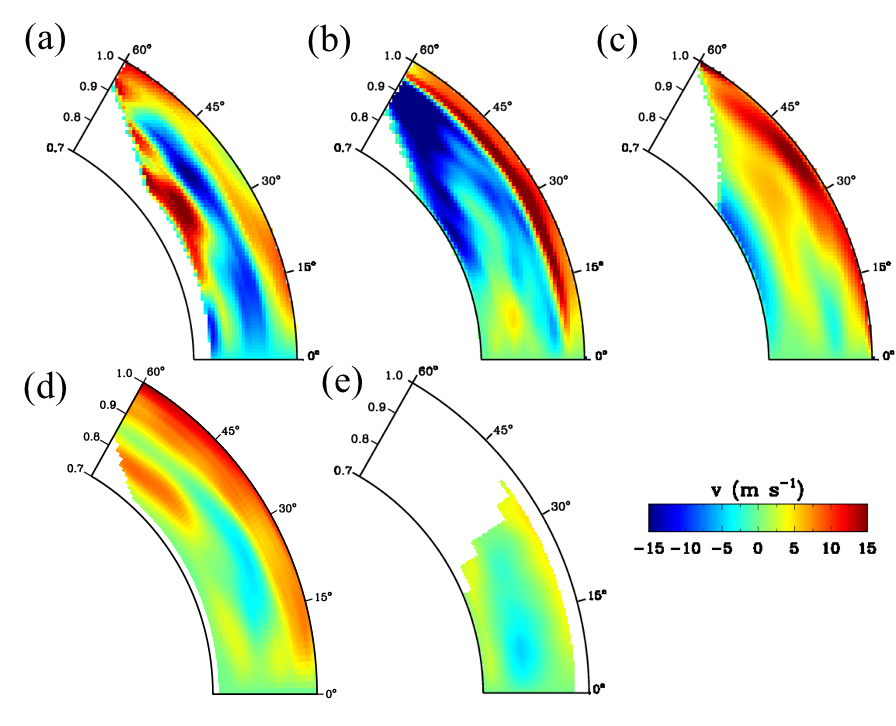}
\caption{Comparison of hemispherically symmetrized meridional circulation for selected results by time-distance method, from (\textit{a}) \citet{Zhao2013}, (\textit{b}) \citet{Kholikov2014, Jackiewicz2015}, (\textit{c}) \citet{Rajaguru2015}, (\textit{d}) \citet{Chen2017}, and (\textit{e}) \citet{Lin2018}. }
\label{MFcomp}
\end{figure}

The physical mechanism of the systematic center-to-limb frequency shift is yet to be established. It could be caused by near-surface convective flows \citep{Baldner2012}, changes in the spectral line formation height \citep{Kitiashvili2015}, or by non-adiabatic effects contributing to the observed Doppler shift \citep{Waidele2023b}.  An empirical method proposed to remove the shifts uses measurements along the East-West direction in the equatorial region \citep{Zhao2013}. After applying the correction, this study showed that the meridional flow has a double-cell structure along the radius in the convection zone. It is directed from the equator to the poles in the top $\sim 65$ Mm layer and at the base of the convection zone, but between (from 65 to 125 Mm beneath the solar surface, or from 0.82 R$_\sun$ to 0.91 R$_\sun$, the meridional flow is equatorward (Figure~\ref{MFcomp}a). 

However, subsequent analyses by different research teams, following similar measurement methods with various modifications, yielded mixed results regarding the detailed circulation profile. A study using GONG observations supported the presence of shallow return flow but reported a single-cell circulation pattern with a weak double-cell structure at low latitudes  \citep[Figure~\ref{MFcomp}b; ][]{Kholikov2014, Jackiewicz2015}. A follow-up inversion using the same observations confirmed the shallow return flow but found the number of cells inconclusive due to the inversion's sensitivity to regularization parameters \citep{Boning2017}. Another measurement using HMI observations and inversions with a mass-conservation constraint reported a single-cell circulation in each hemisphere \citep[Figure~\ref{MFcomp}c; ][]{Rajaguru2015, Mandal2017}. Meanwhile, a study using 15 years of MDI medium-$\ell$ data revealed a double-cell circulation during solar minima and a temporal variation of the circulation pattern with solar cycles in low-latitude regions \citep[Figure~\ref{MFcomp}e; ][]{Liang2015, Lin2018}. All the above measurements used the same proxy method to remove the CtoL effect. This empirical approach was validated by a comprehensive study that measured along all disk radial directions \citep{Chen2017}. The study demonstrated that HMI’s CtoL effect is isotropic with respect to the azimuthal angle, confirming the effectiveness of the simpler proxy method. Using this more robustly determined CtoL effect, the study revealed a double-cell meridional circulation pattern (Figure~\ref{MFcomp}d).

Recent studies have employed multi-instrument measurements to determine the meridional circulation. A forward modeling of combining data from MDI and HMI  favored a single-cell meridional circulation \citep{Liang2018}.
Another study, based on only GONG and MDI data, showed a single-cell meridional circulation throughout Solar Cycles 23 and 24, with a possibility of a weak second cell at low latitudes in the southern hemisphere from 2005–2011 \citep{Gizon2020}. Additional studies that applied a Bayesian inversion method \citep{Jackiewicz2020} on these multi-instrument measurements indicated the meridional circulation in Solar Cycle 23 exhibited a strong single-cell profile, while in Cycle 24, it had a weak poleward flow component reaching nearly zero in the middle of the convection zone, which might indicate a double-celled structure \citep{Herczeg2023}. It was also shown that the meridional flow structure inferred by the inversion of the observed acoustic travel times depends on the flow constraints implemented in the inversion procedures -- applying an angular momentum transport constraint changed the inverted profile from a single-cell to a double-cell configuration \citep{Hatta2024}. Numerical simulations of the travel-time measurements found that the high noise levels of stochastically excited solar oscillations make it difficult to definitively distinguish between single-cell and double-cell structures through observations alone \citep{Stejko2021}. 

\subsubsection{Other Methods}

A different method (Fourier-Legendre decomposition) that included a control measurement designed to assess and remove the center-to-limb effect applied to 88 months of HMI data indicated significant hemispheric differences: the meridional flow found equatorward at approximately 40 Mm below the photosphere in the northern hemisphere while remaining poleward in the southern hemisphere \citep{Braun2021}.  


\subsection{Summary}

The ongoing research continues to focus on understanding these systematics and refining observational methods, showing promise for further advancements in measurements. Simulations and inversions also emphasize the need for improved measurement precision to resolve ambiguities in deep flow inversion. Different techniques and constraints in inversions have yielded varying circulation profiles, and some simulations indicate that current measurement precision may be insufficient to definitively distinguish between single-cell and double-cell circulations. 

\section{Solar Inertial Waves}

\subsection{Introduction}

Waves propagating through the Sun's interior can be classified based on their restoring forces. Acoustic waves (\textit{p}-modes) are restored by pressure gradient forces and excited by turbulent convection. Internal gravity waves (\textit{g}-modes) are waves restored by buoyancy and mainly propagate through the convectively stably-stratified radiative zone. Both acoustic waves and gravity waves are relatively high-frequency modes, oscillating with periods of several minutes and several hours, respectively. In addition to these, inertial waves (including Rossby waves or \textit{r}-modes) that are driven by the Coriolis force and oscillate very slowly with periods of several months were also believed to exist in the Sun \citep[][]{Provost1981,Saio1982,Unno1989,dziembowski87_3}. Detecting such low-frequency modes required long-term observational data with sufficient signal-to-noise ratios spanning many years, which was achieved using the HMI observing series.

 The observations have revealed several types of inertial waves, including equatorial Rossby waves \citep{Loeptien2018}, high-latitude and critical-latitude modes \citep{Gizon2021}, and high-frequency retrograde modes \citep{Hanson2022,Gizon2024}. The frequencies of these modes in the Carrington rotation rate as a function of the azimuthal wavenumber $m$, and examples of the mode eigenfunctions obtained from observations and a theoretical model \citep[][]{Bekki2022a,Bekki2024a} are shown in Figure~\ref{fig:observed_modes}. {Properties of these modes are discussed in detail in Sections~\ref{sec:Inertial-type waves} and \ref{sec:High-latitude modes}.}

\subsection{Equatorial Rossby, Critical-latitude, and High-Frequency Retrograde (HFR) Modes}\label{sec:Inertial-type waves}
A special class of inertial modes is Rossby waves or \textit{r}-modes, whose motions are quasi-toroidal, i.e., the radial motions are very small compared to the horizontal ones \citep[]{Papaloizou1978,Pedlosky1987,Rieutord1997}. They propagate in a retrograde direction (opposite to the rotation) in a co-rotating frame.

\begin{figure} \centering \includegraphics[width=\linewidth]{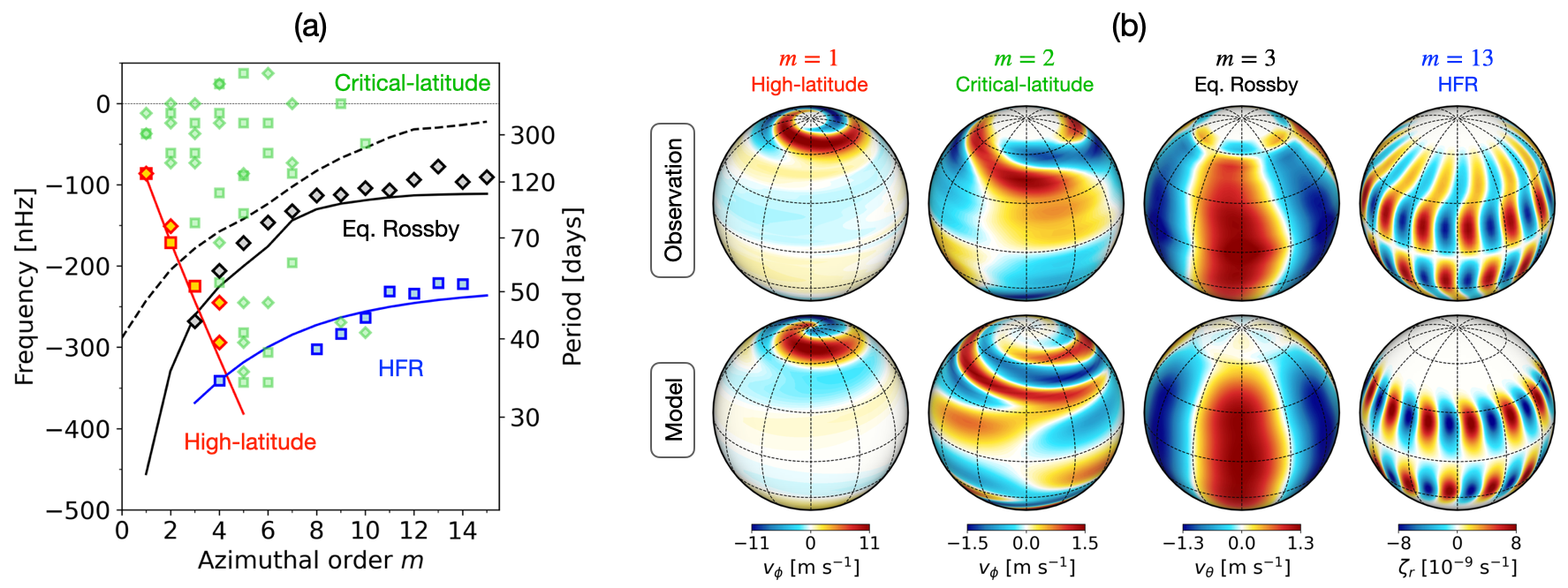} \caption{ Observations and identification of the solar inertial modes. \textit{a}) Frequencies of the solar inertial modes in the Carrington frame ($\Omega_{\rm Car}/2\pi=456.0$~nHz). The \textit{diamonds} and \textit{square symbols} show the frequencies of the observed modes with north-south symmetric and antisymmetric radial vorticity. \textit{Red, green, black, and blue colors} represent the high-latitude modes, the critical-latitude modes, the equatorial Rossby modes, and the high-frequency retrograde modes, respectively. The \textit{curves} show the computed linear dispersion relations from a theoretical model. The \textit{black solid and dashed curves} show the linear dispersion relations of the equatorial Rossby modes with no radial node ($n=0$) and with one radial node ($n=1$), respectively. \textit{b}) Surface eigenfunctions of selected inertial modes deduced from the observations (\textit{top panels}) and the model (\textit{bottom panels}).} \label{fig:observed_modes} \end{figure}


\begin{figure}
	\centering
	\includegraphics[width=\linewidth]{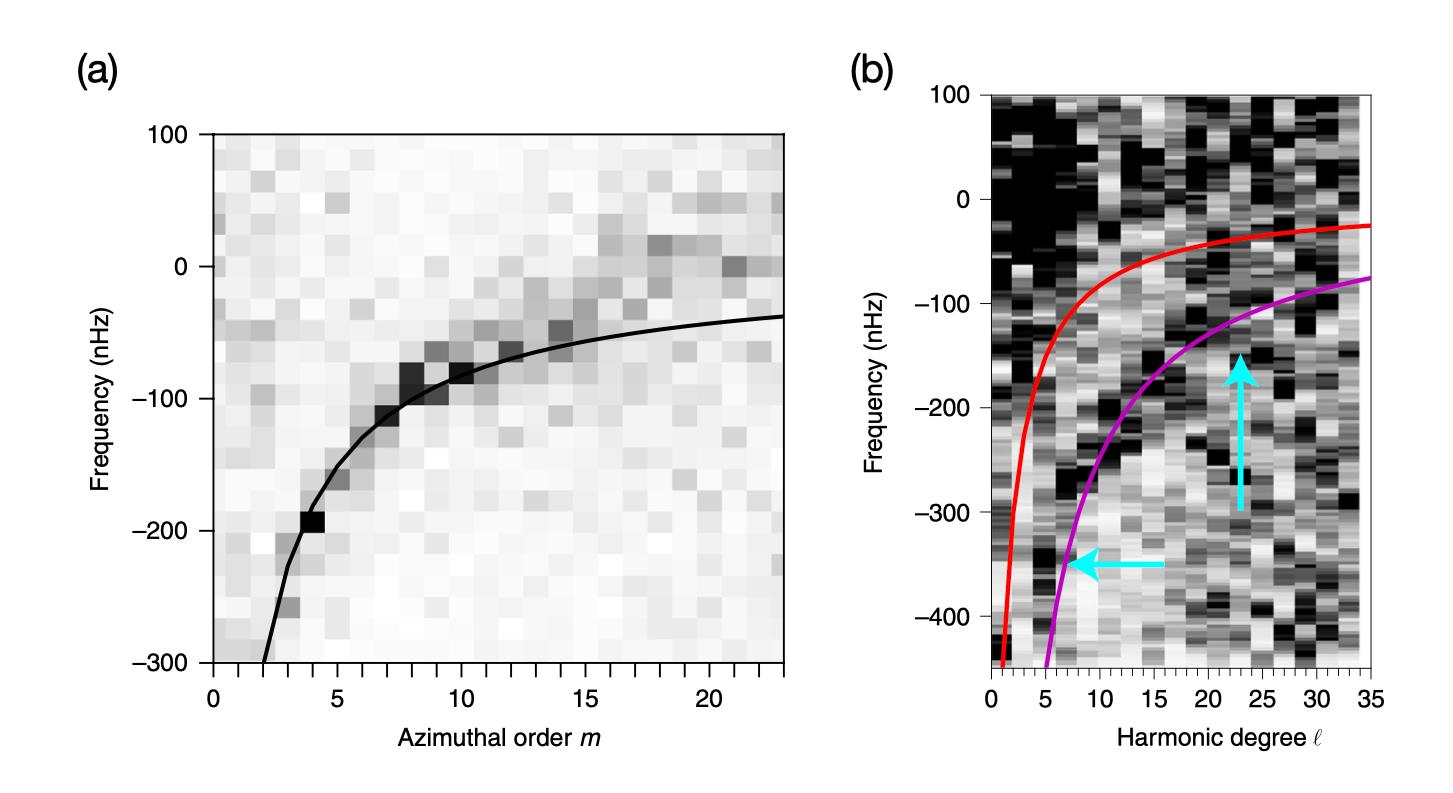}
	\caption{
		Power spectra of radial vorticity near the solar surface.
		\textit{a}) The sectoral ($\ell=m$) component of the radial vorticity power spectrum obtained by granulation tracking. 
		The \textit{black solid curve} shows the theoretical dispersion relation of sectoral Rossby waves  \citep{Loeptien2018}.
		\textit{b}) The $\ell=m+1$ component of the radial vorticity power spectrum obtained by normal mode coupling.
		The \textit{red curve} denotes the theoretical dispersion relation of sectoral Rossby waves.
		The \textit{purple curve} shows the fit to the power ridge of the high-frequency retrograde (HFR) modes, which is approximately three times the frequencies of the sectoral Rossby waves between the \textit{cyan arrows}. 
		In both panels, the frequencies are measured in a co-rotating frame with the Sun's equatorial rotation rate ($\Omega_{0}/2\pi = 453.1$~nHz), and power is normalized at each $m$ or $\ell$ \citep{Hanson2022}.}
	\label{fig:power_radvort} 
\end{figure}

Different methods have been utilized to detect these modes. The initial observation of solar inertial modes was obtained by tracking the motion of granules in intensity images from HMI and also by using subsurface flow maps from the ring-diagram technique. Evidence of the equatorial Rossby waves was found from six years of HMI data (Figure~\ref{fig:power_radvort}a). 
It was found that the sectoral ($\ell=m$) component of radial vorticity power for the range of $3 \leq m \leq 15$ dominantly lies along with the theoretical dispersion relation of the Rossby waves \citep[][]{Papaloizou1978,Provost1981,Saio1982}: \begin{equation} \omega = -\frac{2m \Omega_0}{\ell(\ell+1)}, \label{eq:dispersionRossby} \end{equation} where $\omega$ is the temporal frequency, $m$ is the azimuthal order, $\ell$ is the spherical harmonic degree, and $\Omega_{0}/2\pi = 453.1$~nHz is taken as the equatorial rotation rate of the Sun. Since the sectoral modes have radial vorticity, which peaks at the equator and decreases at higher latitudes, they are also called equatorial Rossby waves. They have typical velocity amplitudes of $1-2$~m~s$^{-1}$ at the surface and lifetimes of several months.

The characteristics of the equatorial Rossby waves have since been corroborated by multiple studies employing various techniques, including the normal mode coupling technique \citep{Hanasoge2019,Mandal2020,Mandal2021}, time-distance helioseismology \citep{Liang2019}, the ring-diagram technique \citep{Hanson2020a,Proxauf2020} and local correlation tracking of features (supergranules) \citep{Hathaway2021}. These studies characterized the frequencies, linewidths, and amplitudes of the Rossby modes for the azimuthal order up to $m=30$ using HMI data products \citep{Bhattacharya2024,Hanson2024}. The Rossby mode with azimuthal order $m = 2$ is not observed; only an upper limit of 0.2~m~s$^{-1}$ on its amplitude has been placed \citep{Mandal2021}.

The measurements of the latitudinal eigenfunctions \citep{Loeptien2018,Proxauf2020,Mandal2024} showed that they differ from the spherical harmonic $Y_m^m$ predicted by theory in the case of uniform rotation \citep{Provost1981}. A more general theory of low-frequency oscillations predicted that the latitudinal eigenfunctions of Rossby (\textit{r}) modes can be represented in terms of eigenfunctions of the Laplace tidal equation \citep{dziembowski87_1}. In addition, the \textit{r}-modes are strongly affected by the differential rotation and, in particular, the critical latitudes where the phase speed of the Rossby waves is equal to the differential rotation speed. At such latitudes, the energy exchange between the flow and waves causes the coupling of the spherical harmonics, resulting in the mode instability and a wave continuum spectrum \citep{dziembowski87_3}. 
\par
In fact, dozens of modes whose eigenfunctions are strongly concentrated near their critical latitudes were found using ten years of HMI data \citep[see, Figure~\ref{fig:observed_modes}: ][]{Gizon2021}. Although the mode power peaks at the critical latitudes, the power excess at the mode frequency can be seen over the entire latitude range. Frequencies of the observed critical-latitude modes cannot be represented by a well-defined dispersion relation.  In such cases, the eigenfunctions are sensitive to the internal properties of the Sun, such as viscosity. It was shown that the latitudinal widths of the critical layers are controlled by turbulent viscosity  \citep{Gizon2020a}. Further, 3D models imply that substantial radial motions are involved, which makes them also sensitive to the superadiabaticity \citep{dziembowski87_1}, particularly, near the critical latitudes \citep{Gizon2021}.

\par
In addition to the equatorial ($\ell=m$) modes,  a distinct power ridge corresponding to the $\ell=m+1$ mode was detected in the near-surface radial vorticity power spectra \citep[Figure~\ref{fig:power_radvort}b; ][]{Hanson2022}. The waves would appear as anti-symmetric vorticity patterns about the equator, in contrast to the equatorially symmetric Rossby modes. The detected power ridge does not follow the theoretical dispersion relation of the classical $\ell=m+1$ Rossby waves given by Equation~(\ref{eq:dispersionRossby}). These modes propagate in a retrograde direction at frequencies about three times faster than those of equatorial Rossby modes. Given these circumstances, the detected modes were called ``high-frequency retrograde (HFR)" modes. 
Theoretical studies imply that these HFR modes are essentially non-toroidal, involving substantial radial motions near the equator \citep{Triana2022,Bhattacharya2023,Bekki2024a,Jain2024,Blume2024}. 

While the latitudinal profile is well constrained by the observations, the variations with depth are more elusive. The eigenfunctions of the equatorial Rossby waves deduced from the ring-diagram analysis up to 15~Mm below the surface \citep{Proxauf2020} deviated from the theoretical eigenfunctions calculated for the uniform rotation and inviscid limit \citep{Provost1981,dziembowski87_2}. A recent theoretical study found that viscous diffusivity has a significantly greater impact on the radial structure of Rossby waves \citep{Bekki2022a}. The mode-coupling technique has the capability to detect the signal up to a depth of $0.83~R_\odot$ \citep{Mandal2024}. However, the radial structure of the mode eigenfunctions has not been established.

\begin{figure} \centering 	\includegraphics[width=\linewidth]{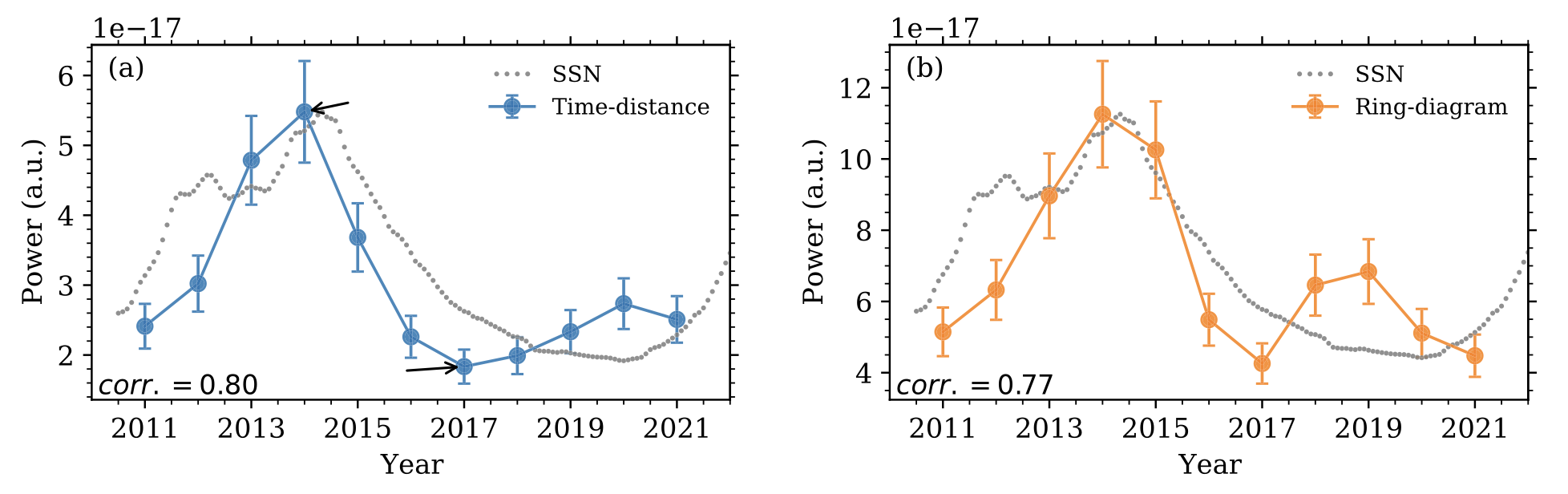}\\ \caption{ The power of the equatorial Rossby modes averaged over $3 \leq m \leq 16$ as a function of time. The \textit{left} and \textit{right panels} show the average power computed from time-distance analysis and ring-diagram analysis, respectively. The \textit{gray dots} show the monthly smoothed sunspot number (SSN) normalized to fit the scale \citep{Waidele2023a}.} \label{fig:cycle_variation_eqRos} \end{figure}

Figure \ref{fig:cycle_variation_eqRos} presents an analysis of 12 years of inferred subsurface velocity fields derived from the SDO/HMI time–distance and ring-diagram pipelines. The results show that the Rossby wave amplitude is higher during the solar activity maximum than during the minimum, while the frequency of Rossby waves is lower during the maximum than during the minimum \citep{Waidele2023a,Lekshmi2024}. This is opposite to the first-order predictions that considered only perturbations due to torsional oscillations \citep{Goddard2020}. 

It raises questions about the excitation mechanism of Rossby waves. It was suggested that, similarly to \textit{p}-modes, Rossby modes are stochastically excited by convection \citep{dziembowski87_2,Philidet2023}. The calculated amplitude is about 1~m~s$^{-1}$, in good agreement with the observations \citep{Gizon2020a}. These modes are also present with significant amplitudes in nonlinear simulations of rotating convection \citep{Bekki2022b,Matilsky2022,Blume2024}. The role of magnetic activity on these modes should be investigated numerically to better understand their variations with the solar cycle.

\subsection{High-Latitude Modes (Polar Spiral Flows)}\label{sec:High-latitude modes}

Global-scale spiral velocity patterns around the poles (above $\pm 60^{\circ}$), which persist for several months and move retrograde with respect to the Carrington rotation rate at high latitudes, were found by tracking the motions of supergranules seen in HMI Doppler observations \citep{Hathaway_etal_2013}. They have a dominant flow component at azimuthal order $m=1$, which corresponds to the high-latitude velocity pattern previously found in Mt. Wilson Observatory (MWO) Dopplergrams \citep{Ulrich1993,Ulrich2001}. Similar polar spiral flow patterns were found in subsurface flows using the ring-diagram analysis \citep{Bogart2015}. 

\begin{figure} \centering \includegraphics[width=0.8\textwidth]{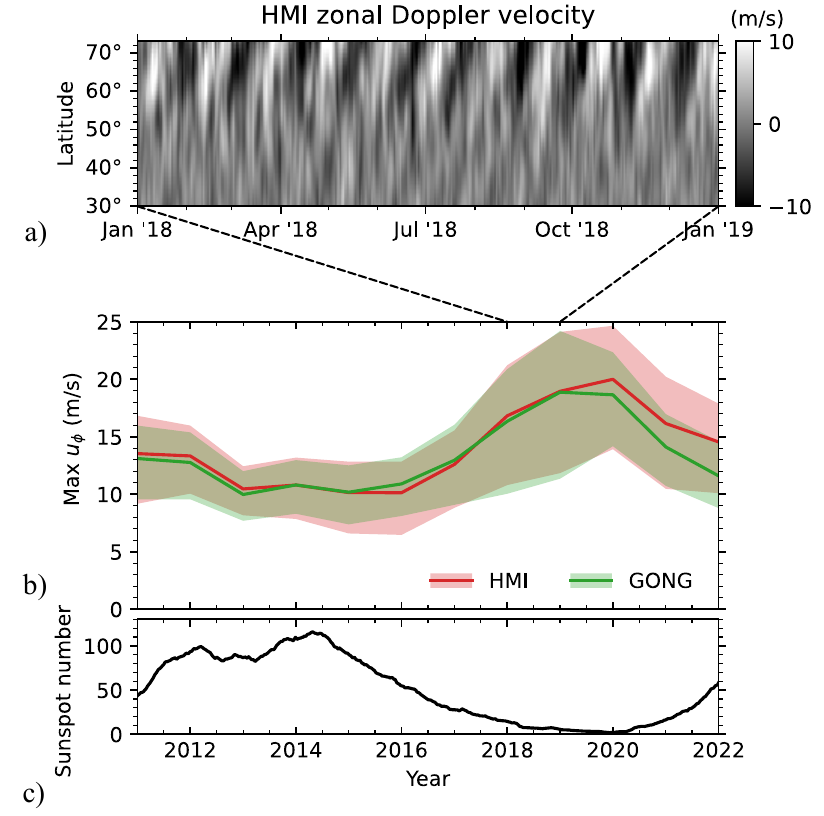} \caption{ Temporal variation of the amplitude of the $m=1$ high-latitude mode. \textit{Top}: Zonal velocity map computed from HMI line-of-sight Doppler data. \textit{Middle}: The amplitude of the $m=1$ mode estimated from the zonal velocity averaged over the latitude range $60^{\circ}-75^{\circ}$. \textit{Bottom}: Sunspot number as a function of time \citep{Gizon2024}. } \label{fig:cycle_variation_HL1} \end{figure}

These spiral flows can represent high-latitude inertial modes because their retrograde propagation frequencies are in the inertial frequency range of the Sun (\textit{red points} in Figure~\ref{fig:observed_modes}a) and the associated power excess characterized by a single frequency can be seen at all latitudes \citep{Gizon2021}. The $m=1$ mode is of particular interest because it has the largest time-averaged amplitudes at the surface ($v_{\phi} \approx 10$~m~s$^{-1}$)  among all the observed solar inertial modes. A recent study using the normal mode coupling technique found that this mode extends (almost uniformly in radius) all the way to the base of the convection zone \citep{Mandal2024}. It was also found that the mode amplitude is stronger during the solar minimum than in the maximum \citep[Figure~\ref{fig:cycle_variation_HL1},][]{Gizon2024, Liang2024}. These new observations imply that the high-latitude inertial modes are substantially influenced by internal dynamo-generated magnetic fields. Theoretical models showed that the high-latitude modes are baroclinically-unstable by the latitudinal entropy gradient and have a significant impact on the Sun's differential rotation \citep{Bekki2024b}. The readers may also refer to a recent review \citep[][]{Gizon2024}. 

\subsection{Conclusion} 

Various inertial modes have been successfully observed on the Sun. Having Coriolis force as their restoring force, these inertial modes have many unique properties that are not possessed by conventional acoustic modes. Theoretical models show that they are sensitive to some currently unknown parameters of the Sun, such as turbulent viscosity and superadiabaticity, both of which are poorly constrained by conventional acoustic-mode helioseismology. The detection of a large number of unique inertial modes hints at the possibility of “inertial-mode helioseismology”, potentially revealing new insights into the deep interior of the Sun. Furthermore, the sensitivity of inertial modes to the global scale magnetic fields may push forward our understanding of the solar dynamo. In the future, further analysis of observational HMI data (in tandem with other instrumental data) and improvement of theoretical models will deepen our understanding of solar inertial modes, which will help us to infer the convective properties and magnetic fields hidden deep in the Sun.

\section{Solar Cycle Variations and Dynamo} \label{sec:dynamo}

A precise knowledge of the variability in the solar interior is important for understanding the origin of magnetic activity and constraining dynamo models.
After the discovery of solar acoustic oscillations in the early 1960s \citep{1962ApJ...135..474L}, the first results on their variability came in the mid-1980s.  \citet{1985Natur.318..449W} reported a  decrease in the frequencies of 0.42 $\mu$Hz between 1980 and 1984 using the low-$\ell$ Active Cavity Radiometer for Irradiance Monitoring data. Since then, a large number of studies have been carried out to improve the characterization of global modes and to explore the changes in the inferred parameters, such as the oscillation frequencies, the mode line widths, and amplitudes, with varying solar activity. In addition, 
helioseismic observations have been widely used to study solar-cycle variations of the internal rotation and large-scale subsurface flows.


In this section, we give an overview of results obtained primarily from the HMI observations and how they compare with the contemporaneous observations of other instruments.

\subsection{Solar-Cycle-Related Changes in Acoustic Mode Parameters and Splittings}

\begin{figure} \centering \includegraphics[width=0.7\linewidth]{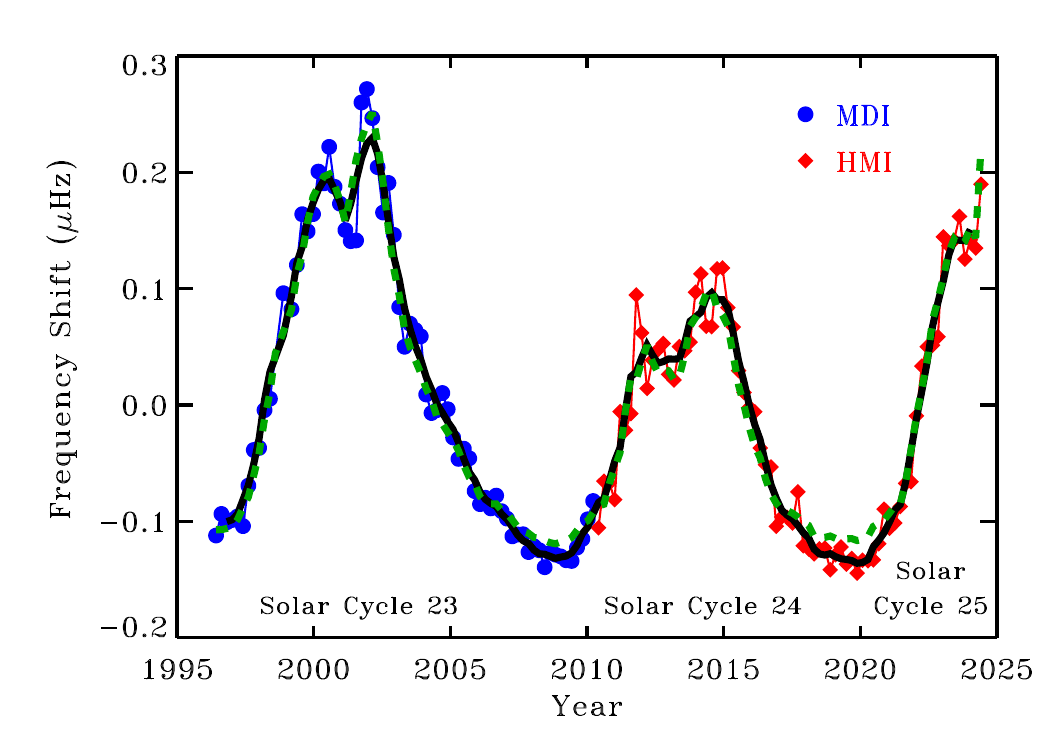} 
\caption{Temporal variation of the mean change in frequencies (\textit{symbols}) for MDI (\textit{blue dots}) and HMI (\textit{red diamonds}) observations. 
A five-point running mean of the frequency shifts is shown by the \textit{solid black line}, and the \textit{green dashed line } depicts the variation of 10.7 cm radio flux averaged over the same time intervals and scaled by assuming a linear relationship between the radio flux and frequency shifts.} \label{fig:delnu} \end{figure}

\begin{figure} \centering \includegraphics[width=0.8\linewidth]{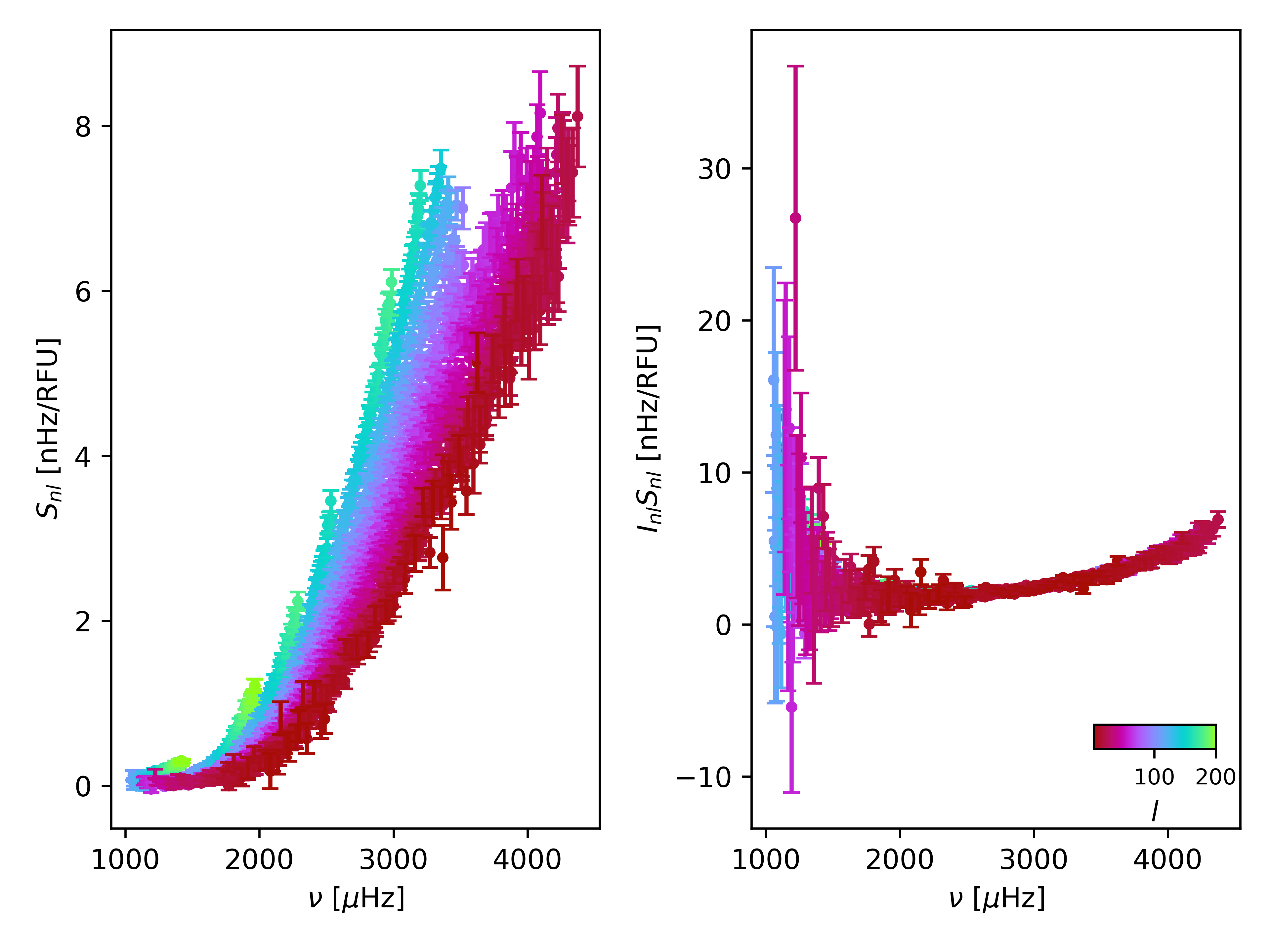} \caption{Frequency change per unit 10.7\,cm radio flux for medium-degree modes from MDI and HMI observed between 1996 and 2024, plotted as a function of frequency and color-coded by degree $\ell$. \textit{Left}: unscaled results. \textit{Right}: results scaled by mode inertia \citep{2018MNRAS.480L..79H}.}     \label{fig:mdisens} \end{figure}

Figure~\ref{fig:delnu} displays the temporal variation in \textit{p}-mode frequency shifts from MDI and HMI observations starting from 1 May 1996. These shifts are calculated using common modes between all 141 datasets in the frequency range of 1500-4500\,$\mu$Hz, each derived from 72~d of observations, with the first 71 datasets from MDI\footnote{jsoc.stanford.edu/ajax/lookdata.html?ds=mdi.vw\_V\_sht\_modes} and the remainder from HMI\footnote{jsoc.stanford.edu/ajax/lookdata.html?ds=hmi.V\_sht\_modes}. The MDI and HMI data are considered together as a continuous set, and the frequency shifts are scaled by mode inertia. The dashed line in the figure depicts the variation of solar activity represented by the 10.7 cm radio flux\footnote{www.spaceweather.gc.ca/forecast-prevision/solar-solaire/solarflux/sx-5-en.php}. It is clear that the frequencies vary in the phase of the solar cycle and show a strong positive linear correlation between them (Pearson's correlation coefficient is 0.99). These findings are consistent with the previous studies \citep[see][and references therein]{2012A&A...545A..73J,2016LRSP...13....2B, 2022ApJ...924L..20J}. Similar results were found for \textit{f}-mode frequencies, which were further utilized to study the changes in solar seismic radius and their implications on the total solar irradiance  \citep{jain_suns_2018,Kosovichev2018}. {In addition, as shown in Figure~\ref{fig:mdisens}, there is a strong $\ell$ dependence in frequency shifts, which is reduced by scaling with the mode inertia, and the obtained frequency shifts provide true variations in frequencies.}

The relative variation in the global oscillation frequencies observed by HMI with the change in solar activity has been studied and compared with those from other instruments, e.g., MDI and GONG, by several authors \citep{2013JPhCS.440a2023J,2016SPD....47.0716J, 2018MNRAS.480L..79H, 2019MNRAS.486.1847R}.  All these studies showed comparable results indicating that the sensitivity of the oscillation frequencies to the change of magnetic activity does not vary significantly with the choice of instrument and is similar for the same phase of solar cycles.  However, the correlation between the frequency variations and the solar radio flux deviates from a linear relation during the very high activity period, suggesting a saturation effect in the oscillation frequency variations \citep{2019MNRAS.486.1847R}.

The progression of the Solar Cycle 24 is  studied by analyzing high-$\ell$ acoustic mode frequencies in localized regions derived from the technique of ring diagrams\footnote{jsoc.stanford.edu/HMI/Rings.html} using HMI full-resolution Dopplergrams \citep{2019AGUFMSH11D3376U}.  The authors confirm that these frequencies are strongly correlated with the changes during the solar cycle and display a strong correlation on spatial as well as temporal scales.  It was found that the frequency variations in the Northern hemisphere peaked before the Southern hemisphere, which is consistent with the findings obtained by analyzing frequencies from GONG \citep{2024ApJ...962..194B} and the measures of magnetic activity.  The temporal variations in the amplitudes and widths of high-degree acoustic modes from GONG, MDI, and HMI measurements during the declining phase of Solar Cycle 23 and the rising phase of Cycle 24 were similar for all three instruments. However, the variations in Cycle 24 were smaller than in Cycle 23 because of the weaker magnetic activity \citep{2013JPhCS.440a2027B}.

\begin{figure} \centering \includegraphics[width=0.75\linewidth]{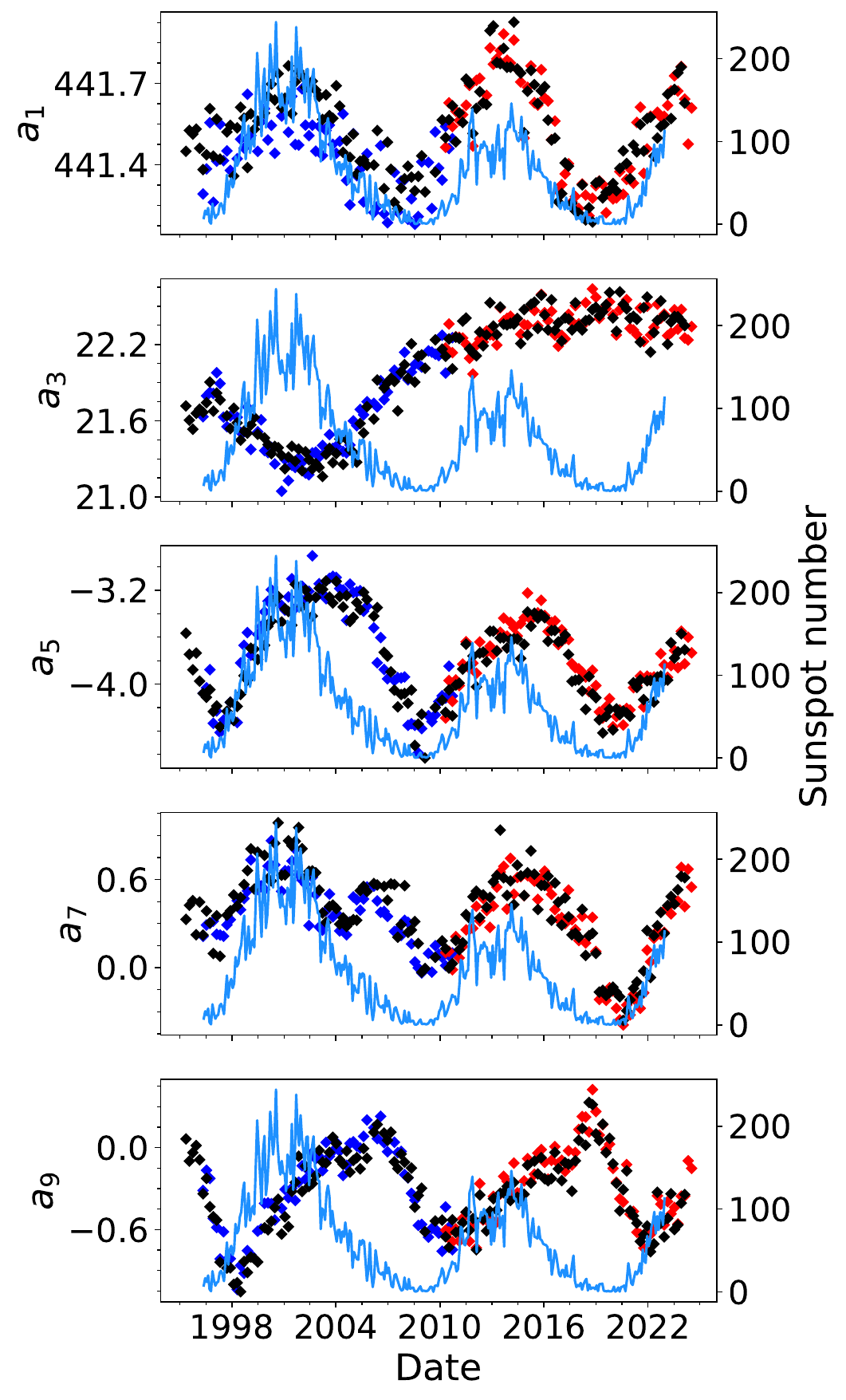} \caption{ Temporal variation of averaged odd-order splitting coefficients (in nHz) obtained by fitting GONG (\textit{black}), MDI (\textit{blue}) and HMI (\textit{red}) data for the oscillation modes of angular degree  $\ell$ = 0\,--\,120 \citep{Korzennik2023FrASS}. The \textit{cyan lines} in all panels depict the variation of sunspot numbers averaged over time intervals that correspond to the GONG sets.}  \label{fig:splittings} \end{figure}

The odd-order splitting coefficients, characterizing latitudinal components of the solar differential rotations, shown in Figure~\ref{fig:splittings} from all three instruments, revealed complex variations with the solar cycles {\citep{Korzennik2023FrASS}}. While the coefficient, $a_1$, describing a latitudinally averaged component, displays an 11-year cyclic pattern in phase with the sunspot number, the other coefficients are either out of phase or do not follow any systematic variations.  Since these coefficients are used to measure rotation rate,  their amplitude and phase variations indicate changes in the internal solar rotation during Solar Cycles 23, 24, and 25 (see Section~\ref{sec:zonal}).

\subsection{Solar-Cycle Variation of Subsurface Flows}

Both the solar rotation  (zonal flow) and the meridional flow vary over the solar cycle. The variation in the rotation rate that manifests in the pattern of migrating bands of faster and slower rotation known as the `torsional oscillations' first discovered in Doppler observations of solar rotation \citep{1980ApJ...239L..33H} is now continuously monitored by helioseismic instruments: GONG (1995 to present), MDI (1996\,--\,2011) and HMI (2010 to present). The flows were reported in early MDI \textit{f}-mode data \citep{1997ApJ...482L.207K}, and they were subsequently shown to penetrate deep into the convection zone \citep[e.g.,][]{2000ApJ...533L.163H, 2002Sci...296..101V}. The near-surface meridional flow has also been found to be modulated by solar activity, with bands of converging flow associated with the magnetic activity belts \citep[e.g.,][]{Zhao2004}. 
The continuously available high-resolution helioseismic observations from HMI have allowed more detailed study of these phenomena during Solar Cycles 24 and 25, revealing the extended 22-year cycle in the variations of the zonal flows and meridional circulation \citep{Komm2018,Getling2021,Mahajan2023}.

In a different approach,  the subsurface zonal and meridional flows were measured by applying the ring-diagram technique to Dopplergrams constructed from the HMI global spherical harmonic coefficients. The preliminary results agree with the traditional ring-diagram analysis near the surface and intermediate depths, but the measured meridional flow in the depth range of $25\pm  7$~Mm is reversed compared to the flows observed near the surface \citep{2023arXiv231211699T}.

\subsubsection{Zonal Flows} \label{sec:zonal} 
\begin{figure}\centering \includegraphics[width=0.8\linewidth,height=5.5cm]{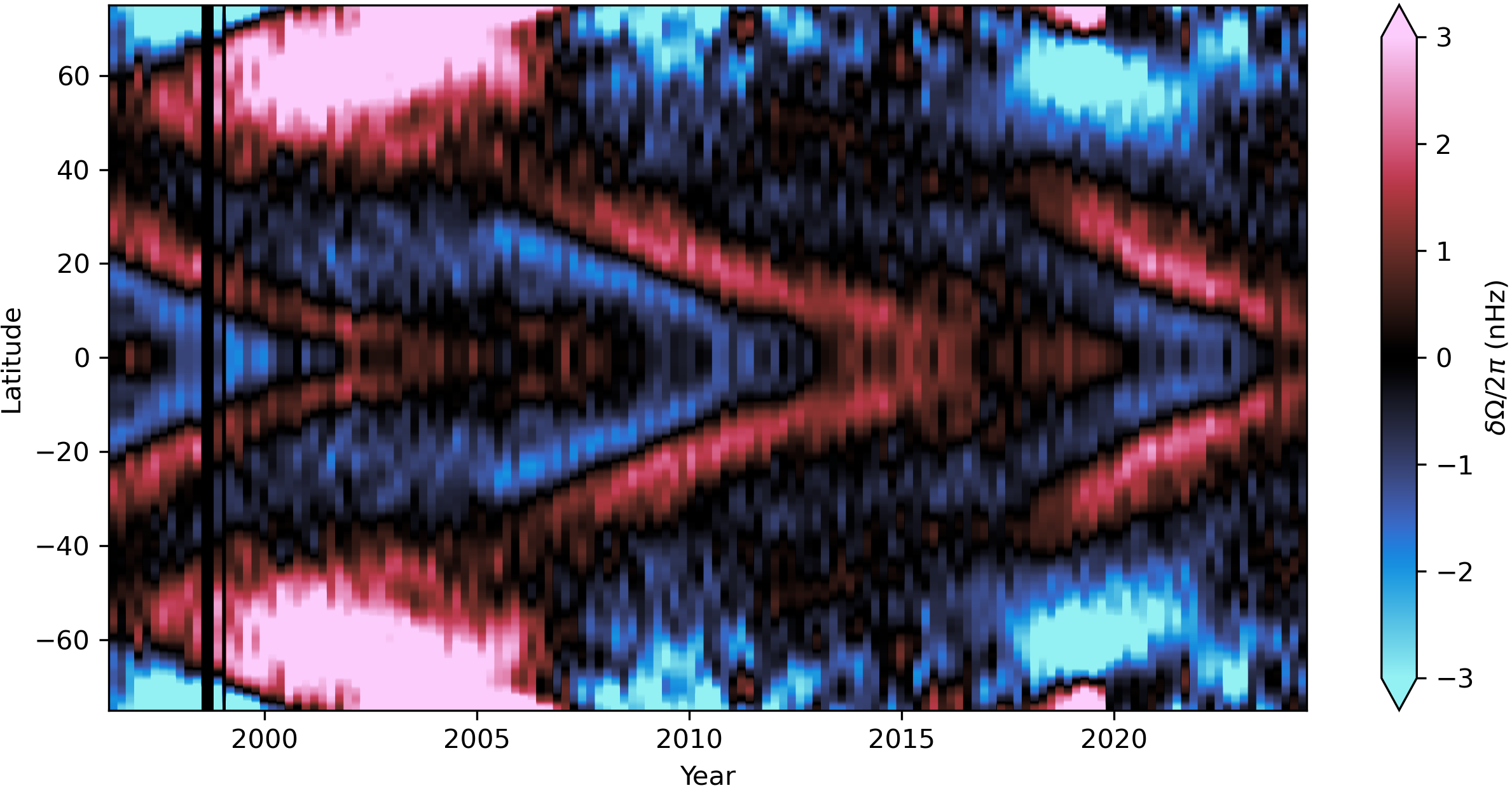} \caption{Time-latitude map of zonal-flow residuals at a target depth of $0.99\,R_{\odot}$ from MDI (1996\,--\,2010) and HMI (2011\,--\,2024) observations.} \label{fig:rhzon1} \end{figure}

\begin{figure}\centering \includegraphics[width=0.95\linewidth]{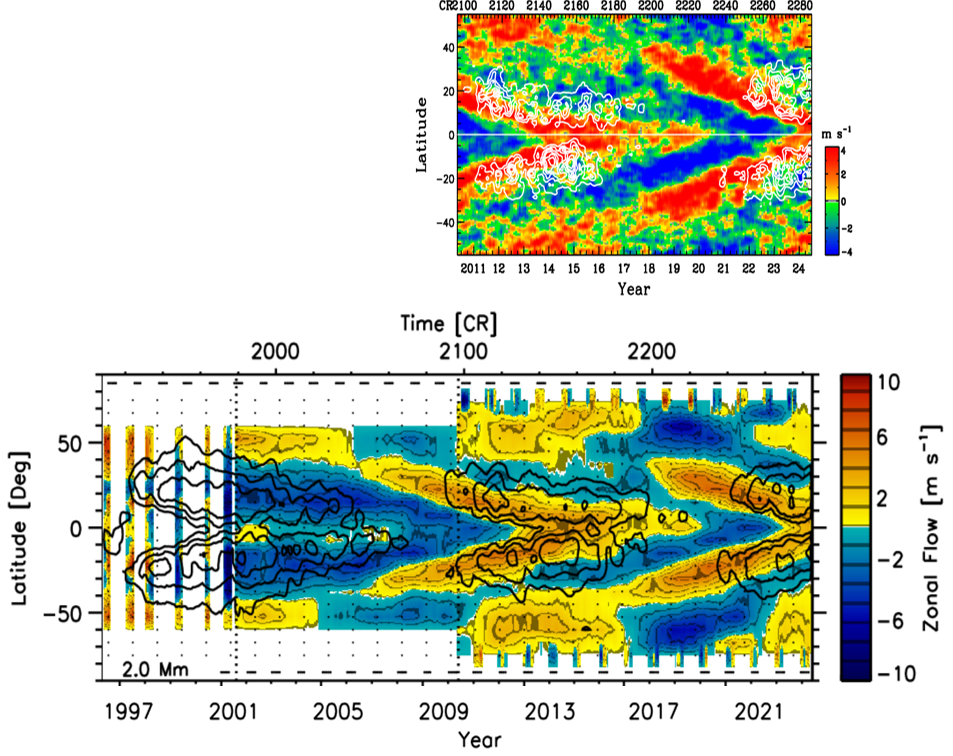} \caption{Time-latitude map of zonal flow residuals (torsional oscillations) from (\textit{Top}) the time-distance method at near the surface layer (0\,--\,1 Mm), and (\textit{Bottom}) the ring-diagram method at a depth 2.0 Mm. The temporal mean has been subtracted at every latitude. The time coordinate is given in years (\textit{bottom x-axis}) and CRs (\textit{top x-axis}). The \textit{vertical dotted lines} at CR 1979 and CR 2097 indicate the starting date of GONG and HMI data. \textit{Black contours} indicate magnetic activity (5, 10, 20, and 40 G) smoothed over five CRs. This figure is an extended version of Figure~1 of \citet{Komm2018}.   } \label{fig:LHZonal} \end{figure}

Zonal flows appear as a modulation of the solar rotation rate and can be inferred by making a data cube of successive two-dimensional rotation-rate inferences and then subtracting a temporal mean at each latitude and depth. When using inversions of global modes, this provides only the equatorially symmetric part of the pattern; to see the anti-symmetric component, we need to use results from local helioseismology. Figure~\ref{fig:rhzon1} shows the zonal-flow measurements at a target depth of $0.99\,R_{\odot}$ from MDI and HMI observations. The equatorward-migrating branches associated with Solar Cycles 23, 24, and 25 are clearly visible, while the behavior at higher latitudes is somewhat less regular. Similar results are obtained from the local helioseismology techniques (Figure~\ref{fig:LHZonal}).

\begin{figure} 
	\begin{center} 
    \includegraphics[width=0.9\linewidth]{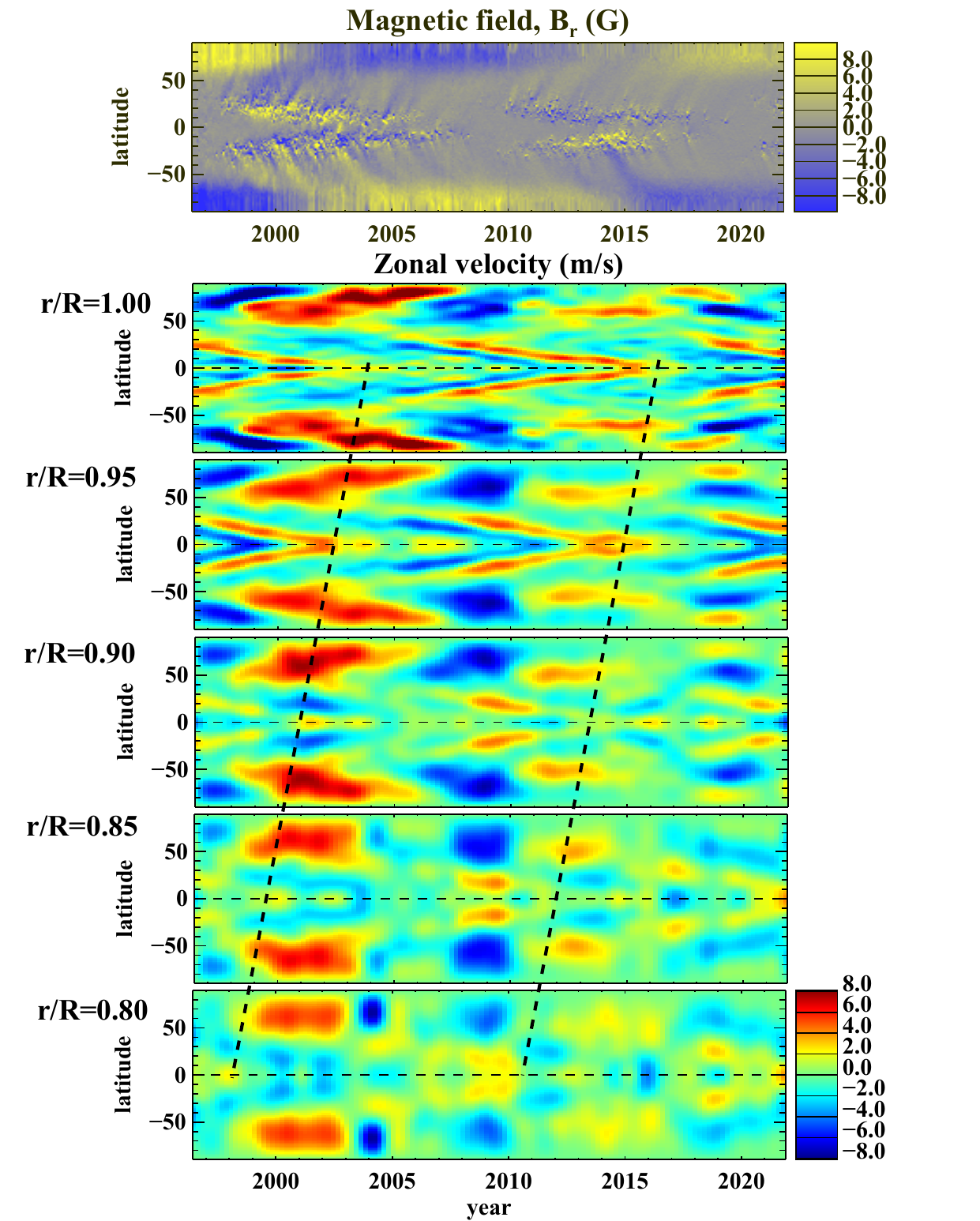} 
		\caption{Time-latitude diagrams for the radial magnetic field in 1996\,--\,2021, obtained from MDI, HMI, and SOLIS data, and the zonal flow velocity calculated from the solar rotation inversion data of MDI and HMI, available from JSOC. The \textit{inclined dashed lines} track the points of convergence of the fast rotating zonal flows at the equator, indicating the end of the extended Solar Cycles 22 and 23 \citep{Kosovichev2023_IAU362}.} \label{fig3} \end{center} \end{figure} 
        
        \begin{figure} 
	\begin{center} 
    \includegraphics[width=0.85\linewidth]{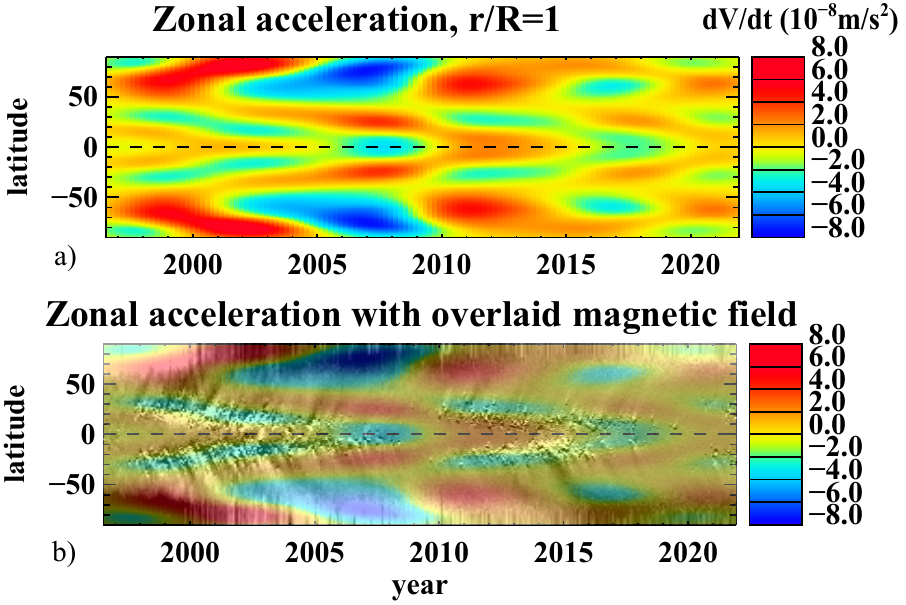}
		\caption{a) Time-latitude diagram of the zonal flow acceleration close to the solar surface from the MDI and HMI data from 1996 to 2022.  b) Overlay of the zonal acceleration and the magnetic butterfly diagram for the same period \citep{2019ApJ...871L..20K,Kosovichev2023_IAU362}.} \label{fig4} \end{center} \end{figure} 
        
        \begin{figure}[ht] 
	\begin{center} 
    \includegraphics[width=0.85\linewidth]{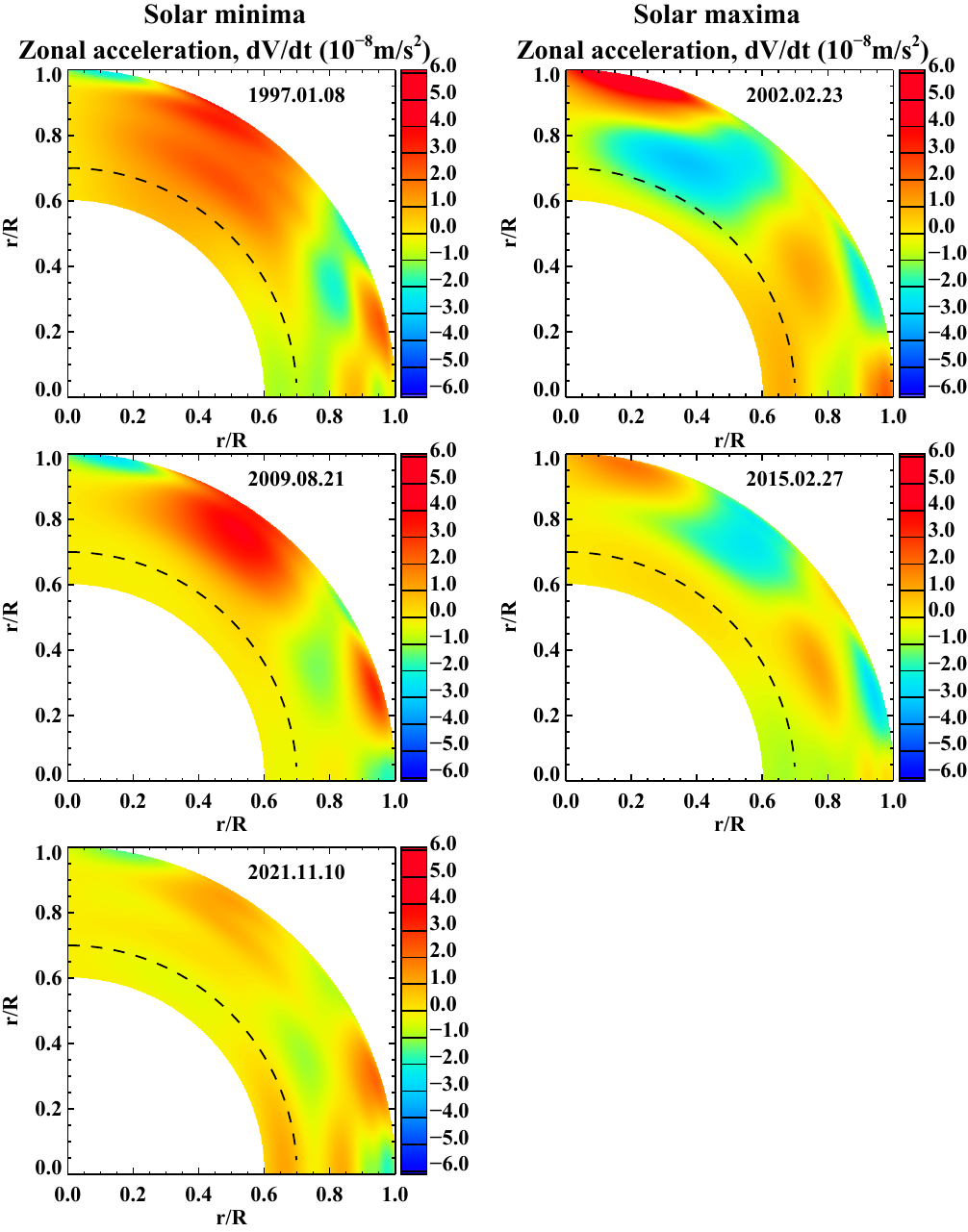} 
		\caption{Distributions of the zonal acceleration in the solar interior during the periods corresponding to the solar minima in 1997, 2009, and 2021 (\textit{left column}) and during the solar maxima in 2002 and 2015 (\textit{right columns}) \citep{2019ApJ...887..215P,Kosovichev2023_IAU362}} \label{fig5} \end{center} \end{figure} 
        
\begin{figure} 
\centering 
\includegraphics[width=0.6\linewidth]{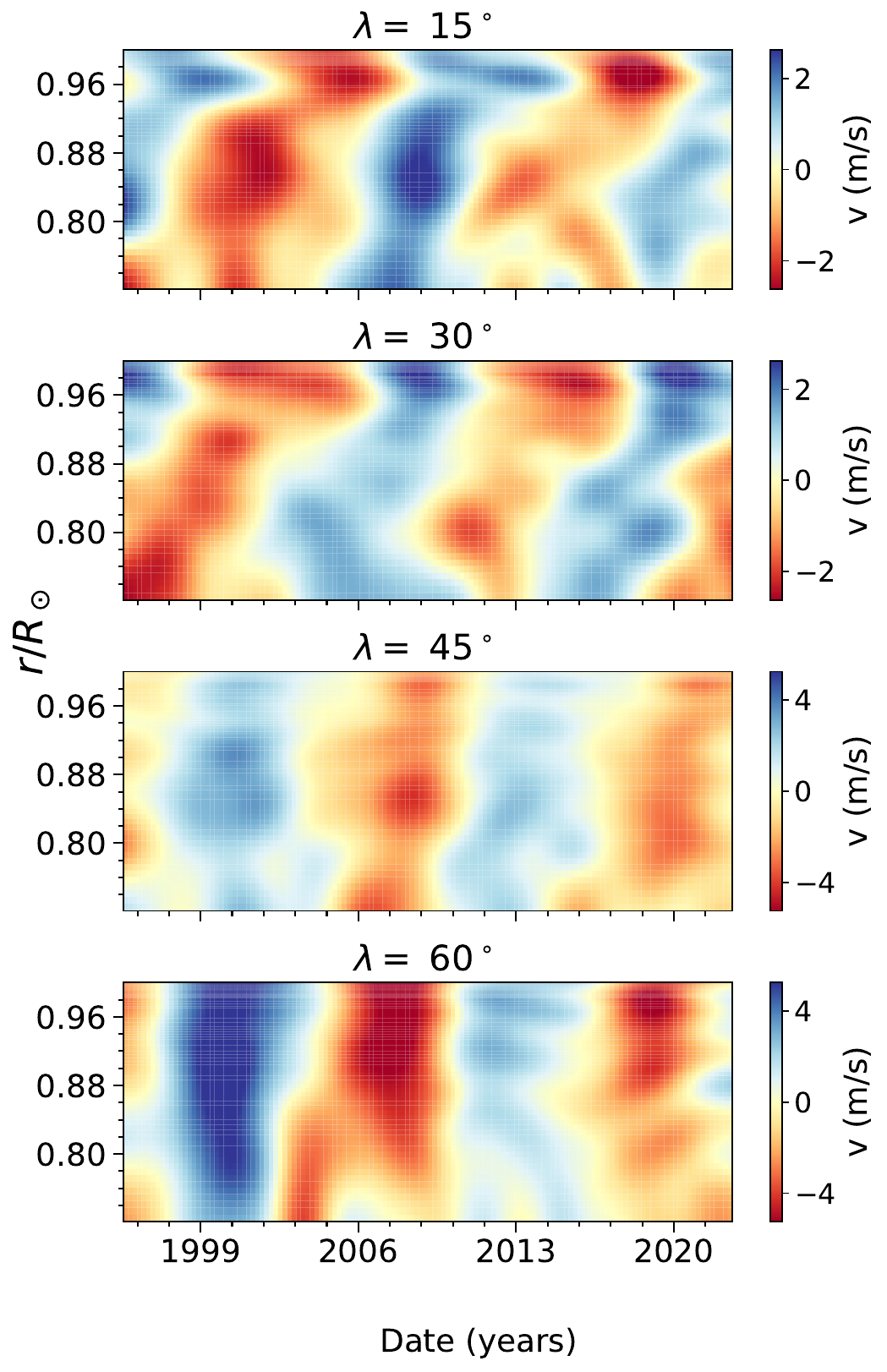} \caption{Zonal flow velocity obtained using combined MDI and HMI datasets is plotted as a function of time and radius for four latitudes, indicated in the title of each panel \citep{2024arXiv240215647M}.} \label{fig:zf_depth_year} \end{figure}

Subtracting a temporal mean over the whole data set, which now covers multiple solar cycles, can cause some problems because the ``underlying'' rotation rate may not be the same as the temporal mean over the available observations. Initial measurements of the torsional oscillations incorporating the HMI data \citep{2013ApJ...767L..20H} found a substantially weaker appearance of the high-latitude branch of the pattern corresponding to Solar Cycle 24 compared to this branch observed in Solar Cycle 23, but much of this appearance can be explained by the slower underlying rotation rate in Cycle 24. This slowing-down of the mean rotation rate at the high latitudes could result from weaker polar fields in Cycle 24, as suggested by a flux-transport model \citep{2012ApJ...750L...8R}, which showed that a reduction in the cycle amplitude leads to a drop in the high-latitude rotation rate. By 2018, with nearly two full cycles of data, the relative slow-down in the higher-latitude rotation for Cycle 24 was even more apparent, while the beginning of the equatorward branch of the torsional oscillation pattern for Cycle 25 was visible at mid-latitudes \citep[e.g.][]{2018ApJ...862L...5H,2019ApJ...883...93B}. 

The hemispheric asymmetry in the solar-cycle evolution of the zonal and meridional flows was initially detected by the time-distance helioseismology analysis using the first 3.7 years of data acquired by SDO/HMI during the rising phase of Cycle 24 \citep{Zhao2014}. A ring-diagram analysis with $30^\circ$ tiles to examine both zonal and meridional flows in HMI observations revealed that the band of faster zonal flow associated with Cycle 25 appeared first in the northern hemisphere. In the range of depths covered by this analysis, down to about 32\,Mm below the surface, the zonal flows appeared almost simultaneously \citep{2021SoPh..296..174K}. However, a global helioseismology analysis that provides measurements of the zonal flows through the whole convection zone (Figure~\ref{fig3}) indicated that the flow bands in the deeper layers appear earlier than in the subsurface layers \citep{2019ApJ...871L..20K} and the time lag of the polar branches is shorter than that of the equatorial branches (indicated by dashed lines in Figure~\ref{fig3}. These flow maps were obtained by subtracting the mean rotational velocities observed by the MDI and HMI instruments from the individual 72-day measurements and then smoothing the variation with a one-year window function.

However, for understanding the physical mechanism of the zonal flows and comparison with models, it is beneficial to consider the zonal acceleration \citep{2019ApJ...871L..20K} or the rate of change of the rotation rate \citep{2024SoPh..299...38M}. Comparison of the time-latitude diagram of the zonal acceleration calculated for a near-surface layer with the photospheric magnetic field map (Figure~\ref{fig4}) showed that the magnetic field regions at low and high latitude coincide with the regions of zonal deceleration. Tracing the evolution of the zonal acceleration with the solar cycle (extracted from the observed variations of the zonal flows by applying a Principal Component Analysis) revealed a pattern of waves originating near the bottom of the convection zone and traveling in radius and latitude \citep[Figure~\ref{fig5}; ][]{2019ApJ...871L..20K}. Two basic wave branches were identified, a relatively fast high-latitude (polar) branch that reaches the surface in 1\,--\,2 years and an equatorial branch that takes about 11 years to reach the surface \citep[Figure~\ref{fig:zf_depth_year}; ][]{2024arXiv240215647M}. It is intriguing that the data in this figure show a special role of the near-surface shear layer, in particular, reverse migration of the flow pattern from the surface to the bottom of this layer, indicating that the solar dynamo generates magnetic field not only in the deep convection zone but also the near-surface layer.

\subsubsection{Temporal Variation of the Near-Surface Shear Layer}

As the amplitude of the zonal flow varies with depth, it might be expected that the gradient of the near-surface shear will also show cycle-related changes as the torsional oscillation moves through it. In fact, however, recent studies suggest that changes in the near-surface shear are related more directly to the latitudes of magnetic activity. The variation of the near-surface shear layer in \textit{f}-mode data from the 360-day MDI and HMI power spectra during Solar Cycle 23 showed an increase in the steepness of the gradient in phase with the solar cycle at low latitudes \citep{2016A&A...595A...8B}. In addition, \citet{2022ApJ...924...19A}, using a combination of \textit{f}-mode and \textit{p}-mode data, found that the radial gradient increases in steepness with increasing solar activity at 0.990~\sunrad\ and decreases in steepness at 0.950~\sunrad\ during Solar Cycles~23 and 24. 
The changes are more pronounced in the active latitudes than at adjoining higher latitudes. Results at high latitudes of about 70\degree\ and poleward are unreliable. The NSSL is deeper at lower latitudes than at high latitudes, with a maximum at the equator. The extent of the NSSL also varies with the solar cycle. More recently, \citet{Rozelot2025} found that the radial gradient of solar rotation is enhanced in the latitudinal zones of sunspot and active region formation. In middle and low latitudes, the gradient enhancements below the leptocline follow the magnetic butterfly diagram. However, in the leptocline, the latitudinal patterns of the enhanced gradient are more complicated, resembling the overlapping `extended' solar cycles of the torsional oscillations.

Local helioseismic techniques, such as ring-diagram analysis, \citep[for example, ][]{2009ASPC..416...99Z} also found that the radial gradient of the rotation rate in the NSSL varies with the solar cycle.  In particular, in regions with strong magnetic flux, the radial gradient decreases (in magnitude) close to the surface at 0.997~\sunrad\ (2.0~Mm) but increases in deeper layers at 0.990~\sunrad\ (7.0~Mm) \citep{2022FrASS...917414K,2022ApJ...924...19A}, suggesting that the solar-cycle effects on the shear were likely to reflect the ``presence or absence of magnetic flux above a certain threshold.'' In addition, the thickness of the NSSL varies with the solar cycle: it extends to about 0.952~\sunrad\ (29.4~Mm) at locations of active regions compared to 0.956~\sunrad\ (35.3~Mm) at those of quiet regions \citep{2022ApJ...924...19A,2023SoPh..298..119K}.

\subsubsection{Meridional Flow Modulation} 
\label{sec:meridional_vary}

The ring-diagram and time-distance analyses of the solar-cycle variation of the meridional flow showed a similar pattern to the torsional oscillation, with bands of converging meridional flow moving from mid- to low latitudes during the solar cycle \citep{2021SoPh..296..174K,Getling2021}. At $45^\circ$ latitude, these bands were found to appear earlier at greater depths, apparently taking about two years to rise through the outer layers of the convection zone from the 32\,Mm depth, but at lower latitudes, they appear almost simultaneously at all depths.

\begin{figure} \includegraphics[width=\linewidth]{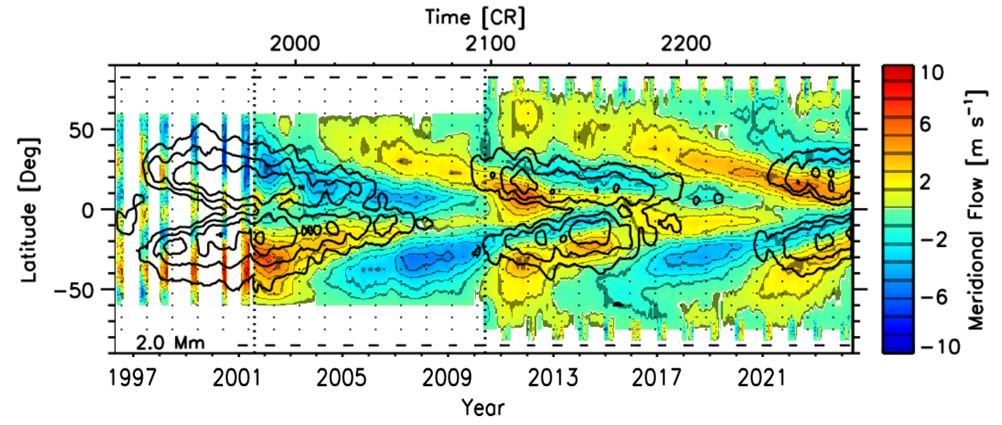} \caption{Time--latitude map of meridional flow residuals from the ring-diagram method at a depth 2.0 Mm. Positive values correspond to northward flows, and negative - to southward flows. The temporal mean has been subtracted at every latitude. The time coordinate is given in years (\textit{bottom x-axis}) and CRs (\textit{top x-axis}). The \textit{vertical dotted lines} at CR 1979 and CR 2097 indicate the starting date of GONG and HMI data. \textit{Black contours }indicate magnetic activity (5, 10, 20, and 40 G) smoothed over five CRs. This figure is an extended version of Figure~2 of \citet{Komm2018}.} \label{fig:LHMeridional} \end{figure}

\begin{figure} 
	\begin{center} \includegraphics[width=\linewidth]{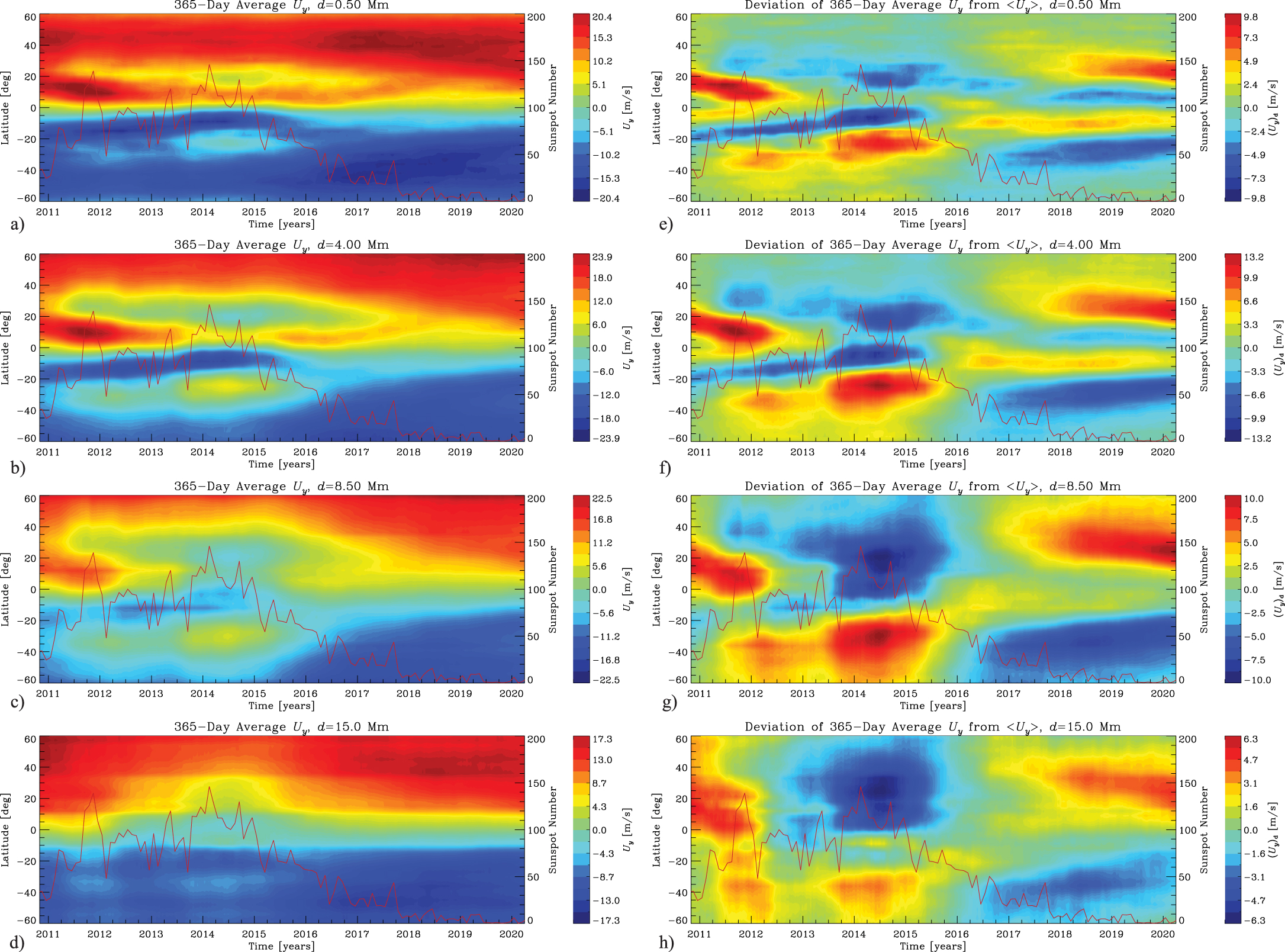} 
		\caption{Measurements of subsurface meridional circulation by time-distance helioseismology. \textit{Left}: time–latitude diagrams representing the 365-day running average of the meridional velocity at depth levels $d$ = 0.5, 4.0, 8.5, 15.0 Mm; \textit{right}: the deviations from the mean over the whole interval at the same depths \citep{Getling2021}.} \label{fig11} \end{center} \end{figure}

Figure~\ref{fig11} displays the time variations of the meridional (South-North) flow and the deviations from the mean revealing migrating bands in meridional flows, obtained by the time-distance method \citep{Getling2021}. Apparently, the meridional flow variations reveal the extended solar-cycle structure originally found in the zonal flows. It is particularly apparent from Figure~\ref{fig:LHMeridional}, which combines the data obtained from GONG, MDI, and HMI by the ring-diagram technique. Similarly to the zonal flows, the near-surface meridional flow showed a hemispheric asymmetry correlated with the asymmetry of magnetic activity, leading the activity by about three years \citep{Lekshmi2019}.

These results show that the meridional flow is significantly disturbed during the solar activity maxima, resulting in a substantial decrease in the flow speed from mid-latitudes to the polar regions and even the appearance of flows across the solar equator \citep{Komm2020}, which may affect magnetic flux transport.

While at least some of the solar-cycle modulation of the near-surface meridional flow can be attributed to inflows into active regions, several studies have found that the active regions cannot account for the full observed effect. 
The contribution from active regions was found to be quite shallow, with maximum strength at 3.1\,Mm depth \citep{Komm2020}, while the solar-cycle variations of the meridional flow extend much deeper. 
In addition, excluding active region areas did not completely remove these variations, leading to a suggestion that the meridional flow is made up of three components: 1) active-region inflows, 2) a variation of about 2 m\,s$^{-1}$ on solar-cycle timescales that cannot be attributed to active regions, and 3) the quiet-Sun background meridional flow \citep{Mahajan2023}.

The surface meridional flow and its variations have been studied by tracking the movement of the Sun's magnetic network during the period May 1996 through May 2022 \citep{hathaway2022}. The results suggested that the meridional flow disappears above about 80$\degree$ latitude and polar counter cells, i.e., equatorward flow from the poles down to about $60\degree$ latitude, appearing from time to time, in general agreement with previous helioseismology results \citep{2013SoPh..287...85K}. However, more work is needed to confirm both findings. 

\subsection{Structural Changes in the Solar Convection Zone}

\begin{figure} \includegraphics[width=\linewidth]{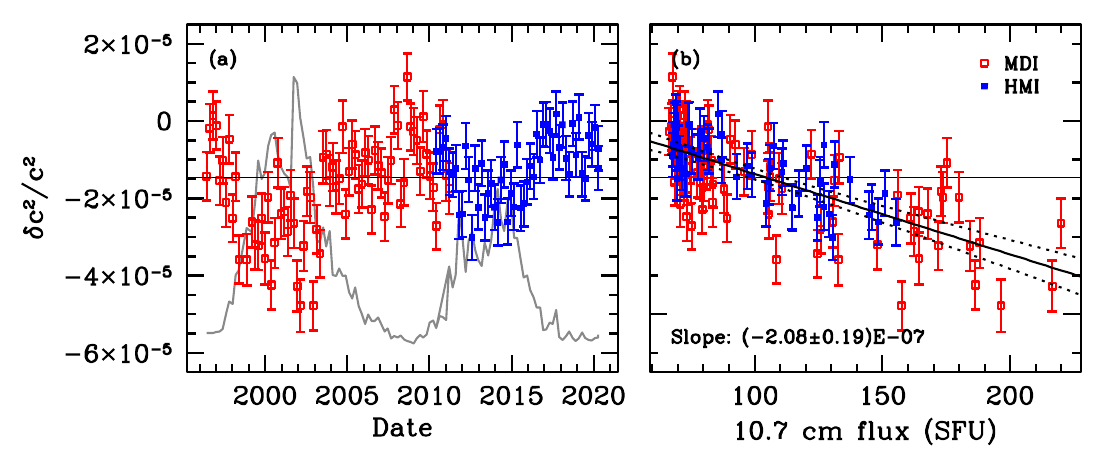} \caption{The changes in the squared sound speed of the Sun at different epochs with respect to the Sun between 27 July 2007 to 16 July 2009. The differences are for a region centered at $0.75$ R$_\odot$. \textit{Panel (a)} shows the differences plotted as a function of time, and \textit{panel (b)} shows the differences as a function of the 10.7cm flux. The \textit{gray curve} in \textit{panel (a)} shows the variation of the 10.7 cm radio flux as a function of time. The \textit{black straight line} in \textit{panel (b)} is the result of a linear regression analysis  \citep{2021ApJ...917...45B}. } \label{fig:czchanges} \end{figure}

Analysis of the sound-speed variations in the solar interior between the activity minimum and maximum of Solar Cycle 23 using global helioseismology data revealed a two-layer configuration in the sound-speed difference, reminiscent of the structure observed beneath active regions \citep[Figure~6 in][]{2012ApJ...745..184R}. 
Near the surface (around 5.5 Mm depth), the sound speed at solar maximum was found to be lower than at solar minimum. This difference decreased until about 7 Mm depth, beyond which the trend reversed. In deeper layers (7-10 Mm), the sound speed at maximum became higher than at minimum, with the relative difference peaking around 10 Mm depth. The magnitude of this deeper positive difference was less than half of the negative difference observed near the surface. These results are consistent with predictions of a magnetic convection model, which considers the magnetic inhibition of convection onset in the presence of a vertical magnetic field \citep{2012ApJ...755...79M}. 
However, it is important to note that the interpretation of these results is not straightforward.  The observed changes in mode frequencies can be attributed to both structural changes in the medium through which the waves propagate and direct effects of magnetic fields on the plasma waves via the Lorentz force \citep{2012ApJ...745..184R} or reflect a combination of structural and non-structural effects of the magnetic fields \citep{2006ESASP.624E..58L}.  

Unlike the very clear changes in solar dynamics, changes in solar structure are small and difficult to detect. Helioseismic data obtained during Solar Cycle~23 hinted at changes in the second helium ionization zone \citep{2004ApJ...617L.155B} and the base of the convection zone \citep{2008ApJ...686.1349B}, but these were not confirmed until HMI data showed that the changes could be seen in Solar Cycle~24 \citep{2020ApJ...903L..29W}. This analysis showed that there is indeed a change in the layers that correspond to the He~II ionization zone, which must be due to temperature changes in that layer. The time variation is easily discernible because the helioseismic data sets used covered two solar cycles.


In addition, small changes, of the order of a few parts in $10^5$ in the of the Sun just below and just above the base of the convection zone (Figure~\ref{fig:czchanges}) were detected by sacrificing spatial resolution in favor of the temporal resolution \citep{2021ApJ...917...45B}. These changes are anti-correlated with solar activity, i.e., the sound speed is higher at the solar minimum than at the solar maximum. There also appears to be a change in the gradient of the sound speed at the convection-zone base, which could be an indication that the position of the base changes with activity. While there is some evidence that the radial location of the base of the convection zone has a slight latitude dependence, evidence of the time variation of this asphericity is marginal \citep{2024ApJ...964....8B}.

\subsection{Quasi-Biennial Oscillations}

In addition to the 11-year cyclic patterns, shorter quasi-periodic variations are also found in the helioseismic data. These variations, commonly known as quasi-biennial oscillations (QBOs), with periods ranging from 1.5 to 4 years, were initially identified in low-$\ell$ frequencies from the integrated-light instruments \citep{2010ApJ...718L..19F}, but later found in the intermediate-$\ell$ frequencies from GONG \citep{2011ApJ...739....6J} too. Subsequent studies using helioseismic data from various sources, including HMI, support these findings \citep[e.g.,][]{2013ApJ...765..100S, 2022MNRAS.515.2415M}, and also reported two different QBO periods, about 2 years in Cycle 23 and 3 years in Cycle 24 \citep{2023ApJ...959...16J}. \citet{2021ApJ...920...49I,Bogart2023} studied the presence and spatio-temporal evolution of QBOs in the rotation-rate residuals in NSSL and showed the QBO-like signals in each latitudinal band and depth; however, their amplitudes increase with increasing depth, suggesting the source region of the QBO to be below 0.78\,\sunrad. 
It is worth mentioning that QBOs are not limited to the helioseismic data only; they are also present in various activity indices and have been widely discussed in numerous papers.

The origin and periods of QBOs have been a subject of debate, and several mechanisms have been suggested.  For instance, two dynamo sources separated in space give rise to two different periods \citep{1998ApJ...509L..49B}. The first source of the dynamo is located near the bottom of the convection zone responsible for the 11-year cycle, while the second source operating near the top in the near-surface shear layer could give rise to short-term periodicities. While some studies based on the helioseismic data suggest that the magnetic field responsible for quasi-periodic variations is anchored within the NSSL \citep{2010ApJ...718L..19F, 2022MNRAS.515.2415M,2023ApJ...959...16J}, there are arguments that QBOs could arise from the beating between two types of magnetic configuration of the solar dynamo \citep{2013ApJ...765..100S}, or from an interplay between the flow and magnetic fields \citep{2019A&A...625A.117I}.

\subsection{Implications for Solar Dynamo Theories}

All manifestations of solar activity, from solar irradiance variations to solar flares, are caused by the Sun's magnetic field. This magnetic field is believed to be generated by a dynamo mechanism operating in the Sun's convection zone \citep{1955ApJ...122..293P}, deep below the solar surface, although proposals exist for the near-surface origin of the Sun's magnetic field \citep{2024Natur.629..769V}. Despite substantial modeling and simulation efforts, our understanding of how the magnetic field is generated, transported to the solar surface, and forms sunspots remains very limited.

Currently, we cannot measure subsurface magnetic fields by helioseismology. Thus, any knowledge about the dynamo process within must be gathered from measuring large-scale subsurface flows, such as differential rotation and meridional flow. It is expected that the 11-year sunspot-cycle evolution of subsurface magnetic fields modulates these flows via magnetic stresses, variations in turbulent heat transport, and inertial forces. Therefore, analyzing the solar-cycle-scale variations in zonal and meridional flows provides us with a way to peek into the heart of the solar dynamo.

Numerous explanations have been proposed for the measured variations in zonal flow (torsional oscillation) ranging from Lorentz forces \citep{1981A&A....94L..17S,2000A&A...360L..21C,2016ApJ...828L...3G} that result in a magnetic torque \citep[e.g.][]{2019PhDT.......170M}, inertial forces that are driven by a combination of meridional flow and momentum conservation \citep{2013SoPh..282..335B}, enhanced thermal cooling around active regions causing geostrophic flows \citep{2003SoPh..213....1S}, and thermally induced deviations of the solar rotation from the cylindrical (Taylor-Proudman) state \citep{2007ApJ...655..651R}. If the torsional oscillation were caused by active regions, then one would expect their amplitude to be closer to active regions and lower amplitude farther away. However, no changes in the amplitude as a function of distance from the centroids of the active regions were detected \citep{Mahajan2023}. It was particularly difficult to explain why the migrating zonal flows formed the extended 22-year cycle (Figure~\ref{fig4}) and remained prominent during the solar minima when there was no significant magnetic flux on the surface.

A recent dynamical mean-field dynamo explained the observed zonal flows, particularly the extended 22-year solar cycle phenomenon \citep{2019ApJ...887..215P}. According to this model, the extended cycle of the solar torsional oscillations (zonal flow variations) is caused by the effects of the overlapped dynamo waves in the solar interior on the angular momentum and heat transport in the solar convection zone.  At high latitudes, the zonal flows are driven by the large-scale Lorentz force with contributions from magnetic quenching of the convective heat flux.  The toroidal magnetic field associated with the dynamo wave propagating towards the equatorial zone results in the quenching of the convective heat flux.  The variations of the meridional circulation affect the amplitude of the zonal flows and produce their effective drifts in radius and latitude, causing the extended 22-year cycles. This model predicted the variations of the meridional flow towards the active latitude and the extended 22-year cycle of these variations in the near-surface layer, which was confirmed by helioseismic observations \citep[Figures~\ref{fig:LHMeridional} and \ref{fig11}; ][]{Komm2018,Getling2021}.

The observed evolution of the zonal acceleration in the solar convection zone in the form of traveling waves \citep[Figures~\ref{fig5} and \ref{fig:zf_depth_year}; ][]{2019ApJ...871L..20K,2024arXiv240215647M} suggested proposed that the variations in zonal flows are a signature of dynamo wave that begins near the base of the convection zone and propagates upwards taking multiple years to reach the surface. This would mean that there is predictive power in observing zonal flows in the interior.

	\subsection{Solar Cycle Variation in Helioseismic Data and its Implications for Asteroseismology}
	
	As discussed above, the magnetic activity cycle influences the central frequency of the solar acoustic modes in all $\ell$ ranges. With the advent of space missions that collected long and continuous high-quality photometric observations, asteroseismic studies of hundreds of solar-like stars were possible. Indeed, with CNES/ESA Convection, Rotation, and exoplanet Transits, \citep[CoRoT: ][]{2006cosp...36.3749B}, NASA {Kepler} \citep{2010Sci...327..977B}, K2 \citep{2014PASP..126..398H}, and the Transit Exoplanet Survey Satellite \citep[TESS:][]{2014SPIE.9143E..20R}, astroseismology has gone through a revolution \citep[see][]{2019LRSP...16....4G}. The ESA mission PLAnetary Transits and stellar Oscillations \citep[PLATO: ][]{2014ExA....38..249R,2024arXiv240605447R}, planned to be launched late 2026, also promises new discoveries with tens of thousands of solar-like stars well characterized by seismology \citep[][]{2024A&A...683A..78G}.
	
	{The CoRoT target, HD\,49933, with individual modes characterized \citep[see][]{2008A&A...488..705A,2009A&A...507L..13B}, was the first solar-like star for which frequency shifts and amplitude variations were observed to be respectively correlated and anti-correlated to the photometric dispersion of the light curves as in the Sun. This work uncovered the existence of a magnetic activity modulation in the star \citep[][]{2010Sci...329.1032G,2011A&A...530A.127S} and opened a new window for studying magnetic activity in other solar-like stars.}
	Nowadays, it is possible to characterize all the seismic parameters of individual modes, including the rotational splittings in main-sequence solar-like stars \citep[e.g.,][]{2011A&A...530A..97B,2013PNAS..11013267G,2015MNRAS.446.2959D,2021NatAs...5..707H}, allowing us to have better information on the structure and dynamics of the stellar interiors in an evolutionary context, as well as providing an estimation of the stellar inclination axis \citep[][]{2003ApJ...589.1009G,2006MNRAS.369.1281B,2018MNRAS.479..391K,2019MNRAS.488..572K}, a key parameter to better characterize exoplanet systems \citep[][and references therein]{2018ASSP...49..119H}. These seismic analysis -- complemented by the determination of the mean stellar surface rotation rate \citep[e.g.,][]{2014ApJS..211...24M,2014A&A...572A..34G,2017A&A...605A.111C,2019ApJS..244...21S,2021ApJS..255...17S,2021ApJ...913...70G,2022ApJ...936..138H} -- have challenged current gyrochronology models \citep[e.g.,][]{1972ApJ...171..565S,2007ApJ...669.1167B,2015Natur.517..589M,2016ApJ...823...16B}. Around the age of the Sun and older, stars seem to be spinning faster than expected leading to weakened magnetic braking \citep[e.g.,][]{2015MNRAS.450.1787A,2016Natur.529..181V,2021NatAs...5..707H,2023MNRAS.523.5947S,2024ApJ...962..138S}. Although the physical origin to this weakened magnetic braking is not yet known \citep[e.g.,][]{2023MNRAS.520..418T}, it would impact the evolution of magnetic cycles and dynamos at intermediate main-sequence ages \citep[e.g.,][]{2019ApJ...871...39M,2021ApJ...921..122M,2022ApJ...933L..17M}.

	Different predictions of detection of frequency shifts due to stellar activity were obtained based on different assumptions \citep[][]{2007MNRAS.377...17C,2007MNRAS.379L..16M}. With the {Kepler} mission, frequency shifts and amplitude variation of acoustic modes were detected in 87 stars \citep[][]{2016A&A...596A..31S,2018A&A...611A..84S,2017A&A...598A..77K,2018ApJS..237...17S}. This larger sample can provide insights into the theoretical understanding of the changes in the acoustic-mode properties. The sample contains stars with different temperatures, metallicities, and ages, allowing us to study the impact of these parameters on the modes \citep{2019ApJ...883...65S} and also on the magnetic activity \citep[e.g.,][]{2023MNRAS.524.5781S,2024MNRAS.533.1290S}.
	
	One very good case is the {Kepler} target KIC~8006161, which is very similar to the Sun in terms of its effective temperature, surface gravity, and age. Thanks to a spectroscopic follow-up of that target for several decades, it is known to have a magnetic activity cycle period of $\sim$\,7.4 years, shorter than the Sun, and with a much stronger chromospheric activity. Differences were also observed for the frequency shifts and amplitude variation \citep[][]{2018ApJ...852...46K}. All these differences compared to the Sun could all be explained by the fact that the star is metal-rich (0.3\,dex), leading to a deeper convective zone and a stronger differential rotation. In another example, \citet{Lombardi2025} compared frequency shifts of a solar-type star KIC~6106415, which has a mass close to the solar mass, is about 20\% bigger and rotates about 50\% faster, with the solar-cycle frequency shift from the Global Oscillations at Low Frequency (GOLF) instrument \citep{Gabriel1995} on the SoHO Mission. They found that the frequency shifts observed for this star are greater than for the Sun, presumably due to stronger starspot number variations.     
    
    These analyses showed how asteroseismology can provide more constraints to better understand magnetic activity in other stars. Moreover, despite the absence of images of stars, it has been shown that it could be possible to have seismic constraints on active latitudes and differential rotation \citep{2018A&A...619L...9B, 2018Sci...361.1231B, 2019A&A...623A.125B, 2019MNRAS.485.3857T, 2023A&A...680A..27B}, both being key ingredients to better understand stellar magnetic activity cycles and magnetic dynamos.
	
	Since higher magnetic activity suppresses the amplitude of the oscillation modes, mode detection will be more difficult for very active stars. Indeed, stars without detection of solar-like oscillations usually have a high level of magnetic activity \citep{2010Sci...329.1032G, 2011ApJ...732L...5C, 2019FrASS...6...46M, 2022ApJ...940...93D}. This needs to be considered when predicting the detection of modes in future space missions.
	
	Understanding the effects of the magnetic cycle on the modes is extremely important as mode frequencies are modified, and they are used to constrain the stellar models to characterize stars \citep{2014A&A...569A..21L}. Any changes in those frequencies would modify the final inferred parameters  \citep[e.g.,][]{2020MNRAS.493L..49H}. There is an angular angular-degree dependence of the \textit{p}-modes with the cycle, which is attributed to the non-spherical nature of the magnetic activity within the stellar convective envelope. These variations in mode frequencies can influence the small frequency separation between low-degree modes (i.e., the frequency difference between consecutive quadrupole and radial modes), which is sensitive to the core structure and, consequently, to its age. Therefore, estimating global stellar parameters, such as age, based on mode frequencies at a specific phase in the magnetic activity cycle could lead to biased results. \citet{2019FrASS...6...41P} estimated that these errors could reach up to 10\% in age and a few percent in mass and radius. Additionally, the frequency dependence of the frequency shifts can also bias the inferred global stellar parameters. In the solar case, this dependence is a smooth function of frequency which is partially masked by the so-called ``surface-effect'' correction \citep[e.g.,][]{2008ApJ...683L.175K,2014A&A...568A.123B, 2017MNRAS.466L..43T}. However, observations of other stars by \citet{2018A&A...611A..84S} suggest the presence of an oscillatory component with a period corresponding to the acoustic depth of the He II zone. This could result in the incorrect estimation of certain global stellar parameters, such as helium abundance. Calculations done by \citet{2019FrASS...6...41P} showed that the uncertainties caused by this effect could be below 3\% for some stars observed by {Kepler}. In the case of the Sun, the error on the age could be up to 6.5\% when comparing observations obtained during minimum or maximum magnetic activity \citep{2024A&A...688L..17B}.
	
	In summary, continuous full-disk observations from HMI and other contemporary instruments continue to advance our understanding of the variability of the solar interior and its connection with the solar atmosphere. The oscillations also reveal small but measurable structural changes with time. The measured flows, e.g., particular torsional oscillations, are proven to be a good precursor for the onset of the next solar cycle. However, the results suggest the complex nature of the solar oscillations and their variability with time that may have implications on the solar dynamo models.   Thus, an improved understanding of the variability below the surface is crucial for constraining the solar dynamo models.
	
\section{Seismology of Sunspots and Active Regions} 

With the advent of helioseismology and the progress made by both global and local helioseismic techniques, much progress has been made in understanding the interior dynamics and rotation profile of the Sun, as described in the earlier sections. However, understanding the structure and dynamics of sunspots and active regions below the surface has been a formidable task due to the complex nature of the interaction between the magnetic field and acoustic waves \citep[for previous reviews see][]{2005LRSP....2....6G, 2012SoPh..279..323K, 2023FrASS..1091777T}.

Solar active regions and sunspots are the manifestation of the solar dynamo generating magnetic field in the solar interior. Thus, the observation of these magnetized regions at the solar surface provides a crucial link between the solar interior and the surface. Although local helioseismic studies have been undertaken for over three decades, significant uncertainty exists in the interpretation and modeling of helioseismic signals due to the complexity of wave interactions with magnetic fields. Recent forward modeling and numerical simulations suggest that an active region works as a window through which seismic waves can leak upwards into the solar atmosphere and downwards into the interior to contaminate the properties of the acoustic waves \citep{2007AN....328..286C, 2013MNRAS.435.2589C}. Here we review some recent results primarily obtained from HMI observations and numerical simulations.

\subsection{Waves and Oscillations in Sunspots}

\begin{figure}[t] \centering \includegraphics[width=0.9\textwidth]{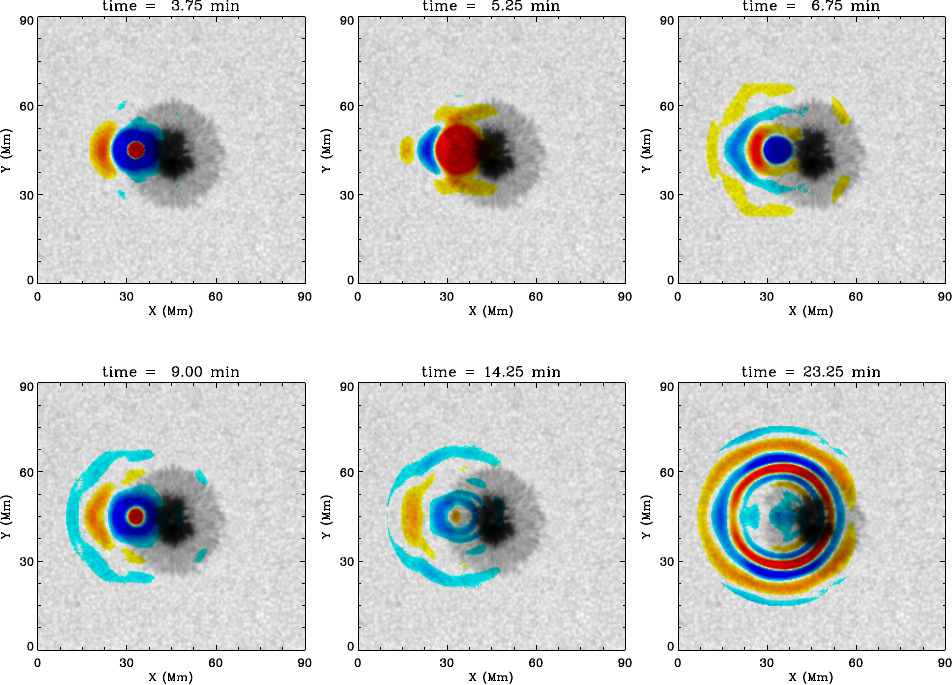} \caption{Selected snapshots of the reconstructed waves propagating away from the sunspot penumbra, shown as \textit{foreground color images}. The \textit{background black-and-white image} shows the continuum intensity of the studied region. The snapshots taken at 3.75, 5.25, 6.75, and 9.00 minutes show the fast-moving wave along the sunspot's radial direction, and the snapshots taken at 14.25 and 23.25 minutes mainly show the typical helioseismic waves expanding in all directions \citep{Zhao2015}.} \label{sunspot_fast_waves} \end{figure}

It is well known that waves of different types and modes are abundant in and above sunspots. Fast waves traveling outward from the sunspot to its vicinity with a phase speed around 45\,km\,s$^{-1}$ were detected through a cross-correlation analysis of oscillation signals observed by HMI in sunspot regions \citep[Figure~\ref{sunspot_fast_waves}: ][]{Zhao2015}. These waves have a typical frequency of 2.5-4.0\,mHz, consistent with \textit{p}-mode waves but much faster than the expected speed of \textit{p}-mode waves in the area. Numerical modeling suggested that these waves were indeed \textit{p}-mode waves, but their excitation source was about 5\,Mm beneath the sunspot's surface.

Deep sources of acoustic waves were found in 3D radiative magnetohydrodynamic (MHD) simulations of solar magnetoconvection with a self-organized pore-like magnetic structure \citep{Kitiashvili2019}. The analysis of more than 600 acoustic events inside and outside the simulated pore-like magnetic structure showed that the depths of acoustic events were mostly located below 1 Mm inside the magnetic region. This is substantially deeper than the acoustic wave sources outside the magnetic region. Similarly, to interpret the ripple-like velocity patterns observed in FeI\,5435\,\AA, \citet{Cho2020} assumed subsurface wave sources and fitted the model to infer a subsurface wave source at a depth of 1-2 Mm. 

The work reviewed above confirmed a fast horizontal wave propagation in and around sunspots. This is consistent with an explanation of subsurface wave sources beneath sunspots, although the detailed depths are still uncertain.

It is also important to investigate whether the photospheric waves are connected to the wave phenomena in the upper atmosphere. A time-distance helioseismic analysis method applied to HMI observations together with chromosphere data observed at the Big Bear Observatory and Sacramento Peak Observatory (IRIS instrument) and coronal observations from the SDO Atmospheric Imaging Assembly (AIA) showed that the photospheric waves can channel up from the photosphere through the chromosphere and transition region into the corona along different wave paths with different phase speeds \citep{Zhao2016}. In addition, direct spectroscopic observations by IBIS characterized waves in the sunspot penumbrae with essentially the same dominant periods and propagation speeds, indicating that the chromospheric running penumbral waves are counterparts of the waves observed in the photosphere \citep{Lohner2015}.

\subsection{Acoustic Halos} 

The enhanced power of high-frequency waves surrounding sunspots and plages, known as ``acoustic halo'', is an intriguing wave-dynamical phenomenon observed in the solar atmosphere \citep{2009A&A...506L...5K}. It was first observed in the early 1990s at photospheric and chromospheric heights \citep{1992ApJ...392..739B,1992ApJ...394L..65B,tonerlabonte93}. This phenomenon is usually observed at frequencies above the photospheric acoustic cut-off of $\approx$ 5.3 mHz, in the range of 5.5\,--\,7~mHz, over regions of weak to intermediate strength (50\,--\,250 G) photospheric magnetic field \citep{1998ApJ...504.1029H,2000ApJ...537.1086T, 2002A&A...387.1092J,2004ApJ...613L.185F,2007A&A...471..961M,2007PASJ...59S.631N, 2009A&A...506L...5K}. It was found that the inclination of the magnetic field plays a prominent role in the production of halos \citep{2011SoPh..268..349S}.

A large number of studies centered around modeling acoustic wave -- magnetic-field interactions over heights from the photosphere to the chromosphere, with relevance to high-frequency power excess observed around sunspots, have been carried out (see \citet{2016ApJ...817...45R} and references therein). A central theme of all the above theoretical studies has been the conversion of acoustic wave modes (from below the photosphere) into magnetoacoustic wave modes (the fast and slow waves) at the magnetic canopy defined by the plasma $\beta$=1 layer.

Utilizing the multi-height capability of HMI and AIA observations, the following properties were found: (i) The halo is present for high-frequency oscillations, beginning at 5.5–6 mHz and up to at least 9–10 mHz. The 6-mHz halo is the strongest in measurements of the Fe I 6173.34\,\AA\, Doppler velocity at $z$ = 140 km; (ii) The halo magnitude is a clear function of height, with no enhancement in intensity continuum ($z = 0$) and in line-wing Doppler velocity ($z$ = 20 km); (iii) In the upper photosphere and chromosphere at the heights sampled by AIA 1600 and 1700\,\AA\, wavelength channels (corresponding approximately to $z$ = 430 and 360 km, respectively), the halo in the 7–10 mHz range is observed to spread out radially with height, and (iv) about 8~mHz, the spatial extent and structure of the halo change to double halo structure with an inner compact halo surrounded by a diffuse, weak halo region, extending radially many megameters into quiet regions \citep{2012SoPh..281..533H,2013SoPh..287..107R,2018AdSpR..61..691T,2020ASSP...57..121T}. 

In addition, a ring-diagram analysis of active regions observed by the HMI during the first five years showed that the frequency at which the mode amplitude changes from attenuation to amplification is around 4.2 mHz in the quiet nearby regions while it is about 5.1 mHz in the active regions. This effect (which has a very weak dependence on the wave propagation direction) corresponds to the amplitude enhancement found in the acoustic power maps \citep{2016ApJ...827..140R}.

\begin{figure} \includegraphics[width=\linewidth]{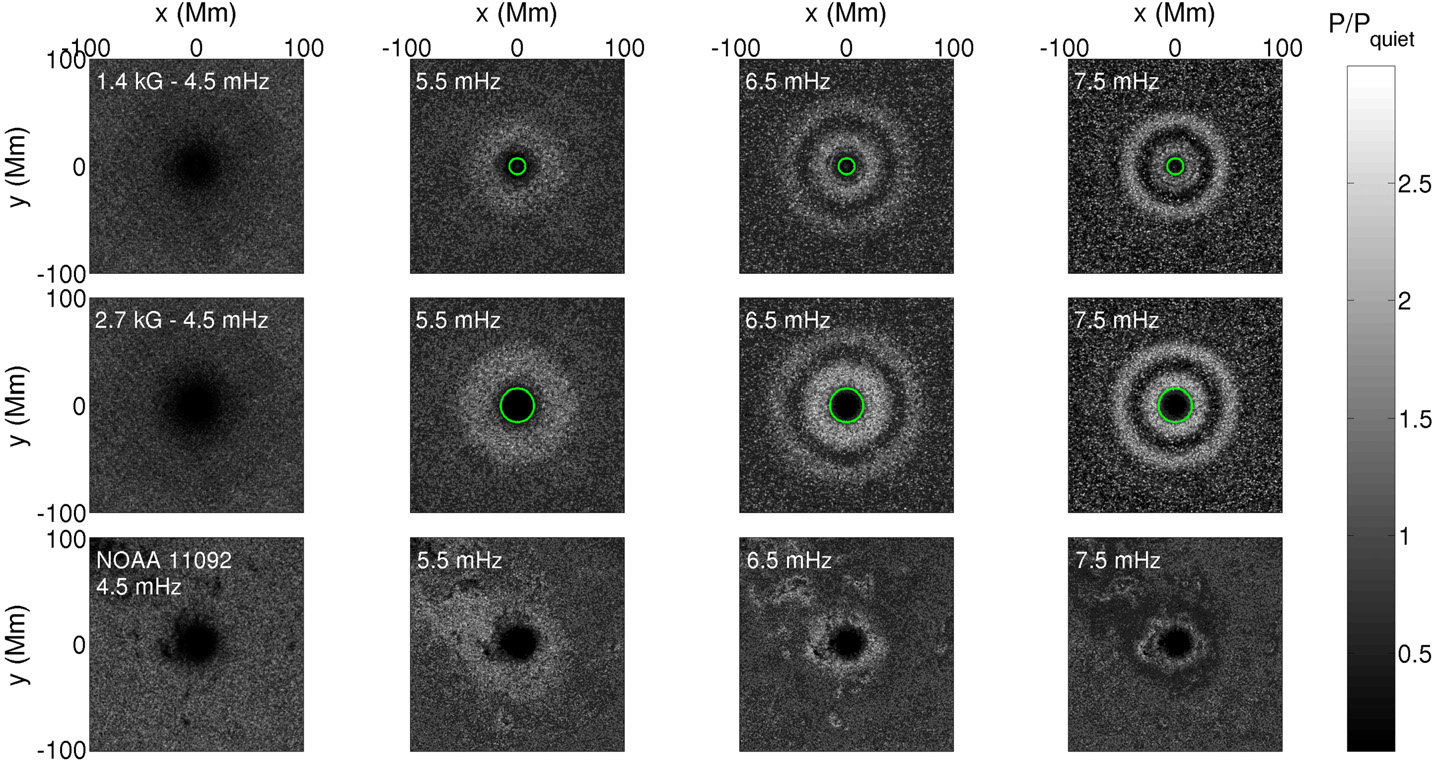} \caption{\textit{Top row}: 6 hr time-averaged vz power maps at the height of formation of the FeI~6173.34~\AA~ line ($z$ = 140 km) for four illustrative frequency ranges for the weak sunspot case (1.4 kG). \textit{Middle row}: the same power maps for the stronger-field case (2.7 kG). \textit{Bottom row}: 14 hr time-averaged observational Doppler velocity power maps of the active region NOAA 11092 
for the same frequency ranges. The \textit{green contour} in rows 1 and 2 is the contour at $z$ = 140 km where the Alfv\'en speed is equal to the sound speed \citep{2016ApJ...817...45R}}.\label{fig:halo}
\end{figure}

From what has been learned so far, from observations as well as theoretical studies, it is clear that transport and conversion of energy between magneto-acoustic wave modes, which are driven by acoustic waves and convection from below the photosphere and mediated by the structured magnetic field in the overlying atmosphere, provide a plausible approach for identifying the exact mechanism. The instruments HMI and AIA onboard SDO, with photospheric Doppler and vector magnetic field information from the former and the upper photospheric and lower chromospheric UV emissions in the 1700~\AA~  and 1600~\AA~ wavelength channels of the latter, have provided interesting possibilities for such studies.  Though none of the proposed theoretical explanations or mechanisms causing the power halos can match all of the observed properties and hence provide an acceptable theory, the mechanism based on MHD fast-mode refraction in the canopy-like structure of strong expanding magnetic field \citep{2016ApJ...817...45R} appears to confirm some major observed features (Figure~\ref{fig:halo}).

\subsection{Subsurface Structure and Dynamics of Sunspots} 

Several helioseismic techniques have been developed to probe the subsurface structure below the sunspots. The most widely used measurement is the phase shifts detected between in and out-bound waves traveling through the active regions and can be obtained through Hankel analysis \citep{1995ApJ...451..859B} or time-distance helioseismology \citep{Duvall1993}. The initial attempt to infer the sound speed structure assumed that only the thermal effects are relevant (the direct effect of the magnetic field was neglected) and found a faster wave speed approximately at a depth of 1~Mm below the sunspot \citep{1995ApJ...451..877F}. Subsequent studies using time-distance helioseismology indicated a two-layer model with reduced wave speed in the top layer and increased wave speed down to 10~Mm below the surface \citep{Kosovichev2000}. Analysis of HMI Dopplergrams through ring-diagram technique also illustrated a two-layer structure \citep{2009SSRv..144..249G}. However, a comparison between ring-diagram, time-distance, and Hankel analysis of the sunspot associated with NOAA AR 9787 showed substantial quantitative differences \citep{2009SSRv..144..249G, 2010SSRv..156..257G, 2011JPhCS.271a2005K} and it was suggested that the structure of this particular sunspot is probably associated with a shallow, positive wave-speed perturbation instead of the standard two-layer model \citep{2010SoPh..267....1M}. Additionally, one cannot rule out the possibility that helioseismic inversions are contaminated by surface effects produced by a strong magnetic field \citep{2007ApJ...661..558C}.

\begin{figure} \centering\includegraphics[width=0.8\linewidth]{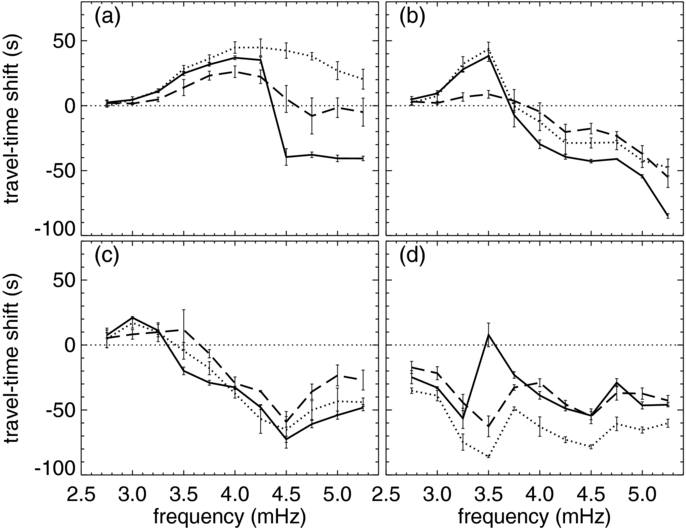} \caption{Averages of the mean travel-time shifts over the umbra of the simulated sunspot (\textit{solid lines}) and AR~11092 (\textit{dotted line}) and AR~10615 (\textit{dashed line}) as a function of the central frequencies measured through helioseismic holography. The panels show the results from four different phase-speed filters \citep[for details see][from where this figure is reproduced]{2012ApJ...744...77B}. }     \label{fig:braun2012} \end{figure}

With the availability of faster high-end computational facilities that led to the generation of more realistic artificial data through numerical modeling and its analysis through various helioseismic techniques, we now have a better understanding of the issues associated with wave propagation in strongly magnetized regions. For example, \citet{2012ApJ...744...77B} compared travel-time shifts measured from a realistic magnetoconvective sunspot simulation and two sunspots observed by HMI. Figure~\ref{fig:braun2012} shows the averages of the mean travel-time shifts over the umbra of the sunspot measured by helioseismic holography. The study found remarkable agreement below 4.5~mHz but significant differences above it.  The most striking difference is seen in the center of the sunspots; the travel-time shifts of the real sunspots are found to be positive, while the corresponding shifts in the simulated spots are negative. Since the hare and hound tests of simulated data considering only sound speed perturbation had shown remarkable agreement between the simulated sunspot and measured shifts \citep{2011SoPh..272...11B}, it is conjectured that the anomaly in the travel-time shifts is caused by the inclusion of the magnetic field.  Later studies focused on understanding the interaction of waves with the magnetic field by analyzing data from simulated sunspots.  \citet{Felipe2016} investigated travel time shifts arising from magnetic and thermal effects and concluded that for certain combinations of phase speeds and frequencies, the travel time shifts could be explained through the thermal effects alone.  These authors later studied the dependence of the travel time on magnetic field strength and Wilson depression \citep{2017A&A...604A.126F} and found that shifts corresponding to frequencies above 3.5~mHz are insensitive to the magnetic field strength \citep[also see][]{2015MNRAS.449.3074M}.  

\subsection{Flows in and around Active Regions}

Apart from strong outflows close to sunspots, active regions exhibit much larger inflows around them in the near-surface layers. These flows first detected by the ring-diagram technique typically have amplitudes of 20\,--\,30 $\rm{m\,s}^{-1}$ and occur up to a depth of 10\,--\,14~Mm below the photosphere and extent up to $30\degree$ ($\sim~300$~Mm) around active regions \citep{2001ESASP.464..209H,2003PhDT.........9G,Zhao2004}.  The ring-diagram analysis using  HMI data confirmed these results \citep{2018ApJ...859....7R,Komm2020}. 

Analysis of the HMI time-distance helioseismology data with a resolution of $0.12\degree$ confirmed these inflows around active regions to have a peak amplitude of $\approx$ 40~m\,s$^{-1}$ for the strongest active regions and about 20 m\,s$^{-1}$ on average, an extent of up to $30\degree$ away for the strongest active regions and an extent of about $15\degree$ to $20\degree$ on average \citep{Kosovichev2016,Mahajan2023}. Helioseismic holography is another technique that has been used to measure inflows around active regions \citep{2019ApJ...873...94B,2024ApJ...972..160B}. It is found that the average inflow profiles around active regions using holography are a factor of two smaller than the ones found by time-distance analysis. It is presently unclear where this difference arises from. Evidence of a deep outflow around sunspots has been observed using Hinode \citep{2010ApJ...708..304Z} and HMI \citep{2018ApJ...859....7R}. This is consistent with a large-scale circular flow in a cylindrical shell around sunspots, with a downflow near the sunspot and an upflow farther away, connecting the near-surface inflow to the deep outflow \citep{2003PhDT.........9G}.
These active-region flows, thus, contribute to longitudinal averages of zonal and meridional flows and, thus, to the time-varying global-scale flows \citep{2019ApJ...873...94B,Getling2021}.

Inflows around active regions impact the solar magnetic cycle in several ways: 1) they may enhance magnetic flux cancellation within active regions as shown by surface flux transport models with inflows \citep{2012A&A...548A..57C,2020JSWSC..10...62N} 2) they may play a role in perturbing the differential rotation of the Sun and hence contribute to the torsional oscillation pattern (Section \ref{sec:zonal}) 3) they play a dominant role in the modulation of the near-surface meridional flow (Section \ref{sec:meridional_vary}) and 4) inflows around active regions combined with hemispheric asymmetry in activity can cause cross-equator flows \citep{2022SoPh..297...99K}.

\begin{figure}
	\begin{center}
		\includegraphics[width=\linewidth]{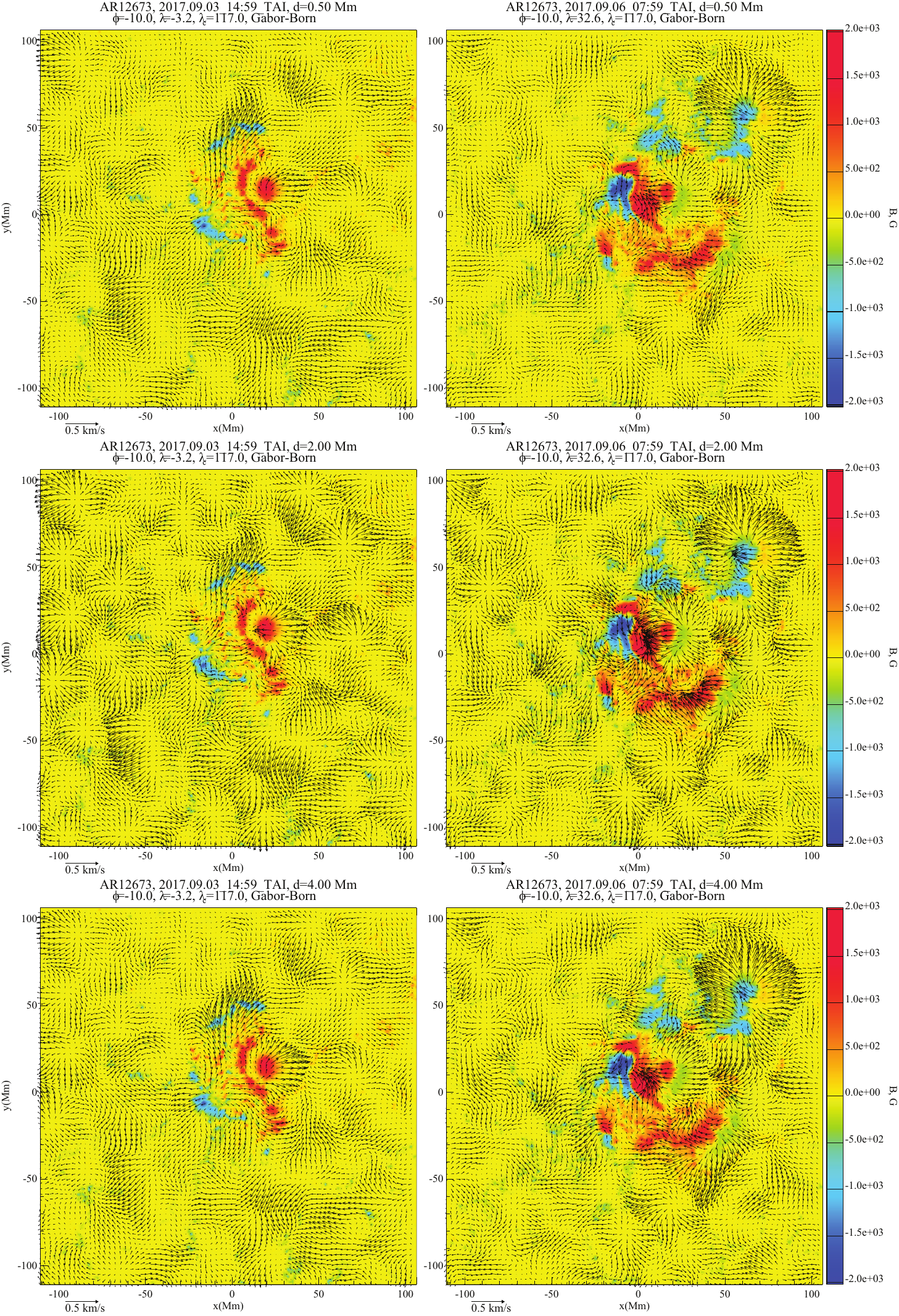} 
		\caption{Flow maps of AR~12673 during the emergence of bipolar magnetic flux in the form of fragmented ribbons near the existing positive polarity sunspot (\textit{left column}) and before the X9.3 flare (\textit{right}) at three depth levels centered at 0.5, 2, and 4~Mm (\textit{from top to bottom}). The background images show the corresponding surface magnetograms \citep{Kosovichev2024}.
		}
		\label{AR12673}
	\end{center}
\end{figure}

A study of subsurface flows in the emerging active region NOAA 11158 by the ring-diagram technique revealed divergent flows, the magnitude of which decreased around the time of major flares, suggesting that the flow kinetic energy contributed to the energy released by flares   \citep{2015ApJ...808...60J}. The subsurface dynamics of emerging active regions inferred from the time-distance helioseismology with higher spatial and temporal resolutions showed that the divergent flow pattern observed at the initial emergence stage changes to predominantly converging flows during the formations of sunspots and flaring activity in the top 4 Mm deep layer while the flows remain diverging in the deeper layers   \citep{Kosovichev2018,Kosovichev2024}. Figure~\ref{AR12673} illustrates the flow patterns in the active region NOAA  AR~12673 during the emergence and before a major X9.3 solar flare.

These studies also indicated a statistically significant correlation between the subsurface flow characteristics and the flare index of active regions. However, a comprehensive study of the complex relationship between subsurface flows, the evolution of active regions, and their magnetic activity has yet to be performed. 

Some attempts are also made to explore the flow characteristics in active regions with peculiar magnetic configurations, e.g., anti-Hale and anti-Joy active regions. Authors reported the anti-clockwise or clockwise flows surrounding the anti-Joy active regions depending on their locations in the Northern or Southern hemispheres, respectively \citep[e.g.,][]{2013JPhCS.440a2050G,2023arXiv231007271J,Kosovichev2024}.

\section{Helioseismic Mapping of the Solar Far-Side Active Regions}
\label{sec_FS}

Active regions observed by HMI on the near side of the Sun are very important for monitoring solar activity and eruptions, and many numerical models using these observations have greatly advanced our understanding of solar eruptions in the last decades.
However, many space-weather prediction models, particularly those dealing with solar wind, coronal holes, and global-scale magnetic properties, need far-side information of active regions for more accurate modeling \citep{Arge2013}. 
Some modelers rely on synoptic maps obtained from magnetograms of the visible surface of the Sun, while others estimate the far-side magnetic field through flux-transport models \citep[e.g.,][]{Schrijver2003, Upton14a}. 
However, new active regions (ARs) may emerge and rapidly evolve on the far side of the Sun. These ARs cannot be modeled by the flux-transport models, causing inaccuracies in the far-side magnetic field data currently used by the global coronal and solar wind models. Local helioseismology has provided the capability of mapping ARs on the far side of the Sun, which was initially developed for the analysis of the SoHO/MDI data.

\subsection{Detection of Far-Side Active Regions Using Helioseismic Holography}
\label{sec_FS_1}

While the Sun’s far side is not directly visible in electromagnetic waves from Earth, it is essentially transparent to acoustic waves. This property was employed in the helioseismic holography method, in which the phase shifts are caused by the scattering of waves from the thermal and magnetic perturbations associated with active regions on the far side \citep{Lindsey2000Farside,Braun2001,Lindsey2017}. This technique allowed us to map large active regions on the Sun’s far side.
The far-side images are routinely obtained twice a day and available online in near-real time \footnote{\url{jsoc.stanford.edu/data/farside/}}.

With the availability of STEREO’s far-side EUV images, it became possible to systematically compared the STEREO/EUVI 304 Å images, in which ARs exhibit enhanced emission with the helioseismic holography far-side images made from observations of both GONG and HMI \citep{Liewer2012, Liewer2014, Liewer2017}. It was found that 95\% of the far-side ARs detected by the holography method correspond to an observed EUV brightening area, but only about 50\% of EUV brightening areas correspond to the holography-detected ARs.  This gives us a sense of the reliability and limitations of the helioseismic holography far-side imaging method. 

Various authors have also tried to improve the detection of active regions in the holographic far-side images. In particular, the far-side ARs were identified with higher sensitivity and better accuracy by applying a machine-learning method \citep{Felipe2019,Broock2022}.
The quality of the individual AR imaging on the far side was improved by defining the ingression and egression wavefields in terms of a Green’s function and using an accurate forward solver \citep{Yang2023}.
A comparison of ARs observed by the SO/PHI near the far-side limb area found reasonable agreement \citep{Yang2023b}.

\subsection{Detection of Far-Side Active Regions Using Time-Distance Helioseismology}
\label{sec_FS_2}

\begin{figure}
	\centering
	\includegraphics[width=0.7\textwidth]{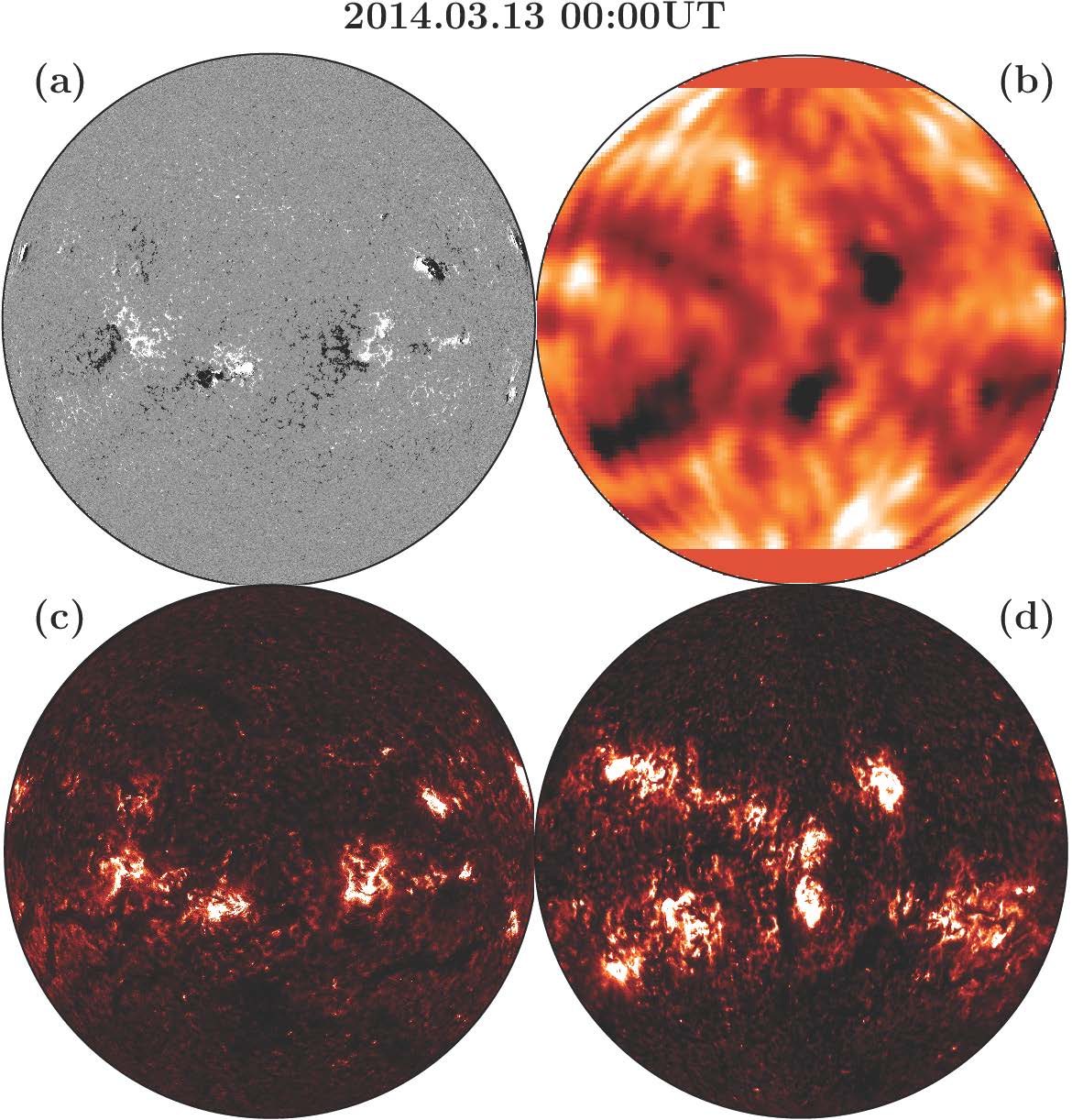}
	\caption{Comparison of near-side and far-side images taken at or near 00:00 UT of 13 March 2014. 
		(\textit{a}) SDO/HMI-observed near-side magnetic field. 
		(\textit{b}) Helioseismic far-side image. 
		(\textit{c}) SDO/AIA-observed near-side 304\,\AA\ image. 
		(\textit{d}) STEREO/EUVI-observed far-side 304\,\AA\ image after merging observations from both STEREO-A and STEREO-B spacecraft. 
		The magnetic field is displayed with a scale of $-200$ to 200 G, the helioseismic image is displayed with a scale of phase shifts from $-0.10$ to 0.10, and the EUV images are displayed in pixel intensity units DN (\textit{Data Number}) from 10 to 2000.
		The STEREO/EUVI 304\,\AA\ image is displayed after its intensity is calibrated to match the SDO/AIA 304\,\AA\ level \citep{Zhao2019}.}
	\label{td_fs}
\end{figure}

Time-distance helioseismology also demonstrated the capability to image the far-side active regions with a deep-focusing measurements scheme with the acoustic ray focus point on the far side of the Sun (a 4-skip scheme) \citep{Duvall2001}. This technique was improved by combining this scheme with 3-5-skip waves \citep{Zhao2007,Ilonidis2009} and applying numerical simulations that helped to identify and remove spurious signals in the far-side images \citep{Hartlep2008}. 
The HMI data allowed to include the 6- and 8-skip waves \citep{Zhao2019}, which increased the reliability and accuracy of the far-side imaging (Figure~\ref{td_fs}). The routine production of the far-side images is available in the near-real time\footnote{\url{jsoc.stanford.edu/data/timed/}}.

The helioseismic images of the solar far side can also be used to study the solar-cycle properties of the Sun. The far-side images obtained by the holography technique from the SOHO/MDI data showed that the mean helioseismic phase shifts associated with far-side ARs varied substantially with the phase of the Solar Cycle 23 \citep{Gonzalez2009}. 
These variations were attributed to the Sun’s seismic radius change with solar cycles. However, using 11 years (a full solar cycle) of HMI time-distance far-side image data, \citet{Zhao2021} concluded that the mean phase shifts (or travel-time shifts) in such images are likely due to a collective effect of surface reflections of the helioseismic waves that unavoidably interact with the surface magnetic fields. 

\subsection{The Space-Weather Connection}
\label{FS_Hel_in_SpaceWeather}

\begin{figure}
	\centering
	\includegraphics[width=0.8\textwidth]{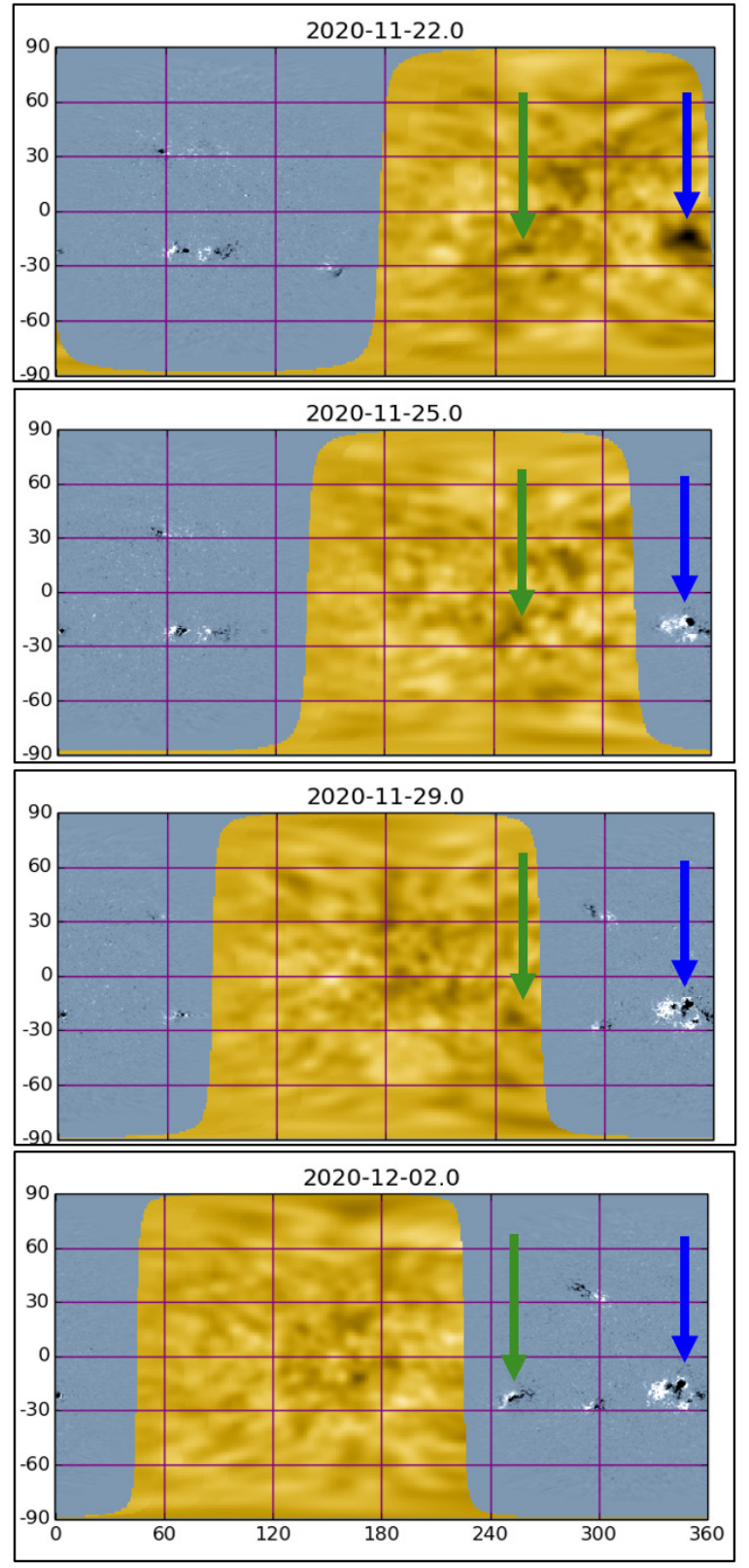}
	\caption{Composite near- and far-side Carrington maps of the Sun composed of HMI magnetograms (\textit{blue-gray}) and helioseismic phase maps (\textit{amber}) solar activity indicating the onset of Solar Cycle 24.
		The active region, to which the \textit{blue arrow} points, emerged in the far hemisphere almost two weeks before it rotated into the direct view from the Earth, is shown in the \textit{third row}.
		The active regions (\textit{green arrow}) points produced the first major flare of Solar Cycle 24.}
	\label{CompFD_Maps}
\end{figure}

The far-side images play an important role in space-weather prediction because they use up to ten days of warning of new large active regions appearing on the Earth's side \citep{2023BAAS...55c.189J}. For example, Figure \ref{CompFD_Maps} shows the helioseismic signature of the strong active-region complex consisting of ARs 12785 and 12786, which crossed the Sun's south-eastern limb on 2020-11-23.

Another active region, AR~12781, indicated by the green arrow, approached the south-eastern solar limb five days later and produced a series of flares.

\subsection{Converting Far-Side Helioseismic Images into Magnetic-Flux Maps}
\label{sec_FS_3}

As pointed out earlier, the far-side magnetic fields are important to global-scale modeling efforts, but the helioseismic far-side images are unsigned measurements of acoustic phase shifts. Converting acoustic signals into magnetic signals is one crucial step toward fully mapping the far-side magnetic fields.  Previous authors have established that the EUV 304\,\AA\ images can be used as a proxy for the magnetic field \citep[e.g.,][]{Ugarte2015,Kim2019}, and the acoustic phase shifts can be used as another proxy for the magnetic field \citep{Gonzalez2007}. 
Figure~\ref{td_fs} shows well the relations among these properties.

\begin{figure}[t]
	\centering
	\includegraphics[width=\textwidth]{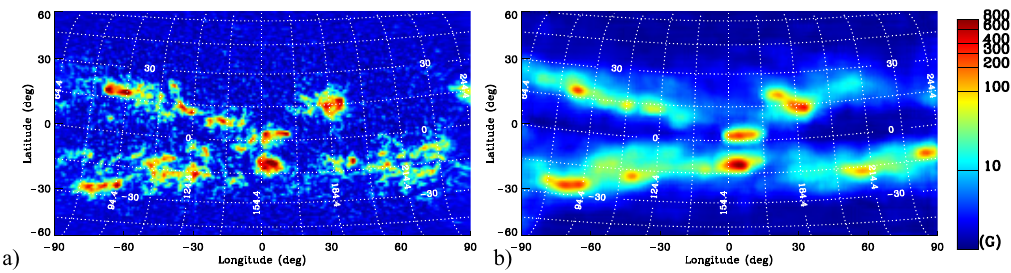}
	\caption{Example showing (\textit{a}) STEREO-converted far-side magnetic-flux map from the machine learning code 1; and (\textit{b}) helioseismic-image-converted far-side magnetic-flux map from the machine learning code 2, both taken at 00:00 UT on 13 March 2014 \citep{2022Chen_Farside}.}
	\label{FS_mag_flux}
\end{figure}

Two machine-learning techniques were used to 
convert the time-distance far-side images combined with the far-side 304\,\AA\ images from STEREO into far-side magnetic-flux maps \citep{2022Chen_Farside}.
Figure~\ref{FS_mag_flux} shows an example of magnetic-flux maps generated by this technique.
Despite the success of these machine-generated far-side magnetic-flux maps, their usefulness is still limited because this method is unable to produce the polarities of the magnetic field. 

\section{Early Detection of Emerging Magnetic Flux}

Detection of emerging active regions (ARs) before they become visible on the solar surface is tightly linked to understanding the coupling of the activity of the Sun on the global and local scales. This understanding is crucial for advancing our capabilities to predict solar activity and its impact on Earth's environment. In particular, theoretical and numerical efforts suggest that ARs form from magnetic flux ropes generated in the solar interior \citep[e.g.,][]{Parker1994,Fan2021} that emerge on the solar surface. However, understanding how and where the magnetic flux is organized in highly magnetized magnetic field bundles, how they evolve during emergence through the convective layers, and how surface and atmospheric layers respond and manifest the upcoming emergence continue to require a detailed theoretical investigation and detailed analysis of observations.

Whole-disk observations with HMI over the solar cycle enable the investigation of emerging ARs from the deep interiors to the photosphere. Since the evolution of subsurface dynamics cannot be observed directly, it can be inferred from photospheric disturbances that are sensitive to structural and dynamical changes through all layers of the convection zone using helioseismic and other methods. In this section, we review advances in the detection of emerging ARs and the analysis of surface and subsurface dynamics to identify precursors of the emergence with different helioseismic techniques (Section~\ref{hseis}) and alternative approaches, including tracking of the photospheric flows and machine-learning approaches (Section~\ref{obs_only}), as well discussion of still-open questions and expectations for future developments (Section~\ref{QA}).

\subsection{Diagnostics of Emerging Magnetic Flux from Helioseismic Analyses of HMI Data}\label{hseis}

The emergence of ARs is usually associated with a significant increase of magnetic flux in the photosphere, accompanied by strong diverging flows, a rapid decrease of continuum intensity, and substantial restructuring in the solar atmosphere \citep[for a review of the topic see][]{vanD2015}. However, before AR emergence, there are generally no obvious signatures of upcoming flux due to slow changes in local thermodynamic properties in the solar interior. 

Most previous helioseismic investigations {by the time-distance technique (Section~\ref{sec:Time-Distance Method})} \citep[e.g.][]{Ilonidis2012,Kholikov2013,Birch2013} have found negative travel time shifts (typically corresponding to an increase in local wave speed), and the regularity with which they have been detected suggests that there is potential for them to be used as an indicator of the magnetic field emergence with the following formation of ARs. Early evidence of this can be found in an increase in the sound speed perturbations inferred by a time-distance inversion technique in the uppermost 18~Mm of the convection zone before the emergence of an AR \citep{Kosovichev2000}. Tracking the evolution of the acoustic wave speed perturbation allowed the authors to obtain one of the earliest observational constraints on flux rise speed, estimating that the speed of emergence is about 1.3 km\,s$^{-1}$.

The capability to detect a signal in the deep convection zone $\sim 24$ hours before the AR emergence was demonstrated by applying a deep-focus time-distance scheme \citep{Ilonidis2011} and a similar approach \citep{Kholikov2013} to the MDI and Global Oscillation Network Group (GONG) data. The follow-up analysis of AR~11158 pre-emergence helioseismic signal from SDO/HMI observations \citep{Ilonidis2012,Ilonidis2013} showed an increase in the local wave speed consistent with the earlier investigations \citep{Kosovichev2000,Ilonidis2011}. 

\begin{figure}
	\centering
	\includegraphics[width=\linewidth]{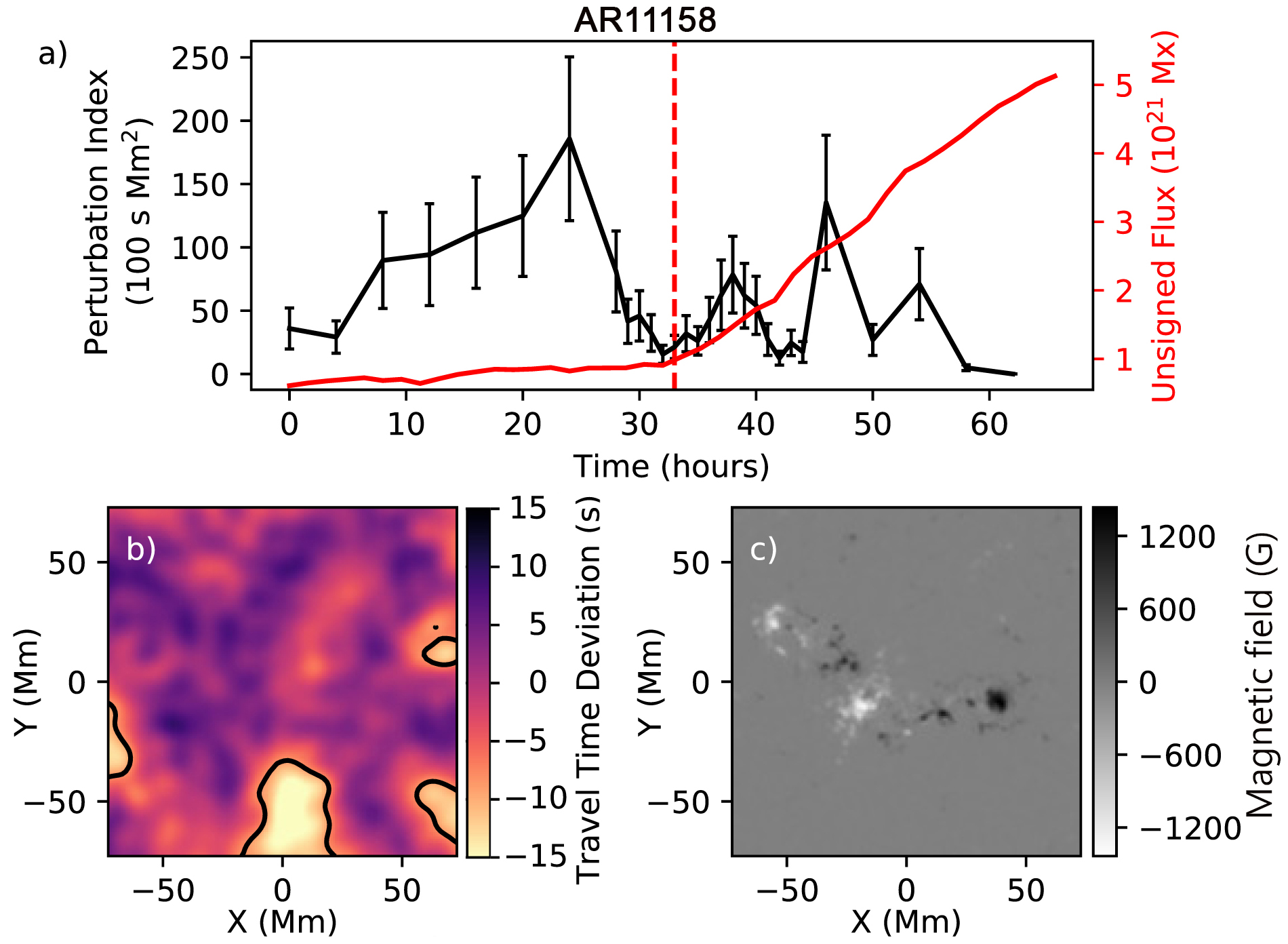}
	\caption{Example of tracking the evolution of AR~11158's emergence from the interior using the time-distance analysis. \textit{Panel (a)} shows the unsigned magnetic flux (\textit{red}), the perturbation index (\textit{black}), and the estimated time of emergence (\textit{vertical dashed red line}). \textit{Panel (b)} shows the travel time map at the greatest perturbation index, where the strong deviations are located adjacent (and not beneath) the eventual location of the AR. \textit{Panel (c)} shows the line-of-sight magnetic field at the end of the analysis \citep{Stefan2021}.}
	\label{fig:AR11158time-dis}
\end{figure}

Subsequent work focused on improving the signal-to-noise ratio by applying an averaging procedure and varying the signal threshold  \citep{Stefan2021}. Testing of this new approach was performed on two simulations with imposed sound speed variations 
showed that the method was most sensitive to deeper perturbations (50~Mm to 60~Mm beneath the simulated photosphere). 
This approach enabled the detection of a robust signal from several ARs up to 26 hours before the emergence. Figure~\ref{fig:AR11158time-dis} shows an example of the successful detection of AR~11158 at least 20 hours before it emerges. Panel (\textit{a}) traces the evolution of the unsigned surface magnetic flux (\textit{red}) and the perturbation index (\textit{black}). The perturbation index is a measure that integrates travel-time perturbations beyond a given threshold to reduce the contribution from the background noise \citep{Ilonidis2011}. Figure~\ref{fig:AR11158time-dis}a displays a gradually increasing perturbation index that peaks most sharply around 8-10 hours prior to the time of emergence (the dashed red line). Figures~\ref{fig:AR11158time-dis}b and c show the corresponding travel-time map at the peak and the configuration of the surface magnetic field at the end of emergence, respectively.  
Later application of this methodology to 46 ARs showed the capability to detect a signal of the emerging flux at least 24 hours before emergence at the photosphere for 28 ARs \citep{Stefan2023}. In particular, a correlation was found between the emergence rate and amplitude in travel time shift variations. Therefore, faster emergence (that typically corresponds to larger magnetic flux due to stronger buoyancy force) allows earlier detection. 

\subsection{Emerging Flux Inferences Based on Tracking of Surface Dynamics}\label{obs_only}

While detection of emerging magnetic flux via helioseismic may be the most direct method for making forecasts of AR emergence, there have also been significant efforts focused on forecasting the emergence of ARs directly from HMI observations. Such works use HMI's observations of the photosphere and search for subtle changes in the surface velocity, brightness, and magnetic field.

\subsubsection{Flow Tracking Techniques}
One of the more robust photospheric signatures is the horizontal diverging flow (HDF) field which forms on the surface just before the flux concentration emerges on the photosphere. Simulations show that HDF results from a ``squeezing'' of plasma above the rising flux concentration and below the stably stratified atmosphere \citep{Cheung2010,Toriumi2013}, and that the magnitude of this diverging flow field is directly related to the flux rise speed \citep{Birch2010,Birch2016}.

Direct measurement of these HDF flows is challenging; however, since they are oriented perpendicular to the photosphere and HMI can measure only line-of-sight (LOS) velocity. Two strategies have been developed to overcome this limitation. First, focusing on observations close to either the West or East solar limb means that the LOS velocity incorporates a greater amount of the horizontal component, and the horizontal diverging flows can be discerned from the background convection based on overall spatial extent (larger than convective cells) and configuration (positive and negative Doppler velocities, generally aligned with magnetic dipole axis). An examination of HMI Dopplergrams centered around the eventual location of AR~11081 ($\sim$25~degrees away from the central longitude) found an HDF that develops around 100 minutes before clear flux emergence \citep{Toriumi2012}. In this case, it was impossible in this case to distinguish between the radial and horizontal components of the flow from the LOS velocity, but tracking the outflow pattern in Doppler images revealed that the positive and negative centroids of these flows drifted apart at speeds between 0.6 and 1.2 km~s$^{-1}$. Further examination of other ARs using a similar strategy also found HDFs in around a third \citep{Rees-Crockford2022} to around half \citep{Toriumi2014} of the AR samples.

An alternative approach is to perform feature tracking based on either HMI Dopplergrams, magnetograms, or intensitygrams that allow for the horizontal velocity to be extracted by differentiating the feature position with respect to time. The advantage of this approach is that HDFs can be measured anywhere on the solar disk, potentially leading to a more general robust forecasting tool. One particular application of this methodology is the Balltracking algorithm \citep{Potts2004,Attie2015}. This method treats the magnitude of the continuum intensity measured by HMI as ``height'', down which simulated balls are allowed to roll, mimicking how a physical object would be pushed around by granular flows. The Balltracking algorithm was applied to the intensitygrams to investigate the evolution of horizontal flows around AR~12673 to search for pre-emergence signatures of the secondary flux, which emerges close to the central, mature sunspot \citep{Attie2018}. The authors find that the radius of the moat flow is perturbed up to 35\% at four different locations, occurring up to 12 hours before the secondary flux emerges. Most notably, the deviation of the moat flow occurs directly over the flux concentration, ``pointing'' in the direction of new flux. Thus, this method has the potential to identify secondary flux emergence, which is difficult to perform using helioseismic methods due to the complicated interaction between \textit{p}-modes and existing magnetic flux concentrations. 

\subsubsection{Machine-Learning Applications}

The prediction of emerging flux using more modern data processing techniques, such as Machine Learning (ML), is still a rather unexplored area of study. Currently, machine learning studies are primarily concentrated on properties and dynamics of already formed active regions, such as automatic identification of active regions on the solar surface \citep[e.g.,][]{quan2021solar,veeramani2023automatic,mourato2024automatic}. A recently developed machine learning approach for the detection of emerging active regions  \citep{kasapis2024predicting} is based on previous demonstrations of a correlation between variations in acoustic power and the presence of strong subsurface magnetic fields or upcoming flux \citep{Hartlep2011}. Utilizing the trained Long Short-Term Memory \citep[LSTM;][]{Sherstinsky2020} -- which is most suitable for the analysis of time-series of HMI/SDO observables -- a series of physics-informed ML models have been developed and tested. During the training phase on 40 ARs, the LSTM models analyze correlations between acoustic power in four frequency ranges and the unsigned magnetic flux and corresponding continuum intensity variations that allowed to predict a decrease of the continuum intensity associated with an AR emergence  \citep{kasapis2024solar}. To evaluate the performance of the developed ML models, the time criterion of AR emergence was defined as the time that follows a sustained (for more than 3 hours) decrease in continuum intensity with a rate of 0.01.  

\begin{figure}
\centering
\includegraphics[width=0.9\linewidth]{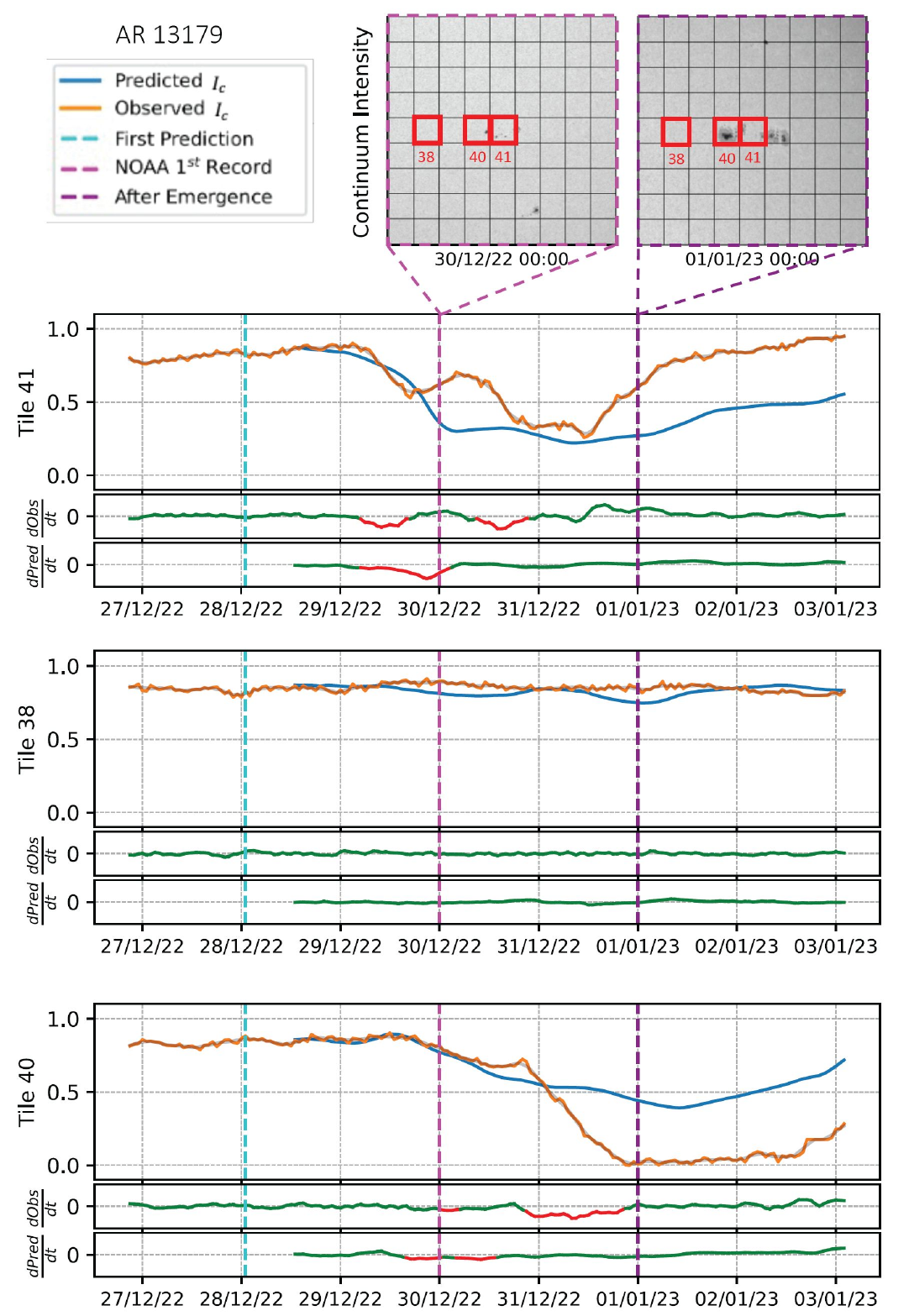}
\caption{Example of the 12-hours-ahead prediction of AR ~13179 emergence using an LSTM model~\citep[Model 8;][]{kasapis2024solar}. The graph panels show variations of observed (\textit{orange curves}) and predicted (\textit{blue}) continuum intensity for the subareas (\textit{tiles}) identified in \textit{red} on the \textit{upper left} continuum intensity images. The time derivative of predicted and observed intensity color codes are \textit{green} if there is no activity in this area and \textit{red} if the decrease in intensity, corresponding to the emergence of magnetic flux, is predicted or observed according to the emergence criterion. The \textit{vertical cyan dashed line} marks the time when the model begins predicting intensity. NOAA's first record of the AR is indicated by the \textit{magenta dashed line}. The \textit{dark magenta dashed line} indicates the moment of time two days after NOAA's first record of AR ~13179, as shown in the corresponding intensity map.}
\label{fig:AR_Pred_Using_ML}
\end{figure}

The developed model can predict continuum intensity in both subareas (tiles) where the emergence of an active region is expected (e.g., Tile 40 and 41 in Figure~\ref{fig:AR_Pred_Using_ML}) and tiles where photospheric dynamics remain quiet (Tile 38). Magnetic flux can occupy more than one tile or be transported from one tile to another during the emergence process, and the model can capture these changes. For instance, Tile 40 shows a delay in the predicted and observed continuum intensity decrease compared to Tile 41. In general, the LSTM model predictions are in reasonable agreement with observations of the changing continuum intensity before and during AR emergence. However, as emerging flux forms an AR, these deviations to the continuum intensity can be significant due to the complexity of the magnetic field structure, which is not considered during the model training. Testing of the LSTM model demonstrated the capability to predict its emergence for the following active regions and time scales: 5 hours in advance for AR~13179, 10 hours for AR~11726, and 29 hours for AR~13165. This new capability is an initial attempt to create a path towards efficient processing of observations and a future application of this capability for the operational forecast of AR emergence. 

In contrast, a machine learning method applied to detect signatures in super-granular scale flow patterns associated with magnetic flux emergence in the quiet Sun found no systematic features \citep{2024ApJ...965..186M}. It was concluded that emergence was neither systematically influenced by nor affected supergranular-scale flows in their vicinity.

\subsection{Open Questions}\label{QA}

Work in the pre-HMI era laid a strong foundation for the forecasting of AR emergence, and many successful efforts have used HMI data to build upon these foundations. Still, many significant open questions remain regarding the detectability and physical origins of emerging magnetic flux. Perhaps the most important open questions are those regarding the mechanism by which the helioseismic precursors of AR emergence are generated. For example, several investigations have found mean phase travel-time perturbations that precede the emergence of ARs that are consistent with increases in the local wave speed. Yet, the direct contribution of the magnetic field to wave speed in the deeper ($ > 20$~Mm) convection zone is expected to be completely undetectable, producing travel-time variations on the order of $10^{-1}$ to $10^{-2}$~s \citep{Braun2012b}. Therefore, the mechanism that generates the observed travel-time perturbations is still unclear.

\begin{figure}
\centering
\includegraphics[width=\textwidth]{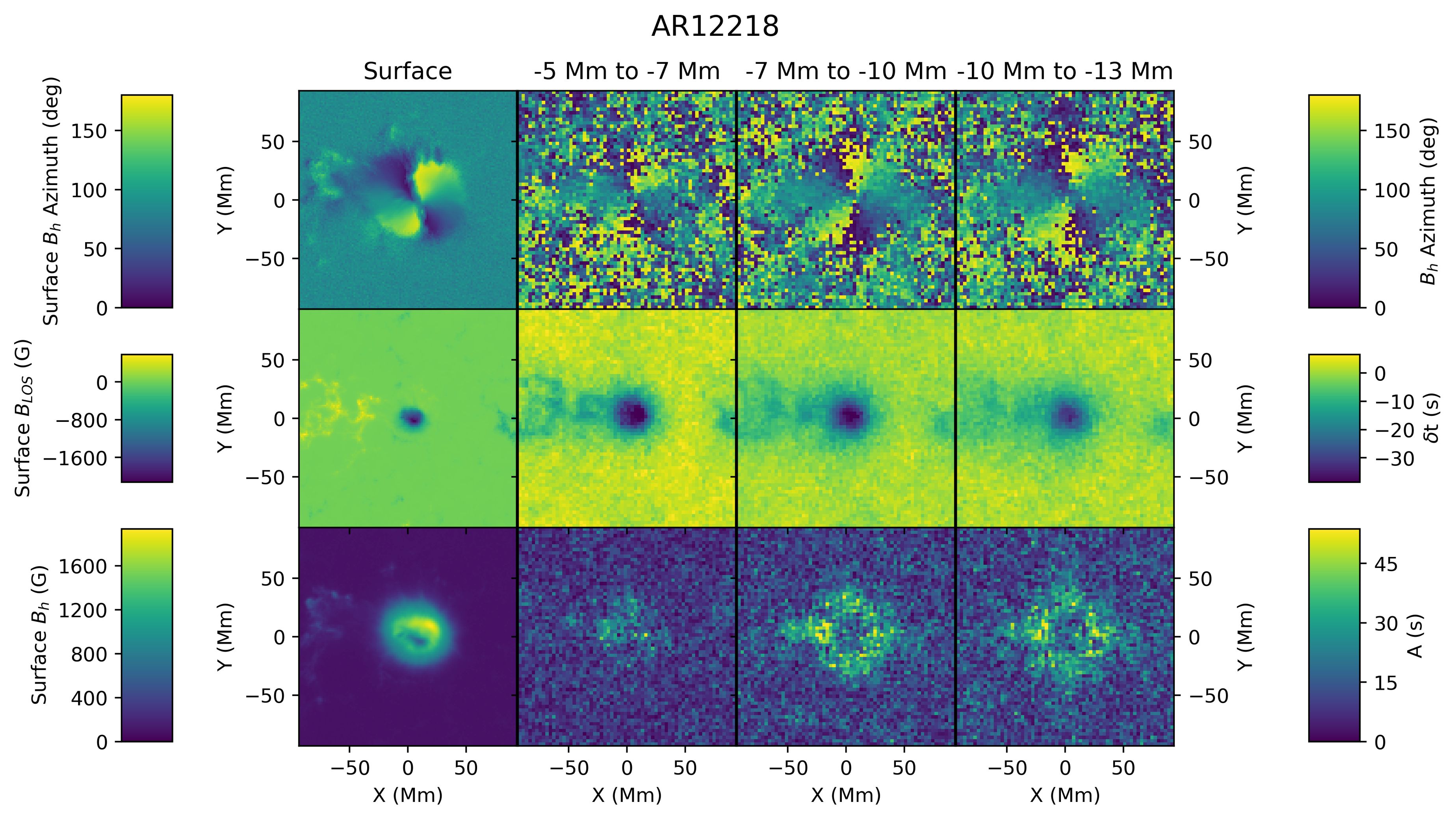}
\caption{Helioseismic investigation of the subsurface horizontal magnetic field in AR~12218. The \textit{first column} shows the surface field azimuth, surface line-of-sight magnetic field, and the surface field's horizontal component in descending order. \textit{Subsequent columns} show the helioseismic measurements in progressively deeper layers, from -5 to -7 Mm, -7 to -10 Mm, and finally -10 to -13 Mm, from \textit{left} to \textit{right}. These helioseismic measurements include the subsurface azimuth (\textit{top row}), mean travel-time deviation (\textit{middle row}), and the travel-time anisotropy caused by the subsurface horizontal magnetic field. The middle row is included as a proxy for the radial magnetic field, though normal perturbations to the sound speed also contribute to this signal \citep{Stefan2022a}.}\label{fig:magfield}
\end{figure}

Despite this, there has been some success in measuring horizontal magnetic fields in the shallow convection zone (down to $\sim$15~Mm) of existing, stable sunspots \citep{Stefan2022a}. Since the speed of magnetoacoustic wave propagation depends on the direction of wave propagation relative to the magnetic field, measuring the travel time of acoustic waves in several different directions allows for the magnetic field's contribution to be isolated. This methodology has been applied to several simulations and, most notably, several ARs observed by HMI. Figure \ref{fig:magfield} shows the results of applying this methodology to HMI Dopplergrams centered on AR~12218, where changes to the configuration of the magnetic field over depth can be tracked. An inversion for the magnetic field has not yet been attempted, so the magnitude of the horizontal magnetic field is quantified in terms of a travel-time shift; however, the direction of the horizontal field can be directly measured from the travel times. However, how this methodology can be applied to the forecasting of AR emergence is unclear since it requires relatively long integration times ($\sim$24~hours). Additionally, the uncertainties for this methodology are significant for weak magnetic fields; this sets a minimum detection threshold on the order of several hundred gauss.

Additionally, purely acoustic waves (\textit{p}-modes) are not the only mode to have been investigated as a probe of interior magnetic fields. In particular, changes to the power of the surface gravity mode (or \textit{f}-mode) have been identified as a potential precursor to AR emergence. Analyses of idealized MHD simulations have found that the \textit{f}-mode tends to ``fan out'' in the Fourier space in the presence of horizontally non-uniform magnetic fields, where the \textit{f}-mode power at any particular wavenumber is spread over a broader range of frequency space than in the unperturbed state \citep{Singh2014}. The exact mechanism of the \textit{f}-mode fanning and the associated increase in the \textit{f}-mode power is not well understood, but a scattering of \textit{p}-modes by the magnetic flux and subsequent leaking into the \textit{f}-mode has been suggested as a possible explanation. 

Furthermore, investigations of changes to the \textit{f}-mode on the Sun have produced mixed results. Early work examined how the \textit{f}-mode power derived from HMI Dopplergrams in six pre-emergence regions changes with respect to quiet regions at the opposite latitude and identical longitude. Ideally, such a comparison will control for variations in the \textit{f}-mode power related to the distance from the disk center. It was found that all six isolated pre-emergence regions show an increase in the \textit{f}-mode power at high wavenumbers (corresponding to $\ell$=1200\,--\,2000) one to two days before emergence \citep{Singh2016}, and in most cases, this increase in the \textit{f}-mode power exceeds that observed in the opposite hemisphere. The results of this particular investigation were called into question by later work, however, where an improvement in the procedure used to calibrate the \textit{f}-mode power against longitudinal variations reduced the changes in \textit{f}-mode power in ARs 11105 and 11130 to within the quiet Sun uncertainty range \citep{Korpi2022}.

\begin{figure}[h!]
\centering
\includegraphics[width=\textwidth]{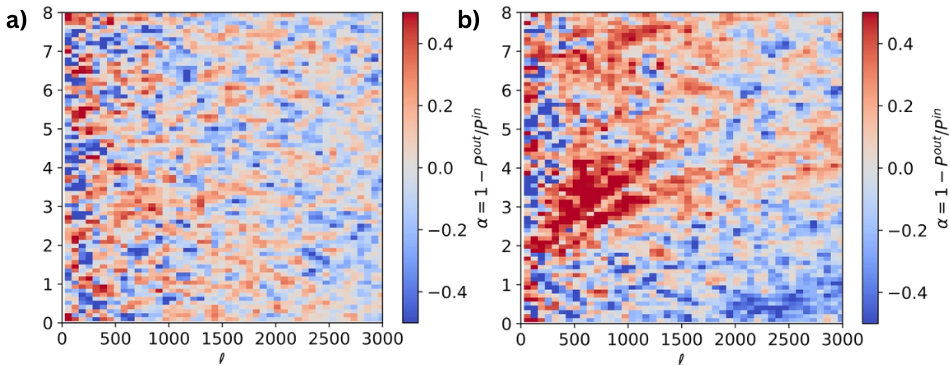} 
\caption{Absorption coefficient 24 hours before (\textit{a}) and after (\textit{b}) the emergence of AR~11158 derived from a series of Doppler velocity images centered on the AR \citep{Waidele2023} There is strong absorption for the \textit{f}-mode and low-order \textit{p}-modes after emergence, characteristic of the depressed acoustic power in the presence of surface magnetic fields. However, there is no coherent absorption or emission of the same modes during the \textit{f}-mode power enhancement.} 
\label{fig:fmode}
\end{figure}

A re-analysis of the same ARs by using a Fourier-Hankel decomposition \citep{Waidele2023} showed that there is indeed a statistically significant increase in the \textit{f}-mode power in ARs 11158, 11768, 11130, and 11072 which is consistent with the previous results. However, the measurement of the absorption coefficient showed no absorption or emission during \textit{f}-mode power enhancement \citep[Figure \ref{fig:fmode},][]{Waidele2023}. This result questioned the mechanism of \textit{p}-mode to \textit{f}-mode conversion via scattering, which was hypothesized in earlier work. 

The lack of a precise description of changes in the \textit{f}-mode power and subsurface magnetic field raises a broader question: what is the relationship between wave propagation and magnetic fields? Many efforts have investigated various aspects of this relationship, for example, disentangling the magnetic effects on wave travel times from thermal effects \citep{Felipe2016}, conversion of acoustic waves to MHD waves \citep{Khomenko2006}, and the influence of surface magnetic fields on helioseismic measurements \citep{Lindsey2005}. A comprehensive understanding of the interaction between magnetic fields and acoustic waves remains unclear, though, especially for complicated configurations of the magnetic field such as those found in solar ARs.

\subsection{Summary}
To conclude, there have been many investigations using HMI data to search for pre-emergence signatures of ARs. Time-distance helioseismology has revealed the presence of mean phase travel time deviations which occur before the AR emergence, in some cases more than 24 hours before the flux concentration reaches the surface. The mechanism that generates these deviations remains to be determined, though, as the direct effect of the magnetic field is expected to be small. Other, more straightforward methods have also found precursors of active region emergence, such as measuring surface horizontal diverging flows (HDFs) that develop within 1-2 hours before emergence. Enhancement to the \textit{f}-mode power has also been suggested as a precursor, though the methodology required to measure these changes is highly sensitive to calibration and normalization procedures and is still somewhat controversial. Finally, the recent work using a type of neural network has been able to predict changes to the continuum intensity that is associated with surface magnetic flux.

\section{Sunquakes}

\subsection{Introduction}

	The potential connection between solar flares and acoustic oscillations has been a subject of interest since the early days of helioseismology. \citet{Wolff1972} initially proposed that flares could stimulate high-order modes of solar whole-body oscillations. While some studies reported changes in \textit{p}-mode parameters associated with flares \citep{2009ApJ...706L.235M,2014A&A...561A.123M}, other studies did not find correlations between high-frequency \textit{p}-modes and flares \citep[e.g.][]{2012SoPh..281...21R,2017MNRAS.471.4677K}. A comprehensive analysis using ring-diagram techniques on MDI and HMI data found no amplification in mode amplitudes due to flaring activity larger than a 10 percent uncertainty \citep{2021MNRAS.505..293R}. As a result, Wolff's original suggestion that flares could excite whole-Sun oscillations remains an open question in solar physics.
	
	However, helioseismic observations performed in 1996 by the MDI instrument detected local helioseismic waves excited by a solar flare \citep{Kosovichev1998}. These waves formed concentric ripples traveling from the flare source, resembling seismic waves during earthquakes. Therefore, this phenomenon was dubbed `sunquake'. 
	The primary interests in studying sunquakes are, first, understanding the physical mechanism of flare impacts that cause helioseismic waves of a relatively high amplitude that reaches several hundred m/s, exceeding the amplitude of background solar convection and oscillations, and, second, understanding how these waves interact with magnetic fields and plasma flows in active regions. 
	
	Observations of several powerful sunquake events by MDI during Solar Cycle 23 showed that sunquake sources are associated with flare hard X-ray emission, both spatially and temporally. It was also found that the wavefronts are often anisotropic, perhaps due to the wave interaction with flows and magnetic fields and also because of the multiple flare impacts usually distributed along the magnetic field polarity inversion line in active regions. The close association with the hard X-ray sources indicated that the sunquake may be caused by energetic electrons penetrating deep into the solar atmosphere, in accordance with the `thick-target' model of solar flares. However, the radiative hydrodynamic simulations of this model showed that the particle beams do not reach the solar photosphere and thus can heat only the upper and middle chromosphere, causing chromospheric evaporation. It was suggested that sunquake impacts could be caused by shocks traveling downward from the heated area. However, these shocks quickly decay and do not produce the required energy and momentum transfer in the photosphere. Therefore, additional mechanisms have been discussed, such as the Lorentz force associated with coronal mass ejections, proton beams, and correlation with convective downdrafts. One of the difficulties was that the MDI observations with sufficient resolution (2 arcsec/pixel) were performed only two months a year, and, therefore, the statistics of sunquakes were incomplete.  
	
	The uninterrupted high-resolution (0.5 arcsec/pixel) observations with the HMI instrument have opened new perspectives for studying sunquakes. It turns out that sunquakes are more common than assumed before and can be initiated by relatively weak flares (in terms of the flare X-ray class), while many strong X-class flares do produce sunquakes. The high temporal cadence of the HMI observations allows us to discover high-frequency acoustic signals of sunquakes. The exact timing of the photospheric impacts relative to the X-ray signals can be measured by using the HMI filtergram data taken every 3.75 seconds. In addition, the links between the sunquake sources, white-light emission kernels, and rapid magnetic field variations are investigated by analyzing the Stokes profiles of the photospheric Fe~I~6173~\AA~ line observed by the HMI. 
	
	This section overviews the observational results from HMI and theoretical models motivated by the observations.
	
	\subsection{Detection Methods}
	
	The detection of sunquakes depends on the amplitude of the excited wave relative to the background solar oscillations and convective flows. Because of the impulsive excitation source, the power of the flare-excited waves peaks in the high-frequency range of solar acoustic oscillations, typically above the acoustic cut-off frequency of the solar atmosphere, which is about 5.2~mHz \citep{2012SoPh..280..335A}. In this frequency range, the background convection power diminishes. Sunquake waves are most easily detected by applying a high-frequency filter centered at about 5\,--\,6~mHz, and sometimes even at higher frequencies \citep{Zharkov2011a,Matthews2015,Lindsey2020}.
	
\begin{figure}
		\begin{center}
			\includegraphics[width=\linewidth]{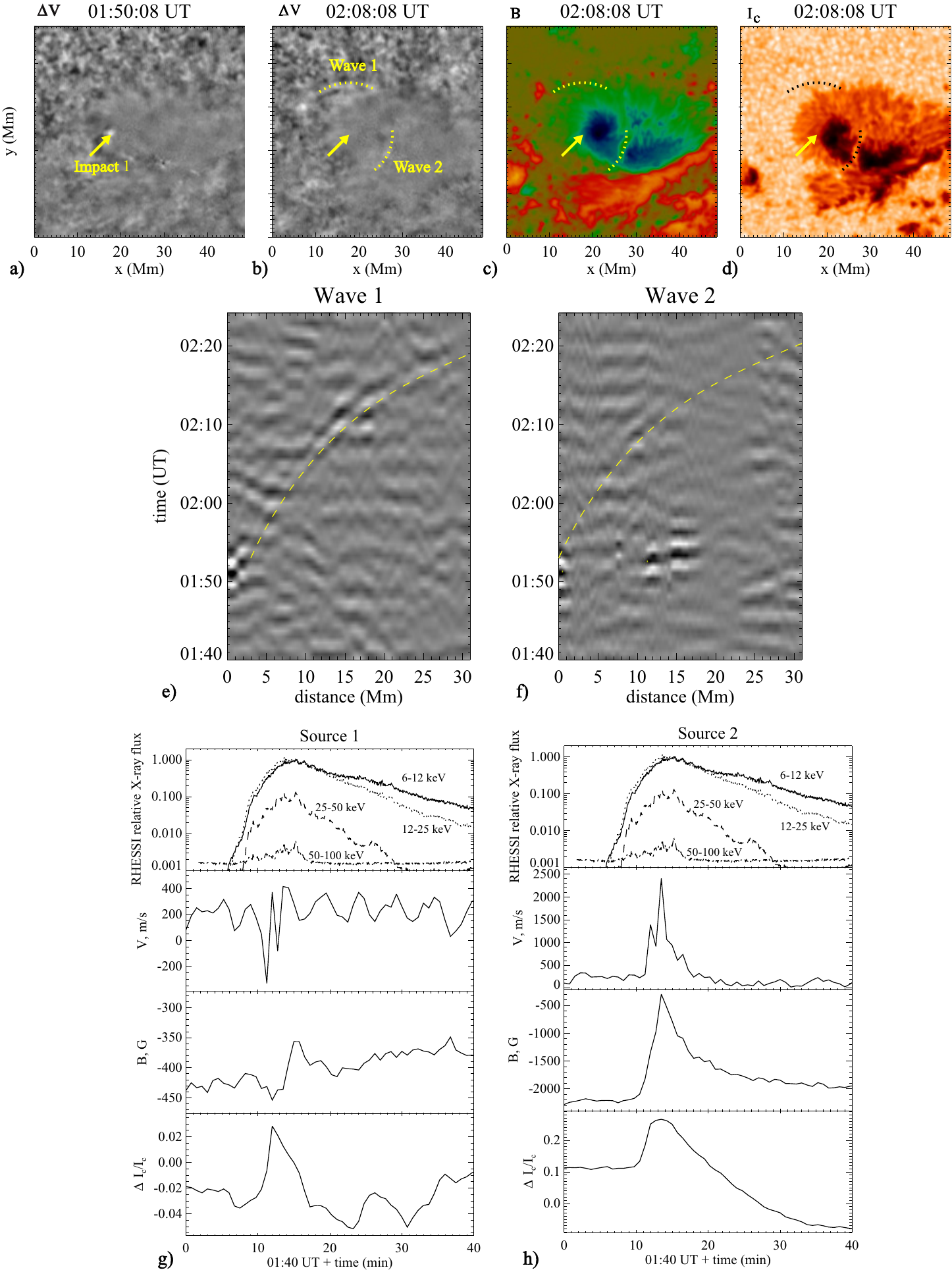}
			\caption{Sunquakes produced by X2.2 flare of 15 February 2011. The \textit{top images} show the sunquake source (\textit{a}) and the helioseismic wavefront (\textit{b}) in the photospheric Dopplergrams and the location of the source in the HMI magnetogram (\textit{c}) and intensity image (\textit{d}). \textit{Middle panels} \textit{(e-f}) show the time-distance diagrams of two sunquake waves. \textit{Bottom panels} (\textit{g-h}) show the X-ray intensities from the RHESSI and Fermi spacecraft and the variations in the Doppler velocity, line-of-sight magnetic field, and intensity in the sunquake sources \citep{Kosovichev2011b}.}\label{sunquakes_fig_01}
		\end{center}
	\end{figure}

\begin{figure}
		\begin{center}
			\includegraphics[width=\linewidth]{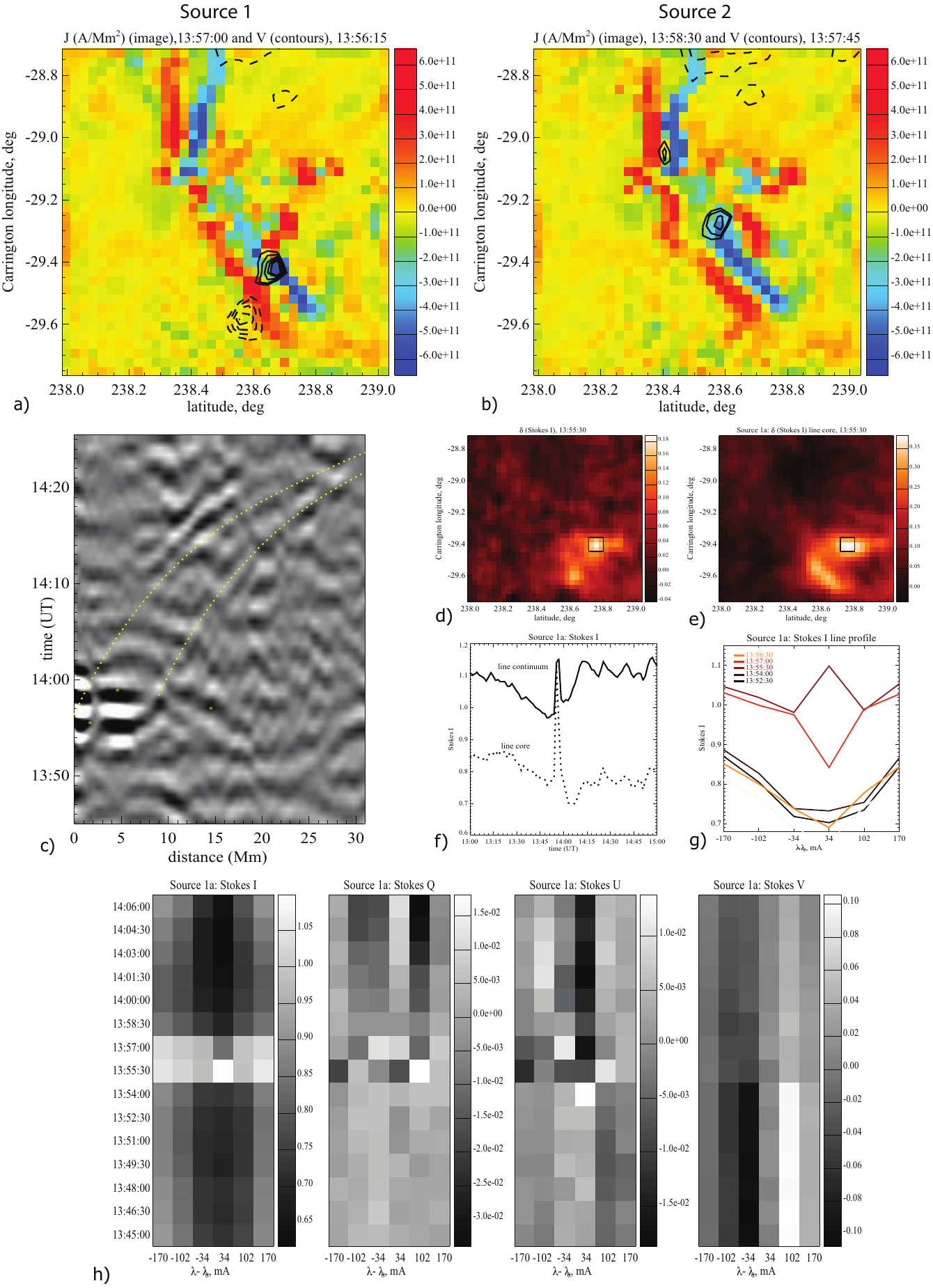}
			\caption{Sunquakes initiated by X1.1 flare of 10 May 2022:  (\textit{a-b}) the distribution of vertical electric currents and location of sunquake sources; (\textit{c}) the time-distance diagram of the helioseismic waves; (\textit{d-e}) White-light emission kernels; (\textit{f}) variations of the HMI continuum and line core intensity and (\textit{g}) the line profiles; (\textit{h}) variations of the Stokes profiles of the Fe~I~6173\AA~ line \citep{Kosovichev2023}.}\label{sunquakes_fig_02}
		\end{center}
	\end{figure}

	The most unambiguous evidence of a sunquake is the visual detection of the expanding ring-shaped ripples from the flare site in the filtered Dopplergram images (e.g., top panels of Figure~\ref{sunquakes_fig_01}b). The ripples are usually best observed starting about 20 minutes after the impulsive flare impacts observed in the HMI Dopplergrams as a high-velocity signal Figure~\ref{sunquakes_fig_01}a) and also in the continuum intensity images as white-light flare kernels (Figure~\ref{sunquakes_fig_02}d-e). Because of the fast and strong variations in the HMI Fe~I 6173~\AA~ line profile (Figure~\ref{sunquakes_fig_02}g), the physical properties of these impacts are difficult to assess \citep{Sadykov2020,Kosovichev2023}.
	
	While expanding ripples are visible in some strong sunquakes, they can be challenging to observe in weaker sunquakes obscured by various background oscillations. A more effective method for detecting these sunquake waves is through a time-distance diagram, which involves averaging time sequences along thin annuli or arcs around the flare footpoints and stacking them by distance. This approach enhances the signal-to-noise ratio by averaging along the wavefront of the ripples, making the sunquake waves more visible against the background. As shown in Figures~\ref{sunquakes_fig_01}(e-f) and \ref{sunquakes_fig_02}(c), the propagation of the sunquake wave is clearly visible and aligns well with the theoretical time-distance relationship (yellow-dashed curve). 
	
	\begin{figure}
		\begin{center}
			\includegraphics[width=0.7\linewidth]{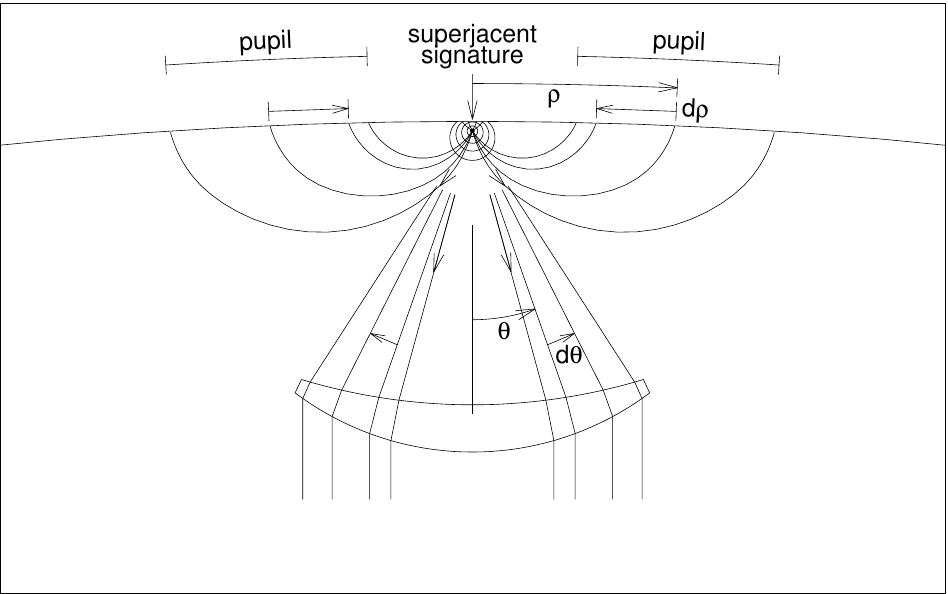}
		\end{center}
		\caption{Diagram illustrating parallels between (1) optical imaging by a lens (\textit{bottom}) of a compact electromagnetic source of waves that travel along straight rays through a uniform medium and (2) computational imaging of an acoustic source of waves propagating along curved rays bent back to the Sun's surface by refraction \citep{Lindsey2020}.}
		\label{holog_diagram}
	\end{figure}

\begin{figure}
		\begin{center}
			\includegraphics[width=0.85\linewidth]{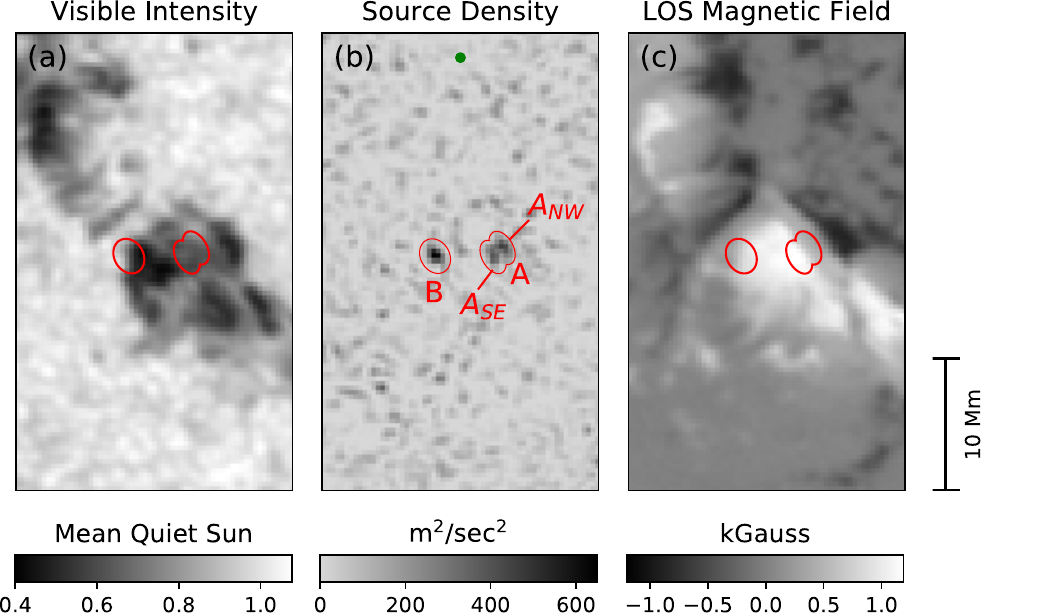}
		\end{center}
		\caption{Computational acoustic holography of impulsive acoustic emission emanating from the M9.3-class flare of 30 July 2011 hosted by NOAA AR~11261 during the early impulsive phase of the flare. Panel ({\it b}) maps the source density of acoustic radiation released into the 2-mHz spectral band centered at 10~mHz at 02:09:45~TAI. This image is focused on the Sun's surface.  Frames ({\it a}) and ({\it c}) show cospatial visible-intensity and LOS-magnetic maps, respectively.  The \textit{filled green circle} at the top of the middle frame indicates the FWHM of the source density profile of an artificial 10-mHz point source at the base of the photosphere as imaged by the diagnostic that generated the source-density map \citep{Lindsey2020}. }\label{holog_image}
	\end{figure}

	Another detection method is the helioseismic holography method \citep[e.g.][]{Lindsey1997, dbl1999, Lindsey2000a, dl2005}. Helioseismic holography can be thought of as a computational implementation of basic principles of wave optics, running a numerical simulation of the job done in familiar electromagnetic optics by lenses in regressing the wave field to the neighborhood of its source (Figure~\ref{holog_diagram}). Among its utilities is mapping the source-power density (hence, just ``source density'') of acoustic transients emitted by acoustically active flares.  In this capacity, the diagnostic delivers diffraction-limited acoustic images of the acoustic sources it focuses on.  Figure \ref{holog_image} shows an example of the diagnostic applied to helioseismic observations of surface ripples emanating outward from the M9.3-class flare of 2011-07-30 in NOAA AR~11261 \citep{Martinez2020}.  Both sources, compact to within the acoustic diffraction limit of 10-mHz acoustic waves from the source region, appear on conspicuous boundaries, the left on the boundary separating a sunspot umbra from the adjoining quiet Sun, and the right split along a penumbral magnetic neutral line.
	
	Applied to outgoing surface ripples from a transient acoustic source, the computational task of helioseismic holography is to reverse the acoustic waves in the solar interior, of which the ripples are the surface manifestation, to the neighborhood source location and time, by applying Green's functions, which describe how waves propagate in the medium. Theoretical Green's functions, which are noise-free and calculated from a quiet-Sun model, are commonly used in holography studies and generally work well for reconstructing sunquake sources \citep{Chen2021}. This enables a statistical analysis of the temporal relationship between flares and the reconstructed sunquake velocities.

	A recent study introduced a novel approach using ML techniques to enhance the detection of sunquakes \citep{mercea2023machine}. 
	This method is still in the prototype stage but has demonstrated the potential to identify weak acoustic emissions that may be difficult to detect using conventional methods.
	
\subsection{Theoretical Interpretations and Models} 

Mechanisms so far considered as contributors to the seismic emissions of flares can be summarized as follows.

{\textit{Shocks}.} Flares exhibit strong particle acceleration, as reflected in their hard X-ray and $\gamma$-ray spectra. The high-energy particles producing these radiations can penetrate into the solar atmosphere, resulting in ``thick-target'' heating of the chromosphere. This heating ejects some of the material up into the corona. The intense heating in the thick target, including the dynamic effects of the ejection, excites downward shock waves that carry some of the flare energy to the photosphere, potentially becoming the source of the seismic emission  \citep[e.g.][]{1975SvA....18..590K,kz1995,1984ApJ...279..896N,1985ApJ...289..434F,2005ApJ...630..573A}.

{\textit{Back warming}.} Much of the downward radiation from a heated chromosphere (the visible continuum, specifically) should penetrate deeper into the atmosphere and be absorbed \citep[e.g.][]{1972SoPh...24..414H}. The resulting photospheric heating, known as ``back-warming''  \citep{2005AdSpR..35.1743G}, if it is sufficiently compact, can generate acoustic transients propagating into the underlying subphotosphere \cite{detal2006}.

{\textit{Penetrating particles}.} High-energy protons can penetrate into the low photosphere and heat it directly \citep{1970SoPh...15..176N,2005AdSpR..35.1743G}. This mechanism requires large particle beam intensities at high energies of order 100~MeV  for protons. Very energetic particles can indeed penetrate below the photosphere, thus optimally transferring momentum and energy between the corona and interior. However, a seismically active flare (SOL2005-01-15T00:43) without appropriate $\gamma$-ray emission led to the arguments that sunquakes may not be caused by high-energy protons \citep{metal2007,Kosovichev2007}.  
Recent radiative hydrodynamics (RADYN) models of the atmosphere heated by non-thermal protons showed that these models could explain the continuum (white-light) enhancements during the flare impulsive phase and also explain the generation of sunquakes \citep{Sadykov2024}. The continuum enhancement was particularly prominent in the models with the low energy cut-off greater than 500 keV. In contrast, the strongest helioseismic impact is found in the models with an energy cut-off of less than 100 keV. In this model, the sunquake momentum was delivered directly by the low-energy (deka- and hecto-keV) protons, suggesting their important role in sunquake generation.

\textit{Lorentz Force Transients}. Flares are dynamic events that cause appreciable changes to the magnetic field configuration in the corona, with corresponding changes in the photosphere itself \citep{anetal1993,detal2006,Hudson2008}. This mechanism differs radically from the hydrodynamic mechanisms involving chromospheric or photospheric heating and gains credence because of the general consensus that identifies flares and  CMEs as magnetic in origin. Observations of Si~I and He~1 infrared lines in the X1-class flare, SOL2014-03-29T17:48, in NOAA AR~12017 \citep{Judge2014}
detected a transient photospheric blue shift in the acoustic source region deduced by acoustic holography applied to the corresponding HMI observation.  The spectral lines indicated a transient shift in the magnetic field orientation. However, extrapolations of the acoustic energy injected into the subphotosphere of the source region by the associated Lorentz-force transient were insufficient to account for the energy carried away by the acoustic transient.

\begin{figure}
		\begin{center}
			\includegraphics[width=0.75\linewidth]{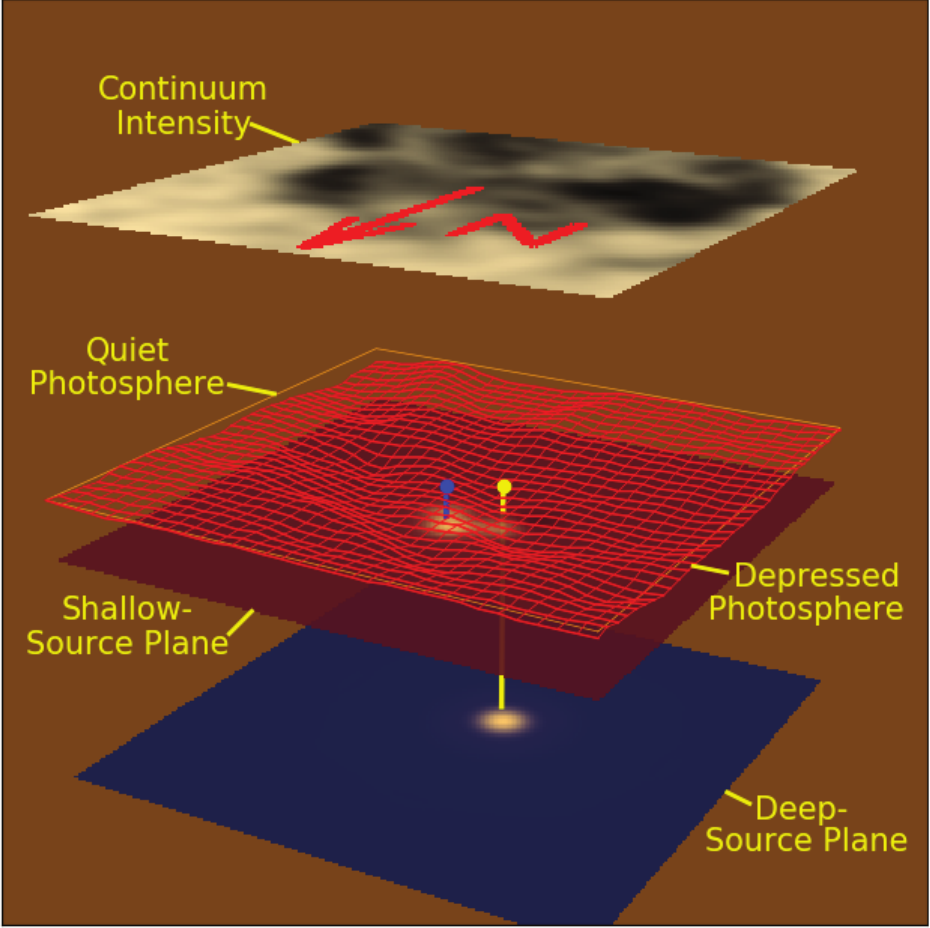}
		\caption{Projection-rendering of `Model 1' that best fits the 10-mHz acoustic source densities mapped by distributions of dipole emission over planes 200 km (\textit{magenta}) and 1150 km (\textit{deep blue}) beneath the quiet photosphere. Each horizontal \textit{square panel}, obliquely projected to the viewer, is 11.13~Mm across. Vertical displacements are stretched by a factor of five with respect to the horizontal scale. The north-western `Shallow Source' is marked by a \textit{blue peg} tacked into the plane 200~km beneath the quiet photosphere. The `Deep Source' is marked by a \textit{yellow peg} extending just short of the deeper plane 1,150~km beneath the quiet photosphere \citep{Lindsey2020}.}\label{model}
        \end{center}
	\end{figure}

{\textit{Transient Emission from Submerged Sources}.} All of the contributors to flare acoustic emission proposed above are confronted, at least to some degree, by occasional instances of acoustic sources that are devoid of any known significant transient photospheric disturbances (continuum-intensity, magnetic, and Doppler) within 5~Mm or more of the acoustic source coherently reconstructed by helioseismic holography.  All of those contributors would disturb the source photosphere conspicuously in quantifiable ways.  A strong emission at 10 mHz in the M9.3 flare of 30 July 2011, suggested that it was potentially caused by a deep source located about 1 Mm below the solar surface \citep{Martinez2020}.  A sunquake associated with the M1.9 of 2 July  2014, flare shows the signature of a high-frequency compact acoustic source at a depth of 2 Mm \citep{Perez-Piel2022}. This has raised the question of whether, at least in some instances, the driver of the emission that emanated, while it must be triggered by MHD above the photosphere, was itself submerged. A model that best fits the acoustic emission of the 30 July 2011 flare (`Model 1')  is shown in projection in Figure \ref{model} \citep{Lindsey2020}.

{\textit{Enhancement by Background Oscillations}.} In addition to these mechanisms, it was hypothesized that a sunquake is more likely to occur if, during the flare's impulsive phase, the background oscillation at the flare footpoint is directed downward in the same direction as the impulse from above, thereby enhancing it \citep{Chen2021}.
Analysis of reconstructed velocity fields of 41 sunquake sources found that over 80\% of them exhibited net downward velocities, supporting this hypothesis.

{\textit{Numerical Models}.} Two types of excitation mechanisms were tested using a 3D hydrodynamic model: 1) an instantaneous momentum transfer at various heights in the atmosphere and beneath the surface, modeling a process that is analogous to the shock excited by the thermalization of the electron beam, and 2) a more gradual energy transfer caused by an external force \citep{Stefan2020}. The models showed the wave amplitude is substantially reduced in the model with the mode damping derived from observations of solar oscillations in magnetic active regions. In the successfully analyzed events of momentum mechanisms, the energy required to excite the quake with the damping in active regions is on the order of $10^{29}$ erg. In contrast, with the damping in nonmagnetic (quiet-Sun) regions, the energy estimate is $10^{28}$ erg, which is consistent with recent estimates using acoustic holography methods \citep{Chen2021}. 

None of the mechanisms has been ruled out, and indeed, some combination of these effects may apply, though not necessarily the same ones in all flares.

	\subsection{Case Studies}

	The first significant sunquake events observed with the HMI instrument were produced on 15 February 2011, by the X2.2-class solar flare \citep{Kosovichev2011b,Zharkov2011a,Zharkov2013}. The initial flare impacts were observed in the form of compact and rapid variations of the Doppler velocity, line-of-sight magnetic field, and continuum intensity. These variations formed a typical two-ribbon flare structure with two sunquake sources and are believed to be associated with the thermal and hydrodynamic effects of high-energy particles heating the lower atmosphere. The analysis of the SDO/HMI and X-ray data from RHESSI showed that the helioseismic waves might be associated with the photospheric impacts in the early impulsive phase observed before the hard X-ray (50-100 keV) impulse.
	An acoustic holography analysis applied to the same HMI data deduced that in addition to these (eastern) sunquake sources, there was another weaker (western) source \citep{Zharkov2011a,Zharkov2013}.  The estimated acoustic energy released was about $1.18\times 10^{28}$ ergs for the eastern source and $6.08\times 10^{27}$ ergs for the western source. These estimates are higher than the Lorentz-force energy estimates \citep{Hudson2008,Alvarado-Gomez2012}.  These observations left open the question of whether the Lorentz force can be a possible mechanism for quake excitation.
	
	The observation of a strong sunquake caused by a compact C7.0 solar flare on 17 February 2013 suggested that the sunquake source might be associated with rapid current dissipation or a localized impulsive Lorentz force in the lower layers of the solar atmosphere because it did not spatially correspond to the strongest hard X-ray emission source \citep{Sharykin2015}. It was the weakest X-ray class flare that produced a significant sunquake. These authors presented another example of a modest X-ray class (M9.3) flare of 30 July 2011, which made a strong photospheric impact and produced a sunquake \citep{Sharykin2015a}. The absence of CME ruled out magnetic rope eruption as a mechanism of helioseismic waves in this flare. Also, the sunquake impact did not coincide with the strongest HXR source, contradicting the standard beam-driven sunquake generation mechanism. However, the sunquake initiation might be related to the rapid dissipation of electric currents and associated high-speed plasma flows in the lower solar atmosphere during the flare energy release.
	
	Investigations of two homologous X-class flares occurred in the same location with similar morphologies on 6 and 7 September 2011, of which only the first flare produced a sunquake \citep{Zharkov2013a}, suggested that the generation of sunquakes may be explained by a unified model involving simultaneously Lorentz force, hydrodynamic shocks and white light emission, all produced by the magnetic energy released during the flare and its conversion into energetic particles (electrons) with their further effects on a flaring atmosphere and the interior \citep{Xu2014,Macrae2018}. However, the question of why some flares produce strong photospheric impacts that generate the helioseismic response and others do not is still unresolved. 
	
	Fast and complex changes in the magnetic field strength and topology complicate the interpretation of observational data. The reconstruction of magnetic field changes in the solar atmosphere during the M5.7 flare of May 2012 using a non-linear force-free extrapolation and comparison with other flares in the same active region suggested that the sunquake initiation might be associated with the rearrangement of magnetic field lines and strong currents at low altitudes above the polarity boundary line are transformed into the currents along the system of loops oriented at wide angles to the neutral line \citep{Livshits2016,Grigoreva2017}. This process may explain the apparent fast supersonic motion of the sunquake sources, originally noticed with the SoHO MDI observations \citep{Kosovichev2007} and confirmed by the analysis of sunquakes observed by HMI \citep{Kosovichev2011b,Zharkov2011a}. The sequential impacts that move along the magnetic arcade can explain the observed anisotropy of the sunquake wavefronts \citep{Kosovichev2006,Kosovichev2006a}.
	
	The X9.3 flare of  6 September 2017, the most powerful flare of Solar Cycle 24, generated strong white-light emission and multiple sunquakes. The photospheric flare impacts started to develop in compact regions near the magnetic polarity inversion line (PIL) in the pre-impulsive phase before detection of the HXR emission. The initial photospheric disturbances were localized in the region of the strong horizontal magnetic field of the PIL and, thus, are likely associated with a compact sheared magnetic structure elongated along the PIL. The acoustic egression power maps revealed two primary sources of sunquake generation, which were associated with places of the strongest photospheric impacts in the pre-impulsive phase and the early impulsive phase \citep{Sharykin2018}. Analysis of the high-cadence HMI filtergrams suggests that the flare energy release developed in the form of sequential involvement of compact low-lying magnetic loops that were sheared along the PIL \citep{Livshits2016}. In addition, a second bounce of the generated helioseismic waves detected in this flare was explained by a stronger momentum delivered by the shock generated by a mixed electron-proton beam in the flaring atmosphere and at deeper depths of the interior where this shock was deposited \citep{Zharkova2020}.
	
	The sunquake waves are usually best observed in the frequency range 5-6 mHz as their amplitude rapidly decreases at high frequencies \citep{Zharkov2013}. Nevertheless,  a strong emission at 10 mHz detected in the M9.3 flare of 30 July 2011, \citep{Martinez2020} was potentially caused by a deep source located about 1 Mm below the solar surface \citep{Lindsey2020}. 
	This result suggested that sunquakes might be caused by perturbations in magnetic subphotospheric layers without producing a compact source signature on the surface (Figure~\ref{model}).  Numerical modeling of deep excitation sources showed the estimated source amplitude is too low to produce the observed wavefront \citep{Stefan2022}.

	\subsection{Statistical Relationship Among Sunquakes, Magnetic Field Variations, White-Light Emission, and X-ray Emission}
	
	Sunquakes represent a critical piece in the broader puzzle of understanding solar flare physics. These helioseismic events provide unique insights into the dynamic processes occurring during some large solar flares, particularly connecting what happens in the low corona and chromosphere with the photosphere and upper sub-photosphere. While sunquakes are invaluable for probing the surface and subsurface seismic impacts of flares, they are only one component of a much broader, complex physical system. A comprehensive understanding of seismically active solar flares requires that sunquakes be studied alongside other key observational phenomena, each offering complementary perspectives on flare dynamics.
	
	Nonthermal hard X-rays, for example, serve as proxies for the dynamics of accelerated electron beams, revealing the distribution and energy of these particles as they precipitate into the lower atmosphere. Magnetic field variations offer another vital piece of the puzzle, providing insights into the magnetic reconnection processes that drive flares and potentially induce Lorentz forces capable of triggering sunquakes. White-light emission, which can account for up to 70\% of the total radiated energy during a flare, is thought to be produced through the hydrogen recombination continuum in the upper chromosphere. This emission not only marks intense energy release but also provides clues about the mechanisms behind flare energy dissipation.
	
	Investigating the relationships between sunquakes and other solar flare phenomena can help achieve a more holistic understanding of flare energy transfer processes. When analyzed in conjunction with nonthermal hard X-rays, magnetic field changes, and white-light emission, sunquakes can reveal the intricate interplay between various layers of the solar atmosphere around the photosphere. This multi-observational approach not only deepens our knowledge of sunquakes themselves but also enhances our overall understanding of the physics underlying solar flares. 
	
	Before HMI, statistical studies of sunquakes with other flare phenomena were often constrained by small sample sizes, primarily limited to events detected by MDI. The advent of HMI observations has enabled the detection of more sunquakes, and the expanded catalogs allow for more robust statistical studies,
	providing deeper insights into the complex relationships between sunquakes and other flare-related phenomena. For instance, the relationship between sunquakes, white-light flares, and hard X-ray emissions showed a general correlation between higher electron energy deposition and sunquake occurrence, with seismically active flares often exhibiting harder hard X-ray spectra and more compact source sizes \citep{pedram2012survey}. 
	This study found that the link between white-light emission and sunquakes was inconsistent, suggesting that other factors, such as energy distribution and timing, may play crucial roles. 
	
	A statistical study of 75 solar flares during Solar Cycle 24, focusing on events with significant hard X-ray (HXR) emissions observed by RHESSI, identified 18 sunquakes, all occurring in flares that exhibited these white-light enhancements. This finding underscores the importance of white-light emission as a marker for sunquake activity and suggests a tight coupling between the energy release processes in the chromosphere and the generation of seismic waves in the photosphere. A later survey of 60 strong flares detected 41 sunquakes, all exhibiting continuum emissions, confirming the association between sunquakes and continuum enhancement  \citep{Chen2021}.
	
\begin{figure}
		\begin{center}
			\includegraphics[width=\linewidth]{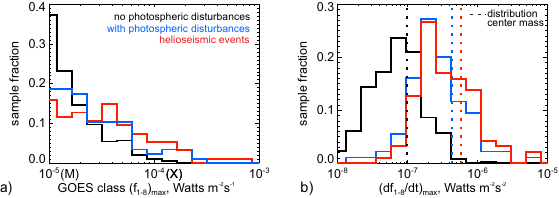}
			\caption{ Statistics of sunquakes observed in Solar Cycle 24 \citep{Sharykin2020}.}\label{sunquakes_fig_03}
		\end{center}
	\end{figure}

	A more comprehensive search for sunquakes from 500 M-X class flares during Solar Cycle 24 identified 94 helioseismic events confirmed by at least one detection method \citep{Sharykin2020}. This analysis revealed that many moderate-class flares can produce strong sunquakes, while some powerful X-class flares may not be helioseismically active. Notably, the total energy of sunquakes exhibits a stronger correlation with the maximum value of the soft X-ray time derivative than with the X-ray class, indicating that sunquake generation is closely related to the impulsiveness of the energy release. Figure~\ref{sunquakes_fig_03} shows the distributions of the seismic and nonseismic flares, between which more evident differences appear in terms of the
	maximum values of the flare-energy release rate (panel \textit{b}) than in the GOES class (panel \textit{a}). 
	
	While sunquakes are consistently associated with Doppler transient impacts (sudden perturbations sometimes reaching extreme values in the HMI observables), not all flares with Doppler transients lead to sunquakes \citep{2015arXiv150207798B}. This suggests that Doppler transients are necessary but not sufficient on their own as a proxy to detect sunquakes, reinforcing the need for a multi-parameter analysis when studying these events \citep{Sharykin2020,Chen2021}. The bottom panels of Figure~\ref{sunquakes_fig_01} show examples of Doppler transients at two sunquake source locations, occurring simultaneously with continuum and magnetic field enhancements.

	\subsection{Imaging Spectro-Polarimetry of Sunquake Sources}
	
	The two cameras of the HMI instrument record filtergrams in the linear and circular polarizations in six positions across the spectral line Fe~I 6173.34~\AA. The filtergrams are used to generate maps of the line width, line depth, Doppler shift, and line-of-sight magnetic field with a 45-sec cadence. In addition, the full Stokes line profiles are calculated with a 90-sec cadence, using the filtergrams from both HMI cameras. 
	
	A study of a flare event (SOL-2010-06-12T00:57) found significant blueshifts at both footpoints, while the line remained in absorption \citep{2011SoPh..269..269M}. The blueshifts can be explained by the motion of the photospheric medium toward the observer, which may be caused by various physical phenomena, such as backwarming in the photosphere. 
	The quantitative interpretation of the transient perturbations quantitatively is complicated by the irregular Fe I line shape during flares and requires a detailed radiative transfer modeling taking into account the HMI instrumental characteristics \citep{Sadykov2020}.
	
	A spectro-polarimetric analysis of the first significant sunquake of Solar Cycle 25, observed during the X1.5 flare of 10 May 2022,  revealed a transient emission in the line core in three of the four sunquake sources, indicating intense, impulsive heating in the lower chromosphere and photosphere. In addition, the observed variations of the Stokes profiles reflected transient and permanent changes in the magnetic field strength and geometry in the sunquake sources  \citep[Figure~\ref{sunquakes_fig_02}; ][]{Kosovichev2023}.  A recent analysis of the HMI spectro-polarimetric observations for the X9.3 flare of 6 September 2017, reveals one or more locations within the sunspot umbra of the associated active region where the Fe I 6173~\AA~line goes into full emission, indicating significant heating of the photosphere and lower chromosphere. Modeling the HMI observables for ad-hoc heating of the initial empirical umbra model indicated that the line emission in the white-light flare kernels, which coincided with sunquake sources, could be explained by the strong heating of initially cool photospheric regions \citep{Granovsky2024}. However, the current flare models do not explain such strong heating.
	
	\subsection{Starquakes}
	
	Stellar flares are powerful bursts of energy emitted from the surfaces of stars, possibly caused by the sudden release of magnetic energy in the star's outer atmosphere. These flares are analogous to solar flares on the Sun but can be far more energetic, especially on stars with more intense magnetic fields, such as red dwarfs, where flares can be particularly powerful and frequent. The sunquakes we observe with HMI do not have sufficient energy to elicit a significant manifestation in individual \textit{p}-modes \citep{2021MNRAS.505..293R}.  
	Stellar flares, due to the enormous energy they emit, could be different. The question arises whether the energy they release is sufficient to make the star rings.  However, a comparison of the {solar-type} \textit{p}-mode spectra of eight stars for periods before and after large flares did not find conclusive evidence that the flares induced global acoustic oscillations \citep[starquakes; ][]{Balona_Oscillationsinstellar_2015}.  
	
	\subsection{Key Questions for Future Studies}
	
	Despite the significant progress made in understanding sunquakes and their relationship with various solar flare phenomena, a comprehensive statistical study specifically focusing on the interplay between sunquakes and solar flare magnetic field dynamics is still lacking. Only a handful of studies have examined this relationship in the context of individual flare events, providing valuable but limited insights into the broader picture. 
	These studies, while informative, are not sufficient to establish a robust statistical framework. Moving forward, it is crucial to conduct a systematic statistical study that rigorously investigates the relationship between sunquakes and the magnetic field dynamics of solar flares. Such a study would significantly enhance our understanding of the conditions under which sunquakes are generated and how magnetic field variations influence these seismic events.
	
	The possibility that the sources of flare acoustic transients are submerged opens new mechanisms of acoustic emission, transients that, while triggered by disturbances that originate above the photosphere, are possibly driven by magnetic energy that, like the source, is submerged.  If the magnetic energy that is eventually released into the corona causes the flares we see above the photosphere, is it possible for some of this energy to be released into the acoustic field while it has yet to break the surface?   This would represent a noteworthy connection with the advent of HMI of two areas of research, flare physics and the solar interior, that, in some ways, have formerly been hermetically separate.
	
	A  realistic understanding of the physics of seismic wave generation would in principle teach us a great deal about how flares work, in a context that now extends their mechanics significantly into the solar interior.

\section{Solar Irradiance}
\label{sec:irradiance}

When the magnetic field generated in the solar interior emerges at the surface of the Sun in a concentrated form, it modifies the radiative properties of the corresponding regions such that some areas on the Sun appear darker (sunspots or pores) or brighter (faculae or network) than the surroundings.
This modulates the total solar brightness or irradiance.
Thus, analysis and understanding of the variations of solar irradiance provide information on the processes under the solar surface and can, in some ways, be considered complementary to helioseismology. 

Solar irradiance is the Sun's radiative energy flux, measured at the top of Earth's atmosphere and normalized to the mean Sun-Earth distance.
The spectrally-resolved and wavelength-integrated solar irradiance are termed spectral and total solar irradiance (SSI and TSI), respectively.

Due to wavelength- and time-dependent atmospheric absorption and dispersion on the way to the Earth's surface, measuring irradiance with accuracy sufficient to detect its variability is only possible from space. 
Direct measurements of TSI and SSI variability exist since 1978 \citep[]{frohlich_total_2012} (top panel of Figure~\ref{fig:tsirecsatires}) and 1967 \citep[though at the beginning over a very limited spectral range only;][]{ermolli_recent_2013}, respectively.
They revealed that solar irradiance varies on all observable timescales with a typical amplitude in
TSI of about 0.1\% over the solar activity cycle and solar rotation timescales \citep[e.g.][]{kopp_magnitudes_2016}. 
In the UV, the variations are significantly stronger, reaching several percent in the near- and mid-UV, several tens of percent in the far-UV, and up to about 100\% around Ly-$\alpha$ \citep{rottman_observations_1988,floyd_11_2003}.

Determining the source of solar irradiance changes and pinpointing its variation in the past (amplitude and timescales of variation) is particularly important for evaluating the solar influence on Earth's climate. 
Solar energy is produced in the core of the Sun through nuclear reactions.
The chemical evolution of the Sun's core is important for brightness variations on timescales greater than $10^6$ years.
However, on shorter timescales, the energy production in the Sun's core is considered to be rather stable.
Thus, the variations in solar irradiance we observe arise due to changes in how the energy produced in the core reaches the surface of the Sun and how it is radiated from there.
The most viable explanation for irradiance variations on timescales from days to millennia lies with changes in the solar surface magnetic field \citep[] {shapiro_nature_2017,yeo_solar_2017}. 
These changes are primarily modulated by the passage of dark sunspots and bright facular and network regions across the solar disk. 
On shorter timescales, \textit{p}-mode oscillations peaking at around 5 minutes and granulation are the main drivers of the irradiance fluctuations.

The effect of solar surface magnetic field on irradiance variations has been hypothesized since the very early measurements of irradiance \citep[e.g.][]{oster_solar_1982,sofia_solar_1982}.
Sunspots block energy, which is redistributed in the solar convection zone.
This stored heat is then re-emitted over the Kelvin-Helmholtz timescale of the convection zone \citep{spruit_flow_1982,domingo_solar_2009,solanki_solar_2013}. This can lead to small variations in the solar radius, which has been suggested to be connected to irradiance variations \citep[]{sofia_solar_1979,sofia_solar_1998,pap_relation_2001}. A surface radius change of about 0.06 arcseconds (about 45 km) is considered sufficient to account for the $\sim$0.1\% variation in TSI values over the solar cycle. 
However, more recent estimates based on SoHO/MDI and SDO/HMI data found significantly smaller changes \citep{emilio_constancy_2000,kuhn_constancy_2004,jain_suns_2018}. This suggests that the radius changes, although interesting in themselves, cannot explain all the TSI variations. 

\subsection{Irradiance Reconstructions with SDO/HMI Magnetograms}
\label{sec:hmi_irradiance}
To compute changes in the irradiance caused by magnetic features at the solar surface, models require knowledge about the surface distribution of these features at a given time as well as their brightness or contrast with respect to their quiet surroundings, i.e., regions free of measurable magnetic field. 
Obviously, the most direct information on the surface distribution of the magnetic field can be retrieved from solar full-disc magnetograms. This information is used by the Spectral and Total Solar irradiance Reconstruction \citep[SATIRE-S, ``S'' for satellite-era;][]{krivova_reconstruction_2003} model to reconstruct total and spectral solar irradiance, just as its name suggests.
SATIRE-S segments the solar surface into four components: faculae, sunspot umbrae, sunspot penumbrae, and quiet Sun.
The brightness of each component is considered time-independent and was computed with the radiative transfer code ATLAS9 \citep{kurucz_atlas_1970}   from the appropriate semi-empirical model atmospheres \citep{unruh_spectral_1999,tagirov_readdressing_2019}.
The brightness depends on the position of the features on the visible solar disc (center-to-limb variation) and the wavelength.
In principle, this information can also be extracted from observations. For example, SDO/HMI observations have been used to derive the contrasts of faculae and the network with respect to the quiet Sun as a function of the heliocentric angle and the magnetic field strength \citep{yeo_intensity_2013,yeo_intensity_2019}.
However, such observation-based contrasts are limited to individual wavelengths and are less reliable close to the solar limb.
The latter can be improved by combining observations from different vantage points \citep{albert_intensity_2023} by complementing SDO/HMI observations with data gathered by the Polarimetric and Helioseismic Imager (PHI) \citep{solanki_polarimetric_2020} onboard Solar Orbiter (SoLo) \citep{muller_solar_2020}.
Such information can be used to better constrain the semi-empirical model atmospheres.

\begin{figure}
\centering
		\includegraphics[width=\linewidth]{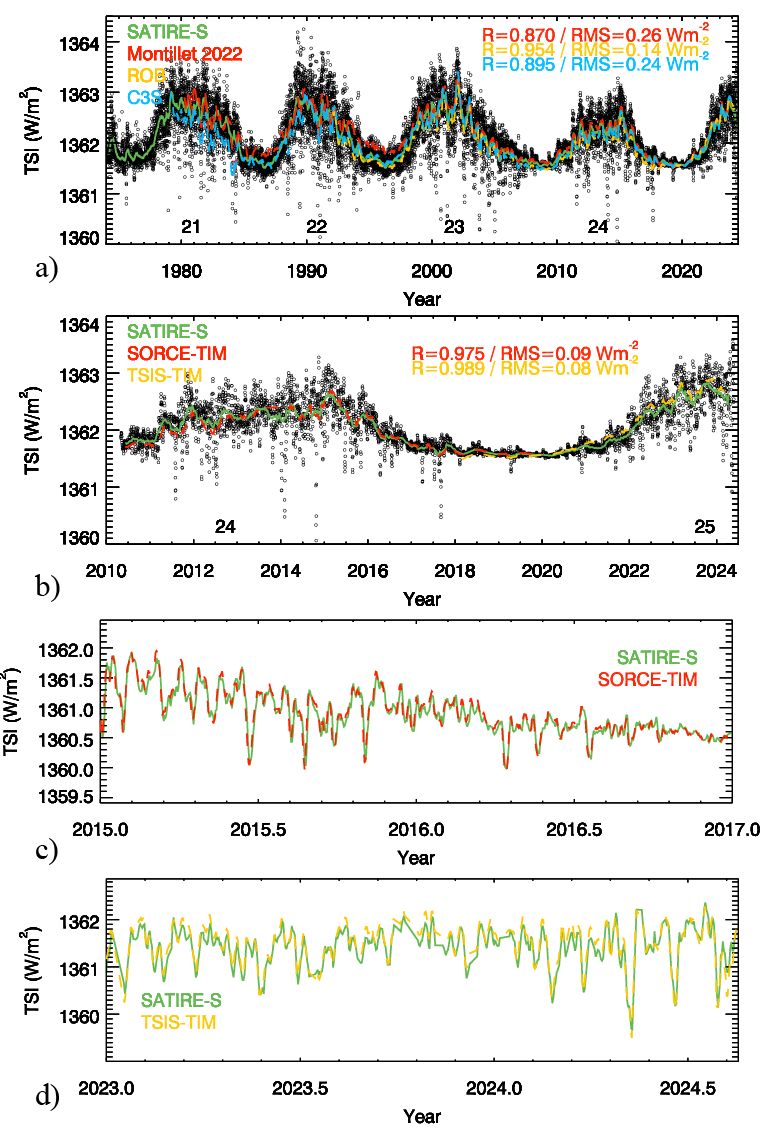}
\caption{TSI reconstruction with the SATIRE-S model (\textit{black} for daily and \textit{green} for 81-day running means) compared to direct measurements. 
	The \textit{top panel} (\textit{a}) shows the complete SATIRE-S composite series, while only the reconstruction with SDO/HMI magnetograms is shown in the other panels. 
	The \textit{lines} in \textit{panels} (\textit{a}) and (\textit{b}) are 81-day running means while they represent daily values in the other panels. Also listed within the first two panels are the linear correlation coefficient, $R$, and the RMS differences between SATIRE-S and the TSI measurements. }\label{fig:tsirecsatires}
	\end{figure}

The first version of SATIRE-S \citep[][]{fligge_modelling_2000,krivova_reconstruction_2003} used SoHO/MDI magnetograms and co-temporal continuum images 
to derive the distribution of the various features on the surface and their evolution with time. 
Later it was updated to incorporate ground-based magnetograms from the Kitt Peak National Observatory \citep{wenzler_comparison_2004,wenzler_can_2005,wenzler_reconstruction_2006,ball_reconstruction_2012}, followed by SDO/HMI magnetograms and continuum images \citep{yeo_reconstruction_2014}, and most recently also Mt Wilson data. 
To our knowledge, these are the only irradiance reconstructions employing SDO/HMI magnetograms.
Figure \ref{fig:tsirecsatires} shows the complete, most recent SATIRE-S TSI reconstruction along with the one using just SDO/HMI magnetograms and compares them to direct measurements of TSI.
In particular, it shows the ROB \citep[named after Royal Observatory of Belgium;][]{dewitte_total_2016,montillet_data_2022}, and the Copernicus Climate Change Service (C3S)\footnote{Available at \url{confluence.ecmwf.int/pages/viewpage.action?pageId=304239361}} TSI composites as well as the Total Irradiance Monitor (TIM) onboard the Solar Radiation and Climate Experiment \citep[SORCE/TIM;][covering 2003\,--\,2020]{kopp_total_2005} and The Total and Spectral Solar Irradiance Sensor TIM \citep[TSIS1/TIM;][covering 2018\,--\,2024]{pilewskie_tsis-1_2018} direct measurements.
The high quality of SDO/HMI magnetograms allows recovering the TSI variations very accurately, reaching linear correlation coefficients (RMS differences) of 0.98 and 0.99 (0.09 and 0.09 Wm$^{-2}$) when comparing the model with SORCE/TIM and TSIS1/TIM direct TSI measurements, as highlighted in Figure~\ref{fig:tsirecsatires}.

More recently, co-temporal magnetograms from SDO/HMI and SoLo/PHI were used to reconstruct irradiance variations as seen from two different vantage points \citep{yeo_reconstruction_2023}.
Thus, SDO/HMI data also helped to check and calibrate the process of reconstructing irradiance with SO/PHI data, which are obtained from different perspectives than the Sun-Earth line and will eventually also reflect the Sun as seen from outside ecliptic plane, from heliolatitudes of up to 33$^\circ$.

\subsection{Using SDO/HMI Magnetograms to Verify the Magnetic Origin of Irradiance Variations}
\label{sec:satire3d}

A limitation of nearly all existing irradiance models is that they have free parameters they need to fix by comparing to irradiance measurements. The first and, until now, the only irradiance model that does not require any tuning to the measured irradiance variations was produced by employing three-dimensional MHD   MURaM \citep{vogler_simulations_2005} simulations of the solar atmosphere \citep{yeo_solar_2017}.
The model produced bolometric images based on the active region distribution found in SDO/HMI magnetograms and achieved an excellent agreement to measured TSI, reaching linear correlation coefficients (RMS differences) of 0.99 (0.07 and 0.09 Wm$^{-2}$) compared to SORCE/TIM and TSIS1/TIM direct measurements \citep{chatzistergos_long-term_2023}. It demonstrated that solar surface magnetism accounts for almost all irradiance variations on time scales of days to the solar cycle, leaving very little room for other mechanisms \citep{shapiro_nature_2017}.

\subsection{Irradiance Variations on Longer Timescales}

While covering only the last 1.5 solar cycles (Solar Cycle 24 and the first half of Cycle 25), SDO/HMI observations are also useful for constraining, even if indirectly, the longer-term changes in the irradiance, which is one of the key open questions in irradiance reconstructions.
Solar full-disc magnetograms have only been available for roughly half a century.
Assessment of the Sun's impact on the Earth's climate requires significantly longer records.
Reconstructions of solar irradiance variations at earlier times are mostly based on sunspot observations or cosmogenic isotope data \citep{solanki_reconstruction_1999,krivova_reconstruction_2007,krivova_reconstruction_2010,vieira_evolution_2011,wu_solar_2018,chatzistergos_discussion_2024}.
These data, however, do not provide direct information on the evolution of the faculae and the network, which are thought to be the main drivers of the variability on time scales of the solar cycle and longer.
This limitation led to significant uncertainty, by about a factor of 10, in the magnitude of the secular irradiance change \citep[e.g.,][]{solanki_solar_2013,chatzistergos_long-term_2023}.
This problem can be addressed by using historical archives of full-disc solar photographs in the \ca line.
Such observations constitute the second-longest photographic archive of solar observations after white-light observations.
Observations in the \ca line started in 1892 and continue to this day at various sites around the world \citep{chatzistergos_analysis_2020,chatzistergos_full-disc_2022}.
It has been shown that solar brightness in the \ca line is an excellent tracer of the Sun's magnetic activity \citep[e.g.][]{schrijver_relations_1989,loukitcheva_relationship_2009,kahil_brightness_2017}.

The relationship between the solar surface magnetic field strength and the brightness in the \ca line core derived from near-co-temporal Rome \ca \; observations \citep{ermolli_rome_2022} and SDO/HMI magnetograms was then used to recover unsigned magnetograms \citep{chatzistergos_recovering_2019} from which solar irradiance was reconstructed  \citep{chatzistergos_reconstructing_2021, chatzistergos_understanding_2024}.
For example, three \ca archives (Meudon, Rome, and San Fernando) were used in the SATIRE-S composite series shown in the top panel of Figure~\ref{fig:tsirecsatires} to fill gaps in the reconstruction left by the direct magnetogram sources. This paved the way to reconstruct unsigned magnetograms from historical \ca observations back to the late 19th century.
Solar irradiance can sometimes be reconstructed before directly measured magnetograms become available. Since TSI reconstructions from the combination of sunspots and such reconstructed magnetograms have been far more reliable than reconstructions based on sunspots alone, we expect that this work will help settle the issue of the increase in TSI since the Maunder minimum.

\section*{Conclusions, Challenges, and Unresolved Problems}

The high-resolution uninterrupted data from the Helioseismic and Magnetic Imager substantially advanced our knowledge of the solar structure and activity.  They have stimulated the development of new helioseismic techniques, such as high-resolution mapping of subsurface flows by local helioseismology, accurate detection of active regions in the solar interior and far side of the Sun, imaging of the zonal and meridional flows in the whole convection zone, high-precision measurements of the solar tachocline and the near-surface shear layer, and analysis of large-scale complex flows using the novel mode coupling technique, which led to discovery of global Rossby waves.  

As a result of these investigations, a new, highly dynamic picture of the Sun is emerging. It is characterized by strong multi-scale interaction of convective flows, differential rotation, global circulation, and magnetic fields. Although the first attempts to measure subsurface magnetic fields have been made, most of our knowledge about the physical processes inside the Sun comes from the flow measurements. In particular, the measurements of the migrating zonal and meridional flows provided evidence for dynamo waves and showed the Near-Surface Shear Layer (NSSL) plays an important role in the solar dynamo and activity cycles. The local helioseismology analysis of subsurface convection revealed the co-existence of multi-scale flows, from supergranulation to giant cells, challenging the classical mixing-length theory of stellar convection. It appears that large-scale convection is linked to Rossby-type waves, creating complicated global flow patterns in the equatorial and polar regions.  

The data analysis and supporting numerical simulation revealed the `leptocline', a highly dynamical shallow substructure of the NSSL in the top 5-10 Mm of the convection zone. It is found that both the zonal and meridional flows vary with the `extended' 22-year solar cycle, and the branches of these flows continue to migrate towards the equator even during the deep minimum of solar activity when no significant magnetic field is observed on the surface, meaning that these variations are driven by subsurface processes, associated with the solar dynamo. During the solar maxima, the zonal and meridional flows interact with converging flows formed around magnetic active regions in the leptocline affecting the meridional flux transport and creating complicated `Solar Subsurface Weather' (SSW) patterns. The data suggest that the subsurface converging and shearing flows in active regions may play a critical role in their flaring and CME activity. Measuring characteristics of these flows (such as divergence and vorticity) and properties of solar oscillations (distribution of the surface power and mode amplitude) may help predict the emergence of sunspot regions and their activity. 

The interaction of subsurface flows with magnetic fields is also critical for understanding the energy flow inside the Sun. 
Current state-of-the-art physics-based irradiance models (e.g., SATIRE-3D;  Section \ref{sec:irradiance}) can explain the measured solar irradiance variations from the surface magnetograms, leaving no room for any additional contributions.
At this stage, however, the influence of subsurface dynamics over timescales significantly longer than those covered by irradiance measurements cannot be ruled out.

In terms of acoustic energy transport, the helioseismology data revealed enhanced power of high-frequency acoustic waves ('acoustic halos') around sunspot regions. In particular, the HMI observations detected a double-ring structure of the power enhancement, which is explained by wave transformation in high-intensity magnetic fields. The halos can contribute to coronal heating in active regions. It was known that solar acoustic oscillations are excited in intergranular lanes close to the surface, but the HMI observations revealed deep acoustic sources beneath sunspot regions at a depth of about 5 Mm. Numerical simulations showed that these sources may be caused by the interaction of magnetic field bundles. 

It is intriguing that the acoustic holography method detected deep (`submerged') acoustic sources associated with helioseismic waves excited during solar flares ('sunquakes'). The HMI showed that sunquakes are more frequent than was thought before and can be produced by relatively weak flares (in terms of the X-ray flux). It was found that the sunquakes are produced by compact impulsive flares at the beginning of their impulsive phase and that their sources are associated with continuum (white-light) emission, Doppler velocity transients, and permanent or transient magnetic field variations. Although the physical mechanism of sunquake sources has not been established, recent radiative hydrodynamic and acoustic models indicate that sunquakes may be caused by deeply penetrating proton beams. Thus, the origin of sunquakes is closely related to the fundamental problem of the flare energy release and transport.

The HMI advances raised new challenges, and many problems remain unsolved. In particular, it is important to implement high-degree helioseismology data analysis and inversion procedures, which will substantially improve the spatial resolution of the solar structure and rotation in the near-surface shear layer. In addition, it is critical for solar dynamics and dynamo modeling to measure the spectrum of solar convection with higher accuracy to solve `the convective conundrum' problem - a large discrepancy among various local helioseismology measurement techniques and numerical models. The deep structure of the meridional circulation and its variations with the solar cycle (particularly at the base of the solar convection zone and high latitudes), which is still a highly debated problem, is also a critical element of dynamo theories. Its solution depends on improving our understanding of the center-to-limb variations of acoustic travel times. The recently discovered Rossby waves potentially play an important role in the solar structure, dynamics, and magnetic activity.
Another unsolved problem that may significantly improve space weather forecasts is the helioseismic diagnostics of emerging and evolving active regions on the front and far sides of the Sun. In addition, the challenges in understanding solar composition, variations in solar irradiance, and sunquake sources are deeply connected to fundamental problems of physics and astrophysics. The future development of this field requires a synergy of new sophisticated data analysis techniques and realistic modeling and simulations. 

The HMI instrument continues providing high-quality data that undoubtedly will lead to new discoveries. We also anticipate that new out-of-ecliptic observations of solar polar regions from Solar Orbiter, which is currently in operation, and the planned Solaris mission will complement the HMI observations and allow us to complete the global dynamic picture of the solar interior.

\vspace*{0.5cm}
{\bf Acknowledgments.} This work utilizes data from HMI onboard NASA's SDO spacecraft, courtesy of NASA/SDO and the HMI Science Teams.  It also utilizes data from the MDI onboard SoHO. SoHO is a mission of international cooperation between ESA and NASA.  This work also utilizes GONG data obtained by the NSO Integrated Synoptic Program, managed by the National Solar Observatory, which is operated by the Association of Universities for Research in Astronomy (AURA), Inc. under a cooperative agreement with the National Science Foundation (NSF) and with contribution from the National Oceanic and Atmospheric Administration (NOAA). The GONG network of instruments is hosted by the Big Bear Solar Observatory, High Altitude Observatory, Learmonth Solar Observatory, Udaipur Solar Observatory, Instituto de Astrof\'{\i}sica de Canarias, and Cerro Tololo Interamerican Observatory. 10.7 cm radio flux data is provided by the National Resources Canada (NRC) Space Weather. Resources supporting this work were provided by the NASA High-End Computing (HEC) Program through the NASA Advanced Supercomputing (NAS) Division at Ames Research Center.
This research has used NASA's Astrophysics Data System (ADS: \url{ui.adsabs.harvard. edu/}) Bibliographic Services.

  This work is partially supported  by the NASA Contract NAS5-02139, NASA grants  80NSSC20K0602, 80NSSC22M0162, 80NSSC20K0194, 80NSSC21K0735,  80NSSC23K0563, 80NSSC23K0404, 80NSSC22K0516, 80NSSC19K0268, 80NSSC20K1320, 80NSSC19K1436,  80NSSC23K0097, and NSF grant 1916509.  This project has received funding from the European Research Council (ERC) under the European Union's Horizon 2020 research and innovation program (grant agreement No. 101097844 — project WINSUN), the UK Science and Technology Facilities Council (STFC) through grant ST/V000500/1,  and the GOLF and PLATO Centre National D'{\'{E}}tudes Spatiales grants.   

%
%
%
%
%
%
%

%
%
\bibliographystyle{spr-mp-sola}


%

%
%

\end{document}